\documentclass[a4paper,11pt]{article}
\pdfoutput=1 

\usepackage{jheppub} 

\usepackage[T1]{fontenc} 
\usepackage[numbers]{natbib}
\usepackage{filecontents}
\usepackage{dynkin-diagrams}
\usepackage{tikz}
\usepackage{placeins}
\usepackage{stackengine}
\tikzset{Rightarrow/.style={double equal sign distance,>={Implies},->},
triple/.style={-,preaction={draw,Rightarrow}},
quadruple/.style={preaction={draw,shorten >=0pt},shorten >=1pt,-,double,double
distance=0.2pt}}
\usetikzlibrary{positioning}
\usetikzlibrary{hobby}  
\definecolor{gold}{rgb}{1.0, 0.84, 0.0}
\definecolor{goldenrod}{rgb}{0.72, 0.53, 0.04}
\definecolor{goldenrod}{rgb}{0.85, 0.65, 0.13}
\usepackage{graphicx}
\usepackage{pdflscape}
\usepackage{caption}
\usepackage{subcaption}
\usepackage{comment}
\usepackage{todonotes}
\usepackage{subfiles}
\newcommand{\Figref}[1]{Figure~\ref{#1}}
\newcommand{\Quiver}[1]{$\mathcal Q_{\ref{#1}}$}
\newcommand{\surm}{\mathrm{SU}}
\newcommand{\urm}{\mathrm{U}}

\newcommand{\sprm}{\mathrm{Sp}}
\newcommand{\hs}{\mathrm{HS}}

\arxivnumber{2308.05853} 

\title{Quotient Quiver Subtraction}

\author[a]{Amihay Hanany,}
\author[a]{Rudolph Kalveks,}
\author[a]{and Guhesh Kumaran}

\affiliation[a]{Theoretical Physics Group, The Blackett Laboratory, Imperial College London, Prince Consort Road, SW7 2AZ, UK}

\emailAdd{a.hanany@imperial.ac.uk}
\emailAdd{rudolph.kalveks09@imperial.ac.uk}
\emailAdd{guhesh.kumaran18@imperial.ac.uk}

\abstract{We develop the diagrammatic technique of quiver subtraction to facilitate the identification and evaluation of the $\surm(n)$ hyper-Kähler quotient (HKQ) of the Coulomb branch of a $3d$ $\mathcal N=4$ unitary quiver theory. The target quivers are drawn from a wide range of theories, typically classified as “good” or “ugly”, which satisfy identified selection criteria. Our subtraction procedure uses quotient quivers that are “bad”, differing thereby from quiver subtractions based on Kraft-Procesi transitions. The simple diagrammatic procedure identifies one or more resultant quivers, the union of whose Coulomb branches corresponds to the desired HKQ. Examples include quivers whose Coulomb branches are moduli spaces of free fields, closures of nilpotent orbits of classical and exceptional type, and slices in the affine Grassmanian. We calculate the Hilbert Series and Highest Weight Generating functions for HKQ examples of low rank. For certain families of quivers, we are able to conjecture HWGs for arbitrary rank. We examine the commutation relations between quotient quiver subtraction and other diagrammatic techniques, such as Kraft-Procesi transitions, quiver folding, and discrete quotients.}

\begin{document} 
\preprint{Imperial/TP/23/AH/02}
\maketitle
\flushbottom

\section{Introduction}
\label{sec:intro2}

A motivation for studying quiver gauge theories is that their graphical nature allows for simple diagrammatic and combinatorial operations that represent actions on the moduli spaces of vacua. As a consequence, information about the structure of these moduli spaces can often be found in an efficient way.

The notion of \textit{magnetic quivers} has been applied to quantum field theories with eight supercharges in three, four, five, and six dimensions with great success \cite{Cabrera:2019izd,Bourget:2019aer,Bourget:2019rtl,Cabrera:2019dob,Grimminger:2020dmg,Bourget:2020gzi,Bourget:2020asf,Bourget:2020xdz,Beratto:2020wmn,Closset:2020scj,Akhond:2020vhc,vanBeest:2020kou,Bourget:2020mez,VanBeest:2020kxw,Giacomelli:2020ryy,Akhond:2021knl,Carta:2021whq,Arias-Tamargo:2021ppf,Bourget:2021xex,Gledhill:2021cbe,vanBeest:2021xyt,Carta:2021dyx,Sperling:2021fcf,Nawata:2021nse,Akhond:2022jts,Giacomelli:2022drw,Kang:2022zsl,Hanany:2022itc,Gu:2022dac,Fazzi:2022hal,Bourget:2022tmw,Gledhill:2022hrz,Fazzi:2022yca,Bhardwaj:2023zix,Bourget:2023uhe,Bourget:2023cgs,DelZotto:2023myd,DelZotto:2023nrb,Hanany:2023uzn,Lawrie:2023uiu,Bourget:2023dkj,Benvenuti:2023qtv,Mansi:2023faa,Fazzi:2023ulb,Bourget:2024mgn}. This demonstrates the power of simple diagrammatic techniques on quivers in solving various problems within physics and is hence a motivation to develop further diagrammatic techniques.

Our focus is on moduli spaces constructed using the Coulomb branches of $3d$ $\mathcal N=4$ unitary quiver gauge theories. These can be described by Hilbert series (HS) computed using the monopole formula \cite{Cremonesi:2013lqa}. The monopole formula treats the Coulomb branch as a moduli space of dressed monopole operators and constructs a HS that captures how the generators of the Coulomb branch are charged under $\surm(2)_R$ $R$-symmetry and their $\urm(1)_J$ topological symmetries. Relations between these charges encoded in the quiver can lead to enhanced global symmetries of the moduli space. We refer to a quiver gauge theory as a "magnetic quiver" for the moduli space, $X$, if the Coulomb branch of the magnetic quiver is $X$ \cite{Cabrera:2019izd}.

From a refined HS one can derive the Highest Weight Generating function (HWG) that enumerates the irreps (identified by Dynkin labels) of the global symmetry that appear at each order in the HS \cite{Hanany:2014dia}. Sometimes, an HS or HWG can be expressed in a compact form using the Plethystic Exponential \cite{Feng:2007ur}.

Diagrammatic operations on quivers include \textit{folding}, \textit{discrete gauging}, and \textit{quiver subtraction}. These have been the subject of various studies. For example, the discrete actions of folding and discrete gauging of magnetic quivers were studied in \cite{Hanany:2018vph,Hanany:2018dvd,Bourget:2020bxh}, and quiver subtractions representing Kraft-Procesi (KP) transitions \cite{Kraft1982OnGroups} between related moduli spaces were studied in \cite{Cabrera:2017njm,Cabrera:2016vvv}. Our aim herein is to build on ideas introduced in  \cite{Hanany:2022itc} and \cite{Bourget:2021qpx} to systematise a way of using quiver subtraction to take an $\surm(n)$ HKQ from the Coulomb branch of a unitary magnetic quiver. This in turn allows us to explore the relationships between such diagrammatic HKQs and other operations.

The \textit{folding} of a magnetic quiver corresponds to an action on the Coulomb branch which reduces it to a subspace that is invariant under some discrete action. This can be represented diagrammatically as the folding of $k$ identical legs in a quiver around a common pivot node, with the introduction of a non-simply laced edge of multiplicity $k$ \cite{Cremonesi:2014xha}, so that the pivot node is associated to long roots of the algebra of the global symmetry. Such folding reduces the dimension of the moduli space and may manifest in the HWG via identifications between highest weight fugacities. Examples of folded magnetic quivers appear in Sections \ref{sec:freetheories},  \ref{sec:BtypeHKQ}, \ref{sec:ExceptionalHKQ}, and \ref{sec:Conclusions}.

The \textit{discrete gauging} of a Coulomb branch involves the identification of points in the moduli space under a discrete action. The dimension of the space is not reduced, but its volume is reduced by a factor of the order of the discrete group. As a special case, the discrete gauging by $S_k$ of the Coulomb branch of a magnetic quiver with a bouquet of $n\geq k$ $\urm(1)$ gauge nodes, can be implemented diagrammatically by collecting $k$ such nodes into a single $\urm(k)$ node with an adjoint hypermultiplet \cite{Hanany:2018dvd}. This action was extended to quivers with a \textit{complete graph} of $\urm(1)$ gauge nodes in \cite{Hanany:2023uzn}.

Discrete gauging may manifest in the HWG (after assigning fugacities under the discrete action to its coefficients) as a finite group average over its terms according to the Burnside lemma \cite{burnside_2012}. Note that the characters of discrete symmetry groups commute with highest
weight fugacities, and not Cartan sub-algebra (CSA) fugacities, so the HWG is essential for this analysis. Examples of magnetic quivers with discrete gauging appear in Sections \ref{sec:BtypeHKQ}, \ref{sec:ExceptionalHKQ}, \ref{sec:miscellaneous}, and \ref{sec:Conclusions}.

The aforementioned examples draw not only upon magnetic quivers for free field theories, which present a simple background against which to develop our methods, but also upon magnetic quivers for the closures of nilpotent orbits, which are more intricate and merit some introductory comments.

The closures of nilpotent orbits of semi-simple Lie algebras ("nilpotent orbits" or "orbits") have deep connections to the Coulomb branches of $3d$ $\mathcal N=4$ theories and many have constructions from magnetic quivers \cite{Hanany:2016gbz,Hanany:2017ooe}.

The theorem of Namikawa \cite{2016arXiv160306105N}, applied to moduli spaces of $3d$ $\mathcal N=4$ theories, states that the chiral ring with generators at spin $1$ of $\surm(2)_R$ is a nilpotent orbit. This has been verified for the Coulomb branches of many magnetic quivers for nilpotent orbits by applying the monopole formula.

Nilpotent orbits, along with other Coulomb branches, are symplectic singularities and enjoy a stratification into a finite number of symplectic leaves \cite{2003math.....10186K}. This structure can be presented in a poset or Hasse diagram \cite{Grimminger:2020dmg}, which shows the partial order between (closures of) symplectic leaves in the moduli space. In the mathematics literature, Kraft and Procesi \cite{Kraft1982OnGroups, Kraft1980MinimalGLn} computed Hasse diagrams for nilpotent orbits of Classical groups and identified the transverse spaces, or KP transitions, between adjacent (closures of) symplectic leaves.

Prior studies \cite{Hanany:2016gbz} have shown that all orbits of A-type algebras, as well as all orbits with characteristic height \cite{Panyushev1999}\footnote{"Characteristic height" is defined as the dot product between the Characteristic of the orbit and the Coxeter Labels of the algebra.} equal to two, being located near the origin of such Hasse diagrams, have magnetic quivers, and magnetic quivers have also been identified for some higher orbits of non A-type algebras. Examples of unitary magnetic quivers for nilpotent orbits appear in Sections  \ref{sec:AType}, \ref{sec:BtypeHKQ}, \ref{sec:DtypeHKQ} and \ref{sec:ExceptionalHKQ}.

Discrete actions on the closures of nilpotent orbits of semi-simple Lie algebras were studied and classified mathematically by Kostant and Brylinski \cite{1992math......4227B} and Kobak and Swann \cite{Kobak1996CLASSICALNO}. These studies found relationships between orbits through discrete actions, and a variety of their results have been replicated using magnetic quivers, through the \textit{discrete gauging} of their Coulomb branches, including by diagrammatic methods \cite{Bourget:2020bxh}.

In both \textit{folding} and \textit{discrete gauging} there exist diagrammatic operations on magnetic quivers, which represents their action on the Coulomb branch in a simple and efficient way without resorting to explicit formulae.

The usual diagrammatic method of \textit{quiver subtraction} \cite{Cabrera:2018ann}, which may relate multiple quivers, is more complicated, and draws on the irreducible nature of KP transitions. Importantly, KP transitions, such as $A_k$ Kleinian singularities and $g_k$ minimal nilpotent orbits, can be represented by unitary magnetic quivers, and \textit{quiver subtraction} permits the deconstruction of a given unitary magnetic quiver into a partially ordered Hasse diagram or poset of magnetic quivers.

The \textit{quiver subtraction} implementation of the KP transitions makes it very simple to compute the Hasse diagrams of moduli spaces which are symplectic singularities (but not necessarily orbits) providing their magnetic quivers are known.

KP transitions also have a physical realisation in Type IIB string theory in NS5-D5-D3 brane systems \cite{Cabrera:2016vvv, Cabrera:2017njm}. This involves the movement of D3 branes, such as coinciding them with 5-branes in particular ways and brane creation/annihilation \cite{Hanany:1996ie}. The net result of the manipulation of the branes matches that obtained by quiver subtraction.

Let us return now to our central topic of hyper-Kähler quotients. The Coulomb branch of a magnetic quiver is always a hyper-Kähler cone, so we can consider its hyper-Kähler quotient (HKQ) by some subgroup of its global symmetry. Physically, such an HKQ is a gauging \cite{Bourget:2021qpx} that incorporates the action of a moment map \cite{Hanany:2017ooe}.

The result of an HKQ of some moduli space $X$ with global symmetry $G_{\mathcal C}$  by a continuous symmetry $G\subset G_\mathcal C$ is denoted $X///G$ and depends on the explicit choice of embedding of $G$ into $G_\mathcal C$. When all the generators of $G$ are involved in the gauging, the resulting (quaternionic) dimension is $|X///G|= |X| - |G|$. However, if the subgroup $G$ is "too large" (in terms of dimensions and structure) relative to $X$, then the resulting moduli space will have a (quaternionic) dimension greater than $|X| - |G|$, and we say that \textit{incomplete Higgsing} has occurred.

Kobak and Swann \cite{Kobak1996CLASSICALNO} showed that the HKQs of classical orbits by continuous symmetry groups are hyper-Kähler, and provide several such examples, based mainly on HKQs by low rank unitary groups, where the resulting space is also an orbit. The $\urm(1)$ HKQs appearing in Kobak-Swann's work can be replicated graphically using magnetic quivers and elementary \textit{quiver subtraction} involving a $\urm(1)$ gauge node, as demonstrated in Appendix \ref{sec:U1HKQ}.

Given the feasibility of implementing discrete actions and $\urm(1)$ HKQs on the Coulomb branch using magnetic quivers and diagrammatic techniques, and the consistency with established mathematical results, it is natural to ask whether classical group HKQs can also be realised diagrammatically.

In \cite{Hanany:2022itc}, the $\surm(3)$ HKQ of the minimal orbit of $E_8$ was computed to give a double cover of the 21-dimensional orbit of $E_6$. The observation was made that this HKQ could have been represented by the subtraction of the quiver $(1)-(2)-(3)-(2)-(1)$ from the \textit{unframed} unitary magnetic quiver for the minimal orbit of $E_8$, to produce a magnetic quiver for this double cover. See Section \ref{sec:ExceptionalHKQ}.

Another work \cite{Bourget:2021qpx} includes an example of the implementation of a $\urm(n)$ HKQ on a unitary magnetic quiver, using $3d$ mirror symmetry and Higgs-Coulomb branch dualities. This $\urm(n)$ HKQ can also be realised diagrammatically as the subtraction of the quiver $(1)-(2)-\cdots-(n)-\cdots-(2)-(1)$ from the framed unitary quiver, as we show in Appendix \ref{sec:unex}.

In this paper we generalise this type of quiver subtraction into a precise procedure for an $\surm(n)$ HKQ, providing a list of selection rules for the starting quivers from which such an HKQ is permissible. A full list of our rules for $\surm(n)$ HKQs via \textit{quotient quiver subtraction} will be provided in Section \ref{sec:rules}, but we can preview some key features.

In particular, we only work with those magnetic quivers where there is a \textit{Dynkin type} embedding of $\surm(n)$ into the global symmetry $\mathcal G_{\mathcal C}$, and our diagrammatic selection rules avoid those magnetic quivers where \textit{incomplete Higgsing} occurs, so that the resulting moduli spaces have the correct integer (quaternionic) dimension.

Notably, our rules involve the subtraction of a quotient quiver with gauge nodes of the form $(1)-(2)-\cdots-(n)-\cdots-(2)-(1)$. Such a quiver is "bad" in the sense of \cite{Gaiotto:2008ak}, and its Coulomb branch cannot therefore be computed using the monopole formula, differing thereby from the quivers for KP transitions, which are always "good". Nevertheless, such a quotient quiver can often be subtracted from a "good" or "ugly" magnetic quiver that has a corresponding "external leg" to produce valid magnetic quiver(s), whose Coulomb branches have the correct dimensions for an HKQ.  We refer to our procedure as "\textit{quotient quiver subtraction}" to avoid confusion with quiver subtraction for KP transitions.

A significant feature of our rules is that, in certain cases, they identify that the HKQ of a Coulomb branch is a union of Coulomb branches. We draw on methods similar to \cite{Bourget:2023cgs} to evaluate the HS of such unions and to illuminate their underlying structures.

To demonstrate our $\surm(n)$ quotient quiver subtraction rules, we show how they can be applied to many examples of unitary magnetic quivers, with Coulomb branches drawn, inter alia, from free fields, classical or exceptional nilpotent orbits, and slices in the affine Grassmannian \cite{Bourget:2021siw}. We find many relationships between such spaces, along with several results that have not, to the best of our knowledge, appeared in the physics or mathematics literature. We validate the results by computing HKQs through analytic methods using Weyl integration.

We find that diagrammatic HKQs provide a means of identifying relationships involving magnetic quivers that lie outside the realms of brane systems and class $\mathcal S$ theories, and which have not been well studied in the literature.

\paragraph{Organisation of the paper}
{In Section \ref{sec:rules} we recap the usual method of taking an HKQ via Weyl integration and present our diagrammatic rules for $\surm(n)$ quotient quiver subtraction, which simplify this calculation.

We also recap briefly on relevant aspects of quiver theory, including the method of (un)framing quivers, and the relationships between magnetic quivers, nilpotent orbits, Slodowy slices and slices in the affine Grassmannian. We also define the notation that we use.} 

In Section \ref{sec:freetheories} we warm-up by testing our rules of $\surm(n)$ quotient quiver subtraction on magnetic quivers for moduli spaces of free fields.

In Sections \ref{sec:AType}, \ref{sec:BtypeHKQ}, \ref{sec:CtypeHKQs}, \ref{sec:DtypeHKQ} and \ref{sec:ExceptionalHKQ} we apply the rules of $\surm(n)$ quotient quiver subtraction to find HKQs of some classical and exceptional nilpotent orbits. We obtain many conjectures for relationships between orbits and/or slices in the affine Grassmannian. For low rank cases we check these relationships by computing HS and/or HWGs.

In Section \ref{sec:ExceptionalAGHKQ} we apply the rules of $\surm(n)$ quotient quiver subtraction to some slices in the affine Grassmannians of $E_{6}$ and $E_{7}$, thereby testing that our rules remain valid when applied to complicated quivers.

In Section \ref{sec:miscellaneous} we apply our rules to some miscellaneous families of magnetic quivers, in order to generalise results from earlier sections.

Finally, in the concluding Section \ref{sec:Conclusions}, we compile some results and conjectures obtained by quotient quiver subtraction (see Tables \ref{tab:FreeSU2}, \ref{tab:OrbSU2}, \ref{tab:OrbSU3} and \ref{tab:OrbSU4}). We also discuss how the diagrammatic technique of $\surm(n)$ \textit{quotient quiver subtraction} typically commutes with \textit{folding} and \textit{discrete gauging}, but not with quiver subtraction for KP transitions. Finally we identify some open problems.

The Appendices contain supplementary materials, including a discussion of situations where precise application of the selection rules is necessary to avoid pathological results. Additionally, we include examples of certain $\urm(n)$ HKQs that follow as a natural extension of the rules for $\surm(n)$ presented in Section \ref{sec:rules}.

\section{Method of quotient quiver subtraction for $\surm(n)$ HKQs}
\label{sec:rules}
\subsection{Analytic computation of $\surm(n)$ HKQ}
The refined Hilbert series $\hs_{G_{\mathcal C}}$  of the Coulomb branch of a magnetic quiver is first computed using the unitary monopole formula, as outlined in \cite{Cremonesi:2013lqa}. This formula counts monopole operators graded by their charges under $\surm(2)_R$ $R$-symmetry and $\urm(1)_J$ topological symmetry. The fugacities associated to the topological symmetry can be mapped to the Cartan sub-algbera (CSA) fugacities of the global symmetry $G_{\mathcal C}$ of the Coulomb branch using the Cartan matrix. When expanded perturbatively, such an HS enumerates the characters of irreps of the global symmetry at each value of $R$-charge.

Suppose the global symmetry of the Coulomb branch is $G_{\mathcal C}=\prod_i G_i$. If there exists an embedding of $\surm(n)$ into some $G_j$, $G_j\hookleftarrow G_j'\times \surm(n)$, then the general form for its hyper-Kähler quotient by $\surm(n)$ is given by \cite{Hanany:2016gbz}: 
\begin{equation}
     \hs_{G_\mathcal C///\surm(n)}(x_1,\cdots, x_r;t)=\int_{\surm(n)}d\mu_{\surm(n)}\frac{\hs_{G_\mathcal C}(x_1,\cdots,x_r;y_1,\cdots,y_{n-1};t)}{PE[\chi([1,0,\cdots,0,1]_{\surm(n)})t^2]},
     \label{eq:general_HKQ}
\end{equation}
where the $x_i$ are fugacities for $G_j'\times \prod_{i\neq j} G_i$, and the $y_i$ are fugacities for $\surm(n)$. The denominator is a PE of the character of the adjoint of $\surm(n)$, graded by the $R$-charge counting fugacity $t$.

The physical interpretation of this formula is the gauging, against a background of relations, of an $\surm(n)$ subgroup of the global symmetry $G_{\mathcal C}$ of the Coulomb branch \cite{Bourget:2021qpx}. Explicitly, the formula starts with a Hilbert series, takes its quotient by symmetric products of the $\surm(n)$ adjoint representation, and projects out $\surm(n)$ singlets. 

It follows from \eqref{eq:general_HKQ} that the quaternionic dimension of $\hs_{G_{\mathcal C}}$ is reduced by at most $n^2-1$. There are two contributions, the first from the denominator of \eqref{eq:general_HKQ}, which reduces the dimension by $(n^2-1)/2$, and the second from the Haar measure and Weyl integration, which affect the dimension up to a further reduction of $(n^2-1)/2$. We refer to \textit{complete Higgsing} whenever $|\hs_{G_\mathcal C///\surm(n)}|=|\hs_{G_\mathcal C}|-(n^2-1)$. Sometimes, however, $\hs_{G_{\mathcal C}}$ fails to explore enough of the lattice of $\surm(n)$ to saturate this dimensional reduction, and this is referred to as \textit{incomplete Higgsing}.

Providing complete Higgsing occurs, the formula \eqref{eq:general_HKQ} is valid; otherwise it cannot be used, and a construction from first principles using moment maps is required. We focus herein on HKQs with complete Higgsing.


\subsection{Rules for $\surm(n)$ HKQ by quotient quiver subtraction}

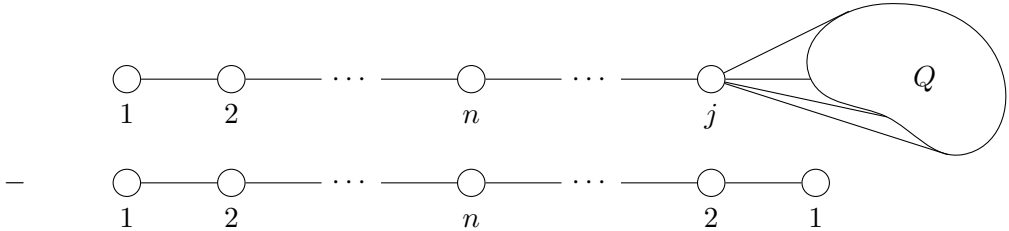
\begin{figure}[h!]
    \centering
    \begin{tikzpicture}[main/.style={draw,circle}]

    \node[main, label=below:$1$] (1){};
    \node[main, label=below:$2$] (2) [right=of 1]{};
    \node[draw=none,fill=none] (cdotsL) [right=of 2]{$\cdots$};
    \node[main, label=below:$n$] (n) [right=of cdotsL]{};
    \node[draw=none,fill=none] (cdotsR) [right=of n]{$\cdots$};
    \node[main,label=below:$j$] (blank) [right=of cdotsR]{};

x     \node (Q) at (10.5,0) {$Q$};
    \path[draw,use Hobby shortcut,closed=true]
    (9,0) ..  (10,1).. (11,-1) .. (10,-.5);

    \draw[-] (1)--(2)--(cdotsL)--(n)--(cdotsR)--(blank);
    \draw[-] (blank)--(9,0);
    \draw[-] (blank)--(9.5,0.9);
    \draw[-] (blank)--(10.8,-1);
    \draw[-] (blank)--(10,-.5);

    \node[main, label=below:$1$] (1subL) [below=of 1]{};
    \node[main, label=below:$2$] (2subL) [right=of 1subL]{};
    \node[] (cdotsresL) [right=of 2subL]{$\cdots$};
    \node[main, label=below:$n$] (nsub) [right=of cdotsresL]{};
    \node[] (cdotsresR) [right=of nsub]{$\cdots$};
    \node[main, label=below:$2$] (2subR) [right=of cdotsresR]{};
    \node[main, label=below:$1$] (1subR) [right=of 2subR]{};

    \draw[-] (1subL)--(2subL)--(cdotsresL)--(nsub)--(cdotsresR)--(2subR)--(1subR);

    \node[] (minus) [left=of 1subL] {$-$};
    
    \end{tikzpicture}
    \caption{Schematic drawing of a valid target quiver (top) having a junction and an external leg. The target quiver for the quotient quiver subtraction must have a leg that starts as $(1)-\cdots-(n)$. If the quotient quiver (bottom) overlaps a junction node of rank $j$ on the target quiver, this must align with a node of rank $2 \leq j$ on the quotient quiver. The remainder of the target quiver, denoted $Q$, can take any "good" or "ugly" form.}
    \label{fig:blobmodeljunction}
\end{figure}
When carrying out $\surm(n)$ HKQ by quotient quiver subtraction, we find it convenient to work with unframed quivers as intermediates, and to subtract between unframed magnetic quivers to obtain a framed result.

We define the $\urm(n)$ \textit{quotient quiver} as the unframed quiver $(1)-(2)-\cdots-(n)-\cdots-(2)-(1)$ and the \textit{target quiver} as the unitary magnetic quiver for which the $\surm(n)$ hyper-Kähler quotient of its Coulomb branch is desired. 
The procedure for performing an $\surm(n)$ HKQ by quotient quiver subtraction is then as follows. At each stage of the procedure there are selection rules (in bold) that should be applied before continuing.

\begin{enumerate}
     \item Select a "good" or "ugly" unitary magnetic quiver as the target.
     \begin{description}
    \item[Long Framing Rule] \hypertarget{rule:LongFraming}
    {The target quiver must be framed only on long nodes, otherwise unframing will give an ambiguous result.}
    \end{description}
    \item Transform the target quiver into its unframed form.
    \begin{description}
    \item[External Leg Rule] \hypertarget{rule:ExternalLeg}
    {The unframed target quiver must have at least one external leg with gauge nodes of $(1)-(2)-\cdots-(n)$. For a non-simply laced target quiver, this leg must correspond to long roots. Multiple legs may give a choice of HKQs.}
    \end{description}
    
    \item Align the quotient quiver against this external leg and subtract its gauge nodes from the target.
    \begin{description}
        \item[Single Edge Rule] \hypertarget{rule:SingleEdge}{The edges of the quotient quiver are simply laced and can only be subtracted from a section of the target quiver with single edges.}
        \item[Junction Rule] \hypertarget{rule:Junction}{If the quotient quiver extends beyond a junction in the target quiver, the junction must align with a trailing node of 2 in the quotient quiver, as drawn in \Figref{fig:blobmodeljunction}.}
        \item[Union Rule] \hypertarget{rule:Union}{If the quotient quiver extends past the junction, then all possible alignments must lead to valid diagrams.}
         \item[Adjoint Hypers Rule] \hypertarget{rule:Adjoint}{Any nodes with an adjoint hypermultiplet must survive the quotient quiver subtraction.}
    \end{description}
    
    \item Restore the original balance of all surviving gauge nodes by attaching flavours.
    \begin{description}
        \item[Rebalancing Rule] \hypertarget{rule:Rebalancing}{The resulting framed quiver(s) must contain no node (gauge or flavour) with negative rank. }
    \end{description}

      \item If a particular alignment of the quotient quiver breaks the target into disconnected pieces, then its contribution to the moduli space is the Cartesian product of the Coulomb branches of these disjoint magnetic quivers.
    
    \item If the target quiver contains a junction, and there are multiple possible alignments of the quotient quiver, each alignment is taken as yielding a different quiver, and the desired moduli space is the union of the Coulomb branches of these magnetic quivers.
    

\end{enumerate}
By way of explanatory comments:
\begin{itemize}

 \item Recall that the Hilbert series from applying the monopole formula to a simply laced magnetic quiver is insensitive to the choice of framing, and that for a non-simply laced quiver, the HS is insensitive to choice of framing amongst long nodes \cite{Hanany:2020jzl}. Quiver subtraction requires compatible framing, so that the framing on the target quiver and the quotient quiver need to be aligned before subtraction, if this is possible, by judicious shifts in framing. Alternatively, one can work with unframed quivers in both cases, and this is the diagrammatic approach that we find simpler for $\surm(n)$ HKQs. 

     \item The \hyperlink{rule:ExternalLeg}{External Leg Rule} requires that an external leg of $(1)-(2)-\cdots-(n)$ must be available for subtraction of the quotient quiver. By construction, this leg has a chain of $n-1$ balanced nodes and so contains the Dynkin diagram of $A_{n-1}$, thereby ensuring that there is a Dynkin type embedding of $\surm(n)$, and permitting comparison with the result from Weyl integration.
    The sum of the ranks of the (unframed) quotient quiver is $n^2$ and so, when it is subtracted, the (quaternionic) dimension of the framed target quiver's Coulomb branch is reduced by $n^2-1$ (allowing for the dimension 1 increase from the initial unframing). This matches with the expectation from Weyl integration for an $\surm(n)$ HKQ whenever there is complete Higgsing.

    \item The unframing of the target quiver re-introduces the center-of-mass $\urm(1)$ to the global symmetry. Thus the subtraction of the $\urm(n)$ quotient quiver (of dimension $n^2$) from the unframed target quiver is consistent with an $\surm(n)$ HKQ.
    
     \item The \hyperlink{rule:ExternalLeg}{External Leg Rule} also requires the external leg to lie within the long roots of the target quiver. This is because the quotient quiver corresponds to long roots and so cannot be subtracted from short roots.

     \item Empirically, we find that quotient quiver subtraction does not match Weyl integration under violation of the \hyperlink{rule:Adjoint}{Adjoint Hypers Rule}.

     \item The \hyperlink{rule:SingleEdge}{Single Edge Rule} forbids the quotient quiver being aligned against a part of the target quiver with non-simply laced edges or multiple hypers. This is because quiver subtraction is undefined in these cases.

     \item The \hyperlink{rule:Rebalancing}{Rebalancing Rule} ensures that the resulting quiver will not have any nodes (gauge or flavour) with negative rank. This is because magnetic quivers with negative rank nodes do not have an interpretation.

     \item The \hyperlink{rule:Junction}{Junction Rule} is an empirical rule for target quivers where the quotient quiver extends past a junction. If this rule is violated, then we have found that incomplete Higgsing occurs under Weyl integration and we are unable to validate the result from quiver subtraction. Conversely, the requirement that a junction should align with a node of at most rank 2 on the quotient quiver leads to any alternative resulting quivers having an intersection related by A-type Kleinian singularities; and this appears to be sufficient to ensure complete Higgsing.

     \item The \hyperlink{rule:Union}{Union Rule} ensures that the union of the Coulomb branches of the alternative magnetic quivers can be constructed.



  \item The interpretation of the Coulomb branch of disconnected quivers is the natural one of a Cartesian product of Coulomb branches. In terms of HS, it is the usual multiplicative product.

     \item In a magnetic quiver each gauge node is associated to a different root fugacity, and the requirement of a Dynkin type embedding fixes a specific labelling of the nodes by fugacities. So even if alternative alignments of the quotient quiver are related by outer automorphism, they represent different moduli spaces, and have to be treated as distinct components of the union.

     \item A trivial comment is that if multiple $\surm(n)$ HKQ are to be computed then one simply does quiver subtraction for each $\surm(n)$ HKQ in turn. However, as we will discuss in Section \ref{sec:Conclusions}, the order in which these are carried out may matter.

\end{itemize}
Some examples that illustrate pathological outcomes from violating the selection rules are contained in Appendices \ref{sec:JunctionRule} and \ref{sec:ExternalLeg}.

Whenever alternative magnetic quivers result from quotient quiver subtraction, it remains to find their union. We compute the Hilbert series for the union of the Coulomb branches of a set of quivers $\{Q_i\}$, where $i=1,\cdots,n$, using the \textit{Unions of Cones} formula \cite{Bourget:2023cgs}:
\begin{align}
\hs\left(\mathcal C\left(Q_1\cup Q_2\cup\cdots\cup Q_n\right)\right)&=\sum_{i}\hs\left(\mathcal C\left(Q_i\right)\right)-\sum_{1\leq i<j\leq n}\hs\left(\mathcal C\left(Q_i\cap Q_j\right)\right)\nonumber\\&+\cdots+\sum_{1\leq i_1<i_2<\cdots<i_p\leq n}(-1)^{p-1} \hs\left(\mathcal C\left(Q_{i_1}\cap\cdots\cap Q_{i_p}\right)\right)+\cdots.
\label{eq:UnionsofCones}
\end{align}
The magnetic quivers for the intersections can be found by carrying out Kraft-Procesi quiver subtractions from each component of the union down to a common quiver.

This decomposition of the union as a signed sum of Coulomb branches is the reason quotient quiver subtraction provides additional insight into the structure of an HKQ compared with Weyl integration. While it is a challenge to decompose a moduli space into its constituent parts from an HS alone, with quotient quiver subtraction, each alignment directly identifies the magnetic quiver for each constituent part of the moduli space.

We test our conjecture that the $\surm(n)$ HKQ of the Coulomb branch of the target quiver corresponds to the union of Coulomb branches from different possible alignments with many examples in the course of this work.
%
To validate a quotient quiver subtraction, we compare the results to those from an $\surm(n)$ HKQ by Weyl integration. We do this using the Dynkin type embedding of $\surm(n)$ into $G_{\mathcal{C}}$ associated with the external leg, all as described further in each example.

\subsection{A physical interpretation of the $\urm(n)$ quotient quiver}
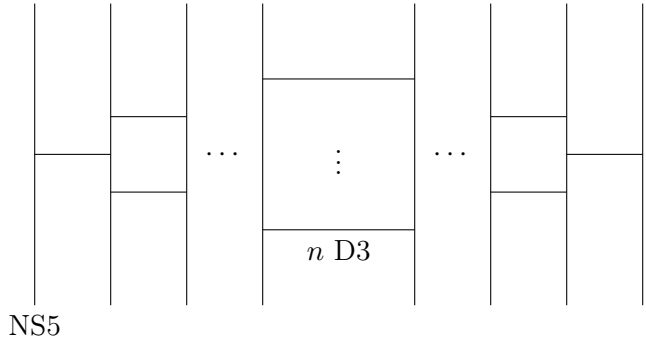
\begin{figure}
    \centering
    \begin{tikzpicture}[main/.style={draw,circle}]
    \draw[-] (0,0)--(0,4)node[pos=0,below]{NS5};
    \draw[-] (1,0)--(1,4);
    \draw[-] (2,0)--(2,4);
    \node[] (cdotsL) at (2.5,2) {$\cdots$};
    \draw[-] (3,0)--(3,4);
    \draw[-] (5,0)--(5,4);
    \node[] (cdotsR) at (5.5,2) {$\cdots$};
    \draw[-] (6,0)--(6,4);
    \draw[-] (7,0)--(7,4);
    \draw[-] (8,0)--(8,4);

    \draw[-] (0,2)--(1,2);
    \draw[-] (1,1.5)--(2,1.5);
    \draw[-] (1,2.5)--(2,2.5);
    \node[] (vdots) at (4,2) {$\vdots$};
    \draw[-] (3,3)--(5,3);
    \draw[-] (3,1)--(5,1)node[pos=0.5,below]{$n$ D3};
    \draw[-] (6,1.5)--(7,1.5);
    \draw[-] (6,2.5)--(7,2.5);
    \draw[-] (7,2)--(8,2);
        
    \end{tikzpicture}
    \caption{Type IIB brane system for the $\urm(n)$ quotient quiver. There are $2n$ NS5 branes with $2n-1$ intervals on which D3 brane branes are suspended.}
    \label{fig:SUNQuotBrane}
\end{figure}

As mentioned, the $\urm(n)$ quotient quiver is ``bad", so its Coulomb branch is not computable with the monopole formula. However, we can go to the Type IIB Hanany-Witten brane system \cite{Hanany:1996ie} for the $\urm(n)$ quotient quiver which is drawn in \Figref{fig:SUNQuotBrane}. It is clear that from the point of view of the NS5 brane that this is the moduli space of an $\surm(2n)$ monopole with magnetic charges $(1,2,\cdots,n,\cdots,2,1)$. The moduli space of such a monopole has been found in the mathematics literature as $T^*\mathrm{SL}(n)$ \cite{2006math.....12365B,Braverman:2017ofm}. The moduli space $T^*\mathrm{SL}(n)$ is smooth and compact and it is at present unclear how this is connected to the $\surm(n)$ HKQ, a singular operation, of Coulomb branches which are also singular and non-compact.

An alternative viewpoint is that the $\urm(n)$ quotient quiver can be framed as a $4d$ $\mathcal N=2$ Class $\mathcal S$ theory consisting of two maximal punctures of $A_{n-1}$\footnote{We thank Tudor Dimofte for discussions about this point.}. 

\subsection{Aspects of Quiver Theory}
\label{sec:quiverology}
It is helpful to summarise key aspects of the theory and notational conventions that surround the magnetic quivers appearing in this paper. These include the method of \textit{unframing}, as well as some canonical moduli spaces, such as nilpotent orbits, Slodowy slices, and Affine Grassmannian slices, many of which have magnetic quivers.





\subsubsection{Unframed quivers}

In any quiver gauge theory there exists an overall center-of-mass $\urm(1)$ symmetry that is gauged away by the choice of framing i.e. specification of flavours. In the monopole formula this is reflected in the choice of origin for the system of monopole charges, and such a choice is indeed necessary for the monopole formula to evaluate without divergence.

However, as observed in \cite{Hanany:2020jzl}, this choice of origin, providing it is made amongst monopole charges associated with long roots, does not affect the Coulomb branch calculation in a fundamental manner, since the impact on the refined HS can be compensated by a suitable choice of map between node and CSA fugacities. This in turn makes it possible to treat framed unitary magnetic quivers within equivalence classes represented by an \textit{unframed} quiver that only contains gauge nodes, as discussed in \cite{Crawley-Boevey2001GeometryQuivers}.

A framed unitary quiver can be converted to its \textit{unframed} representative as follows. 
\begin{enumerate}
\item Replace every flavour node of $n$ by a flavour node of 1, increasing the multiplicity of direct links between it and other nodes by a factor of $n$ to compensate.
\item Redraw the diagram to superimpose all the flavour nodes of 1.
\item Convert the flavour node of 1 to a $\urm(1)$ gauge node.
\end{enumerate}
This introduction of an overall $\urm(1)$ gauge node in place of the flavour nodes is the crucial first step in the procedure for $\surm(n)$ HKQ by quotient quiver subtraction in Section \ref{sec:rules}. A framed quiver that is equivalent to the original can be recovered from the unframed quiver by setting any one of the monopole charges chosen among the long roots to zero.





\subsubsection{"Good", "Bad", and "Ugly" Quivers}
The terms "good", "bad", and "ugly", coined and defined in \cite{Gaiotto:2008ak}, are related to the $R$-charges of monopole operators. These follow directly from the structure of a quiver by computation of the "balance" of each gauge node. The balance $b_i$ of a $\urm(n_i)$ gauge node, having adjacent nodes (gauge or flavour) with ranks $\{r_j\}$ and link multiplicities $\{m_j\}$, is defined as:
\begin{equation}
    b_i=\sum_{j \in adjacent}m_j r_j-2n_i.
    \label{eq:balance}
\end{equation}
A quiver is termed "good" if all gauge nodes have non-negative balance, "ugly" if any gauge node has balance of $-1$ (but not less), or "bad" if any gauge node has balance less than $-1$. The monopole formula will usually converge for a "good" or "ugly" theory, but not for a "bad" theory. A node is termed "balanced" if its balance is 0. In a "good" or "ugly" theory, if a (sub)set of balanced gauge nodes forms a Dynkin diagram, then a factor of this group appears in the Coulomb branch global symmetry \cite{Gledhill:2021cbe}. This is due to the coincidence that additional monopole operators appear on the Coulomb branch when the balance condition is satisfied.

\subsubsection{Nilpotent Orbits and Slodowy Slices}

The nilpotent orbits of a Lie algebra $\mathfrak{g}$ are in one-to-one correspondence with the different possible embeddings $G\hookleftarrow \surm(2)$, 
due to theorems of Jacobson-Morozov \cite{Jacobson1951CompletelyTransformations,Morozov1942OnAlgebra} and Kostant \cite{Collingwood1993NilpotentAlgebras}. Knowledge of all of the embeddings of $\surm(2)$ into a Lie group $G$, as classified by Dynkin \cite{Dynkin1957SemisimpleAlgebras}, is then sufficient to find all of its nilpotent orbits. 

Each such embedding specifies the construction of a nilpotent orbit of $G$ and is easily represented using characters. We define a map $\rho$ which takes the simple root fugacities of $G$, $\{z_1,\cdots,z_{rank(G)}\}$, and the CSA fugacities of $G$, $\{x_1,\cdots,x_{rank(G)}\}$, into the CSA fugacity $x$ of $\surm(2)$:
\begin{align}
    \rho:\{x_1,\cdots,x_{rank(G)}\}\rightarrow \{x^{w_1},\cdots,x^{w_{rank(G)}}\},\\
    \rho:\{z_1,\cdots,z_{rank(G)}\}\rightarrow \{x^{q_1},\cdots,x^{q_{rank(G)}}\},
\end{align}
where the exponents $q_i$ and $w_j$ are necessarily related by the Cartan matrix of $G$:
\begin{equation}
    q_i=\sum_j A_{ij}w_j.
\end{equation}
These maps can be used to decompose irreps of $G$ into irreps of $\surm(2)$ and this permits the labelling of nilpotent orbits in any one of a number of equivalent ways \cite{Collingwood1993NilpotentAlgebras}:
\begin{enumerate}
    \item By the \textit{Characteristic}, or \textit{root map}, set of ordered exponents $\{q_1,\cdots,q_{rank(G)}\}$,
     \item By the \textit{weight map} set of ordered exponents $\{w_1,\cdots,w_{rank(G)}\}$,
    \item By the \textit{partition}\footnote{ $(\ldots,n_i^{m_i},\ldots)$ denotes a partition where the superscript $m_i$ gives the multiplicity of the dimension $n_i$ in the partition.} which enumerates the dimensions and multiplicity of $\surm(2)$ irreps in the vector/fundamental of $G$ under the map $\rho$.
\end{enumerate}
A typical way of labelling the closures of nilpotent orbits is by group and partition data as $\overline{{\mathcal O}}^{G}_{\rho}$, where the overline clarifies that we are refering to the closure of the orbit. The partition data allows the nilpotent orbits of $G$ to be organised as a Hasse diagram, and certain orbits in this poset carry canonical names. These include the minimal orbit $\overline{min. G}$, "next to" minimal orbits $\overline{n. min. G}$ (where we can use $n.$ multiple times to indicate how far an orbit is from the minimal), the sub-regular orbit $\overline{sub. reg. G}$ and the maximal or regular orbit $\overline{max. G}$ (also termed the nilcone ${\mathcal N}$).


Nilpotent orbits have been studied extensively in the quiver gauge theory literature \cite{Hanany:2016gbz,Hanany:2017ooe}.  Notably, there exist unitary magnetic quivers for any orbit of A-type or any orbit of $G$ that has a Characteristic height of 2.
These are all balanced quivers of Dynkin type. A magnetic quiver for a height 2 orbit can be constructed by taking the Dynkin diagram of $G$, numbering each node in the usual way following \cite{Yamatsu:2015npn}, then using the weight map to assign the rank of the $\urm(w_i)$ gauge node at position $i$, and the root map to assign the attached flavours $q_i$.

Nilpotent orbits have transverse spaces formed by the generators of the Lie algebra $\mathfrak{g}$ which are not included in the embedding $\rho$. These spaces are termed Slodowy slices, or Slodowy intersections if restricted by a superior orbit in the poset. Each such transverse space within $\mathfrak{g}$ is defined by a pair of orbits and is notated ${{\mathcal S}^{G}_{\rho,\sigma}}$, where the slice is taken from $\sigma$ to $\rho$. Many Slodowy slices and intersections possess unitary magnetic quivers of Dynkin type, but these are not generally balanced. For further background on Slodowy slices and intersections see \cite{Cabrera:2018ldc,Hanany:2019tji}.

\subsubsection{The affine Grassmannian}
While we do not attempt a systematic account of the relationship between quivers and affine Grassmannian spaces, for which the reader is referred to \cite{Bourget:2021siw}, it is noteworthy that many of the quivers that result from the quotient quiver subtraction examples herein turn out to be slices in an affine Grassmannian. Indeed, any framed unitary magnetic quiver, whose gauge nodes take the form of the Dynkin diagram of some Lie group $G$, and which is "good" in the sense of \cite{Gaiotto:2008ak} has a Coulomb branch that is a slice in the affine Grassmanian of $G$. This is denoted $\left[\overline{\mathcal W_\mathfrak{g}}\right]^{\bf f}_{\bf b}$, where the vectors ${\bf f}\equiv [f_1,\ldots,f_{rank(G)}]$ and ${\bf b}\equiv [b_1,\ldots,b_{rank(G)}]$ represent flavour nodes and the balances of gauge nodes, respectively.

The flavour and balance vectors contain the same information as the flavour and gauge nodes of a unitary magnetic quiver. The ranks of the quiver gauge nodes $\urm(r_i)$ can thus be recovered from the affine Grassmanian slice $\left[\overline{\mathcal W_\mathfrak{g}}\right]^{\bf f}_{\bf b}$ by introducing the vector ${\bf r} \equiv [r_1,\ldots,r_{rank(G)}]$, specialising \eqref{eq:balance} to the case where the quiver is of Dynkin type, and rearranging to obtain:
\begin{equation}
        {\bf r}=A_{\mathfrak g}^{-1}\cdot(\bf f -b),
\end{equation}
where $A_\mathfrak g$ is the Cartan matrix for $G$.

Without going into too many technical details about the affine Grassmannian, in order to ensure that the corresponding magnetic quiver is "good", it is necessary that the vector $\bf f-b$ lies in the positive coroot lattice of $G$, where the lattice is written in a basis of fundamental weights of $G$.

The global symmetry of an Affine Grassmannian slice $\left[\overline{\mathcal W_\mathfrak{g}}\right]^{\bf f}_{\bf b}$ is generally some proper subgroup of $G$, matching $G$ only in the case where the balance vector is trivial. The magnetic quivers for nilpotent orbits of A-type and/or of height 2 constitute such examples where $\bf b = 0$.

The ability of the affine Grassmannian to construct a wide range of "good" theories entails that some moduli spaces known by different names also have a description as an affine Grassmannian slice. Thus, the affine Grassmannian contains magnetic quivers for some nilpotent orbits, Slodowy slices and Slodowy intersections, amongst others. We deploy these various descriptions as appropriate.




\section{Moduli Space of Free Fields}
\label{sec:freetheories}
We begin with magnetic quivers for some of the moduli spaces of free fields and apply the rules presented in Section \ref{sec:rules} to compute the $\surm(n)$ HKQ of the Coulomb branch using $\surm(n)$ quotient quiver subtraction. Additionally, we compare $\surm(n)$ quotient quiver subtraction where the explicit embedding used will be given when studying each example.
\subsection{Moduli Space of Free Fields $\surm(2)$ HKQ}
\subsubsection{$\mathbb H^4///\surm(2)$}
\label{sec:H4SU2}
\begin{figure}[h!]
    \centering
    \begin{tikzpicture}[main/.style={draw,circle}]
    \node[main, label=below:$1$] (black) {};
    \node[main, label=below:$2$] (2) [right=of black]{};
    \node[main, label=right:$1$,blue,fill] (blue) [above right=of 2]{};
    \node[main, label=right:$1$,pink,fill] (pink) [below right=of 2]{};

    \draw[-] (black)--(2)--(blue);
    \draw[-] (2)--(pink);
    \end{tikzpicture}
    \caption{Unframed magnetic quiver \Quiver{fig:4DimFree} for $\mathbb H^4$.}
    \label{fig:4DimFree}
\end{figure}

The magnetic quiver \Quiver{fig:4DimFree} for $\mathbb H^4$ is drawn in \Figref{fig:4DimFree}, where we have introduced colours to distinguish nodes which have different fugacities. The quiver \Quiver{fig:4DimFree} can also be identified as the $T_3\left[\surm(2)\right]$ theory. The global symmetry of its Coulomb branch is $\surm(2)^3$. 

We carry out quiver subtraction by aligning the $\urm(2)$ quotient quiver along the uncoloured nodes and ending on either the \textcolor{blue}{blue} or \textcolor{pink}{pink} node. These alternatives give the following quivers: \begin{equation}
    \mathcal Q_{\ref{fig:4DimFree}\textcolor{blue}{a}}=\resizebox{!}{30pt}{\begin{tikzpicture}[baseline=(current bounding box.center), main/.style = {draw, circle},scale=0.25]

    \node[main, label=below:$1$,pink,fill] (pink) {};
    \node[draw, label=right:$2$] (2) [above=of pink]{};

    \draw[-] (pink)--(2);

    \end{tikzpicture}};\quad  \mathcal Q_{\ref{fig:4DimFree}\textcolor{blue}{b}}=\resizebox{!}{30pt}{\begin{tikzpicture}[baseline=(current bounding box.center), main/.style = {draw, circle},scale=0.25]

    \node[main, label=below:$1$,blue,fill] (blue) {};
    \node[draw, label=right:$2$] (2) [above=of blue]{};

    \draw[-] (blue)--(2);

    \end{tikzpicture}}.
\end{equation} 
To calculate the union of $\mathcal Q_{\ref{fig:4DimFree}\textcolor{blue}{a}}$ and $\mathcal Q_{\ref{fig:4DimFree}\textcolor{blue}{b}}$ diagrammatically, we note that their intersection is at the origin, so the HWG of the union follows from standard results for $A_1$ magnetic quivers \cite{Hanany:2015hxa}:
\begin{align}
    HWG[ \mathcal{C}( {\mathcal Q_{\ref{fig:4DimFree}\textcolor{blue}{a}} \cup \mathcal Q_{\ref{fig:4DimFree}\textcolor{blue}{b}}})]&=PE[\mu^2t^2]+PE[\nu^2t^2]-1\\&
    =PE[(\mu^2 + \nu^2) t^2 - \mu^2 \nu^2 t^4],
\end{align}
where $\mu$ and $\nu$ are highest weight fugacities for each of the two $\surm(2)$ subgroups. This is the $D_2$ minimal orbit.

The unrefined HS evaluates as: 
\begin{align}
    \hs[\mathcal{C}( {\mathcal Q_{\ref{fig:4DimFree}\textcolor{blue}{a}} \cup \mathcal Q_{\ref{fig:4DimFree}\textcolor{blue}{b}}})]&=\frac{1-t^4}{(1-t^2)^3}+\frac{1-t^4}{(1-t^2)^3}-1\\&=\frac{1+4t^2-4t^4}{(1-t^2)^2},
\end{align}
which is of quaternionic dimension 1.

Considering the outer automorphism symmetry, the HKQ may be computed explicitly using Weyl integration by choosing any factor of $\surm(2)$ to implement the quotient. This replicates the results HS and HWG above and so we find that the diagrammatic and analytic integration methods lead to the same result:
\begin{equation}
  \hs[\mathcal{C}({\mathcal Q}_{\ref{fig:4DimFree}})///\surm(2)]= \hs[\mathcal{C}( {\mathcal Q_{\ref{fig:4DimFree}\textcolor{blue}{a}} \cup \mathcal Q_{\ref{fig:4DimFree}\textcolor{blue}{b}}})]
\end{equation}

\subsubsection{$\mathbb{H}^9///\surm(2)$}
\begin{figure}[h!]
    \centering
    \begin{tikzpicture}[main/.style={draw,circle}]
    \node[main,label=below:$1$] (1L){};
    \node[main, label=below:$2$] (2L) [right=of 1L]{};
    \node[main, label=below:$3$] (3) [right=of 2L]{};
    \node[main, label=below:$2$] (2R) [right=of 3]{};
    \node[main, label=below:$1$] (1R) [right=of 2R]{};
    \node[main, label=right:$1$] (1T) [above=of 3]{};

    \draw[-] (1L)--(2L)--(3)--(2R)--(1R);
    \draw[-] (3)--(1T);
        
    \end{tikzpicture}
    \caption{Unframed magnetic quiver $\mathcal Q_{\ref{fig:9DimFree}}$ for $\mathbb H^9$.}
    \label{fig:9DimFree}
\end{figure}
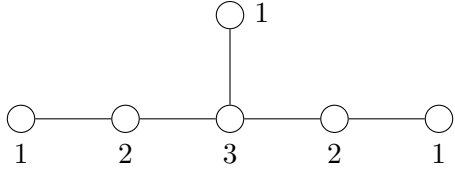
\begin{figure}[h!]
    \centering
    \begin{tikzpicture}[main/.style={draw,circle}]
    \node[main,label=below:$1$] (1L){};
    \node[main, label=below:$2$] (2L) [right=of 1L]{};
    \node[main, label=below:$3$] (3) [right=of 2L]{};
    \node[main, label=below:$2$] (2R) [right=of 3]{};
    \node[main, label=below:$1$] (1R) [right=of 2R]{};
    \node[main, label=right:$1$] (1T) [above=of 3]{};

    \draw[-] (1L)--(2L)--(3)--(2R)--(1R);
    \draw[-] (3)--(1T);

    \node[main, label=below:$1$] (1subL) [below=of 1L]{};
    \node[main, label=below:$2$] (2sub) [right=of 1subL]{};
    \node[main, label=below:$1$] (1subR) [right=of 2sub]{};

    \draw[-] (1subL)--(2sub)--(1subR);

    \node[draw=none,fill=none] (minus) [left=of 1subL]{$-$};

    \node[draw, label=left:$1$] (1fl) [below=of 2sub]{};
    \node[main, label=below:$1$] (1l) [below=of 1fl]{};
    \node[main, label=below:$2$] (2l) [right=of 1l]{};
    \node[main, label=below:$2$] (2r) [right=of 2l]{};
    \node[main, label=below:$1$] (1r) [right=of 2r]{};
    \node[draw, label=right:$1$] (1fr) [above=of 2r]{};

    \draw[-] (1fl)--(1l)--(2l)--(2r)--(1r);
    \draw[-] (1fr)--(2r);
    \node[draw=none,fill=none] (topghost) [right=of 1R]{};
    \node[draw=none,fill=none] (bottomghost) [right=of 1r]{};  
    \draw [->] (topghost) to [out=-30,in=30,looseness=1] (bottomghost);    
    \end{tikzpicture}
    \caption{Subtraction of the $\urm(2)$ quotient quiver from $\mathcal Q_{\ref{fig:9DimFree}}$ to produce $\mathcal Q_{\ref{fig:9DimFreeA1Sub}}$.}
    \label{fig:9DimFreeA1Sub}
\end{figure}
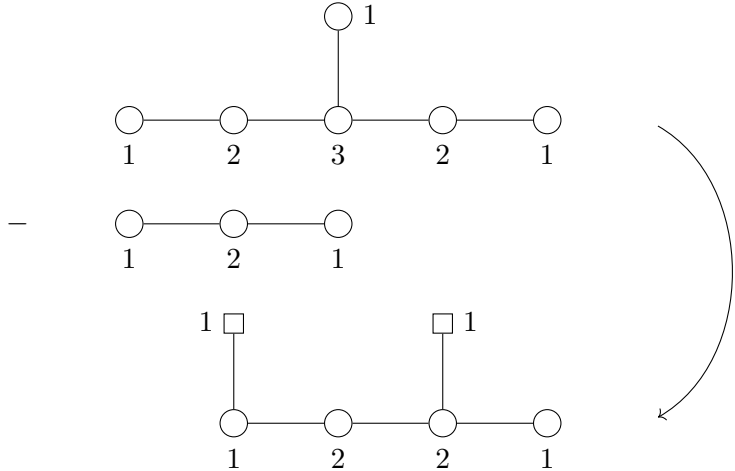
The magnetic quiver $\mathcal Q_{\ref{fig:9DimFree}}$ for the $\mathbb H^9$ is shown in \Figref{fig:9DimFree}. 

The Coulomb branch has an $\surm(3)^2\times \urm(1)$ global symmetry, with the two $\surm(3)$ being equivalent under outer automorphism. The quiver subtraction of the $\urm(2)$ quotient quiver is determined by aligning it along one of the factors of $\surm(3)$, as shown in \Figref{fig:9DimFreeA1Sub}. Applying the monopole formula to $\mathcal Q_{\ref{fig:9DimFreeA1Sub}}$ yields the HKQ from quiver subtraction, which we present as an unrefined HS:
\begin{equation}
    \hs[\mathbb H^9///\surm(2)]=HS\left[\mathcal C\left(\mathcal Q_{\ref{fig:9DimFreeA1Sub}}\right)\right]
    =\frac{(1 + t^2) (1 + 8 t^2 + t^4)}{(1 - t)^{6} (1 - t^2)^6}
\end{equation}
This HS can be identified as the product $ a_3 \times  \mathbb{H}^3$ of the $\surm(4)$ minimal orbit with a 3 (quaternionic) dimensional moduli space of free fields.

To confirm this using Weyl integration, we use the embedding of $A_2\hookleftarrow A_1\times \urm(1)$ which decomposes the fundamental as:
\begin{align}
    [1,0]_{A_2}&\rightarrow [1]_{A_1}(1)+[0]_{A_1}(-2),
\end{align}
where the integer in the round brackets is the $\urm(1)$ charge. The result of the Weyl integration matches the quiver subtraction.

\subsubsection{$\mathbb H^{2k}///\surm(2), k\geq 2$}

\begin{figure}[h!]
    \centering
    \begin{tikzpicture}[main/.style={draw,circle}]
    \node[main, label=below:$1$] (1) []{};
    \node[main, label=below:$2$] (2l) [right=of 1]{};
    \node[main, label=below:$2$] (2m) [right=of 2l]{};
    \node[] (cdots) [right=of 2m]{$\cdots$};
    \node[main, label=below:$2$] (2r) [right=of cdots]{};
    \node[main, label=right:$1$] (1t) [above right=of 2r]{};
    \node[main, label=right:$1$] (1b) [below right=of 2r]{};

    \draw[-] (1)--(2l)--(2m)--(cdots)--(2r)--(1t);
    \draw[-] (2r)--(1b);

    \draw [decorate, 
    decoration = {brace,
        raise=15pt,
        amplitude=5pt}] (2r) --  (2l) node[pos=0.5,below=20pt,black]{$k-1$};

    \node[main, label=below:$1$] (1subL) [below =of 1]{};
    \node[main, label=below:$2$] (2sub) [right=of 1subL]{};
    \node[main, label=below:$1$] (1subR) [right=of 2sub]{};

    \draw[-] (1subL)--(2sub)--(1subR);

    \node[] (minus) [left=of 1subL]{$-$};

    \node[main, label=right:$1$] (1rest) [below =of 1b]{};
    \node[main, label=below:$2$] (2rres) [below left=of 1rest]{};
    \node[main, label=right:$1$] (1resb) [below right=of 2rres]{};
    \node[] (cdotsres) [left=of 2rres]{$\cdots$};
    \node[main, label=below:$2$] (2resm) [left=of cdotsres]{};
    \node[main, label=below:$1$] (1res) [left=of 2resm]{};
    \node[draw, label=left:$1$] (1f) [above=of 2resm]{};

    \draw[-] (1rest)--(2rres)--(1resb);
    \draw[-] (1res)--(2resm)--(cdotsres)--(2rres);
    \draw[-] (1f)--(2resm);

    \draw [decorate, 
    decoration = {brace,
        raise=10pt,
        amplitude=5pt}] (2rres) --  (2resm) node[pos=0.5,below=15pt,black]{$k-3$};

    \node[] (topghost) [right=of 2r]{};
    \node[] (bottomghost) [right=of 2rres]{};
    
    \draw [->] (topghost) to [out=-30,in=30,looseness=1] (bottomghost);

    \end{tikzpicture}
    \caption{One subtraction of the $\urm(2)$ quotient quiver from the magnetic quiver \Quiver{fig:free2kp2} for $\mathbb H^{2k}$ to produce the magnetic quiver \Quiver{fig:quiv_min_Dk} for $\overline{min. D_{k}}$.}
    \label{fig:free2kp2}
\end{figure}

The unframed magnetic quiver \Quiver{fig:free2kp2} for $\mathbb H^{2k}$ is drawn in the top of \Figref{fig:free2kp2}. This quiver takes the form of the finite $D_{k+2}$ Dynkin diagram. The Coulomb branch has a $SO(2k)\times \surm(2)$ global symmetry. Although there are two ends from which a $\urm(2)$ quotient quiver may be subtracted in accordance with the rules in Section \ref{sec:rules}, consider the possibility drawn in \Figref{fig:free2kp2}, which results in the magnetic quiver \Quiver{fig:quiv_min_Dk} for $\overline{min. D_{k}}$. We conjecture that \begin{equation}
    \mathbb H^{2k}///\surm(2)=\overline{min. D_{k}},\quad k\geq 2.
    \label{eq:minDKConj}
\end{equation}

We have tested the conjecture in \eqref{eq:minDKConj} for $k=2,\cdots,6$, finding agreement with Weyl integration. The first member of the family, $k=2$, was evaluated in Section \ref{sec:H4SU2}. The Weyl integration is performed with respect to the fugacity for the $\surm(2)$ in the Coulomb branch global symmetry of \Quiver{fig:free2kp2}.

The result \eqref{eq:minDKConj} is in accordance with gauging the $\surm(2)$ flavour symmetry of the quiver $[\surm(2)]-[D_k]$ to give the electric quiver $(\surm(2))-[D_k]$ and comparing their Higgs branches.

\subsubsection{$\mathbb H^{2k+1}///\surm(2), k\geq 2$}

\begin{figure}[h!]
    \centering
    \begin{tikzpicture}[main/.style={draw,circle}]
    \node[main, label=below:$1$] (1) []{};
    \node[main, label=below:$2$] (2l) [right=of 1]{};
    \node[main, label=below:$2$] (2m) [right=of 2l]{};
    \node[] (cdots) [right=of 2m]{$\cdots$};
    \node[main, label=below:$2$] (2r) [right=of cdots]{};
    \node[main, label=below:$1$] (1r) [right=of 2r]{};
    
    \draw[-] (1)--(2l)--(2m)--(cdots)--(2r);

    \draw [decorate, 
    decoration = {brace,
        raise=15pt,
        amplitude=5pt}] (2r) --  (2l) node[pos=0.5,below=20pt,black]{$k$};

     \draw [line width=1pt, double distance=3pt,
             arrows = {-Latex[length=0pt 3 0]}] (2r) -- (1r);

    \node[main, label=below:$1$] (1subL) [below =of 1]{};
    \node[main, label=below:$2$] (2sub) [right=of 1subL]{};
    \node[main, label=below:$1$] (1subR) [right=of 2sub]{};

    \draw[-] (1subL)--(2sub)--(1subR);

    \node[] (minus) [left=of 1subL]{$-$};

    \node[main, label=below:$1$] (1resl) [below=of 1subR]{};
    \node[main, label=below:$2$] (2resl) [right=of 1resl]{};
    \node[] (cdotsres) [right=of 2resl]{$\cdots$};
    \node[main, label=below:$2$] (2resr) [right=of cdotsres]{};
    \node[main, label=below:$1$] (1resr) [right=of 2resr]{};
    \node[draw, label=left:$1$] (1fres) [above=of 2resl]{};

    \draw[-] (1resl)--(2resl)--(cdotsres)--(2resr);
    \draw[-] (1fres)--(2resl);
    
    \draw [line width=1pt, double distance=3pt,
             arrows = {-Latex[length=0pt 3 0]}] (2resr) -- (1resr);
    
    \draw [decorate, 
    decoration = {brace,
        raise=10pt,
        amplitude=5pt}] (2resr) --  (2resl) node[pos=0.5,below=15pt,black]{$k-2$};

    \node[] (topghost) [right=of 1r]{};
    \node[] (bottomghost) [right=of 1resr]{};
    
    \draw [->] (topghost) to [out=-30,in=30,looseness=1] (bottomghost);

    \end{tikzpicture}
    \caption{Subtraction of the $\urm(2)$ quotient quiver from the magnetic quiver \Quiver{fig:free2kp1} for $\mathbb H^{2k+1}$ to produce the magnetic quiver \Quiver{fig:quiv_min_Bk} for $\overline{min. B_{k}}$.}
    \label{fig:free2kp1}
\end{figure}

The unframed magnetic quiver \Quiver{fig:free2kp1} for $\mathbb H^{2k+1}$ is drawn in the top of \Figref{fig:free2kp1}. This quiver takes the form of the finite $B_{k+2}$ Dynkin diagram. The Coulomb branch has a $SO(2k+1)\times \surm(2)$ global symmetry. There is only one place from which a $\urm(2)$ quotient quiver may be subtracted in accordance with the rules in Section \ref{sec:rules}. This is shown in \Figref{fig:free2kp1} and results in the magnetic quiver \Quiver{fig:quiv_min_Bk} for $\overline{min. B_{k}}$. We conjecture that \begin{equation}
    \mathbb H^{2k+1}///\surm(2)=\overline{min. B_k},\quad k\geq 2.
    \label{eq:minBKConj}
\end{equation}

We have tested the conjecture in \eqref{eq:minBKConj} for $k=1,\cdots,5$, finding agreement with Weyl integration. The Weyl integration is performed with respect to the fugacity for the $\surm(2)$ in the Coulomb branch global symmetry of \Quiver{fig:free2kp1}.

The result \eqref{eq:minBKConj} is in accordance with gauging the $\surm(2)$ flavour symmetry of the quiver $[\surm(2)]-[B_k]$ to give the electric quiver $(\surm(2))-[B_k]$ and comparing their Higgs branches.

\subsection{Moduli Space of Free Fields $\surm(3)$ HKQ}
\subsubsection{$\mathbb H^{16}///\surm(3)$}
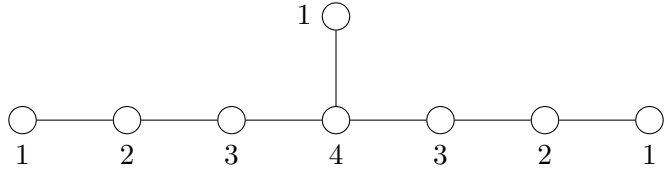
\begin{figure}[h!]
    \centering
    \begin{tikzpicture}[main/.style={draw,circle}]
    \node[main, label=below:$1$] (1L) {};
    \node[main, label=below:$2$] (2L) [right=of 1L]{};
    \node[main, label=below:$3$] (3L) [right=of 2L]{};
    \node[main, label=below:$4$] (4) [right=of 3L]{};
    \node[main, label=below:$3$] (3R) [right=of 4]{};
    \node[main, label=below:$2$] (2R) [right=of 3R]{};
    \node[main, label=below:$1$] (1R) [right=of 2R]{};
    \node[main, label=left:$1$] (1T) [above=of 4]{};

    \draw[-] (1L)--(2L)--(3L)--(4)--(3R)--(2R)--(1R);
    \draw[-] (4)--(1T);
        
    \end{tikzpicture}
    \caption{Unframed magnetic quiver $\mathcal Q_{\ref{fig:16DimFree}}$ for $\mathbb H^{16}$.}
    \label{fig:16DimFree}
\end{figure}
\begin{figure}[h!]
    \centering
    \begin{subfigure}{0.4\textwidth}
    \centering
    \resizebox{0.7\width}{!}{
    \begin{tikzpicture}[main/.style={draw,circle}]
    \node[main, label=below:$1$] (1L) {};
    \node[main, label=below:$2$] (2L) [right=of 1L]{};
    \node[main, label=below:$3$] (3L) [right=of 2L]{};
    \node[main, label=below:$4$] (4) [right=of 3L]{};
    \node[main, label=below:$3$] (3R) [right=of 4]{};
    \node[main, label=below:$2$] (2R) [right=of 3R]{};
    \node[main, label=below:$1$] (1R) [right=of 2R]{};
    \node[main, label=left:$1$] (1T) [above=of 4]{};

    \draw[-] (1L)--(2L)--(3L)--(4)--(3R)--(2R)--(1R);
    \draw[-] (4)--(1T);

    \node[main, label=left:$1$] (1subr) [below=of 4]{};
    \node[main, label=below:$2$] (2subr) [below=of 1subr]{};
    \node[main, label=below:$3$] (3sub) [left=of 2subr]{};
    \node[main, label=below:$2$] (2subl) [left=of 3sub]{};
    \node[main, label=below:$1$] (1subl) [left=of 2subl]{};

    \draw[-] (1subr)--(2subr)--(3sub)--(2subl)--(1subl);

    \node[draw=none,fill=none] (minus) [left=of 1subl]{$-$};

    \node[main, label=below:$2$] (2resl) [below=of 2subr]{};
    \node[main, label=below:$3$] (3res) [right=of 2resl]{};
    \node[main, label=below:$2$] (2resr) [right=of 3res]{};
    \node[main, label=below:$1$] (1res) [right=of 2resr]{};
    \node[draw, label=right:$2$] (2resf) [above=of 3res]{};

    \draw[-] (2resl)--(3res)--(2resr)--(1res);
    \draw[-] (2resf)--(3res);

    \node[draw=none,fill=none] (topghost) [right=of 1R]{};
    \node[draw=none,fill=none] (bottomghost) [right=of 1res]{};
    
    \draw [->] (topghost) to [out=-30,in=30,looseness=1] (bottomghost);
    \end{tikzpicture}}
    
    \caption{}
    
    \label{fig:16DimFreeA2Sub1}
    \end{subfigure}
    \hfill
    \begin{subfigure}{0.5\textwidth}
    \centering\resizebox{0.7\width}{!}{
    \begin{tikzpicture}[main/.style={draw,circle}]
      \node[main, label=below:$1$] (1L) {};
    \node[main, label=below:$2$] (2L) [right=of 1L]{};
    \node[main, label=below:$3$] (3L) [right=of 2L]{};
    \node[main, label=below:$4$] (4) [right=of 3L]{};
    \node[main, label=below:$3$] (3R) [right=of 4]{};
    \node[main, label=below:$2$] (2R) [right=of 3R]{};
    \node[main, label=below:$1$] (1R) [right=of 2R]{};
    \node[main, label=left:$1$] (1T) [above=of 4]{};

    \draw[-] (1L)--(2L)--(3L)--(4)--(3R)--(2R)--(1R);
    \draw[-] (4)--(1T);

    \node[main, label=below:$1$] (1subr) [below=of 3R]{};
    \node[main, label=below:$2$] (2subr) [left=of 1subr]{};
    \node[main, label=below:$3$] (3sub) [left=of 2subr]{};
    \node[main, label=below:$2$] (2subl) [left=of 3sub]{};
    \node[main, label=below:$1$] (1subl) [left=of 2subl]{};

    \draw[-] (1subr)--(2subr)--(3sub)--(2subl)--(1subl);

    \node[draw=none,fill=none] (minus) [left=of 1subl]{$-$};

    \node[draw, label=left:$2$] (2f) [below=of 3sub]{};
    \node[main, label=below:$1$] (1l) [below=of 2f]{};
    \node[main, label=below:$2$] (2l) [right=of 1l]{};
    \node[main, label=below:$2$] (2m) [right=of 2l]{};
    \node[main, label=below:$2$] (2r) [right=of 2m]{};
    \node[main, label=below:$1$] (1r) [right=of 2r]{};
    \node[draw, label=right:$1$] (1f) [above=of 2r]{};

    \draw[-] (2f)--(1l)--(2l)--(2m)--(2r)--(1r);
    \draw[-] (1f)--(2r);

    \node[draw=none,fill=none] (topghost) [right=of 1R]{};
    \node[draw=none,fill=none] (bottomghost) [right=of 1r]{};
    
    \draw [->] (topghost) to [out=-30,in=30,looseness=1] (bottomghost);
    
    \end{tikzpicture}}
    \caption{}
    \label{fig:16DimFreeA2Sub2}
    \end{subfigure}
    \centering
    \begin{subfigure}{\textwidth}
    \centering
    \begin{tikzpicture}[main/.style = {draw, circle}]
    \node[main, label=below:$2$] (2l) []{};
    \node[main, label=below:$2$] (2m) [right=of 2l]{};
    \node[main, label=below:$2$] (2r) [right=of 2m]{};
    \node[main, label=below:$1$] (1) [right=of 2r]{};
    \node[draw, label=left:$1$] (1fl) [above=of 2l]{};
    \node[draw, label=right:$1$] (1fr) [above=of 2r]{};

    \draw[-] (1fl)--(2l)--(2m)--(2r)--(1fr);
    \draw[-] (1)--(2r);
    \end{tikzpicture}
    \caption{}
    \label{fig:16DimFreeA2Int}
    \end{subfigure}

    \caption{Both alignments of the $\urm(3)$ quotient quiver against $\mathcal Q_{\ref{fig:16DimFree}}$ producing $\mathcal Q_{\ref{fig:16DimFreeA2Sub1}}$ and $\mathcal Q_{\ref{fig:16DimFreeA2Sub2}}$. Their intersection, reached via $A_1$ KP transitions, is quiver \Quiver{fig:16DimFreeA2Int}.}
    \label{fig:16DimFreeA2SubBoth}
    
\end{figure}
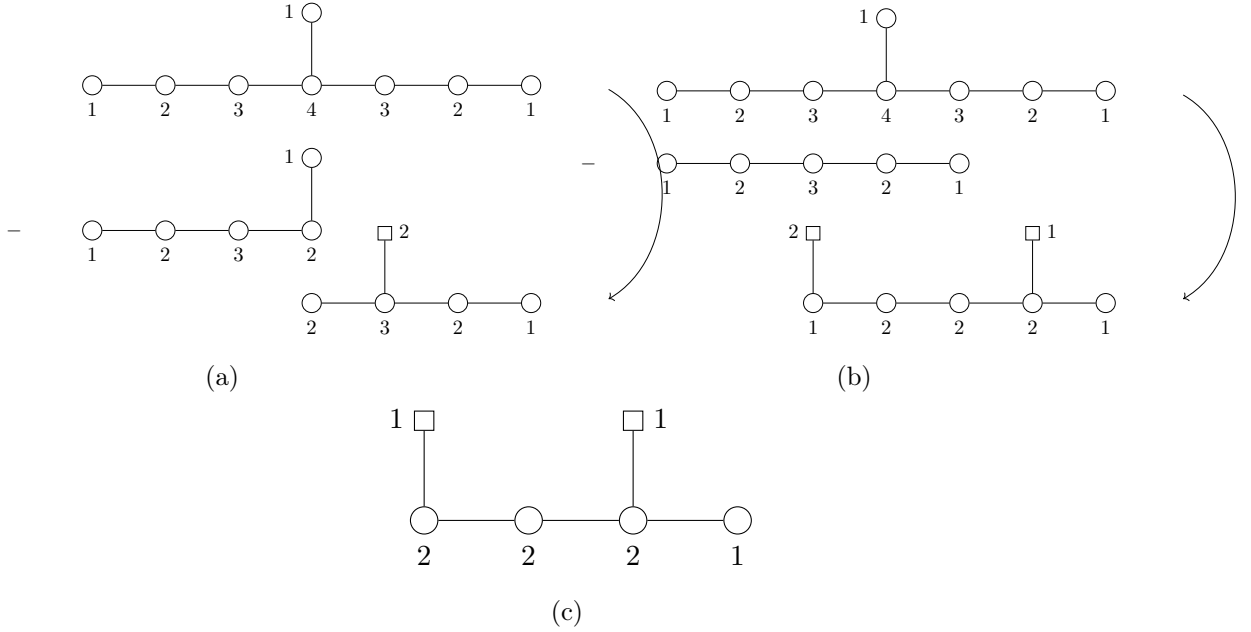

A magnetic quiver $\mathcal Q_{\ref{fig:16DimFree}}$ for $\mathbb H^{16}$ is shown in \Figref{fig:16DimFree}. 

The Coulomb branch has an $\surm(4)^2\times \urm(1)$ global symmetry, with an outer automorphism between the factors of $\surm(4)$. An $\urm(3)$ quotient quiver may be removed from either long leg, but as the quotient quiver extends past a junction, there are two possible alignments as shown in \Figref{fig:16DimFreeA2Sub1} and \Figref{fig:16DimFreeA2Sub2}. These produce quivers $\mathcal Q_{\ref{fig:16DimFreeA2Sub1}}$ and $\mathcal Q_{\ref{fig:16DimFreeA2Sub2}}$ respectively. We expect the HKQ to be the Coulomb branch of the union of $\mathcal Q_{\ref{fig:16DimFreeA2Sub1}}$ and $\mathcal Q_{\ref{fig:16DimFreeA2Sub2}}$.
\begin{equation}
    \mathbb{H}^{16}///\surm(3)=\mathcal C\left(\resizebox{!}{30pt}{\begin{tikzpicture}[baseline=(current bounding box.center),main/.style={draw,circle},scale=0.25]
     \node[main, label=below:$2$] (2resl){};
    \node[main, label=below:$3$] (3res) [right=of 2resl]{};
    \node[main, label=below:$2$] (2resr) [right=of 3res]{};
    \node[main, label=below:$1$] (1res) [right=of 2resr]{};
    \node[draw, label=right:$2$] (2resf) [above=of 3res]{};

    \draw[-] (2resl)--(3res)--(2resr)--(1res);
    \draw[-] (2resf)--(3res);
                
    \end{tikzpicture}}\right)\cup \mathcal C\left(\resizebox{!}{30pt}{\begin{tikzpicture}[baseline=(current bounding box.center),main/.style={draw,circle},scale=0.25]
   \node[draw, label=left:$2$] (2f) {};
    \node[main, label=below:$1$] (1l) [below=of 2f]{};
    \node[main, label=below:$2$] (2l) [right=of 1l]{};
    \node[main, label=below:$2$] (2m) [right=of 2l]{};
    \node[main, label=below:$2$] (2r) [right=of 2m]{};
    \node[main, label=below:$1$] (1r) [right=of 2r]{};
    \node[draw, label=right:$1$] (1f) [above=of 2r]{};

    \draw[-] (2f)--(1l)--(2l)--(2m)--(2r)--(1r);
    \draw[-] (1f)--(2r);
        
    \end{tikzpicture}}\right).
\end{equation}
The intersection of these quivers \Quiver{fig:16DimFreeA2Int} is related to \Quiver{fig:16DimFreeA2Sub1} and \Quiver{fig:16DimFreeA2Sub2} by an $A_1$ Kraft-Procesi transition. The monopole formula can be used to compute the HS of each part of the union. The combination yields the 8 (quaternionic) dimensional HS:
\begin{align}
    HS&=\frac{(1 + t^2)^2 (1 + 5 t^2 + t^4)}{(1 - t)^{8} (1 - t^2)^8} +\frac{\left(\begin{aligned}1 &+ 2 t + 13 t^2 + 28 t^3 + 62 t^4 + 88 t^5 + 128 t^6 + 132 t^7 \\&+ 
 128 t^8 + 88 t^9 + 62 t^{10} + 28 t^{11} + 13 t^{12} + 
 2 t^{13} + t^{14}\end{aligned}\right)}{(1 - t)^{6} (1 - t^2)^6 (1 - t^3)^4}\nonumber\\ &-\frac{(1 + t^2) (1 + 8 t^2 + t^4)}{(1 - t)^{8} (1 - t^2)^6}\label{eq:16DimFreeA2Sub}\\
  &=\frac{\left(\begin{aligned}1 &+ 4 t + 18 t^2 + 56 t^3 + 151 t^4 + 320 t^5 + 581 t^6 + 856 t^7 + 1044 t^8 + 1012 t^9\\
  & + 790 t^{10} + 460 t^{11} + 177 t^{12} + 4 t^{13} - 46 t^{14} - 36 t^{15} - 15 t^{16} - 4 t^{17} - t^{18}\end{aligned}\right)}{(1 - t)^{4} (1 - t^2)^8 (1 - t^3)^4},
\end{align}

where the HS in order in \eqref{eq:16DimFreeA2Sub} are for the Coulomb branches of \Quiver{fig:16DimFreeA2Sub1}, \Quiver{fig:16DimFreeA2Sub2}, and \Quiver{fig:16DimFreeA2Int} respectively.

Explicit computation of the HKQ by Weyl integration may be done with the following embedding of $A_3\hookleftarrow A_2\times \urm(1)$, which decomposes the fundamental
as:
\begin{align}
    [1,0,0]_{A_3}&\rightarrow [1,0]_{A_2}(1)+[0,0]_{A_2}(-3).
\end{align}
The result matches the Hilbert series from the quotient quiver subtraction.

\section{A-type Orbits}
\label{sec:AType}
Here we look at A-type nilpotent orbits and some of the HKQs one can take using $\urm(n)$ quotient quiver subtraction as described in Section \ref{sec:rules}. There are many orbits for which an $\surm(n)$ HKQ can be computed, although we cannot use quiver subtraction to take the $\surm(n)$ HKQ of any A-type orbit. For example, the magnetic quivers for $\overline{min. A_k}$ only contain nodes of rank 1, and so do not permit the subtraction of an $\urm(n)$ quotient quiver – this would violate the \hyperlink{rule:ExternalLeg}{External Leg Rule}.

The examples we use to illustrate the quotient quiver subtraction procedure include the $\surm(2)$ HKQ of $\overline{n.min A_k}$, which we generalise to show how to take the $\surm(n)$ quotient of any height 2 orbit of A-type, and the $\surm(2)$ and $\surm(3)$ HKQs of $\overline{max A_k}$, which we generalise for $\overline{max A_{2n-1}}///\surm(n)$. Many other examples are possible.

To check the results of quiver subtraction using Hilbert Series (and/or HWGs) derived from Weyl integration, we employ an embedding of $A_k\hookleftarrow A_{k-n}\times A_{n-1}\times \urm(1)$ which decomposes the fundamental of $A_k$ as:
\begin{equation}
    [1,0,\cdots,0]_{A_{k}}\rightarrow [1,0,\cdots,0]_{A_{k-n}}(n)+
    [1,0,\cdots,0]_{A_{n-1}}(n-k-1),\label{eq:AkAnEmbed}
\end{equation}
where we have used $()$ to denote $\urm(1)$ charges and suppressed any singlets.

\subsection{A-type Orbit $\surm(2)$ HKQ}
\label{sec:AtypeSU2HKQ}
\subsubsection{$\overline{n. min A_k}///\surm(2)$}
\label{sec:nminAkA1}

The magnetic quivers for $\overline{n.min A_k}$, with $k\geq 4$, meet the criteria for an $\surm(2)$ quiver subtraction. (We note in passing that the unframed magnetic quiver for $\overline{n. min. A_3}$, shown in \Figref{fig:nminA3}, has a double link, and so not all alignments of the $\urm(2)$ quotient quiver are consistent with the \hyperlink{rule:SingleEdge}{Single Edge Rule}, and the quiver subtraction fails.) 

\begin{figure}[h!]
    \centering
    \begin{tikzpicture}[main/.style={draw,circle}]
        \node[main, label=below:$1$] (1L) {};
        \node[main, label=below:$2$] (2) [right=of 1L]{};
        \node[main, label=below:$1$] (1R) [right=of 2]{};
        \node[main, label=left:$1$] (1T) [above=of 2]{};
        \draw[-] (1L)--(2)--(1R);
        \draw[double,double distance=3pt,line width=0.4pt] (1T)--(2);    
    \end{tikzpicture}
    \caption{Unframed magnetic quiver \Quiver{fig:nminA3} for $\overline{n.min. A_3}$.}
    \label{fig:nminA3}
\end{figure}

\begin{figure}[h!]
    \centering
     \begin{tikzpicture}[main/.style = {draw, circle}]
        \node[main, label=below:$1$] (1) []{};
        \node[main, label=below:$2$] (2L) [right=of 1]{};
        \node[draw=none,fill=none] (2M) [right=of 2L]{$\cdots$};
        \node[main, label=below:$2$] (2R) [right=of 2M]{};
        \node[main, label=below:$1$] (1R) [right=of 2R]{};
        \node[main, label=above:$1$] (1T) [above= of 2M]{};
        \draw[-] (1)--(2L)--(2M)--(2R)--(1R);
        \draw[-] (2L)--(1T)--(2R);
        \draw [decorate, decoration = {brace, raise=10pt, amplitude=5pt}] (2R) --  (2L) node[pos=0.5,below=15pt,black]{$k-2$};
    \end{tikzpicture}
    \caption{Unframed magnetic quiver \Quiver{fig:nminAk} for $\overline{n. min. A_k}$ for $k\geq4.$}
    \label{fig:nminAk}
\end{figure}

A magnetic quiver \Quiver{fig:nminAk} for $\overline{n.min. A_k}$ is drawn in \Figref{fig:nminAk}. We wish to carry out quiver subtraction to find the $\surm(2)$ HKQ. There is one special case for $k=4$ to consider before generalising. There are two possible alignments of the $\urm(2)$ quotient quiver and so the result is a union of Coulomb branches. The quiver subtractions (once an end node has been chosen) are as shown in \Figref{fig:minA4A1SubBoth}, and result in the framed quivers \Quiver{fig:nminA4A1Sub1} and \Quiver{fig:nminA4A1Sub2}.

\begin{figure}[h!]
    \centering
\begin{subfigure}{0.4\textwidth}
    \centering
    \begin{tikzpicture}[main/.style = {draw, circle}]
        \node[main, label=below:$1$] (1) []{};
        \node[main, label=below:$2$] (2L) [right=of 1]{};
        \node[main, label=below:$2$] (2R) [right=of 2L]{};
        \node[main, label=below:$1$] (1R) [right=of 2R]{};
        \node[main, label=above:$1$] (1T) [above right= 1cm and 0.5cm of 2L]{};
        \draw[-] (1)--(2L)--(2R)--(1R);
        \draw[-] (2L)--(1T)--(2R);
        \node[main, label=below:$1$] (1subL) [below right= 1cm and 0.5cm of 2L]{};
        \node[main, label=below:$2$] (2sub) [below right=1cm and 0.5cm of 1subL]{};
        \node[main, label=below:$1$] (1subR) [right=of 2sub]{};
        \node[draw=none,fill=none] (minus) [left=of 2sub]{$-$};
        \draw[-] (1subL)--(2sub)--(1subR);
        \node[draw, label=left:$3$] (threeF) [below left= 1cm and 0.5cm of 2sub]{};
        \node[main, label=below:$2$] (two) [below=of threeF]{};
        \node[main, label=below:$1$] (one) [left=of two]{};
        \draw[-] (threeF)--(two)--(one);
        \node[draw=none,fill=none] (topghost) [right=of 1R]{};
        \node[draw=none,fill=none] (bottomghost) [right=of two]{};
        \draw [->] (topghost) to [out=-30,in=30,looseness=1] (bottomghost);
    \end{tikzpicture}
    \caption{}
    \label{fig:nminA4A1Sub1}
\end{subfigure}
\hfill
\begin{subfigure}{0.4\textwidth}
    \centering
    \begin{tikzpicture}[main/.style = {draw, circle}]
    \node[main, label=below:$1$] (1) []{};
    \node[main, label=below:$2$] (2L) [right=of 1]{};
    \node[main, label=below:$2$] (2R) [right=of 2L]{};
    \node[main, label=below:$1$] (1R) [right=of 2R]{};
    \node[main, label=above:$1$] (1T) [above right= 1cm and 0.5cm of 2L]{};
    
    \draw[-] (1)--(2L)--(2R)--(1R);
    \draw[-] (2L)--(1T)--(2R);
    
    \node[main, label=below:$1$] (1subL) [below= of 2L]{};
    \node[main, label=below:$2$] (2sub) [right=of 1subL]{};
    \node[main, label=below:$1$] (1subR) [right=of 2sub]{};
    \node[draw=none,fill=none] (minus) [left=of 1subL]{$-$};
    
    \draw[-] (1subL)--(2sub)--(1subR);
    
    \node[draw, label=left:$3$] (threeF) [below=of 2sub]{};
    \node[main, label=below:$1$] (oneR) [below=of threeF]{};
    \node[main, label=below:$1$] (oneM) [left=of oneR]{};
    \node[main, label=below:$1$] (oneL) [left=of oneM]{};
    \node[draw, label=left:$1$] (oneF) [above=of oneL]{};
    
    \draw[-] (threeF)--(oneR)--(oneM)--(oneL)--(oneF);
    
    \node[draw=none,fill=none] (topghost) [right=of 1R]{};
    \node[draw=none,fill=none] (bottomghost) [right=of oneR]{};
    
    \draw [->] (topghost) to [out=-30,in=30,looseness=1] (bottomghost);
    
    \end{tikzpicture}
   \caption{}
    \label{fig:nminA4A1Sub2}
    \end{subfigure}
    \centering
    \begin{subfigure}{\linewidth}
    \centering
        \begin{tikzpicture}[main/.style = {draw, circle}]

        \node[draw,label=left:$1$] (1FL) []{};
        \node[main, label=below:$1$] (1l) [below=of 1FL]{};
        \node[main, label=below:$1$] (1r) [right=of 1l]{};
        \node[draw, label=right:$1$] (1fr) [above=of 1r]{};

        \draw[-] (1FL)--(1l)--(1r)--(1fr);
            
        \end{tikzpicture}
    \caption{}
    \label{fig:nminA4A1SubInt}
    \end{subfigure}
    \caption{Both alignments of the $\urm(2)$ quotient quiver against the unframed magnetic quiver \Quiver{fig:nminAk} for $\overline{min. A_4}$ to give the quivers \Quiver{fig:nminA4A1Sub1} and \Quiver{fig:nminA4A1Sub2}. Their intersection, reached via $A_2$ Kraft-Procesi transitions, is quiver \Quiver{fig:nminA4A1SubInt}.}
    \label{fig:minA4A1SubBoth}
\end{figure}

We can identify the Coulomb branches of these quivers as the orbit $\overline{\mathcal O}^{A_2}_{(3)}$ and the affine Grassmannian slice $\overline{[\mathcal W_{D_3}]}^{[0,1,3]}_{[0,0,2]}$.
These are 3 (quaternionic) dimensional moduli spaces, with global symmetries $\surm(3)$ and $\surm(3)\times \urm(1)$, respectively.\footnote{This slice in the affine Grassmannian of $D_3$ can also be identified as the Slodowy intersection $\mathcal S^{A_5}_{(4,1^2),(2^3)}$.} We conjecture that:
\begin{equation}
    \overline{n. min. A_4}///\surm(2)=\overline{\mathcal O}^{A_2}_{(3)}\cup\overline{[\mathcal W_{D_3}]}^{[0,1,3]}_{[0,0,2]}.
    \label{eq:n2mA4hkq}
\end{equation}
To compute this union, we first find the intersection of \Quiver{fig:nminA4A1Sub1} and  \Quiver{fig:nminA4A1Sub2}. This is a magnetic quiver \Quiver{fig:nminA4A1SubInt} for $\overline{min. A_2}$, reached in both cases by an $A_2$ Kraft-Procesi transition.

Thus, the conjecture leads to an HWG for the union as the signed sum of HWGs:
\begin{align}
    HWG\left[\overline{\mathcal {O}}^{A_2}_{(3)}\cup\overline{[\mathcal W_{D_3}]}^{[0,1,3]}_{[0,0,2]}\right]&=PE\left[\mu_1\mu_2t^2+\mu_1\mu_2t^4+\left(\mu_1^3+\mu_2^3\right)t^6-\mu_1^3\mu_2^3t^{12}\right]\nonumber\\&+PE\left[(1+ \mu_1 \mu_2)t^2 + \left(\mu_2 q^2+\frac{\mu_1}{q^2}\right) t^4 - \mu_1\mu_2 t^8\right]\nonumber\\&-PE\left[\mu_1\mu_2t^2\right],
    \label{eq:n2mA4hwg}
\end{align}
where $\mu_{1,2}$ are highest weight fugacities for $A_2$, $q$ is a fugacity for $\urm(1)$ charges, and the three PEs encode the HWGs for $\overline{{\mathcal O}}^{A_2}_{(3)}$, $\overline{[\mathcal W_{D_3}]}^{[0,1,3]}_{[0,0,2]}$, and the intersection $\overline{min. A_2}$, respectively. The results \eqref{eq:n2mA4hkq} and \eqref{eq:n2mA4hwg} are confirmed by Weyl integration.

\begin{figure}[h!]
    \centering
    
\begin{subfigure}{0.4\textwidth}
    \centering
    \begin{tikzpicture}[main/.style = {draw, circle}]
    \node[main, label=below:$1$] (1) []{};
    \node[main, label=below:$2$] (2L) [right=of 1]{};
    \node[draw=none,fill=none] (2M) [right=of 2L]{$\cdots$};
    \node[main, label=below:$2$] (2R) [right=of 2M]{};
    \node[main, label=below:$1$] (1R) [right=of 2R]{};
    \node[main, label=above:$1$] (1T) [above= of 2M]{};
    
    \draw[-] (1)--(2L)--(2M)--(2R)--(1R);
    \draw[-] (2L)--(1T)--(2R);
    \draw [decorate, 
    decoration = {brace,
        raise=10pt,
        amplitude=5pt}] (2R) --  (2L) node[pos=0.5,below=15pt,black]{$k-2$};
    
    \node[main, label=below:$1$] (1subL) [below = of 2M]{};
    \node[main, label=below:$2$] (2sub) [below right= of 1subL]{};
    \node[main, label=below:$1$] (1subR) [right=of 2sub]{};
    \node[draw=none,fill=none] (minus) [left=of 2sub]{$-$};
    
    \draw[-] (1subL)--(2sub)--(1subR);
    
    \node[draw, label=left:$2$] (twoF) [below= of 2sub]{};
    \node[main, label=below:$2$] (twoR) [below=of twoF]{};
    \node[draw=none,fill=none] (cdots2) [left=of twoR]{$\cdots$};
    \node[main, label=below:$2$] (twoL) [left=of cdots2]{};
    \node[main, label=below:$1$] (one) [left=of twoL]{};
    \node[draw, label=left:$1$] (oneF) [above=of twoL]{};
    
    \draw[-] (twoF)--(twoR)--(cdots2)--(twoL)--(oneF);
    \draw[-] (one)--(twoL);
    \draw [decorate, 
    decoration = {brace,
        raise=10pt,
        amplitude=5pt}] (twoR) --  (twoL) node[pos=0.5,below=15pt,black]{$k-3$};
    
    \node[draw=none,fill=none] (topghost) [right=of 1R]{};
    \node[draw=none,fill=none] (bottomghost) [right=of twoR]{};
    
    \draw [->] (topghost) to [out=-30,in=30,looseness=1] (bottomghost);
    
    \end{tikzpicture}
    \caption{}
    \label{fig:nminAkA1Sub1}
\end{subfigure}
\hfill
\begin{subfigure}{0.4\textwidth}
    \centering
    \begin{tikzpicture}[main/.style = {draw, circle}]
    \node[main, label=below:$1$] (1) []{};
    \node[main, label=below:$2$] (2L) [right=of 1]{};
    \node[draw=none, fill=none] (2M) [right=of 2L]{$\cdots$};
    \node[main, label=below:$2$] (2R) [right=of 2M]{};
    \node[main, label=below:$1$] (1R) [right=of 2R]{};
    \node[main, label=above:$1$] (1T) [above=of 2M]{};
    
    \draw[-] (1)--(2L)--(2M)--(2R)--(1R);
    \draw[-] (2L)--(1T)--(2R);
    \draw [decorate, 
    decoration = {brace,
        raise=10pt,
        amplitude=5pt}] (2R) --  (2L) node[pos=0.5,below=15pt,black]{$k-2$};
    
    \node[main, label=below:$1$] (1subL) [below= of 2M]{};
    \node[main, label=below:$2$] (2sub) [right=of 1subL]{};
    \node[main, label=below:$1$] (1subR) [right=of 2sub]{};
    \node[draw=none,fill=none] (minus) [left=of 1subL]{$-$};
    
    \draw[-] (1subL)--(2sub)--(1subR);
    
    \node[draw, label=left:$2$] (twoF) [below left=of 1subL]{};
    \node[main, label=left:$1$] (oneT) [below=of twoF]{};
    \node[main, label=below:$2$] (twoL) [below =of oneT]{};
    \node[main, label=below:$1$] (oneL) [left=of twoL]{};
    \node[draw=none,fill=none] (cdots2) [right=of twoL]{$\cdots$};
    \node[main, label=below:$2$] (twoR) [right=of cdots2]{};
    \node[main, label=below:$1$] (oneR) [right=of twoR]{};
    \node[draw,label=right:$1$] (oneF) [above=of twoR]{};
    
    \draw[-] (oneL)--(twoL)--(cdots2)--(twoR)--(oneR);
    \draw[-] (twoF)--(oneT)--(twoL);
    \draw[-] (oneF)--(twoR);
    \draw [decorate, 
    decoration = {brace,
        raise=10pt,
        amplitude=5pt}] (twoR) --  (twoL) node[pos=0.5,below=15pt,black]{$k-4$};
    
    \node[draw=none,fill=none] (topghost) [right=of 1R]{};
    \node[draw=none,fill=none] (bottomghost) [right=of oneR]{};
    
    \draw [->] (topghost) to [out=-30,in=30,looseness=0.7] (bottomghost);
    
    \end{tikzpicture}
    \caption{}
    \label{fig:nminAkA1Sub2}
\end{subfigure}
\centering
\begin{subfigure}{\textwidth}
    \centering
    \begin{tikzpicture}[main/.style = {draw, circle}]
    
    \node[draw, label=left:$1$] (oneT) []{};
    \node[main, label=below:$2$] (twoL) [below =of oneT]{};
    \node[main, label=below:$1$] (oneL) [left=of twoL]{};
    \node[draw=none,fill=none] (cdots2) [right=of twoL]{$\cdots$};
    \node[main, label=below:$2$] (twoR) [right=of cdots2]{};
    \node[main, label=below:$1$] (oneR) [right=of twoR]{};
    \node[draw,label=right:$1$] (oneF) [above=of twoR]{};
    
    \draw[-] (oneL)--(twoL)--(cdots2)--(twoR)--(oneR);
    \draw[-] (oneT)--(twoL);
    \draw[-] (oneF)--(twoR);
    \draw [decorate, 
    decoration = {brace,
        raise=10pt,
        amplitude=5pt}] (twoR) --  (twoL) node[pos=0.5,below=15pt,black]{$k-4$};
    \end{tikzpicture}
    \caption{}
    \label{fig:nminAkA1SubInt}
\end{subfigure}
    \caption{Both alignments of the $\urm(2)$ quotient quiver against the unframed magnetic quiver \Quiver{fig:nminAk} for $\overline{min. A_k}$ for $k\geq 5$ to give the quivers \Quiver{fig:nminAkA1Sub1} and \Quiver{fig:nminAkA1Sub2}. Their intersection, reached via $A_1$ Kraft-Procesi transitions, is quiver \Quiver{fig:nminAkA1SubInt}.}
    \label{fig:nminAkA1SubBoth}
\end{figure}


Now consider the general case for $\overline{n. min. A_k}, k \geq 5$. There are two possible alignments of the $\urm(2)$ quotient quiver, as shown in \Figref{fig:nminAkA1SubBoth}.

These yield the quivers \Quiver{fig:nminAkA1Sub1} and \Quiver{fig:nminAkA1Sub2} respectively. We identify their Coulomb branches as the orbit $\overline{\mathcal O}_{(3,1^{k-4})}^{A_{k-2}}$ and the affine Grassmannian slice $\overline{\left[\mathcal W_{D_{k-1}}\right]}^{[0,1,\cdots,2]}_{[0,0,\cdots,2]}$, which have global symmetries $A_{k-2}$ and $A_{k-2}\times \urm(1)$, respectively.  
So, based on quiver subtraction, we conjecture that:
\begin{equation}
    \overline{n. min. A_k}///\surm(2)=\overline{\mathcal O}_{(3,1^{k-4})}^{A_{k-2}}\cup \overline{\left[\mathcal W_{D_{k-1}}\right]}^{[0,1,\cdots,2]}_{[0,0,\cdots,2]},\quad k\geq5.
    \label{eq:n2mAkhkq}
\end{equation}

In order to evaluate the union, we note that quivers \Quiver{fig:nminAkA1Sub1} and \Quiver{fig:nminAkA1Sub2} are both related by an $A_1$ KP transition to the magnetic quiver \Quiver{fig:nminAkA1SubInt} for $\overline{n.min. A_{k-2}}$. So, we evaluate the HWG as:
\begin{align}
    HWG&\left[\overline{\mathcal O}_{(3,1^{k-4})}^{A_{k-2}}\cup \overline{\left[\mathcal W_{D_{k-1}}\right]}^{[0,1,\cdots,2]}_{[0,0,\cdots,2]}\right]\nonumber\\&=PE\left[\sum_{i=1}^2\mu_i\mu_{k-1-i}t^{2i}+\mu_1\mu_{k-2}t^4+\left(\mu_1^2\mu_{k-3}+\mu_2\mu_{k-2}^2\right)t^6-\mu_1^2\mu_2\mu_{k-3}\mu_{k-2}^2t^{12}\right]\nonumber\\&+PE\left[t^2+\sum_{i=1}^2\mu_i\mu_{k-1-i}t^{2i}+\left(\mu_2q^2+\frac{\mu_{k-3}}{q^2}\right)t^4-\mu_2\mu_{k-3}t^8\right]\nonumber\\&-PE\left[\sum_{i=1}^2\mu_i\mu_{k-1-i}t^{2i}\right],\quad k\geq 5,
    \label{eq:n2mAkhwg}
\end{align}
where $\mu_{1,2,\cdots,k-2}$ are highest weight fugacities for $A_{k-2}$ and $q$ is a fugacity for $\urm(1)$ charges. 

In principle, the HWG \eqref{eq:n2mAkhwg} can be checked for any given value of $k$ using Weyl integration. We have evaluated the cases for $k \leq 8$ and found these to be consistent with the conjecture \eqref{eq:n2mAkhkq}. However, the nice decomposition of the HWG as a sum of polynomial PEs is only manifest under the quiver subtraction approach.

\subsubsection{$\overline{max. A_k}///\surm(2)$}
\label{sec:MaxAkA1}
\begin{figure}[h!]
    \centering
    \begin{tikzpicture}[main/.style={draw,circle}]
    \node[main, label=below:$1$] (1) []{};
    \node[main, label=below:$2$] (2)[right=of 1]{};
    \node[draw=none,fill=none] (cdots) [right=of 2]{$\cdots$};
    \node[main, label=below:$k$] (k) [right=of cdots]{};
    \node[main, label=right:$1$] (1T) [above=of k]{};
    \node[draw=none,fill=none] (empty) [above left= 0.25cm and 0.05cm of k]{$k+1$};

    \draw[-] (1)--(2)--(cdots)--(k);
     \draw[double,double distance=3pt,line width=0.4pt] (1T)--(k);
        
    \end{tikzpicture}
    \caption{Unframed magnetic quiver \Quiver{fig:maxAk} for $\overline{max. A_k}$ for $k\geq 1$.}
    \label{fig:maxAk}
\end{figure}

A magnetic quiver \Quiver{fig:maxAk} for $\overline{max. A_k}$ is shown in \Figref{fig:maxAk}. The label $k+1$ next to the multiple edges indicates that there are $k+1$ hypermultiplets transforming in the bifundamental of the $\urm(k)$ and $\urm(1)$ gauge groups. For $k\geq 3$, this magnetic quiver meets the criteria in Section \ref{sec:rules} for quiver subtraction by an $\surm(2)$ HKQ.

\begin{figure}[h!]
    \centering
    \begin{tikzpicture}[main/.style = {draw, circle} ,quadruple/.style={double distance=6pt, thin,
        postaction={draw,thin, double distance=1.5pt}},
    quadruple/.default=8pt]
    \node[main, label=below:$1$] (1) []{};
    \node[main, label=below:$2$] (2)[right=of 1]{};
    \node[main, label=below:$3$] (3) [right=of 2]{};
    \node[main, label=right:$1$] (1T) [above=of 3]{};
    
    \draw[-] (1)--(2)--(3);
    \draw[quadruple] (3)--(1T);

    \node[main, label=below:$1$] (1subL) [below= of 1]{};
    \node[main, label=below:$2$] (2sub) [right=of 1subL]{};
    \node[main, label=below:$1$] (1subR) [right=of 2sub]{};
    \node[draw=none,fill=none] (minus) [left=of 1subL]{$-$};
    
    \draw[-] (1subL)--(2sub)--(1subR);

    \node[main, label=right:$1$] (one) [below=of 1subR]{};
    \node[draw, label=left:$4$] (4F) [left=of one]{};
    \node[main, label=below:$2$] (two) [below=of one]{};

    \draw[-] (4F)--(one);
    \draw[quadruple] (one)--(two);
     \node[draw=none,fill=none] (topghost) [right=of 3]{};
    \node[draw=none,fill=none] (bottomghost) [right=of one]{};
    \draw [->] (topghost) to [out=-30,in=30,looseness=1] (bottomghost);
    \end{tikzpicture}
    \caption{Subtraction of the $\urm(2)$ quotient quiver from the magnetic quiver \Quiver{fig:maxAk} for $\overline{max. A_3}$ to give quiver \Quiver{fig:maxA3A1Sub}.}
    \label{fig:maxA3A1Sub}
\end{figure}

There is a special case for $k=3$. \Figref{fig:maxA3A1Sub} shows the subtraction of the $\urm(2)$ quotient quiver from the unframed magnetic quiver for $\overline{max. A_3}$, which results in the framed quiver \Quiver{fig:maxA3A1Sub}.

In order to decipher the Coulomb branch of \Quiver{fig:maxA3A1Sub}, we note that this quiver is simply laced, so we can choose to shift the framing to the gauge node of rank 1, such that the quiver decouples into a pair of quivers, $(2)-[4]$ and $(1)-[4]$. We can identify their Coulomb branches as the Slodowy slices $\mathcal{S}^{A_3}_{\mathcal N,(2^2)}$ \cite{Cabrera:2018ldc} and $ \mathbb{C}^2/\mathbb{Z}_4$  \cite{Grimminger:2020dmg}, respectively. Thus, we obtain the result (verified by Weyl integration):

\begin{equation}
    \overline{max. A_3}///\surm(2)=\mathcal C\left(\resizebox{!}{30pt}{\begin{tikzpicture}[baseline=(current bounding box.center),main/.style={draw,circle},quadruple/.style={double distance=6pt, thin,
        postaction={draw,thin, double distance=1.5pt}},
    quadruple/.default=8pt]
        \node[draw, label=left:$4$] (4F) {};
    \node[main, label=below:$1$] (one) [below=of 4F]{};
    \node[main, label=below:$2$] (two) [left=of one]{};

    \draw[-] (4F)--(one);
    \draw[quadruple] (one)--(two);
    \end{tikzpicture}}\right)=\mathcal C\left(\resizebox{!}{30pt}{\begin{tikzpicture}[baseline=(current bounding box.center),main/.style={draw,circle}]
        \node[main, label=below:$2$] (2) {};
        \node[draw, label=left:$4$] (4) [above=of 2]{};
        \node[main, label=below:$1$] (1) [right=of 2]{};
        \node[draw, label=right:$4$] (four) [above=of 1]{};
        \node[draw=none,fill=none] (otimes) [above right= 0.3cm and 0.3cm of 2]{$\otimes$};

        \draw[-] (2)--(4);
        \draw[-] (1)--(four);
    \end{tikzpicture}}\right)=\mathcal{S}^{A_3}_{\mathcal N,(2^2)}\otimes\mathbb{C}^2/\mathbb{Z}_4 \label{eq:maxA3A1HKQ}
\end{equation}
 
\begin{figure}[h!]
    \centering
    \begin{tikzpicture}[main/.style = {draw, circle}]
    \node[main, label=below:$1$] (1) []{};
    \node[main, label=below:$2$] (2)[right=of 1]{};
    \node[draw=none,fill=none] (cdots) [right=of 2]{$\cdots$};
    \node[main, label=below:$k$] (k) [right=of cdots]{};
    \node[main, label=right:$1$] (1T) [above=of k]{};
    \node[draw=none,fill=none] (empty) [above left= 0.25cm and 0.05cm of k]{$k+1$};

    \draw[-] (1)--(2)--(cdots)--(k);
     \draw[double,double distance=3pt,line width=0.4pt] (1T)--(k);

    \node[main, label=below:$1$] (1subL) [below= of 1]{};
    \node[main, label=below:$2$] (2sub) [right=of 1subL]{};
    \node[main, label=below:$1$] (1subR) [right=of 2sub]{};
    \node[draw=none,fill=none] (minus) [left=of 1subL]{$-$};
    
    \draw[-] (1subL)--(2sub)--(1subR);

    \node[draw, label=left:$1$] (one) [below=of k]{};
    \node[main, label=below:$4$] (four) [below=of one]{};
    \node[main, label=below:$2$] (two) [left=of four]{};
    \node[main, label=below:$5$] (five) [right=of four]{};
    \node[draw=none,fill=none] (cdots2) [right=of five]{$\cdots$};
    \node[main, label=below:$k$] (kay) [right=of cdots2]{};
    \node[main, label=left:$1$] (kayplus1) [above=of kay]{};
    \node[draw=none,fill=none] (empty) [above left=0.25cm and 0.05cm of kay]{$k+1$};

    \draw[-] (two)--(four)--(five)--(cdots2)--(kay);
    \draw[double,double distance=3pt,line width=0.4pt] (kay)--(kayplus1);
    \draw[-] (four)--(one);

    \node[draw=none,fill=none] (topghost) [right=of k]{};
    \node[draw=none,fill=none] (bottomghost) [above right=0.2cm and 0.2cm of four]{};
    
    \draw [->] (topghost) to [out=-30,in=30,looseness=1] (bottomghost);
        
    \end{tikzpicture}
    \caption{Subtraction of the $\urm(2)$ quotient quiver from the magnetic quiver \Quiver{fig:maxAk} for $\overline{max. A_k}$ to give quiver \Quiver{fig:maxAKA1Sub}.}
    \label{fig:maxAKA1Sub}
\end{figure}

Now consider the general case for $k>3$. The subtraction of the $\urm(2)$ quotient quiver from the unframed magnetic quiver for $\overline{max. A_k}$ is shown in \Figref{fig:maxAKA1Sub}. The Coulomb branch of the resulting quiver \Quiver{fig:maxAKA1Sub} is the affine Grassmannian slice $\left[\overline{\mathcal W_{D_{k-1}}}\right]^{[k+1,0,\cdots,0]}_{[0,0,\cdots,2]}$, (as can be seen by shifting the framing to the gauge node of rank 1 with $k+1$ links). This has (quaternionic) dimension $k(k+1)/2-3$ and a global symmetry $\surm(k-1)\times \urm(1)$.

Thus, we have obtained the general conjecture:
\begin{equation}
     \overline{max. A_k}///\surm(2)=\mathcal C\left(\resizebox{!}{30pt}{\begin{tikzpicture}[baseline=(current bounding box.center),main/.style = {draw, circle}]
    \node[main, label=below:$2$] (two) []{};
    \node[main, label=below:$4$] (four) [left=of two]{};
    \node[main, label=below:$5$] (five) [left=of four]{};
    \node[draw=none,fill=none] (cdots2) [left=of five]{$\cdots$};
    \node[main, label=below:$k$] (kay) [left=of cdots2]{};
    \node[draw, label=right:$k+1$] (kayplus1) [above=of kay]{};
    \node[main, label=right:$1$] (one) [above=of four]{};

    \draw[-] (two)--(four)--(five)--(cdots2)--(kay)--(kayplus1);
    \draw[-] (four)--(one);
    \end{tikzpicture}}
    \right)=\left[\overline{\mathcal W_{D_{k-1}}}\right]^{[k+1,0\cdots,0]}_{[0,0,\cdots,2]},\quad k\geq4. \label{eq:maxAKA1HKQ}
\end{equation}
We have checked \eqref{eq:maxA3A1HKQ} and \eqref{eq:maxAKA1HKQ} explicitly for $k \leq 7 $, using Weyl integration, with the embeddings \eqref{eq:AkAnEmbed}.

\subsection{A-type Orbit $\surm(3)$ HKQ}
\label{sec:AtypeSU3HKQ}
\subsubsection{$\overline{max. A_k}///\surm(3)$}
\label{sec:MaxAkA2}
For $k\geq 5$, the magnetic quivers \Quiver{fig:maxAk} for $\overline{max. A_k}$ permit subtraction of an $\urm(3)$ quotient quiver.

There is a special case for $k=5$. Upon performing the quiver subtraction and shifting the framing, as before, we obtain:
\begin{equation}
    \overline{max. A_5}///\surm(3)=\mathcal{C}\left( \resizebox{!}{30pt}{\begin{tikzpicture}[baseline=(current bounding box.center), main/.style = {draw, circle},scale=0.25]
    \node[main, label=below:$2$] (2) []{};
    \node[main, label=below:$4$] (4) [right=of 2]{};
    \node[main, label=below:$1$] (1) [right=of 4]{};
    \node[draw, label=right:$6$] (6F) [above=of 1]{};
    \node[draw=none,fill=none] (empty) [above left= 0.05cm and 0.25cm of 1]{$6$};

    \draw[-] (2)--(4);
    \draw[-] (1)--(6F);
    \draw[double,double distance=3pt,line width=0.4pt] (4)--(1);
    
    \end{tikzpicture}}\right)=\mathcal C\left(\resizebox{!}{40pt}{\begin{tikzpicture}[baseline=(current bounding box.center), main/.style = {draw, circle},scale=0.25]
    \node[main, label=below:$1$] (1)[]{};
    \node[draw,label=above:$6$] (6) [above=of 1]{};
    \node[draw=none,fill=none] (times) [above right= 0.3cm and 0.45cm of 1]{$\otimes$};
    \node[main, label=below:$2$] (2) [right=of 1]{};
    \node[main, label=below:$4$] (4) [right=of 2]{};
    \node[draw,label=above:$6$] (6F) [above=of 4]{};
    \draw[-] (1)--(6);
    \draw[-] (2)--(4)--(6F);\end{tikzpicture}}\right),
\end{equation}
where the index above the multiple edge connecting the nodes of rank 4 and 1 indicates an edge multiplicity of 6. We observe that the $\urm(3)$ quotient quiver saturates the magnetic quiver for $\overline{max. A_5}$, with the result that the latter decouples as the product of two moduli spaces. The Coulomb branch of $(1)-[6]$ is the $A_5$ Kleinian singularity $\mathbb{C}^2/\mathbb{Z}_6$ and the Coulomb branch of $(2)-(4)-[6]$ is  $\mathcal{S}^{\mathcal N}_{A_5,(2^3)}$, the Slodowy slice to the orbit of $A_5$ labelled by partition $(2^3)$. The combined moduli space is 7 (quaternionic) dimensional and has an $\surm(3) \times \urm(1)$ global symmetry. Thus, we have obtained the result that: 
\begin{equation}
    \overline{max. A_5}///\surm(3)= \mathcal{S}^{\mathcal N}_{A_5,(2^3)}\otimes \mathbb{C}^2/\mathbb{Z}_6.
\end{equation}

The next case is that for $k=6$. Performing the quiver subtraction, we find the magnetic quiver for $\left[\overline{\mathcal W_{A_4}}\right]^{[0,0,7,0]}_{[0,0,0,4]}$ whose Coulomb branch is 13 (quaternionic) dimensional and has an $\surm(4)\times \urm(1)$ global symmetry. We conclude that 
\begin{equation}
    \overline{max. A_6}///\surm(3)=\mathcal{C}\left( \resizebox{!}{30pt}{\begin{tikzpicture}[baseline=(current bounding box.center), main/.style = {draw, circle},scale=0.25]
    \node[main, label=below:$2$] (2) []{};
    \node[main, label=below:$4$] (4) [right=of 2]{};
    \node[main, label=below:$6$] (6) [right=of 4]{};
    \node[main, label=below:$1$] (1) [right=of 6]{};
    \node[draw, label=left:$7$] (7F) [above=of 6]{};
    \draw[-] (2)--(4)--(6)--(1);
    \draw[-] (7F)--(6);
    
    \end{tikzpicture}}\right)=\left[\overline{\mathcal W_{A_4}}\right]^{[0,0,7,0]}_{[0,0,0,4]}
\end{equation}
Finally, for $k\geq7$ we obtain the general result 
\begin{equation}
    \overline{max. A_k}///\surm(3)=\mathcal{C}\left( \resizebox{!}{30pt}{\begin{tikzpicture}[baseline=(current bounding box.center), main/.style = {draw, circle},scale=0.25]
    \node[main, label=below:$2$] (2) []{};
    \node[main, label=below:$4$] (4) [right=of 2]{};
    \node[main, label=below:$6$] (6) [right=of 4]{};
    \node[main, label=below:$7$] (7) [right=of 6]{};
    \node[main, label=left:$1$] (1) [above=of 6]{};
    \node[draw=none, fill=none] (cdots) [right=of 7]{$\cdots$};
    \node[main, label=below:$k$] (k) [right=of cdots]{};
    \node[draw, label=left:$k+1$] (kplus1) [above=of k]{};
    \draw[-] (2)--(4)--(6)--(7)--(cdots)--(k)--(kplus1);
    \draw[-] (1)--(6);
    
    \end{tikzpicture}}\right),\quad k\geq7,
\end{equation} 
where we have made the choice to frame the node of $k+1$, instead of the node added for rebalancing. The resulting quiver has a global symmetry of $\surm(k-2)\times \urm(1)$ on its Coulomb branch. For the cases $k=7,8,9,10$, these Coulomb branches are slices in the affine Grassmannians $\overline{\left[\mathcal W_{D_5}\right]}^{[0,0,0,8,0]}_{[0,0,0,0,4]}$, $\overline{\left[\mathcal W_{E_6}\right]}^{[0,0,0,0,9,0]}_{[0,0,0,0,0,4]}$, $\overline{\left[\mathcal W_{E_7}\right]}^{[0,0,0,0,0,10,0]}_{[0,0,0,0,0,0,4]}$, and $\overline{\left[\mathcal W_{E_8}\right]}^{[0,0,0,0,0,0,11,0]}_{[0,0,0,0,0,0,0,4]}$, respectively. 

Two things must be noted; the first is that all of the balance vectors are of the form $[0,\cdots,0,4]$ and the flavour vectors for $k$ is of the form $[0,\cdots,0,k+3,0]$. The second point to note is that the gauge nodes in the resulting quiver for $\overline{max. A_k}///\surm(3)$ forms the $E_{k-2}$ Dynkin diagram and so this particular HKQ gives an $E$-sequence.

\subsection{A-type Orbit $\surm(n)$ HKQ}
\label{sec:AtypeSUNHKQ}
\subsubsection{$\overline{max. A_{2k-1}}///\surm(k)$}
\label{sec:MaxAkSUk}
The maximal orbits $\overline{max. A_{2k-1}}$ are saturated by HKQs of type $\surm(k)$, and this gives rise to decoupling into a product space. The cases $\overline{max. A_3}///\surm(2)$ and $\overline{max. A_5}///\surm(3)$ analysed earlier are the lowest rank members of this family, which generalises to all $k \geq 2$.

We carry out the subtraction of an $\urm(k)$ quotient quiver from the magnetic quiver for the maximal orbit $\overline{max. A_{2k-1}}$ using the rules, followed by a frame shift to manifest the global symmetry. We obtain the product of two magnetic quivers:
\begin{equation}
    \overline{max. A_{2k-1}}///\surm(k)=\mathcal C\left(\resizebox{!}{40pt}{\begin{tikzpicture}[baseline=(current bounding box.center), main/.style = {draw, circle},scale=0.25]
    \node[main, label=below:$1$] (1)[]{};
    \node[draw,label=above:$2k$] (6) [above=of 1]{};
    \node[draw=none,fill=none] (times) [above right= 0.3cm and 0.45cm of 1]{$\otimes$};
    \node[main, label=below:$2$] (2) [right=of 1]{};
    \node[main, label=below:$4$] (4) [right=of 2]{};
    \node[draw=none,fill=none] (cdots) [right=of 4]{$\cdots$};
    \node[main, label=below:$2k-2$] (2mminus2) [right=of cdots]{};
    
    \node[draw,label=above:$2k$] (2M) [above=of 2mminus2]{};
    \draw[-] (1)--(6);
    \draw[-] (2)--(4)--(cdots)--(2mminus2)--(2M);\end{tikzpicture}}\right).
\end{equation}
The Coulomb branch of the first quiver is the $A_{2k-1}$ Kleinian singularity or $\mathbb{C}^2/\mathbb{Z}_{2k}$, which has the global symmetry $\urm(1)$. (It can also be identified as the Slodowy slice $\mathcal{S}^{A_{2k-1}}_{\mathcal N,(2^k)}$.) The second quiver is also a Slodowy slice to an orbit of $A_{2k-1}$, labelled this time by the partition $(2^{k})$, which has the global symmetry $A_{k-1}$. Thus we obtain the result:
\begin{equation}
    \overline{max. A_{2k-1}}///\surm(k)=\mathbb C^2/\mathbb Z_{2k}\otimes\mathcal{S}^{A_{2k-1}}_{\mathcal N,(2^k)}.
\end{equation}
The combined moduli space has quaternionic dimension $k^2-k+1$ and its global symmetry is $A_{k-1} \times \urm(1)$. The Hilbert series of this product space can be found using the formulae in \cite{Cabrera:2019izd}.

\subsubsection{Height 2 Orbits $\overline{\mathcal O}^{A_k}_{(2^p,1^{k-2p+1})}///\surm(n)$}
\label{sec:H2AType1}
We consider the two-parameter family of orbits of $A_k$ that are given by a partition of $k+1$ of the form $(2^p,1^{k-2p+1})$, for some integer $p$. These are all height 2 orbits, have the property that their HWGs are PEs of a polynomial, and take a known form derivable from the orbit partition data \cite{Hanany:2016gbz}.

The magnetic quiver \Quiver{fig:2k1lmin1Quiv} for this two-parameter family is drawn in \Figref{fig:2k1lmin1Quiv}.
\begin{figure}[h!]
    \centering
    \begin{tikzpicture}[main/.style = {draw, circle}]
    \node[main, label=below:$1$] (1L) [] {};
    \node[main, label=below:$2$] (2L) [right= of 1L]{};
    \node[draw=none,fill=none] (cdotsL) [right=of 2L]{$\cdots$};
    \node[main, label=below:$p-1$] (kminL) [right=of cdotsL]{};
    \node[main, label=below:$p$] (kL) [right=of kminL]{};
    \node[draw=none,fill=none] (cdotsM) [right=of kL]{$\cdots$};
    \node[main, label=below:$p$] (kR) [right=of cdotsM]{};
    \node[main, label=below:$p-1$] (kminR) [right=of kR]{};
    \node[draw=none,fill=none] (cdotsR) [right=of kminR]{$\cdots$};
    \node[main, label=below:$2$] (2R) [right=of cdotsR]{};
    \node[main, label=below:$1$] (1R) [right=of 2R]{};
    \node[main, label=above:$1$] (1T) [above=of cdotsM]{};
    \draw[-] (1L)--(2L)--(cdotsL)--(kminL)--(kL)--(cdotsM)--(kR)--(kminR)--(cdotsR)--(2R)--(1R);
    \draw[-] (kL)--(1T)--(kR);
     \draw [decorate, 
    decoration = {brace,
        raise=10pt,
        amplitude=5pt}] (kR) --  (kL) node[pos=0.5,below=15pt,black]{$k-2p+2$};
     \end{tikzpicture}
    \caption{Unframed magnetic quiver \Quiver{fig:2k1lmin1Quiv} for the orbit of $A_{k}$ with partition $(2^p,1^{k-2p+1})$ of $k+1$.}
    \label{fig:2k1lmin1Quiv}
    \end{figure}
All these quivers are candidates for the subtraction of an $\urm(n)$ quotient quiver – up to some maximum value for $n$, determined as follows.

There are two cases to consider: $p$ even and $p$ odd. The $\urm(n)$ quotient quiver is of length $2n-1$, which is odd. So, if $p$ is even we can subtract the quotient quiver providing $n \leq (p+2)/2$; if $p$ is odd then the condition is $n \leq (p+1)/2$. If $n$ is greater than the critical values which saturate these inequalities then the \hyperlink{rule:Junction}{Junction Rule} is violated.

In either case, if $n$ is less than these critical values, $n<\lfloor p/2\rfloor$, then quiver subtraction yields the following result:
\begin{align}
    \overline{\mathcal O}^{A_k}_{(2^p,1^{k-2p+1})}///\surm(n)&=\mathcal C\left(\resizebox{!}{27pt}{\begin{tikzpicture}[baseline=(current bounding box.center), main/.style = {draw, circle},scale=0.25]
    \node[main, label=below:$2$] (2L) []{};
    \node[main, label=below:$4$] (4L) [right=of 2L]{};
    \node[draw=none,fill=none] (cdotsL) [right=of 4L]{$\cdots$};
    \node[main, label=below:$2n-2$] (2m) [right=of cdotsL]{};
    \node[main, label=below:$2n-1$] (2mplus1) [right=of 2m]{};
    \node[draw=none,fill=none] (cdotsM) [right=of 2mplus1]{$\cdots$};
    \node[main, label=below:$p$] (pL) [right=of cdotsM]{};
    \node[draw=none,fill=none] (cdotsR) [right=of pL] {$\cdots$};
    \node[main, label=below:$p$] (pR) [right=of cdotsR]{};
    \node[main, label=below:$p-1$] (pminus1R) [right=of pR]{};
    \node[draw=none,fill=none] (cdotsRR) [right=of pminus1R]{$\cdots$};
    \node[main, label=below:$2$] (2R) [right=of cdotsRR]{};
    \node[main, label=below:$1$] (1R) [right=of 2R]{};
    \node[draw,label=left:$1$] (2F) [above=of 2m]{};
    \node[main, label=above:$1$] (1T) [above=of cdotsR]{};
    \draw[-] (2L)--(4L)--(cdotsL)--(2m)--(2mplus1)--(cdotsM)--(pL)--(cdotsR)--(pR)--(pminus1R)--(cdotsRR)--(2R)--(1R);
    \draw[-] (2F)--(2m);
    \draw[-] (pL)--(1T)--(pR);
    \draw [decorate, 
    decoration = {brace,
        raise=10pt,
        amplitude=5pt}] (pR) --  (pL) node[pos=0.5,below=15pt,black]{$k-2p+2$};
        \end{tikzpicture}}\right).        
\end{align}

This magnetic quiver has the global symmetry
$\surm(k-n+1)\times \urm(1)$; the balanced gauge nodes along the bottom form the $A_{k-n}$ Dynkin diagram, and the top gauge node of 1 is overbalanced with balance $2p-2$. We conjecture the Coulomb branch is the hyper-Kähler quotient of the orbit by $\surm(n)$.

If the inequality is saturated then we need to consider $p$ odd and $p$ even separately. In the case of $p$ odd we have:

\begin{align}
    \overline{\mathcal O}^{A_k}_{(2^p,1^{k-2p+1})}///SU\left((p+1)/2\right)&=\mathcal C\left(\resizebox{!}{33pt}{\begin{tikzpicture}[baseline=(current bounding box.center), main/.style = {draw, circle},scale=0.25]
     \node[main, label=below:$2$] (2L) []{};
    \node[main, label=below:$4$] (4L) [right=of 2L]{};
    \node[draw=none,fill=none] (cdotsL) [right=of 4L]{$\cdots$};
    \node[main, label=below:$p-1$] (pL) [right=of cdotsL]{};
    \node[main, label=below:$p$] (2qplus1L)[right=of pL]{};
    \node[draw=none,fill=none] (cdotsR) [right=of 2qplus1L] {$\cdots$};
    \node[main, label=below:$p$] (pR) [right=of cdotsR]{};
    \node[main, label=below:$p-1$] (pminus1R) [right=of pR]{};
    \node[draw=none,fill=none] (cdotsRR) [right=of pminus1R]{$\cdots$};
    \node[main, label=below:$2$] (2R) [right=of cdotsRR]{};
    \node[main, label=below:$1$] (1R) [right=of 2R]{};
    \node[main, label=above:$1$] (1T) [above=of cdotsM]{};
    \node[draw,label=left:$1$] (1FM) [above= 0.25cm of cdotsM]{};
    \draw[-] (2L)--(4L)--(cdotsL)--(pL)--(2qplus1L)--(cdotsR)--(pR)--(pminus1R)--(cdotsRR)--(2R)--(1R);
    \draw[-] (pL)--(1T)--(pR);
    \draw[-] (2qplus1L)--(1FM)--(1T);
    \draw [decorate, 
    decoration = {brace,
        raise=20pt,
        amplitude=5pt}] (pR) --  (pL) node[pos=0.5,below=25pt,black]{$k-2p+2$};
\end{tikzpicture}}\right).
\end{align}
The Coulomb branch of this quiver has an $SU\left(k-(p-1)/2\right)\times \urm(1)$ global symmetry; the balanced gauge nodes along the bottom form the $A_{k-(p+1)/2}$ Dynkin diagram and the top node of 1 is overbalanced with balance $2p-3$.

The $p$ even case is given by: 

\begin{align}
    \overline{\mathcal O}^{A_k}_{(2^p,1^{k-2p+1})}///SU\left((p+2)/2\right)&=\mathcal C\left(\resizebox{!}{30pt}{\begin{tikzpicture}[baseline=(current bounding box.center), main/.style = {draw, circle},scale=0.25]
     \node[main, label=below:$2$] (2L) []{};
    \node[main, label=below:$4$] (4L) [right=of 2L]{};
    \node[draw=none,fill=none] (cdotsL) [right=of 4L]{$\cdots$};
    \node[main, label=below:$p-2$] (pL) [right=of cdotsL]{};
    \node[main, label=below:$p-1$] (2qplus1L)[right=of pL]{};
    \node[main, label=below:$p$] (2qL)[right=of 2qplus1L]{};
    \node[draw=none,fill=none] (cdotsR) [right=of 2qL] {$\cdots$};
    \node[main, label=below:$p$] (pR) [right=of cdotsR]{};
    \node[main, label=below:$p-1$] (pminus1R) [right=of pR]{};
    \node[draw=none,fill=none] (cdotsRR) [right=of pminus1R]{$\cdots$};
    \node[main, label=below:$2$] (2R) [right=of cdotsRR]{};
    \node[main, label=below:$1$] (1R) [right=of 2R]{};
    \node[main, label=above:$1$] (1T) [above=of cdotsM]{};
    \node[draw,label=right:$1$] (1FM) [above= 0.3cm of 2qL]{};
    \draw[-] (2L)--(4L)--(cdotsL)--(pL)--(2qplus1L)--(2qL)--(cdotsR)--(pR)--(pminus1R)--(cdotsRR)--(2R)--(1R);
    \draw[-] (pL)--(1T)--(pR);
    \draw[-] (1FM)--(2qL);
     \draw[double,double distance=3pt,line width=0.4pt] (1T)--(1FM);
     \draw [decorate, 
    decoration = {brace,
        raise=20pt,
        amplitude=5pt}] (pR) --  (pL) node[pos=0.5,below=25pt,black]{$k-2p+2$};
    \end{tikzpicture}}\right)\nonumber\\&\cup\mathcal C\left(\resizebox{!}{25pt}{\begin{tikzpicture}[baseline=(current bounding box.center), main/.style = {draw, circle},scale=0.25]
     \node[main, label=below:$2$] (2L) []{};
    \node[main, label=below:$4$] (4L) [right=of 2L]{};
    \node[draw=none,fill=none] (cdotsL) [right=of 4L]{$\cdots$};
    \node[main, label=below:$p-2$] (pL) [right=of cdotsL]{};
    \node[main, label=below:$p$] (2qL)[right=of pL]{};
    \node[draw=none,fill=none] (cdotsR) [right=of 2qL] {$\cdots$};
    \node[main, label=below:$p$] (pR) [right=of cdotsR]{};
    \node[main, label=below:$p-1$] (pminus1R) [right=of pR]{};
    \node[draw=none,fill=none] (cdotsRR) [right=of pminus1R]{$\cdots$};
    \node[main, label=below:$2$] (2R) [right=of cdotsRR]{};
    \node[main, label=below:$1$] (1R) [right=of 2R]{};
    \node[draw, label=left:$2$] (2F) [above=of 2qL]{};
    \node[draw, label=right:$1$] (1F) [above=of pR]{};
    \draw[-] (2L)--(4L)--(cdotsL)--(pL)--(2qL)--(cdotsR)--(pR)--(pminus1R)--(cdotsRR)--(2R)--(1R);
    \draw[-] (2F)--(2qL);
    \draw[-] (1F)--(pR);
     \draw [decorate, 
    decoration = {brace,
        raise=20pt,
        amplitude=5pt}] (pR) --  (pL) node[pos=0.5,below=25pt,black]{$k-2p+2$};
        \end{tikzpicture}}\right)\label{eq:OrbAk22pSub2}.
\end{align}
The Coulomb branch of the first quiver in this union has a global symmetry of $\surm(k-p/2)\times \urm(1)$; the balanced gauge nodes along the bottom form the Dynkin diagram of $A_{k-(p+2)/2}$ and the top gauge node of 1 is overbalanced with balance $2p-2$. The Coulomb branch of the second quiver in the union is indeed the orbit $\overline{\mathcal O}^{A_{k-p/2-1}}_{(3^{p/2},1^{k-2p})}$. 

The limiting cases for $n=2,p=2$ were treated above in Section \ref{sec:nminAkA1}.
\subsubsection{Height 2 Orbits $\overline{\mathcal O}^{A_{2k-1}}_{(2^k)}///\surm(n)$}
\label{sec:H2AType2}
\begin{figure}[h!]
    \centering
    \begin{tikzpicture}[main/.style = {draw, circle}]
    \node[main, label=below:$1$] (1L)[]{};
    \node[main, label=below:$2$] (2L)[right=of 1L]{};
    \node[draw=none, fill=none] (cdotsL)[right=of 2L]{$\cdots$};
    \node[main, label=below:$p$] (k)[right=of cdotsL]{};
    \node[draw=none, fill=none] (cdotsR)[right=of k]{$\cdots$};
    \node[main, label=below:$2$] (2R)[right=of cdotsR]{};
    \node[main, label=below:$1$] (1R)[right=of 2R]{};
    \node[main, label=above:$1$] (1T) [above=of k]{};
    \draw[-] (1L)--(2L)--(cdotsL)--(k)--(cdotsR)--(2R)--(1R);
     \draw[double,double distance=3pt,line width=0.4pt] (1T)--(k);
    \end{tikzpicture}
    \caption{Unframed magnetic quiver \Quiver{fig:A2k} for the orbit of $A_{2p-1}$ labelled by partition $(2^p)$.}
    \label{fig:A2k}
\end{figure}
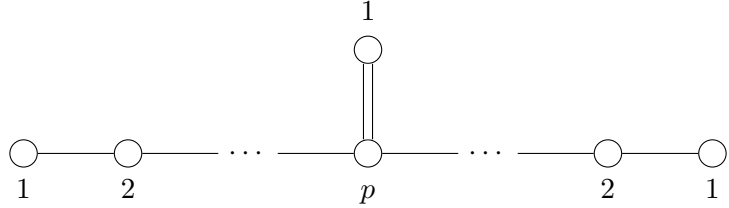

The magnetic quiver \Quiver{fig:A2k} for $\overline{\mathcal O}^{A_{2k-1}}_{(2^k)}$ is shown in \Figref{fig:A2k}.

Consider the subtraction of the $\urm(n)$ quotient quiver to give the $\surm(n)$ HKQ of the Coulomb branch. The presence of the double hyper means that the \hyperlink{rule:Junction}{Junction Rule} is violated if the $\urm(n)$ quotient quiver goes past the node of rank $p$. So from consideration of the length of the $\urm(n)$ quotient quiver, it is clear that $n$ is bounded by $n\leq (p+1)/2$, and that we can only saturate the bound when $p$ is odd.

In the case where $n$ does not saturate the bound,
\begin{align}
    \overline{\mathcal O}^{A_{2p-1}}_{(2^p)}///\surm(n)&=\mathcal C\left(\resizebox{!}{30pt}{\begin{tikzpicture}[baseline=(current bounding box.center), main/.style = {draw, circle},scale=0.25]
    \node[main, label=below:$2$] (2L) []{};
    \node[main, label=below:$4$] (4L)[right=of 2L]{};
    \node[draw=none,fill=none] (cdotsL) [right=of 4L]{$\cdots$};
    \node[main, label=below:$2n-2$] (2minus2) [right=of cdotsL]{};
    \node[main, label=below:$2n-1$] (2minus1) [right=of 2minus2]{};
    \node[draw=none,fill=none] (cdotsM) [right=of 2minus1]{$\cdots$};
    \node[main, label=below:$p$] (k) [right=of cdotsM]{};
    \node[draw=none,fill=none] (cdotsR) [right=of k]{$\cdots$};
    \node[main, label=below:$2$] (2R) [right=of cdotsR]{};
    \node[main, label=below:$1$] (1R) [right=of 2R]{};
    \node[draw, label=left:$1$] (1F) [above=of 2minus2]{};
    \node[main, label=left:$1$] (1T) [above=of k]{};
    \draw[-] (2L)--(4L)--(cdotsL)--(2minus2)--(2minus1)--(cdotsM)--(k)--(cdotsR)--(2R)--(1R);
    \draw[-] (1F)--(2minus2);
     \draw[double,double distance=3pt,line width=0.4pt] (1T)--(k);
    \end{tikzpicture}}\right),\nonumber\\
    & n<(p+1)/2\label{eq:OrbA2pSub}.
\end{align}
The Coulomb branch of the resulting quiver $\mathcal Q_{\eqref{eq:OrbA2pSub}}$, has the expected global symmetry of $\surm(2p-n)\times \urm(1)$, since the balanced gauge nodes along the bottom form the Dynkin diagram of $A_{2p-n-1}$ and the top gauge node of 1 is overbalanced with balance $2p-2$. 

Now consider the case $n=(p+1)/2$, for odd $p$, where $n$ saturates the bound.

\begin{align}
    \overline{\mathcal O}^{A_{2p-1}}_{(2^p)}///SU\left((p+1)/2\right)&=\mathcal C\left(\resizebox{!}{50pt}{\begin{tikzpicture}[baseline=(current bounding box.center), main/.style = {draw, circle},scale=0.25]
    \node[main, label=below:$2$] (2L) []{};
    \node[main, label=below:$4$] (4L)[right=of 2L]{};
    \node[draw=none,fill=none] (cdotsL) [right=of 4L]{$\cdots$};
    \node[main, label=below:$p-1$] (pm1) [right=of cdotsL]{};
    \node[main, label=below:$p-1$] (pm12) [right=of pm1]{};
    \node[main, label=below:$p-2$] (pm2) [right=of pm12]{};
    \node[draw=none,fill=none] (cdotsR) [right=of pm2]{$\cdots$};
    \node[main, label=below:$1$] (1R) [right=of cdotsR]{};
    \node[main, label=left:$1$] (1T) [above=of pm1]{};
    \node[draw, label=left:$2$] (2F) [above=of 1T]{};
    \node[draw,label=right:$1$] (1F) [above=of pm12]{};
    \draw[-] (2L)--(4L)--(cdotsL)--(pm1)--(pm12)--(pm2)--(cdotsR)--(1R);
    \draw[-] (pm12)--(1F);
    \draw[-] (1T)--(2F);
     \draw[double,double distance=3pt,line width=0.4pt] (pm1)--(1T);
    \end{tikzpicture}}\right),\nonumber\\
    &n=(p+1)/2\label{eq:OrbA2pSub2}\end{align}

The Coulomb branch of the resulting quiver $\mathcal Q_{\eqref{eq:OrbA2pSub2}}$ has global symmetry $SU\left((3p-1)/2\right)\times \urm(1)$ as the balanced gauge nodes along the bottom form the Dynkin diagram of $A_{3(p+1)/2}$ and the top gauge node of 1 is overbalanced with balance $2p-2$.

\section{B-type Orbits}
\label{sec:BtypeHKQ}
Here we consider nilpotent orbits of B-type algebras and the possible $\surm(n)$ HKQs that one can take using the quiver subtraction procedure described in Section \ref{sec:rules}.

In particular, we take the examples of magnetic quivers for the orbits $\overline{min. B_k}$ and $\overline{n. min. B_k}$ computing their $\surm(2)$ HKQs. We also study the magnetic quiver for the orbit $\overline{\mathcal O}^{B_k}_{(2^4,1^{2k-7})}$ under an $\surm(3)$ HKQ. These orbits are all height 2 and have magnetic quivers that are flavoured on a long node. Based on this analysis we provide a conjecture that extends to all valid $\surm(n)$ HKQs of such orbits. In each case we specify the conditions on the values of $k$ and $n$ which follow from the quotient quiver subtraction rules in Section \ref{sec:rules}.

The results for HKQs from quiver subtraction can be validated by comparison with explicit computations using Weyl integration. For the $\surm(2)$ HKQs, we use the embedding $B_k\hookleftarrow B_{k-2}\times D_2$, which decomposes the vector of $B_k$ as:

\begin{align}
    [1,0,0]_{B_3}&\rightarrow [2]_{B_1}+[1,1]_{D_2},
    \label{eq:B3A1Embed}\\
    [1,0,\cdots,0]_{B_{k\geq 4}}&\rightarrow [1,0,\cdots,0]_{B_{k-2}}+[1,1]_{D_2}
    \label{eq:BkA1Embed},
\end{align}
where the singlets have been suppressed. In the above branching, the embedding is symmetric in the two factors of $A_1$ inside $D_2$ and so we can perform the HKQ w.r.t either. For the $\surm(3)$ HKQ, we use the embedding of $B_k\hookleftarrow B_{k-3}\times A_2\times \urm(1)$, under which the vector of $B_{k}$ decomposes as:

\begin{align}
    [1,0\cdots,0]_{B_k}&\rightarrow [1,0,\cdots,0]_{B_{k-3}}(0)+[1,0]_{A_2}(1)+[0,1]_{A_2}(-1),
    \label{eq:BkA2Embed}
\end{align}
where the $\urm(1)$ charge is given by $()$.

\subsection{B-type Orbit $\surm(2)$ HKQ}
\label{sec:BtypeSU2HKQ}
\subsubsection{$\overline{min B_k}///\surm(2)$ for $k\geq 4$}
\label{sec:minBkA1}
\begin{figure}[h!]
    \centering
    \begin{tikzpicture}[main/.style = {draw, circle}]
    \node[main, label=below:$1$] (a) {};
    \node[main,label=below:$2$] (b) [right=of a] {};
    \node[main, label=below:$2$] (c) [right=of b]{};
    \node[] (d) [right=of c] {$\cdots$};
    \node[main,label=below:$2$] (e) [right=of d] {};
    \node[main,label=below:$1$] (f) [right=of e] {};
    \node[main, label=left:$1$] (g) [above=of b]{};
    
    \draw[-] (a)--(b)--(c)--(d)--(e)--(f);
    \draw[-] (b)--(g);
    \draw [line width=1pt, double distance=3pt,
             arrows = {-Latex[length=0pt 3 0]}] (e) -- (f);
    \draw [decorate, 
    decoration = {brace,
        raise=15pt,
        amplitude=5pt}] (e) --  (b) node[pos=0.5,below=20pt,black]{$k-2$};
    \end{tikzpicture}
    \caption{Unframed magnetic quiver \Quiver{fig:quiv_min_Bk} for $\overline{min. B_k}$.}
    \label{fig:quiv_min_Bk}
\end{figure}

\begin{figure}[h!]
    \centering
    \begin{subfigure}{0.4\textwidth}
    \centering
    \resizebox{0.75\width}{!}{\begin{tikzpicture}[main/.style = {draw, circle}]
    \node[main, label=below:$1$] (a) {};
    \node[main,label=below:$2$] (b) [right=of a] {};
    \node[main, label=below:$2$] (c) [right=of b]{};
    \node[] (d) [right=of c] {$\cdots$};
    \node[main,label=below:$2$] (e) [right=of d] {};
    \node[main,label=below:$1$] (f) [right=of e] {};
    \node[main, label=left:$1$] (g) [above=of b]{};
    
    \draw[-] (a)--(b)--(c)--(d)--(e)--(f);
    \draw[-] (b)--(g);
    \draw [line width=1pt, double distance=3pt,
             arrows = {-Latex[length=0pt 3 0]}] (e) -- (f);
    \draw [decorate, 
    decoration = {brace,
        raise=15pt,
        amplitude=5pt}] (e) --  (b) node[pos=0.5,below=20pt,black]{$k-2$};

    \node[main, label=left:$1$] (1subT) [below=of b]{};
    \node[main, label=below:$2$] (2sub) [below=of 1subT]{};
    \node[main, label=below:$1$] (1subL) [left=of 2sub]{};
    \node[draw=none,fill=none] (-) [left=of 1subL]{$-$};
    \node[draw=none,fill=none] (ghost) [right=of 2sub]{};
    \draw[-] (1subT)--(2sub)--(1subL);
    
    \node[main,label=left:$2$] (2resF) [below=of ghost]{};
    \node[] (dotsres) [right=of 2resF] {$\cdots$};
    \node[main,label=below:$2$] (eres) [right=of dotsres] {};
    \node[main,label=below:$1$] (fres) [right=of eres] {};
    \node[draw, label=left:$2$] (gres) [above=of 2resF]{};
    
    \draw[-] (gres)--(2resF)--(dotsres)--(eres);
    \draw [line width=1pt, double distance=3pt,
             arrows = {-Latex[length=0pt 3 0]}] (eres) -- (fres);
    \draw [decorate, 
    decoration = {brace,
        raise=10pt,
        amplitude=5pt}] (eres) --  (2resF) node[pos=0.5,below=15pt,black]{$k-3$};

    \node[draw=none,fill=none] (topghost) [right=of f]{};
    \node[draw=none,fill=none] (bottomghost) [right=of fres]{};
    
    \draw [->] (topghost) to [out=-30,in=30,looseness=1] (bottomghost);

   \end{tikzpicture}}
   \caption{}
    \label{fig:minBKA1Sub1}
    \end{subfigure}
\hfill
\begin{subfigure}{0.4\textwidth}
    \centering
    \resizebox{0.75\width}{!}{\begin{tikzpicture}[main/.style = {draw, circle}]
    \node[main, label=below:$1$] (a) {};
    \node[main,label=below:$2$] (b) [right=of a] {};
    \node[main, label=below:$2$] (c) [right=of b]{};
    \node[] (d) [right=of c] {$\cdots$};
    \node[main,label=below:$2$] (e) [right=of d] {};
    \node[main,label=below:$1$] (f) [right=of e] {};
    \node[main, label=left:$1$] (g) [above=of b]{};
    
    \draw[-] (a)--(b)--(c)--(d)--(e)--(f);
    \draw[-] (b)--(g);
    \draw [line width=1pt, double distance=3pt,
             arrows = {-Latex[length=0pt 3 0]}] (e) -- (f);
    \draw [decorate, 
    decoration = {brace,
        raise=15pt,
        amplitude=5pt}] (e) --  (b) node[pos=0.5,below=20pt,black]{$k-2$};

    \node[main, label=left:$1$] (1subT) [below=of b]{};
    \node[main, label=below:$2$] (2sub) [below=of 1subT]{};
    \node[main, label=below:$1$] (1subR) [right=of 2sub]{};
    \node[draw=none,fill=none] (-) [left=of 2sub]{$-$};
    \draw[-] (1subT)--(2sub)--(1subR);
    
    \node[draw,label=left:$2$] (2resF) [below left=of 2sub]{};
    \node[main, label=below:$1$] (1) [below=of 2resF] {};
    
    \draw[-] (2resF)--(1);
    
    \node[draw,label=left:$1$] (1F) [below right=of 1subR]{};
    \node[main, label=below:$2$] (2) [below =of 1F]{};
    \node[main, label=below:$1$] (1) [left=of 2]{};
    \node[draw=none,fill=none] (cdots) [right=of 2]{$\cdots$};
    \node[main,label=below:$2$] (2R) [right=of cdots]{};
    \node[main, label=below:$1$] (1R)[right=of 2R]{};
    
    \draw[-] (1F)--(2);
    \draw[-] (1)--(2)--(cdots)--(2R);
    \draw [line width=1pt, double distance=3pt,
             arrows = {-Latex[length=0pt 3 0]}] (2R) -- (1R);
    \draw [decorate, 
    decoration = {brace,
        raise=15pt,
        amplitude=5pt}] (2R) --  (1) node[pos=0.5,below=15pt,black]{$k-3$};

    \node[draw=none,fill=none] (topghost) [right=of f]{};
    \node[draw=none,fill=none] (bottomghost) [right=of 1R]{};
    
    \draw [->] (topghost) to [out=-30,in=30,looseness=1] (bottomghost);

   \end{tikzpicture}}
   \caption{}
    \label{fig:minBKA1Sub2}
\end{subfigure}
\centering
    \begin{subfigure}{\textwidth}
\centering
\begin{tikzpicture}[main/.style = {draw, circle}]
    \node[draw,label=left:$1$] (1F) []{};
    \node[main, label=below:$2$] (2) [below =of 1F]{};
    \node[main, label=below:$1$] (1) [left=of 2]{};
    \node[draw=none,fill=none] (cdots) [right=of 2]{$\cdots$};
    \node[main,label=below:$2$] (2R) [right=of cdots]{};
    \node[main, label=below:$1$] (1R)[right=of 2R]{};
    
    \draw[-] (1F)--(2);
    \draw[-] (1)--(2)--(cdots)--(2R);
    \draw [line width=1pt, double distance=3pt,
             arrows = {-Latex[length=0pt 3 0]}] (2R) -- (1R);
    \draw [decorate, 
    decoration = {brace,
        raise=15pt,
        amplitude=5pt}] (2R) --  (1) node[pos=0.5,below=15pt,black]{$k-3$};
\end{tikzpicture}
\caption{}
\label{fig:minBKA1SubInt}
    \end{subfigure}
    \caption{Both alignments of the $\urm(2)$ quotient quiver against the unframed magnetic quiver \Quiver{fig:quiv_min_Bk} for $\overline{min. B_k}$, giving the quivers \Quiver{fig:minBKA1Sub1} and \Quiver{fig:minBKA1Sub2}. Their intersection, reached via $A_1$ KP transitions, is quiver \Quiver{fig:minBKA1SubInt}.}
    \label{fig:minBkA1SubBoth}
\end{figure}

An unframed magnetic quiver \Quiver{fig:quiv_min_Bk} for the orbit $\overline{min. B_k}$ is shown in \Figref{fig:quiv_min_Bk}. Only the cases for $k\geq 4$ respect all of the selection rules in Section \ref{sec:rules}. (The case $k=3$ would require the alignment of the $\urm(2)$ quotient quiver with a non-simply laced edge in order to comply with the \hyperlink{rule:Union}{Union Rule}, and this would in turn violate the \hyperlink{rule:SingleEdge}{Single Edge Rule}.)

Considering the permissible cases for $k\geq 4$, there are two possible alignments of the $\urm(2)$ quotient quiver, as shown in \Figref{fig:minBkA1SubBoth}, and these result in the quivers \Quiver{fig:minBKA1Sub1} and \Quiver{fig:minBKA1Sub2}.
The Coulomb branches of \Quiver{fig:minBKA1Sub1} and \Quiver{fig:minBKA1Sub2} are $\overline{n.min. B_{k-2}}$ and $\overline{min. A_1}\otimes\overline{min. B_{k-2}}$, respectively. Thus we conjecture that:
\begin{equation}
    \overline{min. B_k}///\surm(2)=\overline{n.min B_{k-2}}\cup \left(\overline{min. A_1}\otimes \overline{min. B_{k-2}}\right), k\geq 4.
\end{equation}
\Quiver{fig:minBKA1Sub1} and \Quiver{fig:minBKA1Sub2} are both related to their intersection \Quiver{fig:minBKA1SubInt}, which is a magnetic quiver for $\overline{min. B_{k-2}}$, by $A_1$ Kraft-Procesi transitions.

The HWG corresponding to the Hilbert Series of this union follows from the HWGs for its components and intersection, according to the unions of cones formula \ref{eq:UnionsofCones}:
\begin{align}
    HWG\left[\overline{n.min B_{k-2}}\cup \left(\overline{min. A_1}\otimes \overline{min. B_{k-2}}\right)\right]&=PE[\mu_2t^2+\mu_1^2t^4]+PE[\nu^2 t^2]PE[\mu_2t^2]-PE[\mu_2 t^2]\nonumber\\&=PE\left[\left(\mu_2+\nu^2\right)t^2+\mu_1^2t^4-\mu_1^2\nu^2t^6\right],
\end{align}
where $\mu_{1,2}$ are highest weight fugacities for $B_{k-2}$ and $\nu$ is a highest weight fugacity for $A_1$. The first line gives the signed sum of the HWGs for $\overline{n. min. B_{k-2}}$, $\overline{min. A_1}\otimes\overline{min. B_{k-2}}$, and $\overline{min. B_{k-2}}$ respectively. As we see, this form of the HWG is easily reached using the diagrammatic technique of quotient quiver subtraction.  Conveniently, the component HWGs all share a common $PE[\mu_2t^2]$ term and this allows the recombination into a single PE, as in the second line.

We have checked the agreement between quiver subtraction and Weyl integration for $k \leq 8$ (using the embedding in \eqref{eq:BkA1Embed}).

\subsubsection{$\overline{n. min. B_k}///\surm(2)$ for $k\geq 3$}
\label{sec:nminBkA1}
\begin{figure}[h!]
    \centering
    \begin{subfigure}{0.4\textwidth}
    \centering
    \begin{tikzpicture}[main/.style={draw,circle}]
    \node[main, label=below:$2$] (2L){};
    \node[draw=none,fill=none] (cdots) [right=of 2L]{$\cdots$};
    \node[main, label=below:$2$] (2R) [right=of cdots]{};
    \node[main, label=below:$1$] (1) [right=of 2R]{};
    \node[main, label=left:$1$] (2F) [above=of 2L]{};

    \draw[-] (2L)--(cdots)--(2R);
     \draw [line width=1pt, double distance=3pt,
             arrows = {-Latex[length=0pt 3 0]}] (2R) -- (1);
    \draw [decorate, 
    decoration = {brace,
        raise=15pt,
        amplitude=5pt}] (1) --  (2L) node[pos=0.5,below=20pt,black]{$k$};
    \draw[double,double distance=3pt,line width=0.4pt] (2F)--(2L);
   \end{tikzpicture}
   \caption{}
    \label{fig:minBKUsual}
    \end{subfigure}
\hfill
\begin{subfigure}{0.4\textwidth}
    \centering
    \begin{tikzpicture}[main/.style={draw,circle}]
    \node[main, label=below:$2$] (2L){};
    \node[main, label=below:$1$] (1L) [left=of 2L]{};
    \node[draw=none,fill=none] (cdots) [right=of 2L]{$\cdots$};
    \node[main, label=below:$2$] (2R) [right=of cdots]{};
    \node[main, label=left:$1$] (1T) [above=of 2L]{};

    \draw[-] (1L)--(2L)--(cdots)--(2R);
    \draw[-] (1T)--(2L);
    \draw (2R) to [out=135, in=45,looseness=8] (2R);
    \draw [decorate, 
    decoration = {brace,
        raise=15pt,
        amplitude=5pt}] (2R) --  (1L) node[pos=0.5,below=20pt,black]{$k$};
    
   \end{tikzpicture}
   \caption{}
   \label{fig:minBkAdj}
    \end{subfigure}
    \caption{Alternative unframed magnetic quivers \Quiver{fig:minBKUsual} and \Quiver{fig:minBkAdj} for the $\overline{n.min. B_k
}$ for $k\geq 2$. Only \Quiver{fig:minBkAdj} is amenable to the $\surm(n)$ HKQ quiver subtraction rules.}
    \label{fig:twominBk}
\end{figure}
\begin{figure}[h!]
    \centering
    \begin{subfigure}{0.45\textwidth}
    \centering
    \begin{tikzpicture}[main/.style = {draw, circle}]
     \node[main, label=below:$1$] (a) {};
    \node[main,label=below:$2$] (b) [right=of a] {};
    \node[main,label=below:$2$] (f) [right=of b] {};
    \node[main, label=left:$1$] (g) [above=of b]{};
    
    \draw[-] (a)--(b)--(f);
    \draw[-] (b)--(g);
    \draw (f) to [out=135, in=45,looseness=8] (f);

    \node[main, label=left:$1$] (1subT) [below=of b]{};
    \node[main, label=below:$2$] (2sub) [below=of 1subT]{};
    \node[main, label=below:$1$] (1subL) [left=of 2sub]{};
    \node[draw=none,fill=none] (-) [left=of 1subL]{$-$};
    \node[draw=none,fill=none] (ghost) [right=of 2sub]{};
    \draw[-] (1subT)--(2sub)--(1subL);
    
    \node[draw,label=left:$2$] (2resF) [below=of ghost]{};
    \node[main, label=left:$2$] (2res) [below=of 2resF]{};

    \draw[-] (2resF)--(2res);
    \draw (2res) to [out=-45, in=225,looseness=8] (2res);

    \node[draw=none,fill=none] (topghost) [right=of f]{};
    \node[draw=none,fill=none] (bottomghost) [right=of 2res]{};
    
    \draw [->] (topghost) to [out=-30,in=30,looseness=0.6] (bottomghost);

   \end{tikzpicture}
   \caption{}
    \label{fig:nminB3A1Sub1}
\end{subfigure}
\hfill
\begin{subfigure}{0.45\textwidth}
    \centering
    \begin{tikzpicture}[main/.style = {draw, circle}]
    \node[main, label=below:$1$] (a) {};
    \node[main,label=below:$2$] (b) [right=of a] {};
    \node[main,label=below:$2$] (f) [right=of b] {};
    \node[main, label=left:$1$] (g) [above=of b]{};
    
    \draw[-] (a)--(b)--(f);
    \draw[-] (b)--(g);
    \draw (f) to [out=135, in=45,looseness=8] (f);

    \node[main, label=left:$1$] (1subT) [below=of b]{};
    \node[main, label=below:$2$] (2sub) [below=of 1subT]{};
    \node[main, label=below:$1$] (1subR) [right=of 2sub]{};
    \node[draw=none,fill=none] (-) [left=of 2sub]{$-$};
    \draw[-] (1subT)--(2sub)--(1subR);
    
    \node[draw,label=left:$2$] (2resF) [below =of 1subR]{};
    \node[main, label=left:$1$] (1) [below=of 2resF] {};
    
    \draw[-] (2resF)--(1);
    \draw (1) to [out=225, in=-45,looseness=8] (1);

    \node[draw=none,fill=none] (ghost) [left=of 2resF]{};

    \node[draw,label=left:$2$] (2resFtwo) [left =of ghost]{};
    \node[main, label=below:$1$] (1two) [below=of 2resFtwo] {};
    
    \draw[-] (2resFtwo)--(1two);

    \node[draw=none,fill=none] (topghost) [right=of f]{};
    \node[draw=none,fill=none] (bottomghost) [right=of 1]{};
    
    \draw [->] (topghost) to [out=-30,in=30,looseness=1] (bottomghost);

   \end{tikzpicture}
   \caption{}
    \label{fig:nminB3A1Sub2}
\end{subfigure}
\centering
\begin{subfigure}{\textwidth}
\centering
\begin{tikzpicture}[main/.style = {draw, circle}]
    \node[draw,label=left:$2$] (2resF) []{};
    \node[main, label=left:$1$] (1) [below=of 2resF] {};    
    \draw[-] (2resF)--(1);
    \draw (1) to [out=225, in=-45,looseness=8] (1);

\end{tikzpicture}
    \caption{}
    \label{fig:nminB3A1SubInt}
\end{subfigure}

    \caption{Both alignments of the $\urm(2)$ quotient quiver against the unframed magnetic quiver \Quiver{fig:minBkAdj} for $\overline{n. min. B_3}$, giving the quivers \Quiver{fig:nminB3A1Sub1} and \Quiver{fig:nminB3A1Sub2}. Their intersection, reached via $A_1$ KP transitions, is quiver \Quiver{fig:nminB3A1SubInt}.}
    \label{fig:nminB3A1SubBoth}
\end{figure} 
\begin{figure}[h!]
    \centering
    \begin{subfigure}{0.45\textwidth}
    \centering\resizebox{0.75\width}{!}{
    \begin{tikzpicture}[main/.style = {draw, circle}]
    \node[main, label=below:$1$] (a) {};
    \node[main,label=below:$2$] (b) [right=of a] {};
    \node[main, label=below:$2$] (c) [right=of b]{};
    \node[] (d) [right=of c] {$\cdots$};
    \node[main,label=below:$2$] (e) [right=of d] {};
    \node[main,label=below:$2$] (f) [right=of e] {};
    \node[main, label=left:$1$] (g) [above=of b]{};
    
    \draw[-] (a)--(b)--(c)--(d)--(e)--(f);
    \draw[-] (b)--(g);
    \draw (f) to [out=135, in=45,looseness=8] (f);
    \draw [decorate, 
    decoration = {brace,
        raise=20pt,
        amplitude=5pt}] (e) --  (b) node[pos=0.5,below=25pt,black]{$k-2$};

    \node[main, label=left:$1$] (1subT) [below=of b]{};
    \node[main, label=below:$2$] (2sub) [below=of 1subT]{};
    \node[main, label=below:$1$] (1subL) [left=of 2sub]{};
    \node[draw=none,fill=none] (-) [left=of 1subL]{$-$};
    \node[draw=none,fill=none] (ghost) [right=of 2sub]{};
    \draw[-] (1subT)--(2sub)--(1subL);
    
    \node[main,label=left:$2$] (2resF) [below=of ghost]{};
    \node[] (dotsres) [right=of 2resF] {$\cdots$};
    \node[main,label=below:$2$] (eres) [right=of dotsres] {};
    \node[main,label=below:$2$] (fres) [right=of eres] {};
    \node[draw, label=left:$2$] (gres) [above=of 2resF]{};
    
    \draw[-] (gres)--(2resF)--(dotsres)--(eres)--(fres);
    \draw [decorate, 
    decoration = {brace,
        raise=10pt,
        amplitude=5pt}] (eres) --  (2resF) node[pos=0.5,below=15pt,black]{$k-3$};
    \draw (fres) to [out=135, in=45,looseness=8] (fres);

    \node[draw=none,fill=none] (topghost) [right=of f]{};
    \node[draw=none,fill=none] (bottomghost) [right=of fres]{};
    
    \draw [->] (topghost) to [out=-30,in=30,looseness=0.6] (bottomghost);

   \end{tikzpicture}}
   \caption{}
    \label{fig:nminBKA1Sub1}
\end{subfigure}
\hfill
\begin{subfigure}{0.45\textwidth}
    \centering\resizebox{0.75\width}{!}{
    \begin{tikzpicture}[main/.style = {draw, circle}]
    \node[main, label=below:$1$] (a) {};
    \node[main,label=below:$2$] (b) [right=of a] {};
    \node[main, label=below:$2$] (c) [right=of b]{};
    \node[] (d) [right=of c] {$\cdots$};
    \node[main,label=below:$2$] (e) [right=of d] {};
    \node[main,label=below:$2$] (f) [right=of e] {};
    \node[main, label=left:$1$] (g) [above=of b]{};
    
    \draw[-] (a)--(b)--(c)--(d)--(e)--(f);
    \draw[-] (b)--(g);
    \draw [decorate, 
    decoration = {brace,
        raise=20pt,
        amplitude=5pt}] (e) --  (b) node[pos=0.5,below=25pt,black]{$k-2$};
    \draw (f) to [out=135, in=45,looseness=8] (f);

    \node[main, label=left:$1$] (1subT) [below=of b]{};
    \node[main, label=below:$2$] (2sub) [below=of 1subT]{};
    \node[main, label=below:$1$] (1subR) [right=of 2sub]{};
    \node[draw=none,fill=none] (-) [left=of 2sub]{$-$};
    \draw[-] (1subT)--(2sub)--(1subR);
    
    \node[draw,label=left:$2$] (2resF) [below left=of 2sub]{};
    \node[main, label=below:$1$] (1) [below=of 2resF] {};
    
    \draw[-] (2resF)--(1);
    
    \node[draw,label=left:$1$] (1F) [below right=of 1subR]{};
    \node[main, label=below:$2$] (2) [below =of 1F]{};
    \node[main, label=below:$1$] (1) [left=of 2]{};
    \node[draw=none,fill=none] (cdots) [right=of 2]{$\cdots$};
    \node[main,label=below:$2$] (2R) [right=of cdots]{};
    \node[main, label=below:$2$] (1R)[right=of 2R]{};
    
    \draw[-] (1F)--(2);
    \draw[-] (1)--(2)--(cdots)--(2R)--(1R);
    \draw [decorate, 
    decoration = {brace,
        raise=20pt,
        amplitude=5pt}] (2R) --  (1) node[pos=0.5,below=25pt,black]{$k-3$};
    \draw (1R) to [out=135, in=45,looseness=8] (1R);

    \node[draw=none,fill=none] (topghost) [right=of f]{};
    \node[draw=none,fill=none] (bottomghost) [right=of 1R]{};
    
    \draw [->] (topghost) to [out=-30,in=30,looseness=1] (bottomghost);

   \end{tikzpicture}}
   \caption{}
    \label{fig:nminBKA1Sub2}
\end{subfigure}
\centering
\begin{subfigure}{\textwidth}
    \centering
    \begin{tikzpicture}[main/.style = {draw, circle}]
    \node[draw,label=left:$1$] (1F) []{};
    \node[main, label=below:$2$] (2) [below =of 1F]{};
    \node[main, label=below:$1$] (1) [left=of 2]{};
    \node[draw=none,fill=none] (cdots) [right=of 2]{$\cdots$};
    \node[main,label=below:$2$] (2R) [right=of cdots]{};
    \node[main, label=below:$2$] (1R)[right=of 2R]{};
    
    \draw[-] (1F)--(2);
    \draw[-] (1)--(2)--(cdots)--(2R)--(1R);
    \draw [decorate, 
    decoration = {brace,
        raise=20pt,
        amplitude=5pt}] (2R) --  (1) node[pos=0.5,below=25pt,black]{$k-3$};
    \draw (1R) to [out=135, in=45,looseness=8] (1R);
    \end{tikzpicture}
    \caption{}
    \label{fig:nminBkA1SubInt}
\end{subfigure}

    \caption{Both alignments of the $\urm(2)$ quotient quiver against the unframed magnetic quiver \Quiver{fig:minBkAdj} for $\overline{n. min. B_{k\geq 4}}$, giving the quivers \Quiver{fig:nminBKA1Sub1} and \Quiver{fig:nminBKA1Sub2}. Their intersection, reached via $A_1$ KP transitions, is quiver \Quiver{fig:nminBkA1SubInt}.}
    \label{fig:nminBkA1SubBoth}
\end{figure}

A typical unframed magnetic quiver \Quiver{fig:minBKUsual} for $\overline{n.min  B_k}$ is drawn in \Figref{fig:minBKUsual}. However, the presence of the double link means that this violates the \hyperlink{rule:ExternalLeg}{External Leg Rule} for $\surm(2)$ quiver subtraction. Instead, we consider an alternative magnetic quiver \Quiver{fig:minBkAdj} for $\overline{n.min. B_k}$, as shown in \Figref{fig:minBkAdj}; this does permit the subtraction of the $\urm(2)$ quotient quiver while respecting all of the selection rules in Section \ref{sec:rules} for $k\geq3$. Note that the case for $k=2$ violates the \hyperlink{rule:Adjoint}{Adjoint Hypers Rule}.

First, there is a special case for $k=3$. In this case there are two alignments, with the quiver subtractions as shown in \Figref{fig:nminB3A1SubBoth}, producing quivers \Quiver{fig:nminB3A1Sub1} and \Quiver{fig:nminB3A1Sub2}. We identify the Coulomb branches of these as $Sym^2\left(\mathbb C^2/\mathbb Z_2\right)\simeq Sym^2\left(\overline{min. A_1}\right)$ \cite{Bourget:2022tmw} and $\overline{min. A_1}\otimes\overline{min. A_1}$, since the adjoint hypermultiplet on a $\urm(1)$ gauge node makes a trivial contribution to the monopole formula. We find:
\begin{equation}
    \overline{n.min. B_3}///\surm(2)=Sym^2\left(\overline{min. A_1}\right)\cup \left(\overline{min. A_1}\otimes\overline{min. A_1}\right).
\end{equation}

To compute the union, we note that there is an $A_1$ Kraft-Procesi transition from each quiver to their intersection \Quiver{fig:nminB3A1SubInt} which is a magnetic quiver for $\overline{min. A_1}$. The HWG for the components can be computed, and using the union of cones formula \eqref{eq:UnionsofCones} one finds that:
\begin{align}
    HWG\left[Sym^2\left(\overline{min. A_1}\right)\cup \left(\overline{min. A_1}\otimes\overline{min. A_1}\right)\right]&=PE[\mu^2 t^2 + (1 + \mu^4) t^4 + \mu^4 t^6 - \mu^8 t^{12}]\nonumber\\&+PE[\mu^2t^2]PE[\nu^2t^2]-PE[\mu^2t^2],\label{eq:nminB3SU2HWG}
\end{align}where $\mu$ is a highest weight fugacity for the common $A_1$ shared by the Coulomb branches of \Quiver{fig:nminB3A1Sub1} and \Quiver{fig:nminB3A1Sub2}, and $\nu$ is the highest weight fugacity for the $A_1$ in the Coulomb branch of \Quiver{fig:nminB3A1Sub2} only. We have checked this explicitly using the embedding \eqref{eq:B3A1Embed} and Weyl integration.

Now we consider the more general case for $k\geq 4$. As shown in \Figref{fig:nminBkA1SubBoth}, there are two possible alignments of the $\urm(2)$ quotient quiver, which produce quivers \Quiver{fig:nminBKA1Sub1} and \Quiver{fig:nminBKA1Sub2}. The Coulomb branches of these are identified as $\overline{n.min. D_{k-2}}/\mathbb Z_2$ and $\overline{min. A_1}\otimes\overline{n.min B_{k-2}}$, respectively. The HKQ is the union of the Coulomb branches of these quivers. We conjecture:
\begin{equation}
    \overline{n.min. B_k}///\surm(2)=\overline{n. min. D_{k-2}}/\mathbb Z_2\cup \left(\overline{min. A_1}\otimes \overline{n. min. B_{k-2}}\right),\quad k\geq 4.
\end{equation}
 We can perform an $A_1$ KP transition on each quiver to give the intersection \Quiver{fig:nminBkA1SubInt} which is a magnetic quiver for $\overline{n. min. B_{k-2}}$. The HWG for the union can be found as the signed sum \eqref{eq:UnionsofCones} of the component HWGs:
\begin{align}
&HWG[\overline{n. min. D_{k-2}}/\mathbb Z_2\cup \left(\overline{min. A_1}\otimes \overline{n. min. B_{k-2}}\right)]\nonumber\\&=PE\left[\mu_2t^2+\left(2\mu_1^2+1\right)t^4+\mu_1^2t^6-\mu_1^4t^{12}\right]+PE[\nu^2t^2]PE\left[\mu_2t^2+\mu_1^2t^4\right]-PE\left[\mu_2t^2+\mu_1^2t^4\right]
\end{align} where $\mu_{1,2}$ are highest weight fugacities for $B_k$ and $\nu$ is a highest weight fugacity for $A_1$. The component HWGs are those for $\overline{n.min. D_{k-2}}/\mathbb{Z}_2$, $\overline{min. A_1}\otimes\overline{n.min. B_{k-2}}$, and $\overline{n.min. B_{k-2}}$, respectively. We have checked this explicitly using the embedding \eqref{eq:BkA1Embed} and Weyl integration.

\FloatBarrier
\subsection{B-type Orbit $\surm(3)$ HKQ}
\label{sec:BtypeSU3HKQ}
\subsubsection{$\overline{\mathcal O}^{B_k}_{(2^4,1^{2k-7})}///\surm(3)$}
\label{sec:BtypeA2}
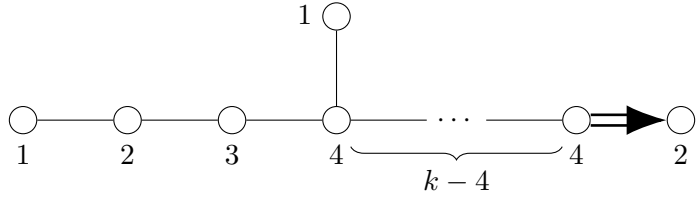
\begin{figure}[h!]
    \centering
    \begin{tikzpicture}[baseline=(current bounding box.center), main/.style = {draw, circle},scale=0.25]
    \node[main, label=below:$1$] (1L) []{};
    \node[main, label=below:$2$] (2L) [right=of 1L]{};
    \node[main, label=below:$3$] (3L) [right=of 2L]{};
    \node[main, label=below:$4$] (4L) [right=of 3L]{};
    \node[draw=none,fill=none] (cdots) [right=of 4L]{$\cdots$};
    \node[main, label=below:$4$] (4R) [right=of cdots]{};
    \node[main, label=below:$2$] (2R) [right=of 4R]{};
    \node[main, label=left:$1$] (1T) [above=of 4L]{};

    \draw[-] (1L)--(2L)--(3L)--(4L)--(cdots)--(4R);
    \draw[-] (1T)--(4L);
     \draw [line width=1pt, double distance=3pt,
             arrows = {-Latex[length=0pt 3 0]}] (4R) -- (2R);
    \draw [decorate, 
    decoration = {brace,
        raise=10pt,
        amplitude=5pt}] (4R) --  (4L) node[pos=0.5,below=15pt,black]{$k-4$};
    \end{tikzpicture}
    \caption{Unframed magnetic quiver \Quiver{fig:Bk24} for the $\overline{\mathcal O}^{B_k}_{(2^4,1^{2k-7})}$.}
    \label{fig:Bk24}
\end{figure}
\begin{figure}[h!]
    \centering
    \begin{subfigure}{0.4\textwidth}
    \centering
    \resizebox{0.75\width}{!}{\begin{tikzpicture}[baseline=(current bounding box.center), main/.style = {draw, circle}]
    \node[main, label=below:$1$] (1L) []{};
    \node[main, label=below:$2$] (2L) [right=of 1L]{};
    \node[main, label=below:$3$] (3L) [right=of 2L]{};
    \node[main, label=below:$4$] (4L) [right=of 3L]{};
    \node[main, label=below:$4$] (4R) [right=of 4L]{};
    \node[main, label=below:$2$] (2R) [right=of 4R]{};
    \node[main, label=left:$1$] (1T) [above=of 4L]{};

    \draw[-] (1L)--(2L)--(3L)--(4L)--(4R);
    \draw[-] (1T)--(4L);
     \draw [line width=1pt, double distance=3pt,
             arrows = {-Latex[length=0pt 3 0]}] (4R) -- (2R);
    
    \node[main, label=left:$1$] (1subT) [below=of 4L]{};
    \node[main, label=below:$2$] (2sub) [below=of 1subT]{};
    \node[main, label=below:$3$] (3sub) [left=of 2sub]{};
    \node[main, label=below:$2$] (2subL) [left=of 3sub]{};
    \node[main, label=below:$1$] (1subL) [left=of 2subL]{};
    \node[draw=none,fill=none] (-) [left=of 1subL]{$-$};
    
    \draw[-] (1subT)--(2sub)--(3sub)--(2subL)--(1subL);

     \node[main, label=below:$2$] (2resL) [below=of 2sub]{};
    \node[main, label=below:$4$] (4res) [right=of 2resL]{};
    \node[main, label=below:$2$] (2resR) [right=of 4res]{};
    \node[draw, label=left:$2$] (2resT) [above=of 4res]{};
    \draw[-] (2resL)--(4res)--(2resT);
    \draw [line width=1pt, double distance=3pt,
             arrows = {-Latex[length=0pt 3 0]}] (4res) -- (2resR);
             
    \node[draw=none,fill=none] (topghost) [right=of 2R]{};
    \node[draw=none,fill=none] (bottomghost) [right=of 2resR]{};
    
    \draw [->] (topghost) to [out=-30,in=30,looseness=1] (bottomghost);
   \end{tikzpicture}}
   \caption{}
    \label{fig:orbB6A2Sub1}
    \end{subfigure}
\hfill
\begin{subfigure}{0.4\textwidth}
    \centering
    \resizebox{0.75\width}{!}{\begin{tikzpicture}[baseline=(current bounding box.center), main/.style = {draw, circle}]
    \node[main, label=below:$1$] (1L) []{};
    \node[main, label=below:$2$] (2L) [right=of 1L]{};
    \node[main, label=below:$3$] (3L) [right=of 2L]{};
    \node[main, label=below:$4$] (4L) [right=of 3L]{};
    \node[main, label=below:$4$] (4R) [right=of 4L]{};
    \node[main, label=below:$2$] (2R) [right=of 4R]{};
    \node[main, label=left:$1$] (1T) [above=of 4L]{};

    \draw[-] (1L)--(2L)--(3L)--(4L)--(4R);
    \draw[-] (1T)--(4L);
     \draw [line width=1pt, double distance=3pt,
             arrows = {-Latex[length=0pt 3 0]}] (4R) -- (2R);
    
    \node[main, label=below:$1$] (1subT) [below=of 4R]{};
    \node[main, label=below:$2$] (2sub) [left=of 1subT]{};
    \node[main, label=below:$3$] (3sub) [left=of 2sub]{};
    \node[main, label=below:$2$] (2subL) [left=of 3sub]{};
    \node[main, label=below:$1$] (1subL) [left=of 2subL]{};
    \node[draw=none,fill=none] (-) [left=of 1subL]{$-$};
    
    \draw[-] (1subT)--(2sub)--(3sub)--(2subL)--(1subL);

    \node[draw,label=left:$2$] (2resT) [below=of 3sub]{};
    \node[main, label=below:$1$] (1resL) [below=of 2resT]{};
     \node[main, label=below:$2$] (2resL) [right=of 1resL]{};
     \node[main, label=below:$3$] (3res) [right=of 2resL]{};
     \node[main, label=below:$2$] (2resR) [right=of 3res]{};
     \node[draw, label=right:$1$] (1resT) [above=of 2resR]{};

     \draw[-] (2resT)--(1resL)--(2resL)--(3res);
     \draw[-] (1resT)--(2resR);
     \draw [line width=1pt, double distance=3pt,
             arrows = {-Latex[length=0pt 3 0]}] (3res) -- (2resR);
             
    \node[draw=none,fill=none] (topghost) [right=of 2R]{};
    \node[draw=none,fill=none] (bottomghost) [right=of 2resR]{};
    
    \draw [->] (topghost) to [out=-30,in=30,looseness=1] (bottomghost);

   \end{tikzpicture}}
   \caption{}
    \label{fig:orbB6A2Sub2}
\end{subfigure}
\centering
\begin{subfigure}{\textwidth}
\centering
\begin{tikzpicture}[main/.style={draw,circle}]

     \node[main, label=below:$2$] (2resL) [below=of 2sub]{};
    \node[main, label=below:$3$] (4res) [right=of 2resL]{};
    \node[main, label=below:$2$] (2resR) [right=of 4res]{};
    \node[draw, label=left:$1$] (1FL) [above=of 2resL]{};
    \node[draw, label=right:$1$] (1FR) [above=of 2resR]{};
    \draw[-] (2resL)--(4res);
    \draw[-] (1FL)--(2resL);
    \draw[-] (1FR) --(2resR);
   \draw [line width=1pt, double distance=3pt,
             arrows = {-Latex[length=0pt 3 0]}] (4res) -- (2resR);
\end{tikzpicture}
    \caption{}
    \label{fig:orbB6A2SubInt}
\end{subfigure}

    \caption{Both alignments of the $\urm(3)$ quotient quiver against \Quiver{fig:Bk24} for $k=6$ giving the quivers \Quiver{fig:orbB6A2Sub1} and \Quiver{fig:orbB6A2Sub2}. Their intersection, reached via $A_1$ KP transitions, is quiver \Quiver{fig:orbB6A2SubInt}.}
    \label{fig:orbB6A2SubBoth}
\end{figure}

The magnetic quiver \Quiver{fig:Bk24} for the orbit $\overline{\mathcal O}^{B_k}_{(2^4,1^{2k-7})}$ is shown in \Figref{fig:Bk24}. This quiver only respects the rules in Section \ref{sec:rules} for $\surm(3)$ quiver subtraction when $k\geq 6$.

The case for $k=6$ is special. There are two alignments, shown in \Figref{fig:orbB6A2SubBoth}, which produce quivers \Quiver{fig:orbB6A2Sub1} and \Quiver{fig:orbB6A2Sub2}. We identify the Coulomb branches of these quivers as $\overline{\left[\mathcal W_{B_3}\right]}^{[0,2,0]}_{[0,0,0]}$ and $\overline{\left[\mathcal W_{B_4}\right]}^{[2,0,0,1]}_{[2,0,0,0]}$, which have global symmetries of $SO(7)$ and $SO(7)\times \urm(1)$ respectively. We conclude:
\begin{equation}
    \overline{\mathcal O}^{B_6}_{(2^4,1^{5})}///\surm(3)=\overline{\left[\mathcal W_{B_3}\right]}^{[0,2,0]}_{[0,0,0]}\cup \overline{\left[\mathcal W_{B_4}\right]}^{[2,0,0,1]}_{[2,0,0,0]}.
    \label{eq:OrbB6A2Sub}
\end{equation}
Once again there is an $A_1$ KP transition from each of the two quivers to their intersection, \Quiver{fig:orbB6A2SubInt} which is a magnetic quiver for $\overline{\left[\mathcal W_{B_3}\right]}^{[1,0,1]}_{[0,0,0]}$, the double cover of the orbit $\overline{\mathcal O}^{B_3}_{(3^2,1)}$ \cite{2011arXiv1108.4999A}. The HWG of the union in \eqref{eq:OrbB6A2Sub} can be computed as:
\begin{align}
&HWG\left(\overline{\left[\mathcal W_{B_3}\right]}^{[0,2,0]}_{[0,0,0]}\cup \overline{\left[\mathcal W_{B_4}\right]}^{[2,0,0,1]}_{[2,0,0,0]}\right)\\&=\left(\begin{aligned}1 &+ \mu_2 t^6 + \mu_1 \mu_3^2 t^8 + \mu_1 \mu_2 \mu_3^2 t^{10} - \mu_1 \mu_2 \mu_3^2 t^{12} \\&-\mu_1 \mu_2^2 \mu_3^2 t^{14} - \mu_1^2 \mu_2 \mu_3^4 t^{16} - 
     \mu_1^2 \mu_2^2 \mu_3^4 t^{22}\end{aligned}\right)PE\left[\mu_2t^2+\left(1+\mu_1^2+\mu_2+\mu_3^2\right)+2\mu_1\mu_3^2t^6+
    \mu_2^2t^8\right]\nonumber\\
   &+\left(\begin{aligned}1 &+ \mu_1 \mu_3^2 t^6 - \mu_1^2 t^8 - 2 \mu_1 \mu_3^2 t^{10} - (
     \mu_1^2 \mu_3^2 t^{10})/q - \mu_1^2 \mu_3^2 q t^{10} - \mu_3^4 t^{12}\\& - (
     \mu_1 \mu_3^4 t^{12})/q - \mu_1 \mu_3^4 q t^{12} + \mu_1^3 \mu_3^2 t^{14} + (
     \mu_1^2 \mu_3^2 t^{14})/q + \mu_1^2 \mu_3^2 q t^{14}\\& + 2 \mu_1^2 \mu_3^4 t^{16} + (
     \mu_1 \mu_3^4 t^{16})/q + \mu_1 \mu_3^4 q t^{16} + \mu_1 \mu_3^6 t^{18} - 
     \mu_1^2 \mu_3^4 t^{20} - 
     \mu_1^3 \mu_3^6 t^26\end{aligned}\right)\nonumber\\&\times PE\left[\left(1+\mu_2\right)t^2+\left(\mu_1^2+\mu_1 q+\mu_1/q+\mu_2+\mu_3^2\right)t^4+\left(\mu_3^2q+\mu_3^2/q+\mu_1\mu_3^2\right)t^6\right] \nonumber\\
   &-PE\left[\mu_2t^2+\left(\mu_1^2+\mu_2+\mu_3^2\right)t^4+2\mu_1\mu_3^2t^6-\mu_1^2\mu_3^4t^{12}\right],
   \label{eq:OrbB6A2Union}
\end{align}
where the constituent HWG in the signed sum \eqref{eq:OrbB6A2Union} are for; $\overline{\left[\mathcal W_{B_3}\right]}^{[0,2,0]}_{[0,0,0]}$, $\overline{\left[\mathcal W_{B_4}\right]}^{[2,0,0,1]}_{[2,0,0,0]}$, and their intersection $\overline{\left[\mathcal W_{B_3}\right]}^{[1,0,1]}_{[0,0,0]}$ respectively. The HS for the HKQ in \eqref{eq:OrbB6A2Sub} can be computed using the embedding \eqref{eq:BkA2Embed} and is matching with the HS of the union in \eqref{eq:OrbB6A2Sub} thereby verifying \eqref{eq:OrbB6A2Sub}, however the HWG of the HKQ in \eqref{eq:OrbB6A2Sub} is difficult to compute and the expression as a signed sum \eqref{eq:OrbB6A2Union} is conjectured to be the correct answer. The advantage of the method of $\urm(n)$ quotient quiver subtraction to aid in computations is clear, as the computation of the HWG for the HKQ \eqref{eq:OrbB6A2Sub} has been broken down into smaller computations.
\begin{figure}[h!]
    \centering
    \begin{subfigure}{0.4\textwidth}
    \centering
    \resizebox{0.6\width}{!}{\begin{tikzpicture}[baseline=(current bounding box.center), main/.style = {draw, circle}]
    \node[main, label=below:$1$] (1L) []{};
    \node[main, label=below:$2$] (2L) [right=of 1L]{};
    \node[main, label=below:$3$] (3L) [right=of 2L]{};
    \node[main, label=below:$4$] (4L) [right=of 3L]{};
    \node[draw=none,fill=none] (cdots) [right=of 4L]{$\cdots$};
    \node[main, label=below:$4$] (4R) [right=of cdots]{};
    \node[main, label=below:$2$] (2R) [right=of 4R]{};
    \node[main, label=left:$1$] (1T) [above=of 4L]{};

    \draw[-] (1L)--(2L)--(3L)--(4L)--(cdots)--(4R);
    \draw[-] (1T)--(4L);
     \draw [line width=1pt, double distance=3pt,
             arrows = {-Latex[length=0pt 3 0]}] (4R) -- (2R);
    \draw [decorate, 
    decoration = {brace,
        raise=10pt,
        amplitude=5pt}] (4R) --  (4L) node[pos=0.5,below=15pt,black]{$k-4$};
    
    \node[main, label=left:$1$] (1subT) [below=of 4L]{};
    \node[main, label=below:$2$] (2sub) [below=of 1subT]{};
    \node[main, label=below:$3$] (3sub) [left=of 2sub]{};
    \node[main, label=below:$2$] (2subL) [left=of 3sub]{};
    \node[main, label=below:$1$] (1subL) [left=of 2subL]{};
    \node[draw=none,fill=none] (-) [left=of 1subL]{$-$};
    
    \draw[-] (1subT)--(2sub)--(3sub)--(2subL)--(1subL);
    
     \node[main, label=below:$2$] (2resL) [below=of 2sub]{};
    \node[main, label=below:$4$] (4resL) [right=of 2resL]{};
    \node[draw=none,fill=none] (cdotsres) [right=of 4resL] {$\cdots$};
    \node[main, label=below:$4$] (4resR) [right=of cdotsres]{};
    \node[main, label=below:$2$] (2resR) [right=of 4resR]{};
    \node[draw, label=left:$2$] (2resT) [above=of 4resL]{};
    
    \draw[-] (2resL)--(4resL)--(cdotsres)--(4resR);
    \draw[-] (2resT)--(4resL);
    \draw [line width=1pt, double distance=3pt,
             arrows = {-Latex[length=0pt 3 0]}] (4resR) -- (2resR);

    \draw [decorate, 
    decoration = {brace,
        raise=10pt,
        amplitude=5pt}] (4resR) --  (4resL) node[pos=0.5,below=15pt,black]{$k-5$};
             
    \node[draw=none,fill=none] (topghost) [right=of 2R]{};
    \node[draw=none,fill=none] (bottomghost) [right=of 2resR]{};
    
    \draw [->] (topghost) to [out=-30,in=30,looseness=1] (bottomghost);

   \end{tikzpicture}}
   \caption{}
    \label{fig:orbBkA2Sub1}
    \end{subfigure}
\hfill
\begin{subfigure}{0.4\textwidth}
    \centering
    \resizebox{0.6\width}{!}{\begin{tikzpicture}[baseline=(current bounding box.center), main/.style = {draw, circle}]
    \node[main, label=below:$1$] (1L) []{};
    \node[main, label=below:$2$] (2L) [right=of 1L]{};
    \node[main, label=below:$3$] (3L) [right=of 2L]{};
    \node[main, label=below:$4$] (4L) [right=of 3L]{};
    \node[draw=none,fill=none] (cdots) [right=of 4L]{$\cdots$};
    \node[main, label=below:$4$] (4R) [right=of cdots]{};
    \node[main, label=below:$2$] (2R) [right=of 4R]{};
    \node[main, label=left:$1$] (1T) [above=of 4L]{};

    \draw[-] (1L)--(2L)--(3L)--(4L)--(cdots)--(4R);
    \draw[-] (1T)--(4L);
     \draw [line width=1pt, double distance=3pt,
             arrows = {-Latex[length=0pt 3 0]}] (4R) -- (2R);
     \draw [decorate, 
    decoration = {brace,
        raise=10pt,
        amplitude=5pt}] (4R) --  (4L) node[pos=0.5,below=15pt,black]{$k-4$};
    
    \node[main, label=below:$1$] (1subL) [below=of 1L]{};
    \node[main, label=below:$2$] (2subL) [right=of 1subL]{};
    \node[main, label=below:$3$] (3sub) [right=of 2subL]{};
    \node[main, label=below:$2$] (2subR) [right=of 3sub]{};
    \node[main, label=below:$1$] (1subT) [right=of 2subR]{};
    \node[draw=none,fill=none] (-) [left=of 1subL]{$-$};
    
    \draw[-] (1subT)--(2subR)--(3sub)--(2subL)--(1subL);

    \node[draw,label=left:$2$] (2resT) [below=of 3sub]{};
    \node[main, label=below:$1$] (1resL) [below=of 2resT]{};
     \node[main, label=below:$2$] (2resL) [right=of 1resL]{};
     \node[main, label=below:$3$] (3res) [right=of 2resL]{};
     \node[main, label=below:$4$] (4resL) [right=of 3res]{};
     \node[draw=none,fill=none] (cdotsres) [right=of 4resL]{$\cdots$};
     \node[main, label=below:$4$] (4resR) [right=of cdotsres]{};
     \node[main, label=below:$2$] (2resR) [right=of 4resR]{};
     \node[draw, label=right:$1$] (1resT) [above=of 4resL]{};

     \draw[-] (2resT)--(1resL)--(2resL)--(3res)--(4resL)--(cdotsres)--(4resR)--(2resR);
     \draw[-] (1resT)--(4resL);
     \draw [line width=1pt, double distance=3pt,
             arrows = {-Latex[length=0pt 3 0]}] (4resR) -- (2resR);

    \draw [decorate, 
    decoration = {brace,
        raise=10pt,
        amplitude=5pt}] (4resR) --  (4resL) node[pos=0.5,below=15pt,black]{$k-6$};
             
    \node[draw=none,fill=none] (topghost) [right=of 2R]{};
    \node[draw=none,fill=none] (bottomghost) [right=of 2resR]{};
    
    \draw [->] (topghost) to [out=-30,in=30,looseness=1] (bottomghost);

   \end{tikzpicture}}
   \caption{}
    \label{fig:orbBkA2Sub2}
\end{subfigure}
\centering
\begin{subfigure}{\textwidth}
\centering
\begin{tikzpicture}[main/.style = {draw, circle}]
    \node[draw, label=left:$1$] (1resFL) []{};
     \node[main, label=below:$2$] (2resL) [below=of 1resFL]{};
     \node[main, label=below:$3$] (3res) [right=of 2resL]{};
     \node[main, label=below:$4$] (4resL) [right=of 3res]{};
     \node[draw=none,fill=none] (cdotsres) [right=of 4resL]{$\cdots$};
     \node[main, label=below:$4$] (4resR) [right=of cdotsres]{};
     \node[main, label=below:$2$] (2resR) [right=of 4resR]{};
     \node[draw, label=right:$1$] (1resT) [above=of 4resL]{};

     \draw[-] (1resFL)--(2resL)--(3res)--(4resL)--(cdotsres)--(4resR)--(2resR);
     \draw[-] (1resT)--(4resL);
     \draw [line width=1pt, double distance=3pt,
             arrows = {-Latex[length=0pt 3 0]}] (4resR) -- (2resR);

    \draw [decorate, 
    decoration = {brace,
        raise=10pt,
        amplitude=5pt}] (4resR) --  (4resL) node[pos=0.5,below=15pt,black]{$k-6$};
\end{tikzpicture}
\caption{}
\label{fig:orbBkA2SubInt}
\end{subfigure}

    \caption{Both alignments of the $\urm(3)$ quotient quiver against \Quiver{fig:Bk24} for $k\geq7$ giving quivers \Quiver{fig:orbBkA2Sub1} and \Quiver{fig:orbBkA2Sub2}. Their intersection, reached via $A_1$ KP transitions, is quiver \Quiver{fig:orbBkA2SubInt}.}
    \label{fig:orbBkA2SubBoth}
\end{figure}
Now we consider the more general case for $k\geq 7$. There are again two possible alignments of the $\urm(3)$ quotient quiver against \Quiver{fig:Bk24} which results in the quivers \Quiver{fig:orbBkA2Sub1} and \Quiver{fig:orbBkA2Sub2}. 
The Coulomb branches of these are $\overline{\left[\mathcal W_{B_{k-3}}\right]}^{[0,2,0,\cdots,0]}_{[0,\cdots,0]}$ and $\overline{\left[\mathcal W_{B_{k-2}}\right]}^{[2,0,0,1,0,\cdots,0]}_{[2,0,\cdots,0]}$, with global symmetries $SO(2k+1)$ and $SO(2k+1)\times \urm(1)$ respectively. Taking an $A_1$ KP transition from each quiver we obtain their intersection, which is the magnetic quiver for  $\overline{\left[\mathcal W_{B_{k-3}}\right]}^{[1,0,1,0,\cdots,0]}_{[0,\cdots,0]}$. We conjecture that:
\begin{equation}
    \overline{\mathcal O}^{B_k}_{(2^4,1^{2k-7})}///\surm(3)=\overline{\left[\mathcal W_{B_{k-3}}\right]}^{[0,2,0,\cdots,0]}_{[0,\cdots,0]}\cup\overline{\left[\mathcal W_{B_{k-2}}\right]}^{[2,0,0,1,0,\cdots,0]}_{[2,0,\cdots,0]},\quad k\geq 7\label{eq:OrbBkA2Sub}
\end{equation}
We have checked this quiver subtraction result \eqref{eq:OrbBkA2Sub} using Weyl integration with the embedding \eqref{eq:BkA2Embed} for $k=7$.
\subsection{B-type Orbit $\surm(n)$ HKQ}
\label{sec:BtypeSUN}
\subsubsection{Height 2 Orbits $\overline{\mathcal O}^{B_k}_{(2^{2p},1^{2k-4p+1})}///\surm(n)$}
\label{sec:H2BType}
\begin{figure}[h!]
    \centering
    \begin{tikzpicture}[baseline=(current bounding box.center), main/.style = {draw, circle}]
    \node[main, label=below:$1$] (1L) []{};
    \node[main, label=below:$2$] (2L) [right=of 1L]{};
    \node[draw=none,fill=none] (cdotsL) [right=of 2L]{$\cdots$};
    \node[main, label=below:$2p$] (2pL) [right=of cdotsL]{};
    \node[main, label=below:$2p$] (2pM) [right=of 2pL]{};
     \node[draw=none,fill=none] (cdotsR) [right=of 2pM]{$\cdots$};
     \node[main, label=below:$2p$] (2pR) [right=of cdotsR]{};
     \node[main, label=below:$p$] (p) [right=of 2pR]{};
     \node[main, label=left:$1$] (1T) [above=of 2pL]{};

     \draw[-] (1L)--(2L)--(cdotsL)--(2pL)--(2pM)--(cdotsR)--(2pR);
     \draw[-] (1T)--(2pL);
      \draw [line width=1pt, double distance=3pt,
             arrows = {-Latex[length=0pt 3 0]}] (2pR) -- (p);
    \draw [decorate, 
    decoration = {brace,
        raise=20pt,
        amplitude=5pt}] (p) --  (1L) node[pos=0.5,below=25pt,black]{$k$}; 
    \end{tikzpicture}
    \caption{Unframed magnetic quiver \Quiver{fig:OrbB22p} for the height 2 orbit of $B_k$ specified by the partition $(2^{2p},1^{2k-4p+1})$ of $2k+1$.}
    \label{fig:OrbB22p}
\end{figure}

We now consider the family of orbits of $B_k$ that are given by a partition $(2^{2p},1^{2k-4p+1})$ of $2k+1$, where $ k \geq 4, p < k/2 $. These orbits are of height 2 and so their HWGs are complete intersections and take a known form derivable from the orbit partition data. The generic magnetic quiver \Quiver{fig:OrbB22p} is drawn in \Figref{fig:OrbB22p}. This magnetic quiver is amenable to quiver subtraction by an $\urm(n)$ quotient quiver, subject to limits on the range of $n$ that satisfies the rules.

In order to satisfy the \hyperlink{rule:Junction}{Junction Rule}, the length of the $\urm(n)$ quotient quiver, $2n-1$, must be at most $2p+1$ which gives the bound $n\leq p+1$. We present the result of the quiver subtraction, separating the different cases according to whether $n$ saturates this bound.

Firstly, when $n<p+1$:
\begin{align}
    \overline{\mathcal O}^{B_k}_{(2^{2p},1^{2k-4p+1})}///\surm(n)&=\mathcal C\left(\resizebox{!}{30pt}{\begin{tikzpicture}[baseline=(current bounding box.center), main/.style = {draw, circle},scale=0.25]
    \node[main, label=below:$2$] (2L) []{};
    \node[main, label=below:$4$] (4L) [right=of 2L]{};
    \node[draw=none,fill=none] (cdotsL) [right=of 4L]{$\cdots$};
    \node[main, label=below:$2n-2$] (2nm2) [right=of cdotsL]{};
    \node[main, label=below:$2n-1$] (2nm1) [right=of 2nm2]{};
    \node[draw=none,fill=none] (cdotsM) [right=of 2nm1]{$\cdots$};
    \node[main, label=below:$2p$] (2pL) [right=of cdotsM]{};
    \node[main, label=below:$2p$] (2pM) [right=of 2pL]{};
    \node[draw=none,fill=none] (cdotsR) [right=of 2pM]{$\cdots$};
    \node[main, label=below:$2p$] (2pR) [right=of cdotsR]{};
    \node[main, label=below:$p$] (p) [right=of 2pR]{};
    \node[main, label=right:$1$] (1T) [above=of 2pL]{};
    \node[draw, label=left:$1$] (1F) [above=of 2nm2]{};
    \draw[-] (2L)--(4L)--(cdotsL)--(2nm2)--(2nm1)--(cdotsM)--(2pL)--(2pM)--(cdotsR)--(2pR);
    \draw[-] (1F)--(2nm2);
    \draw[-] (1T)--(2pL);
    \draw [line width=1pt, double distance=3pt,
             arrows = {-Latex[length=0pt 3 0]}] (2pR) -- (p);
    \draw [decorate, 
    decoration = {brace,
        raise=20pt,
        amplitude=5pt}] (p) --  (2L) node[pos=0.5,below=25pt,black]{$k-n$};
    \end{tikzpicture}}\right)\label{eq:OrbB22pSub1}
\end{align}
The resulting magnetic quiver shown in \eqref{eq:OrbB22pSub1} has a Coulomb branch with global symmetry of $SO(2k-2n+1)\times \urm(1)$, as can be seen by consideration of the balance of the gauge nodes. This quiver does not in general come from a Dynkin diagram of finite type.

Now consider the saturating case where $n=p+1$. Then the $\urm(n)$ quotient quiver has two possible alignments and we obtain a union of Coulomb branches: 
\begin{subequations}
     \label{eq:OrbB22pSub2}
\begin{align}
    \overline{\mathcal O}^{B_k}_{(2^{2p},1^{2k-4p+1})}///\surm(n)&=\mathcal C\left(\resizebox{!}{35pt}{\begin{tikzpicture}[baseline=(current bounding box.center), main/.style = {draw, circle},scale=0.25]
    \node[main, label=below:$2$] (2L) []{};
    \node[main, label=below:$4$] (4L) [right=of 2L]{};
    \node[draw=none,fill=none] (cdotsL) [right=of 4L]{$\cdots$};
    \node[main, label=below:$2p-2$] (2nm2) [right=of cdotsL]{};
    \node[main, label=below:$2p$] (2pL) [right=of 2nm2]{};
    \node[main, label=below:$2p$] (2pM) [right=of 2pL]{};
    \node[draw=none,fill=none] (cdotsR) [right=of 2pM]{$\cdots$};
    \node[main, label=below:$2p$] (2pR) [right=of cdotsR]{};
    \node[main, label=below:$p$] (p) [right=of 2pR]{};
    \node[draw, label=left:$2$] (2T) [above=of 2pL]{};
    \draw[-] (2L)--(4L)--(cdotsL)--(2nm2)--(2pL)--(2pM)--(cdotsR)--(2pR);
    \draw[-] (2T)--(2pL);
    \draw [line width=1pt, double distance=3pt,
             arrows = {-Latex[length=0pt 3 0]}] (2pR) -- (p);
    \draw [decorate, 
    decoration = {brace,
        raise=20pt,
        amplitude=5pt}] (p) --  (2L) node[pos=0.5,below=25pt,black]{$k-p-1$};
    \end{tikzpicture}}\right)\label{eq:OrbB22pSub2a}\\&
    \cup\mathcal C\left(\resizebox{!}{45pt}{\begin{tikzpicture}[baseline=(current bounding box.center), main/.style = {draw, circle},scale=0.25]
    \node[main, label=below:$2$] (2L) []{};
    \node[main, label=below:$4$] (4L) [right=of 2L]{};
    \node[draw=none,fill=none] (cdotsL) [right=of 4L]{$\cdots$};
    \node[main, label=below:$2p-2$] (2nm2) [right=of cdotsL]{};
    \node[main, label=below:$2p-1$] (2nm1) [right=of 2nm2]{};
    \node[main, label=below:$2p$] (2pL) [right=of 2nm1]{};
    \node[main, label=below:$2p$] (2pM) [right=of 2pL]{};
    \node[draw=none,fill=none] (cdotsR) [right=of 2pM]{$\cdots$};
    \node[main, label=below:$2p$] (2pR) [right=of cdotsR]{};
    \node[main, label=below:$p$] (p) [right=of 2pR]{};
    \node[main, label=left:$1$] (1T) [above=of 2nm2]{};
    \node[draw, label=left:$2$] (2T) [above=of 1T]{};
    \node[draw, label=right:$1$] (1F) [above=of 2pL]{};
    \draw[-] (2L)--(4L)--(cdotsL)--(2nm2)--(2nm1)--(2pL)--(2pM)--(cdotsR)--(2pR);
    \draw[-] (2nm2)--(1T)--(2T);
    \draw[-] (1F)--(2pL);
    \draw [line width=1pt, double distance=3pt,
             arrows = {-Latex[length=0pt 3 0]}] (2pR) -- (p);
    \draw [decorate, 
    decoration = {brace,
        raise=20pt,
        amplitude=5pt}] (p) --  (2L) node[pos=0.5,below=25pt,black]{$k-p-1$};
    \end{tikzpicture}}\right)\label{eq:OrbB22pSub2b}
     \end{align} 

\end{subequations}
    The first quiver in the union \eqref{eq:OrbB22pSub2}, \Quiver{eq:OrbB22pSub2a}, is a slice in the affine Grassmannian $\overline{\left[\mathcal W_{B_{k-p-1}}\right]}^{[0,\cdots,0,2,0\cdots,0]}_{[0,\cdots,0]}$ where the flavour is on the $p^{th}$ node. This quiver has a Coulomb branch global symmetry of $SO(2k-2p-2)$. The Coulomb branch of the second quiver \Quiver{eq:OrbB22pSub2b} has a global symmetry of $SO(2k-2p-2)\times \urm(1)$. Again, this quiver does not in general come from a Dynkin diagram of finite type. There is an $A_1$ KP transition from each quiver to their intersection, which is a magnetic quiver for the affine Grassmannian slice $\overline{\left[\mathcal W_{B_{k-p-1}}\right]}^{[0,\cdots,0,1,0,1,0,\cdots,0]}_{[0,\cdots,0]}$, where the flavours are on the $(p-1)^{th}$ and $(p+1)^{th}$ nodes. The Coulomb branch of this quiver has a $SO(2k-2p-2)$ global symmetry.

\section{C-type Orbits}
\label{sec:CtypeHKQs}
There are no relevant orbits to consider here. This is because although there may be a leg of $(1)-(2)-\cdots-(n)-\cdots$ this will always correspond to short roots due to the form of the C-type Dynkin diagram and thus would violate the \hyperlink{rule:ExternalLeg}{External Leg Rule}.

\section{D-type Orbits}
\label{sec:DtypeHKQ}
Here we consider orbits of D-type and the possible $\surm(n)$ HKQs that one can compute using quiver subtraction. These embrace all height 2 orbits, including $\overline{min. D_k}///\surm(2)$, $\overline{n.min. D_k}///\surm(2)$, as well as some $\surm(n)$ HKQs for higher orbits for $k>4$.

In order to check results by Weyl integration, we use the embedding of $D_k\hookleftarrow D_{k-2}\times A_1\times A_1$ which decomposes the vector representation as:
\begin{align}
    [1,0,0,0]_{D_4}&\rightarrow [1;1]_{D_2}+[1;1]_{A_1\times A_1}
    \label{eq:D4A1Embed}\\ [1,0,\cdots,0]_{D_k}&\rightarrow [1,0,\cdots,0]_{D_{k-2}}+[1;1]_{A_1\times A_1},\quad k\geq 5,
    \label{eq:DkA1Embed}
\end{align}
where the cases for $k=4$ and $k\geq 5$ are slightly different and singlets have been suppressed for the latter. In both cases all factors of $A_1$ are isomorphic in the decomposition and the $\surm(2)$ HKQ can be taken w.r.t any of them.
\subsection{D-type Orbit $\surm(2)$ HKQ}
\label{sec:DtypeSU2HKQ}
\subsubsection{$\overline{min. D_k}///\surm(2), k\geq 4$}
\begin{figure}[h!]
    \centering
    \begin{tikzpicture}[main/.style = {draw, circle}]
    \node[main, label=below:$1$] (a) {};
    \node[main,label=below:$2$] (b) [right=of a] {};
    \node[] (c) [right=of b] {$\cdots$};
    \node[main,label=below:$2$] (d) [right=of c] {};
    \node[main,label=below:$1$] (e) [right=of d] {};
    \node[main, label=left:$1$] (f) [above=of b]{};
    \node[main, label=left:$1$] (g) [above=of d]{};
    \draw[-] (a)--(b);
    \draw[-] (e)--(d);
    \draw[-] (f)--(b)--(c)--(d)--(g);
    \draw [decorate, 
    decoration = {brace,
        raise=10pt,
        amplitude=5pt}] (d) --  (b) node[pos=0.5,below=15pt,black]{$k-3$};
    
    \end{tikzpicture}
    \caption{Unframed magnetic quiver \Quiver{fig:quiv_min_Dk} for $\overline{min. D_k}$.}
    \label{fig:quiv_min_Dk}
\end{figure}
\begin{figure}[h!]
    \centering
    \begin{tikzpicture}[main/.style={draw,circle}]
       \node[main, label=below:$2$] (2) {};
       \node[main, label=below:$1$] (1) [below left=of 2]{};
       \node[main, orange, label=left:$1$,fill] (blue) [above left=of 2]{};
       \node[main, purple,label=below:$1$,fill] (red) [below right=of 2]{};
       \node[main, green, label=right:$1$,fill] (green) [above right=of 2]{};

       \draw[-] (1)--(2)--(blue);
       \draw[-] (red)--(2)--(green);
       
    \end{tikzpicture}
    \caption{Unframed magnetic quiver \Quiver{fig:minD4Quiv} for $\overline{min. D_4}$ with three nodes of rank 1 coloured.}
    \label{fig:minD4Quiv}
\end{figure}

The unframed magnetic quiver \Quiver{fig:quiv_min_Dk} for $\overline{min. D_k}$ is drawn in \Figref{fig:quiv_min_Dk}. The cases for $k\geq 4$ are all compatible with the rules for $\surm(2)$ quiver subtraction in \ref{sec:rules}.

We first consider the special case of $k=4$. Since the $\urm(2)$ quotient quiver ends on nodes which are related by outer automorphism, we draw the magnetic quiver \Quiver{fig:minD4Quiv} for $\overline{min. D_4}$ in \Figref{fig:minD4Quiv} using different colours to distinguish the nodes of rank 1. 

The subtraction of the $\urm(2)$ quotient quiver, aligned with the uncoloured nodes and each coloured node in turn, gives the resulting quivers 
$\mathcal{Q}_{\ref{fig:minD4Quiv} \textcolor{blue}{a}}$, $\mathcal{Q}_{\ref{fig:minD4Quiv} \textcolor{blue}{b}}$ and $\mathcal{Q}_{\ref{fig:minD4Quiv} \textcolor{blue}{c}}$:

\begin{equation}
    \mathcal{Q}_{\ref{fig:minD4Quiv} \textcolor{blue}{a}}=\resizebox{!}{30pt}{\begin{tikzpicture}[baseline=(current bounding box.center), main/.style = {draw, circle},scale=0.25]

    \node[main, label=below:$1$,orange,fill] (L) {};
    \node[main, label=below:$1$,purple,fill] (R) [right=of L]{};
    \node[draw=none,fill=none] (times) [above right=0.3cm and 0.3cm of L] {$\otimes$};
    \node[draw,label=right:$2$] (2FL) [above=of L]{};
    \node[draw,label=right:$2$] (2FR) [above=of R]{};

    \draw[-] (L)--(2FL);
    \draw[-] (R)--(2FR);
    \end{tikzpicture}};\quad \mathcal{Q}_{\ref{fig:minD4Quiv} \textcolor{blue}{b}}=\resizebox{!}{30pt}{\begin{tikzpicture}[baseline=(current bounding box.center), main/.style = {draw, circle},scale=0.25]

    \node[main, label=below:$1$,purple,fill] (L) {};
    \node[main, label=below:$1$,green,fill] (R) [right=of L]{};
    \node[draw=none,fill=none] (times) [above right= 0.3cm and 0.3cm of L] {$\otimes$};
    \node[draw,label=right:$2$] (2FL) [above=of L]{};
    \node[draw,label=right:$2$] (2FR) [above=of R]{};

    \draw[-] (L)--(2FL);
    \draw[-] (R)--(2FR);
    
    \end{tikzpicture}};\quad\mathcal{Q}_{\ref{fig:minD4Quiv} \textcolor{blue}{c}}=\resizebox{!}{30pt}{\begin{tikzpicture}[baseline=(current bounding box.center), main/.style = {draw, circle},scale=0.25]

    \node[main, label=below:$1$,green,fill] (L) {};
    \node[main, label=below:$1$,orange,fill] (R) [right=of L]{};
    \node[draw=none,fill=none] (times) [above right= 0.3cm and 0.3cm of L] {$\otimes$};
    \node[draw,label=right:$2$] (2FL) [above=of L]{};
    \node[draw,label=right:$2$] (2FR) [above=of R]{};

    \draw[-] (L)--(2FL);
    \draw[-] (R)--(2FR);
    
    \end{tikzpicture}}
\end{equation}
Thus, we find from quiver subtraction that:
\begin{equation}
    \overline{min. D_4}///\surm(2)=\left(\textcolor{orange}{\overline{min. A_1}}\otimes \textcolor{purple}{\overline{min. A_1}}\right)\cup\left(\textcolor{purple}{\overline{min. A_1}}\otimes \textcolor{green}{\overline{min. A_1}}\right)\cup\left(\textcolor{green}{\overline{min. A_1}}\otimes \textcolor{orange}{\overline{min. A_1}}\right).
\end{equation}
Each pair of quivers in the union is related to its intersection by an $A_1$ KP transition,
\begin{equation}
    \mathcal{Q}_{\ref{fig:minD4Quiv} \textcolor{blue}{a}}\cap\mathcal{Q}_{\ref{fig:minD4Quiv} \textcolor{blue}{b}}=\resizebox{!}{30pt}{\begin{tikzpicture}[baseline=(current bounding box.center), main/.style = {draw, circle},scale=0.25]

    \node[main, label=below:$1$,purple,fill] (1) {};
    \node[draw,label=right:$2$] (2F) [above=of 1]{};

    \draw[-] (1)--(2F);
        
    \end{tikzpicture}};\quad \mathcal {Q}_{\ref{fig:minD4Quiv} \textcolor{blue}{a}}\cap\mathcal{Q}_{\ref{fig:minD4Quiv} \textcolor{blue}{c}}=\resizebox{!}{30pt}{\begin{tikzpicture}[baseline=(current bounding box.center), main/.style = {draw, circle},scale=0.25]

    \node[main, label=below:$1$,orange,fill] (1) {};
    \node[draw,label=right:$2$] (2F) [above=of 1]{};

    \draw[-] (1)--(2F);
        
    \end{tikzpicture}};\quad \mathcal{Q}_{\ref{fig:minD4Quiv} \textcolor{blue}{b}}\cap\mathcal{Q}_{\ref{fig:minD4Quiv} \textcolor{blue}{c}}=\resizebox{!}{30pt}{\begin{tikzpicture}[baseline=(current bounding box.center), main/.style = {draw, circle},scale=0.25]

    \node[main, label=below:$1$,green,fill] (1) {};
    \node[draw,label=right:$2$] (2F) [above=of 1]{};

    \draw[-] (1)--(2F);
        
    \end{tikzpicture}},
\end{equation}
and further $A_1$ KP transitions yield their trivial three-way intersection.
The HWG for this union is thus given by the signed sum of HWGs:
\begin{align}
    HWG&[\left(\textcolor{orange}{\overline{min. A_1}}\otimes \textcolor{purple}{\overline{min. A_1}}\right)\cup\left(\textcolor{purple}{\overline{min. A_1}}\otimes \textcolor{green}{\overline{min. A_1}}\right)\cup\left(\textcolor{green}{\overline{min. A_1}}\otimes \textcolor{orange}{\overline{min. A_1}}\right)]\nonumber\\&=\left(\frac{1}{1-\mu^2t^2}\frac{1}{1-\nu^2t^2}-\frac{1}{1-\rho^2t^2}+\mu\rightarrow\nu\rightarrow\rho\right)+1\label{eq:minD4A1SubHWGTop}\\&=PE\left[\left(\mu^2+\nu^2+\rho^2\right)t^2 -\mu^2\nu^2\rho^2 t^6\right],
    \label{eq:minD4A1SubHWGBottom}
\end{align} 
where $\mu, \nu$, and $\rho$ are highest weight fugacities for each remaining factor of $\surm(2)$. In \eqref{eq:minD4A1SubHWGTop} we have written the result from the union of cones formula \eqref{eq:UnionsofCones} using shorthand notation. Due to cancellations of invariants under the outer automorphism group this simplifies further to \eqref{eq:minD4A1SubHWGBottom}. The result is easily checked by Weyl integration using the embedding \eqref{eq:D4A1Embed} and we find agreement with quiver subtraction.

\begin{figure}[h!]
    \centering
   \begin{subfigure}{0.45\textwidth}
    \centering\resizebox{0.75\width}{!}{
    \begin{tikzpicture}[main/.style = {draw, circle}]
    \node[main, label=below:$1$] (a) {};
    \node[main,label=below:$2$] (b) [right=of a] {};
    \node[] (c) [right=of b] {$\cdots$};
    \node[main,label=below:$2$] (d) [right=of c] {};
    \node[main,label=below:$1$] (e) [right=of d] {};
    \node[main, label=left:$1$] (f) [above=of b]{};
    \node[main, label=left:$1$] (g) [above=of d]{};
    \draw[-] (a)--(b);
    \draw[-] (e)--(d);
    \draw[-] (f)--(b)--(c)--(d)--(g);
    \draw [decorate, 
    decoration = {brace,
        raise=10pt,
        amplitude=5pt}] (d) --  (b) node[pos=0.5,below=15pt,black]{$k-3$};
    
    \node[main, label=left:$1$] (1subT) [below=of b]{};
    \node[main, label=below:$2$] (2sub) [below=of 1subT]{};
    \node[main, label=below:$1$] (1subL) [left=of 2sub]{};
    \node[draw=none,fill=none] (-) [left=of 1subL]{$-$};
    \node[draw=none,fill=none] (ghost) [right=of 2sub]{};
    \draw[-] (1subT)--(2sub)--(1subL);
    
    \node[main,label=left:$2$] (2resF) [below=of ghost]{};
    \node[] (dotsres) [right=of 2resF] {$\cdots$};
    \node[main,label=below:$2$] (eres) [right=of dotsres] {};
    \node[main,label=below:$1$] (fres) [right=of eres] {};
    \node[draw, label=left:$2$] (gres) [above=of 2resF]{};
    \node[main, label=left:$1$] (hres) [above=of eres]{};
    
    \draw[-] (gres)--(2resF)--(dotsres)--(eres)--(fres);
    \draw[-] (hres)--(eres);
    
    \node[draw=none,fill=none] (topghost) [right=of e]{};
    \node[draw=none,fill=none] (bottomghost) [right=of fres]{};
    \draw [decorate, 
    decoration = {brace,
        raise=10pt,
        amplitude=5pt}] (eres) --  (2resF) node[pos=0.5,below=15pt,black]{$k-4$};
    
    \draw [->] (topghost) to [out=-30,in=30,looseness=1] (bottomghost);

   \end{tikzpicture}}
   \caption{}
    \label{fig:minDKA1Sub1}
\end{subfigure}
\hfill
\begin{subfigure}{0.45\textwidth}
    \centering\resizebox{0.75\width}{!}{
    \begin{tikzpicture}[main/.style = {draw, circle}]
    \node[main, label=below:$1$] (a) {};
    \node[main,label=below:$2$] (b) [right=of a] {};
    \node[] (c) [right=of b] {$\cdots$};
    \node[main,label=below:$2$] (d) [right=of c] {};
    \node[main,label=below:$1$] (e) [right=of d] {};
    \node[main, label=left:$1$] (f) [above=of b]{};
    \node[main, label=left:$1$] (g) [above=of d]{};
    \draw[-] (a)--(b);
    \draw[-] (e)--(d);
    \draw[-] (f)--(b)--(c)--(d)--(g);
    \draw [decorate, 
    decoration = {brace,
        raise=10pt,
        amplitude=5pt}] (d) --  (b) node[pos=0.5,below=15pt,black]{$k-3$};
        
    \node[main, label=left:$1$] (1subT) [below=of b]{};
    \node[main, label=below:$2$] (2sub) [below=of 1subT]{};
    \node[main, label=below:$1$] (1subR) [right=of 2sub]{};
    \node[draw=none,fill=none] (-) [left=of 2sub]{$-$};
    \draw[-] (1subT)--(2sub)--(1subR);
    
    \node[draw,label=left:$2$] (2resF) [below left=of 2sub]{};
    \node[main, label=below:$1$] (1) [below=of 2resF] {};
    
    \draw[-] (2resF)--(1);
    
    \node[draw,label=left:$1$] (1F) [below right=of 1subR]{};
    \node[main, label=below:$2$] (2) [below =of 1F]{};
    \node[main, label=below:$1$] (1) [left=of 2]{};
    \node[draw=none,fill=none] (cdots) [right=of 2]{$\cdots$};
    \node[main,label=below:$2$] (2R) [right=of cdots]{};
    \node[main, label=below:$1$] (1R)[right=of 2R]{};
    \node[main, label=left:$1$] (1T) [above=of 2R]{};
    
    \draw[-] (1F)--(2);
    \draw[-] (1)--(2)--(cdots)--(2R)--(1R);
    \draw[-] (2R)--(1T);
    \draw [decorate, 
    decoration = {brace,
        raise=10pt,
        amplitude=5pt}] (2R) --  (2) node[pos=0.5,below=15pt,black]{$k-5$};

    \node[draw=none,fill=none] (topghost) [right=of e]{};
    \node[draw=none,fill=none] (bottomghost) [right=of 1R]{};
    
    \draw [->] (topghost) to [out=-30,in=30,looseness=1] (bottomghost);

   \end{tikzpicture}}
   \caption{}
    \label{fig:minDKA1Sub2}
\end{subfigure}
\centering
\begin{subfigure}{\textwidth}
\centering
    \begin{tikzpicture}[main/.style={draw,circle}]
        \node[draw,label=left:$1$] (1F) []{};
    \node[main, label=below:$2$] (2) [below =of 1F]{};
    \node[main, label=below:$1$] (1) [left=of 2]{};
    \node[draw=none,fill=none] (cdots) [right=of 2]{$\cdots$};
    \node[main,label=below:$2$] (2R) [right=of cdots]{};
    \node[main, label=below:$1$] (1R)[right=of 2R]{};
    \node[main, label=left:$1$] (1T) [above=of 2R]{};
    
    \draw[-] (1F)--(2);
    \draw[-] (1)--(2)--(cdots)--(2R)--(1R);
    \draw[-] (2R)--(1T);
    \draw [decorate, 
    decoration = {brace,
        raise=10pt,
        amplitude=5pt}] (2R) --  (2) node[pos=0.5,below=15pt,black]{$k-5$};
    \end{tikzpicture}
    \caption{}
    \label{fig:minDKA1SubInt}
\end{subfigure}
    \caption{Both alignments of the $\urm(2)$ quotient quiver against \Quiver{fig:quiv_min_Dk} for $k\geq 5$ producing quivers \Quiver{fig:minDKA1Sub1} and \Quiver{fig:minDKA1Sub2}. Their intersection, reached via $A_1$ KP transitions, is quiver \Quiver{fig:minDKA1SubInt}.}
    \label{fig:minDkA1SubBoth}
\end{figure}

Now we consider the more general case for $k\geq 5$. There are two choices of alignment for the $\urm(2)$ quotient quiver against \Quiver{fig:quiv_min_Dk}. These are shown in \Figref{fig:minDkA1SubBoth} and produce quivers \Quiver{fig:minDKA1Sub1} and  \Quiver{fig:minDKA1Sub2}, whose Coulomb branches are $\overline{n.min D_{k-2}}$ and $\overline{min. A_1}\otimes\overline{min. D_{k-2}}$. We conjecture: \begin{equation}
\overline{min. D_k}///\surm(2)=\overline{n.min D_{k-2}}\cup \left(\overline{min. A_1}\otimes \overline{min. D_{k-2}}\right), k\geq 5.
\end{equation}
The intersection of \Quiver{fig:minDKA1Sub1} and  \Quiver{fig:minDKA1Sub2} is a magnetic quiver \Quiver{fig:minDKA1SubInt} for $\overline{min. D_{k-2}}$, reached in both cases by an $A_1$ KP transition. The HWG of the union is thus the signed sum of component HWGs:
\begin{align}
    HWG\left[\overline{n.min D_{k-2}}\cup \left(\overline{min. A_1}\otimes \overline{min. D_{k-2}}\right)\right]&=PE[\mu_2t^2+\mu_1^2t^4]+PE[\mu_2t^2]PE[\nu^2t^2]\nonumber\\&-PE[\mu_2t^2]
\label{eq:minDkA1SubHWGTop}\\&=PE\left[\left(\mu_2+\nu^2\right)t^2+\mu_1^2t^4-\mu_1^2\nu^2t^6\right],
    \label{eq:minDkA1SubHWGBottom}
    \end{align}
where $\mu_{1,2}$ are highest weight fugacities for $D_{k-2}$ and $\nu$ is a highest weight fugacity for $A_1$. The first line \eqref{eq:minDkA1SubHWGTop} follows the union of cones formula \eqref{eq:UnionsofCones}, giving the HWGs for $\overline{n.min. D_{k-2}}$ and $\overline{min. A_1}\otimes \overline{min. D_{k-2}}$, less the HWG for intersection $\overline{min. D_{k-2}}$. Each term has a common $PE[\mu_2t^2]$ term, permitting the simplification into \eqref{eq:minDkA1SubHWGBottom}. This is easily checked by Weyl integration using the embedding \eqref{eq:DkA1Embed} and we find agreement with quiver subtraction.

\subsubsection{$\overline{n. min. D_k}///\surm(2)$ for $k\geq 4$}
\begin{figure}[h!]
    \centering
    \begin{tikzpicture}[main/.style = {draw, circle}]
    \node[main, label=below:$1$] (a) {};
    \node[main, label=below:$2$] (b) [right=of a]{};
    \node[draw=none,fill=none] (cdots) [right=of b]{$\cdots$};
    \node[main, label=below:$2$] (c) [right=of cdots]{};
    \node[main, label=right:$1$] (d)[above right=of c]{};
    \node[main, label=right:$1$] (e)[below right=of c]{};
    
    \draw[-] (e)--(c)--(d);
    \draw[-] (c)--(cdots)--(b);
    \draw[double,double distance=3pt,line width=0.4pt] (a)--(b);
    \draw [decorate, 
    decoration = {brace,
        raise=10pt,
        amplitude=5pt}] (c) --  (b) node[pos=0.5,below=15pt,black]{$k-2$};
    
    \end{tikzpicture}
    \caption{Unframed magnetic quiver \Quiver{fig:quiv_nmin_Dk} for $\overline{n. min D_k}$.}
    \label{fig:quiv_nmin_Dk}
\end{figure}
\begin{figure}[h!]
    \centering
    \begin{subfigure}{0.45\textwidth}
    \centering\begin{tikzpicture}[main/.style = {draw, circle}]
    \node[main, label=below:$1$] (a) {};
    \node[main, label=below:$2$] (b) [right=of a]{};
    \node[main, label=below:$2$] (c) [right=of b]{};
    \node[main, label=right:$1$] (d)[above right=of c]{};
    \node[main, label=right:$1$] (e)[below right=of c]{};
    
    \draw[-] (e)--(c)--(d);
    \draw[-] (c)--(b);
    \draw[double,double distance=3pt,line width=0.4pt] (a)--(b);
    
    \node[main, label=below:$1$] (1subT) [below=of e]{};
    \node[main, label=below:$2$] (2sub) [below left=of 1subT]{};
    \node[main, label=below:$1$] (1subB) [below right=of 2sub]{};
    \node[draw=none,fill=none] (-) [left=of 2sub]{$-$};
    \draw[-] (1subT)--(2sub)--(1subB);
    
    \node[draw, label=left:$2$] (2F) [below left=of 2sub]{};
    \node[main, label=below:$2$] (2R) [below=of 2F]{};
    \node[main, label=below:$1$] (1L) [left=of 2R]{};
    
    \draw[-] (2R)--(2F);
    \draw[double,double distance=3pt,line width=0.4pt] (1L)--(2R);
    \end{tikzpicture}
    \caption{}
    \label{fig:nminD4A1Sub1}\end{subfigure}
    \hfill
    \begin{subfigure}{0.45\textwidth}
    \centering\begin{tikzpicture}[main/.style = {draw, circle}]
   \node[main, label=below:$1$] (a) {};
    \node[main, label=below:$2$] (b) [right=of a]{};
    \node[main, label=below:$2$] (c) [right=of b]{};
    \node[main, label=right:$1$] (d)[above right=of c]{};
    \node[main, label=right:$1$] (e)[below right=of c]{};
    
    \draw[-] (e)--(c)--(d);
    \draw[-] (c)--(b);
    \draw[double,double distance=3pt,line width=0.4pt] (a)--(b);
    
    \node[main, label=below:$1$] (1subT) [below=of e]{};
    \node[main, label=below:$2$] (2sub) [below left=of 1subT]{};
    \node[main, label=below:$1$] (1subB) [left =of 2sub]{};
    \node[draw=none,fill=none] (-) [left=of 1subB]{$-$};
    \draw[-] (1subT)--(2sub)--(1subB);
    
    \node[draw, label=left:$2$] (2F) [below left=of 1subB]{};
    \node[main, label=below:$1$] (1L) [below=of 2F]{};
    \node[main, label=below:$1$] (1R) [right=of 1L]{};
    
    \draw[-] (2F)--(1L);
    \draw[double,double distance=3pt,line width=0.4pt] (1L)--(1R);

    \node[draw, label=right:$2$] (2F2) [below right=of 2sub]{};
    \node[main, label=below:$1$] (1) [below=of 2F2]{};
    \draw[-] (2F2)--(1);
     \end{tikzpicture}
    \caption{}
    \label{fig:nminD4A1Sub2}
    \end{subfigure}
    \centering
    \begin{subfigure}{\textwidth}
    \centering
    \begin{tikzpicture}[main/.style={draw,circle}]
    \node[draw, label=left:$2$] (2F) []{};
    \node[main, label=below:$1$] (1L) [below=of 2F]{};
    \node[main, label=below:$1$] (1R) [right=of 1L]{};
    
    \draw[-] (2F)--(1L);
    \draw[double,double distance=3pt,line width=0.4pt] (1L)--(1R);
        
    \end{tikzpicture}        
    \caption{}
    \label{fig:nminD4A1SubInt}
    \end{subfigure}
    \caption{Both alignments of the $\urm(2)$ quotient quiver against \Quiver{fig:quiv_nmin_Dk} for $\overline{n.min. D_4}$ producing quivers \Quiver{fig:nminD4A1Sub1} and \Quiver{fig:nminD4A1Sub2}. Their intersection, reached via $A_1$ KP transitions, is quiver \Quiver{fig:nminD4A1SubInt}.}
    \label{fig:nminD4A1SubBoth}
\end{figure}
\begin{figure}[h!]
    \centering
    \begin{subfigure}{0.45\textwidth}
    \centering\begin{tikzpicture}[main/.style = {draw, circle}]
    \node[main, label=below:$1$] (a) {};
    \node[main, label=below:$2$] (b) [right=of a]{};
    \node[draw=none,fill=none] (cdots) [right=of b]{$\cdots$};
    \node[main, label=below:$2$] (c) [right=of cdots]{};
    \node[main, label=right:$1$] (d)[above right=of c]{};
    \node[main, label=right:$1$] (e)[below right=of c]{};
    
    \draw[-] (e)--(c)--(d);
    \draw[-] (c)--(cdots)--(b);
    \draw[double,double distance=3pt,line width=0.4pt] (a)--(b);
    \draw [decorate, 
    decoration = {brace,
        raise=10pt,
        amplitude=5pt}] (c) --  (b) node[pos=0.5,below=15pt,black]{$k-2$};
    
    \node[main, label=below:$1$] (1subT) [below=of e]{};
    \node[main, label=below:$2$] (2sub) [below left=of 1subT]{};
    \node[main, label=below:$1$] (1subB) [below right=of 2sub]{};
    \node[draw=none,fill=none] (-) [left=of 2sub]{$-$};
    \draw[-] (1subT)--(2sub)--(1subB);
    
    \node[draw, label=left:$2$] (2F) [below left=of 1subB]{};
    \node[main, label=below:$2$] (2R) [below=of 2F]{};
    \node[draw=none,fill=none] (cdotsRes) [left=of 2R]{$\cdots$};
    \node[main, label=below:$2$] (2L) [left=of cdotsRes]{};
    \node[main, label=below:$1$] (1L) [left=of 2L]{};
    
    \draw[-] (2L)--(cdotsRes)--(2R)--(2F);
    \draw[double,double distance=3pt,line width=0.4pt] (2L)--(1L);
    \draw [decorate, 
    decoration = {brace,
        raise=10pt,
        amplitude=5pt}] (2R) --  (2L) node[pos=0.5,below=15pt,black]{$k-3$};
    
    \end{tikzpicture}
    \caption{}
    \label{fig:nminDkA1Sub1}\end{subfigure}
    \hfill
    \begin{subfigure}{0.45\textwidth}
    \centering\begin{tikzpicture}[main/.style = {draw, circle}]
    \node[main, label=below:$1$] (a) {};
    \node[main, label=below:$2$] (b) [right=of a]{};
    \node[draw=none,fill=none] (cdots) [right=of b]{$\cdots$};
    \node[main, label=below:$2$] (c) [right=of cdots]{};
    \node[main, label=right:$1$] (d)[above right=of c]{};
    \node[main, label=right:$1$] (e)[below right=of c]{};
    
    \draw[-] (e)--(c)--(d);
    \draw[-] (c)--(cdots)--(b);
    \draw[double,double distance=3pt,line width=0.4pt] (a)--(b);
    \draw [decorate, 
    decoration = {brace,
        raise=10pt,
        amplitude=5pt}] (c) --  (b) node[pos=0.5,below=15pt,black]{$k-2$};
    
    \node[main, label=below:$1$] (1subT) [below=of e]{};
    \node[main, label=below:$2$] (2sub) [below left=of 1subT]{};
    \node[main, label=below:$1$] (1subB) [left=of 2sub]{};
    \node[draw=none,fill=none] (-) [left=of 1subB]{$-$};
    \draw[-] (1subT)--(2sub)--(1subB);
    
    \node[draw, label=right:$2$] (2F) [below right=of 2sub]{};
    \node[main, label=below:$1$] (1) [below=of 2F]{};
    \draw[-] (2F)--(1);
    
    \node[main, label=below:$1$] (one) [left=of 1]{};
    \node[main, label=below:$2$] (twoR) [left=of one]{};
    \node[draw, label=left:$1$] (oneF) [above= of twoR]{};
    \node[draw=none, fill=none] (cdotsres) [left=of twoR]{$\cdots$};
    \node[main, label=below:$2$] (twoL) [left=of cdotsres]{};
    \node[main, label=below:$1$] (oneL) [left=of twoL]{};
    
    \draw[-] (one)--(twoR)--(cdotsres)--(twoL);
    \draw[-] (oneF)--(twoR);
    
    \draw[double,double distance=3pt,line width=0.4pt] (oneL)--(twoL);
    \draw [decorate, 
    decoration = {brace,
        raise=10pt,
        amplitude=5pt}] (twoR) --  (twoL) node[pos=0.5,below=15pt,black]{$k-4$};
    
    \end{tikzpicture}
    \caption{}
    \label{fig:nminDkA1Sub2}
    \end{subfigure}
    \centering
    \begin{subfigure}{\textwidth}
    \centering
    \begin{tikzpicture}[main/.style={draw,circle}]
    \node[main, label=below:$1$] (one) []{};
    \node[main, label=below:$2$] (twoR) [left=of one]{};
    \node[draw, label=left:$1$] (oneF) [above= of twoR]{};
    \node[draw=none, fill=none] (cdotsres) [left=of twoR]{$\cdots$};
    \node[main, label=below:$2$] (twoL) [left=of cdotsres]{};
    \node[main, label=below:$1$] (oneL) [left=of twoL]{};
    
    \draw[-] (one)--(twoR)--(cdotsres)--(twoL);
    \draw[-] (oneF)--(twoR);
    
    \draw[double,double distance=3pt,line width=0.4pt] (oneL)--(twoL);
    \draw [decorate, 
    decoration = {brace,
        raise=10pt,
        amplitude=5pt}] (twoR) --  (twoL) node[pos=0.5,below=15pt,black]{$k-4$};
    
    \end{tikzpicture}
    \caption{}
    \label{fig:nminDkA1SubInt}
    \end{subfigure}
    \caption{Both alignments of the $\urm(2)$ quotient quiver against \Quiver{fig:quiv_nmin_Dk} for $k\geq 5$ producing quivers \Quiver{fig:nminDkA1Sub1} and \Quiver{fig:nminDkA1Sub2}. Their intersection, reached via $A_1$ KP transitions, is quiver \Quiver{fig:nminDkA1SubInt}.}
    \label{fig:nminDkA1SubBoth}
\end{figure}

An unframed magnetic quiver \Quiver{fig:quiv_nmin_Dk} for the $\overline{n. min. D_k}$ is shown in \Figref{fig:quiv_nmin_Dk}. We consider the subtraction of a $\urm(2)$ quotient quiver.

First consider the special case for $k=4$. There are two possible alignments, as shown in \Figref{fig:nminD4A1SubBoth}, producing quivers \Quiver{fig:nminD4A1Sub1} and \Quiver{fig:nminD4A1Sub2}. The Coulomb branches of these quivers can be identified as $\overline{n. min. B_2}/\mathbb{Z}_2$ (which has a $D_2$ global symmetry) and $\overline{min. A_1}\otimes \overline{max. D_2}$. The former is suggested by \cite{Hanany:2020jzl} and verified by direct computation using the monopole formula. The latter can be seen by moving the framing on the quiver $[2]-(1)=(1)$. We conjecture that:
\begin{equation}
    \overline{n. min. D_4}///\surm(2)=\overline{n.min. B_2}/\mathbb{Z}_2\cup \left(\overline{min. A_1}\otimes\overline{max. D_2}\right),
\end{equation}
To compute the union, we note that the intersection is \Quiver{fig:nminD4A1SubInt}, a magnetic quiver for $\overline{max. D_2}$, reached in both cases by an $A_1$ KP transition. This yields the HWG as the signed sum:
\begin{align}
    HWG\left[\overline{n.min. B_2}/\mathbb{Z}_2\cup \overline{min. A_1}\otimes\overline{max. D_2}\right]&=PE[(\mu_1^2+\mu_1 \mu_2+\mu_2^2) t^2+(1+\mu_1 \mu_2) t^4-\mu_1^2 \mu_2^2 t^8]\nonumber\\&+PE[(\mu_1^2+\mu_2^2+\nu^2)t^2]-PE[(\mu_1^2+\mu_2^2)t^2],
    \label{eq:nminD4A1HWG}
\end{align}
where the $\mu_{1,2}$ are highest weight fugacities for $D_2$, and $\nu$ is a highest weight fugacity for $A_1$. The PE terms in \eqref{eq:nminD4A1HWG} are HWGs for $\overline{n.min. B_2}/\mathbb{Z}_2$ (expressed in $D_2$ fugacities), $\overline{min. A_1}\otimes\overline{max. D_2}$, and their intersection $\overline{max. D_2}$, respectively. The result from quiver subtraction can be confirmed by using Weyl integration with the embedding \eqref{eq:D4A1Embed}. We point out that this form of the HWG is exactly the same as in \eqref{eq:nminB3SU2HWG} and will explore the relationship between the HKQs of B-type and D-type orbits further in Section \ref{sec:Conclusions}.

Now consider the more general case for $k\geq 5$. There are two alignments of the $\urm(2)$ quotient quiver against \Quiver{fig:quiv_nmin_Dk}, as shown in \Figref{fig:nminDkA1SubBoth}, and these produce quivers \Quiver{fig:nminDkA1Sub1} and \Quiver{fig:nminDkA1Sub2}, whose Coulomb branches can be identified as $\overline{n. min. B_{k-2}}/\mathbb{Z}_2$ and $\overline{n. min. D_{k-2}}$, respectively. We conjecture that:
\begin{equation}
\overline{n. min. D_k}///\surm(2)=\overline{n. min. B_{k-2}}/\mathbb{Z}_2\cup \left(\overline{min. A_1}\otimes \overline{n. min. D_{k-2}}\right)
\end{equation}
The intersection \Quiver{fig:nminDkA1SubInt} is a magnetic quiver for $\overline{n. min. D_{k-2}}$, as can be seen by moving the framing onto the node of rank 1 with the double hyper. The HWG for this hyper-Kähler quotient follows as:
\begin{align}
    HWG[\overline{n. min. D_{k}}///\surm(2)]&=PE\left[\mu_2t^2+\left(2\mu_1^2+1\right)t^4+\mu_1^2t^6-\mu_1^4t^{12}\right]+PE[\nu^2t^2]PE\left[\mu_2t^2+\mu_1^2t^4\right]\nonumber\\&-PE\left[\mu_2t^2+\mu_1^2t^4\right].
\end{align}
where $\mu_{1,2}$ are highest weight fugacities for $D_k$ and $\nu$ is a highest weight fugacity for $A_1$. The first PE term is the HWG for $\overline{n. min. B_{k-2}}/\mathbb{Z}_2$, the second term is the HWG for $\overline{min A_1}\otimes \overline{n.min D_{k-2}}$, and the third is the HWG for $\overline{n.min. D_{k-2}}$.

\subsection{D-type Orbit $\surm(3)$ HKQ}
\label{sec:DtypeSU3HKQ}
\subsubsection{$\overline{n. n. min. D_6}///\surm(3)$}
\begin{figure}[h]
    \centering
    \begin{tikzpicture}[main/.style= {draw,circle}]
    \node[main, label=below:$1$] (1) {};
    \node[main, label=below:$2$] (2) [right=of 1]{};
    \node[main, label=below:$3$] (3) [right=of 2]{};
    \node[main, label=below:$4$] (4) [right=of 3]{};
    \node[main, label=right:$2$,magenta,fill] (2T) [above right=of 4]{};
    \node[main, label=right:$2$,goldenrod,fill] (2B) [below right=of 4]{};
    \node[main, label=left:$1$,teal,fill] (1T) [above=of 4]{};

    \draw[-] (1)--(2)--(3)--(4)--(2T);
    \draw[-] (2B)--(4)--(1T);
        
    \end{tikzpicture}
    \caption{Magnetic quiver \Quiver{fig:nnnminD6Quiv} for the $\overline{n.n. min. D_6}$.}
    \label{fig:nnnminD6Quiv}
\end{figure}
The magnetic quiver \Quiver{fig:nnnminD6Quiv} for the $\overline{n.n.min. D_6}$ is shown in \Figref{fig:nnnminD6Quiv}. We have used different colours for the nodes attached to the node of rank 4. We consider an $\surm(3)$ HKQ  by the subtraction of the $\urm(3)$ quotient quiver. There are three alignments in which the quotient quiver starts on the uncoloured nodes and ends on the \textcolor{teal}{teal}, \textcolor{magenta}{magenta}, and \textcolor{goldenrod}{gold} nodes. The respective resulting quivers $\mathcal{Q}_{\ref{fig:nnnminD6Quiv} \textcolor{blue}{a}}$, $\mathcal{Q}_{\ref{fig:nnnminD6Quiv} \textcolor{blue}{b}}$, and $\mathcal{Q}_{\ref{fig:nnnminD6Quiv} \textcolor{blue}{c}}$ are:
\begin{equation}
    \mathcal{Q}_{\ref{fig:nnnminD6Quiv} \textcolor{blue}{a}}=\resizebox{!}{30pt}{\begin{tikzpicture}[baseline=(current bounding box.center), main/.style = {draw, circle},scale=0.25]

    \node[main, label=below:$2$,magenta,fill] (orange){};
    \node[main, label=below:$2$] (2) [right=of orange]{};
    \node[main, label=below:$2$,goldenrod,fill] (purple) [right=of 2]{};
    \node[draw, label=left:$2$] (2F) [above=of orange]{};
    \node[draw, label=right:$2$] (2FR) [above=of purple]{};

    \draw[-] (2F)--(orange)--(2)--(purple)--(2FR);
    \end{tikzpicture}};\quad \mathcal{Q}_{\ref{fig:nnnminD6Quiv} \textcolor{blue}{b}}=\resizebox{!}{30pt}{\begin{tikzpicture}[baseline=(current bounding box.center), main/.style = {draw, circle},scale=0.25]
    \node[main, label=below:$1$,magenta,fill] (orange) {};
    \node[main, label=below:$2$] (2) [right=of orange]{};
    \node[main, label=below:$2$,goldenrod,fill] (purple) [right=of 2]{};
    \node[main, label=right:$1$,teal,fill] (green) [above=of 2]{};
    \node[draw, label=left:$2$] (2F) [left=of green]{};
    \node[draw, label=right:$2$] (2FT) [above=of purple]{};

    \draw[-] (orange)--(2)--(purple)--(2FT);
    \draw[-] (2)--(green)--(2F);
    \end{tikzpicture}};\quad \mathcal{Q}_{\ref{fig:nnnminD6Quiv} \textcolor{blue}{c}}=\resizebox{!}{30pt}{\begin{tikzpicture}[baseline=(current bounding box.center), main/.style = {draw, circle},scale=0.25]
    \node[main, label=below:$1$,goldenrod,fill] (purple) {};
    \node[main, label=below:$2$] (2) [right=of purple]{};
    \node[main, label=below:$2$,magenta,fill] (orange) [right=of 2]{};
    \node[main, label=right:$1$,teal,fill] (green) [above=of 2]{};
    \node[draw, label=left:$2$] (2F) [left=of green]{};
    \node[draw, label=right:$2$] (2FT) [above=of orange]{};

    \draw[-] (purple)--(2)--(orange)--(2FT);
    \draw[-] (2)--(green)--(2F);
    \end{tikzpicture}.}
\end{equation}
We compute the two-way intersections to be: \begin{equation}
    \mathcal{Q}_{\ref{fig:nnnminD6Quiv} \textcolor{blue}{a}}\cap\mathcal{Q}_{\ref{fig:nnnminD6Quiv} \textcolor{blue}{b}}=\resizebox{!}{30pt}{\begin{tikzpicture}[baseline=(current bounding box.center), main/.style = {draw, circle},scale=0.25]
        \node[main, label=below:$1$,magenta,fill] (orange) {};
        \node[main, label=below:$2$] (2) [right=of orange]{};
        \node[main, label=below:$2$,goldenrod,fill] (purple) [right=of 2]{};
        \node[draw, label=left:$1$] (1F) [above=of 2]{};
        \node[draw, label=right:$2$] (2F) [above=of purple]{};

        \draw[-] (orange)--(2)--(purple)--(2F);
        \draw[-] (2)--(1F);
    \end{tikzpicture}};\quad \mathcal{Q}_{\ref{fig:nnnminD6Quiv} \textcolor{blue}{a}}\cap\mathcal{Q}_{\ref{fig:nnnminD6Quiv} \textcolor{blue}{c}}=\resizebox{!}{30pt}{\begin{tikzpicture}[baseline=(current bounding box.center), main/.style = {draw, circle},scale=0.25]
    \node[main, label=below:$1$,goldenrod,fill] (purple) {};
        \node[main, label=below:$2$] (2) [right=of purple]{};
        \node[main, label=below:$2$,magenta,fill] (orange) [right=of 2]{};
        \node[draw, label=left:$1$] (1F) [above=of 2]{};
        \node[draw, label=right:$2$] (2F) [above=of orange]{};

        \draw[-] (purple)--(2)--(orange)--(2F);
        \draw[-] (2)--(1F);
        
    \end{tikzpicture}};\quad \mathcal{Q}_{\ref{fig:nnnminD6Quiv} \textcolor{blue}{b}}\cap\mathcal{Q}_{\ref{fig:nnnminD6Quiv} \textcolor{blue}{c}}=\resizebox{!}{30pt}{\begin{tikzpicture}[baseline=(current bounding box.center), main/.style = {draw, circle},scale=0.25]
         \node[main, label=below:$1$,magenta,fill] (orange) {};
    \node[main, label=below:$2$] (2) [right=of orange]{};
    \node[main, label=below:$1$,goldenrod,fill] (purple) [right=of 2]{};
    \node[main, label=right:$1$,teal,fill] (green) [above=of 2]{};
    \node[draw, label=left:$2$] (2F) [left=of green]{};
    \node[draw, label=right:$1$] (1F) [above right=of 2]{};

    \draw[-] (orange)--(2)--(purple);
    \draw[-] (2)--(green)--(2F);
    \draw[-] (2)--(1F);
    \end{tikzpicture},}
\end{equation}
and, finally the three-way intersection is:
\begin{equation}
\mathcal{Q}_{\ref{fig:nnnminD6Quiv} \textcolor{blue}{a}}\cap\mathcal{Q}_{\ref{fig:nnnminD6Quiv} \textcolor{blue}{b}}\cap\mathcal{Q}_{\ref{fig:nnnminD6Quiv} \textcolor{blue}{c}}=\resizebox{!}{30pt}{\begin{tikzpicture}[baseline=(current bounding box.center), main/.style = {draw, circle},scale=0.25]
        \node[main, label=below:$1$,magenta,fill] (orange){};
        \node[main, label=below:$2$] (2) [right=of orange]{};
        \node[main, label=below:$1$,goldenrod,fill] (1) [right=of 2]{};
        \node[draw,label=left:$2$] (2F) [above=of 2]{};

        \draw[-] (orange)--(2)--(purple);
        \draw[-] (2F)--(2);
    \end{tikzpicture}}.
\end{equation}
All of the quivers share a global symmetry of $D_3 \times \urm(1)$, although the $\urm(1)$ may be trivial. The quiver subtraction and Weyl integration are found to give consistent results. The HWGs for these quivers are not particularly illuminating, so we present some unrefined HS to illustrate the details of the construction:
\begin{align}
    \hs[\mathcal{Q}_{\ref{fig:nnnminD6Quiv} \textcolor{blue}{a}}]&=\frac{\left(\begin{aligned}1 &+ 9 t^2 + 54 t^4 + 194 t^6 + 471 t^8 + 771 t^{10} + 916 t^{12}\\ & + 
 771 t^{14} + 471 t^{16}+ 194 t^{18} + 54 t^{20} + 
 9 t^{22} + t^{24}\end{aligned}\right)}{(1 - t^2)^{12}(1 + t^2)^6}\\\hs[\mathcal{Q}_{\ref{fig:nnnminD6Quiv} \textcolor{blue}{b}}]=\hs[\mathcal{Q}_{\ref{fig:nnnminD6Quiv} \textcolor{blue}{c}}]&=\frac{\left(\begin{aligned}1 &+ 8 t^2 + 43 t^4 + 128 t^6 + 238 t^8 + 288 t^{10} \\&+ 238 t^{12} + 
 128 t^{14} + 43 t^{16} + 
 8 t^{18} + t^{20}\end{aligned}\right)}{(1 - t^2)^{12}(1 + t^2)^4}\\\hs[\mathcal{Q}_{\ref{fig:nnnminD6Quiv} \textcolor{blue}{a}}\cap\mathcal{Q}_{\ref{fig:nnnminD6Quiv} \textcolor{blue}{b}}]=\hs[\mathcal{Q}_{\ref{fig:nnnminD6Quiv} \textcolor{blue}{a}}\cap\mathcal{Q}_{\ref{fig:nnnminD6Quiv} \textcolor{blue}{c}}]&=\frac{(1 + t^2) (1 + 4 t^2 + 10 t^4 + 4 t^6 + t^8)}{(1 - t^2)^{10}}\\\hs[\mathcal{Q}_{\ref{fig:nnnminD6Quiv} \textcolor{blue}{b}}\cap\mathcal{Q}_{\ref{fig:nnnminD6Quiv} \textcolor{blue}{c}}]&=\frac{\left(\begin{aligned}1 &+ 11 t^2 + 57 t^4 + 170 t^6 + 324 t^8 + 398 t^{10}\\& + 324 t^{12} + 
 170 t^{14} + 57 t^{16} + 11 t^{18} + t^{20}\end{aligned}\right)}{(1 - t^2)^{10} (1 + t^2)^5}\\\hs[\mathcal{Q}_{\ref{fig:nnnminD6Quiv} \textcolor{blue}{a}}\cap\mathcal{Q}_{\ref{fig:nnnminD6Quiv} \textcolor{blue}{b}}\cap\mathcal{Q}_{\ref{fig:nnnminD6Quiv} \textcolor{blue}{c}}]&=\frac{(1 + t^2)^2 (1 + 5 t^2 + t^4)}{(1 - t^2)^8}.
\end{align}Then one can check that \begin{align}
    \hs[HKQ]&=\hs[\mathcal{Q}_{\ref{fig:nnnminD6Quiv} \textcolor{blue}{a}}]+\hs[\mathcal{Q}_{\ref{fig:nnnminD6Quiv} \textcolor{blue}{b}}]+\hs[\mathcal{Q}_{\ref{fig:nnnminD6Quiv} \textcolor{blue}{c}}]-\hs[\mathcal{Q}_{\ref{fig:nnnminD6Quiv} \textcolor{blue}{a}}\cap\mathcal{Q}_{\ref{fig:nnnminD6Quiv} \textcolor{blue}{b}}]-\hs[\mathcal{Q}_{\ref{fig:nnnminD6Quiv} \textcolor{blue}{a}}\cap\mathcal{Q}_{\ref{fig:nnnminD6Quiv} \textcolor{blue}{c}}]\nonumber\\&-\hs[\mathcal{Q}_{\ref{fig:nnnminD6Quiv} \textcolor{blue}{b}}\cap\mathcal{Q}_{\ref{fig:nnnminD6Quiv} \textcolor{blue}{c}}]+\hs[\mathcal{Q}_{\ref{fig:nnnminD6Quiv} \textcolor{blue}{a}}\cap\mathcal{Q}_{\ref{fig:nnnminD6Quiv} \textcolor{blue}{b}}\cap\mathcal{Q}_{\ref{fig:nnnminD6Quiv} \textcolor{blue}{c}}]\\&=\frac{\left(\begin{aligned}1 &+ 10 t^2 + 76 t^4 + 365 t^6 + 1188 t^8 + 2571 t^{10} + 3675 t^{12} + 
 3315 t^{14} \\&+ 1701 t^{16} + 267 t^{18} - 176 t^{20} - 90 t^{22} - 2 t^{24} + 
 6 t^{26} + t^{28}\end{aligned}\right)}{(1 - t^2)^{12} (1 + t^2)^6}
\end{align}

\subsection{D-type Orbit $\surm(n)$ HKQ}
\label{sec:DtypeSUNHKQ}
Quiver subtraction allows us to conjecture a general construction for any permissible hyper-Kähler quotient of a height 2 orbit of $D_k$ by $\surm(n)$. The structure of the quiver subtraction differs depending on whether the spinor nodes are involved. 
\subsubsection{$\overline{\mathcal O}^{D_k}_{(2^{2p},1^{2k-4p})}///\surm(n), n\leq p+1$}
\begin{figure}[h!]
    \centering
    \begin{tikzpicture}[main/.style = {draw, circle}]
    \node[main, label=below:$1$] (1L) []{};
    \node[main, label=below:$2$] (2L) [right=of 1L]{};
    \node[draw=none,fill=none] (cdotsL) [right=of 2L]{$\cdots$};
    \node[main, label=below:$2p$] (2pL) [right=of cdotsL]{};
    \node[draw=none,fill=none] (cdotsR) [right=of 2pL]{$\cdots$};
    \node[main, label=below:$2p$] (2pR) [right=of cdotsR]{};
    \node[main, label=right:$p$] (2T) [above right=of 2pR]{};
    \node[main, label=right:$p$] (2B) [below right=of 2pR]{};
    \node[main, label=right:$1$] (1T) [above =of 2pL]{};
    \draw[-] (1L)--(2L)--(cdotsL)--(2pL)--(cdotsR)--(2pR);
    \draw[-] (1T)--(2pL);
    \draw[-] (2T)--(2pR)--(2B);
    \draw [decorate, 
    decoration = {brace,
        raise=10pt,
        amplitude=5pt}] (2pR) --  (2pL) node[pos=0.5,below=15pt,black]{$k-(2p+1)$};
    \end{tikzpicture}
    \caption{Unframed magnetic quiver \Quiver{fig:Dk22pquiv} for the orbit of $D_k$ labelled by the partition $(2^{2p},1^{2k-4p})$ of $2k$.}
    \label{fig:Dk22pquiv}
\end{figure}

First, consider the orbit of $D_k$ with partition $(2^{2p},1^{2k-4p})$ of $2k$, where $k\geq 2p+2$, for which an unframed magnetic quiver \Quiver{fig:Dk22pquiv} is given in \Figref{fig:Dk22pquiv}. To be consistent with the rules of quiver subtraction, the $\urm(n)$ quotient quiver can only be subtracted from the leg that starts $(1)-(2)-\cdots-(2p) - \cdots$; this puts a constraint on the values of $n$. The length of the $\urm(n)$ quotient quiver is $2n-1$ and the length of the leg $(1)-(2)-\cdots-(2p)$, up to the junction with the rank 1 node, is $2p$. Since the rules permit the quotient quiver to run past this junction by one node at most, the constraint becomes $2n-1\leq2p+1\implies n\leq p+1$. There are two cases; one for $n<p+1$ and a special case for $n=p+1$.

If $n<p+1$, then there is only one alignment of the $\urm(n)$ quotient quiver. The result of the quiver subtraction is:
\begin{equation}
    \overline{\mathcal O}^{D_k}_{(2^{2p},1^{2k-4p})}///\surm(n)=\mathcal{C}\left( \resizebox{!}{32pt}{\begin{tikzpicture}[baseline=(current bounding box.center), main/.style = {draw, circle},scale=0.25]
    \node[main, label=below:$2$] (twoL) []{};
    \node[main, label=below:$4$] (fourL) [right=of twoL]{};
    \node[draw=none,fill=none] (cdotsL) [right=of fourL]{$\cdots$};
    \node[main,label=below:$2n-2$] (2m) [right=of cdotsL]{};
    \node[main, label=below:$2n-1$] (2mplus1) [right=of 2m]{};
    \node[draw=none,fill=none] (cdotsR) [right=of 2mplus1]{$\cdots$};
    \node[main, label=below:$2p$] (2pL) [right=of cdotsR]{};
    \node[draw=none,fill=none] (cdotsRR) [right=of 2pL]{$\cdots$};
    \node[main, label=below:$2p$] (2pR) [right=of cdotsRR]{};
    \node[main, label=right:$p$] (pT) [above right=of 2pR]{};
    \node[main, label=right:$p$] (pB) [below right=of 2pR]{};
    \node[draw,label=left:$1$] (1FL) [above=of 2m]{};
    \node[main, label=left:$1$] (1g) [above=of 2pL]{};
    
    \draw[-] (twoL)--(fourL)--(cdotsL)--(2m)--(2mplus1)--(cdotsR)--(2pL)--(cdotsRR)--(2pR);
    \draw[-] (pT)--(2pR)--(pB);
    \draw[-] (1FL)--(2m);
    \draw[-] (2pL)--(1g);
    \draw [decorate, 
    decoration = {brace,
        raise=10pt,
        amplitude=5pt}] (2pR) --  (2pL) node[pos=0.5,below=15pt,black]{$k-(2p+1)$};      
    \end{tikzpicture}}\right),
    \end{equation}
This magnetic quiver has a global symmetry of $SO(2k-2n)\times \urm(1)$.

In the saturated case of $n=p+1$ this further degenerates to two special cases. The first is for $k>2p+2$, there are two alignments of the $\urm(n)$ quotient quiver and so the $\surm(n)$ HKQ is a union according to the rules in \ref{sec:rules}. We find:
    \begin{align}
    \overline{\mathcal O}^{D_k}_{(2^{2p},1^{2k-4p})}///\surm(p+1)&=\mathcal C\left(\resizebox{!}{40pt}{\begin{tikzpicture}[baseline=(current bounding box.center), main/.style = {draw, circle},scale=0.25]
    \node[main, label=below:$2$] (twoL) []{};
    \node[main, label=below:$4$] (fourL) [right=of twoL]{};
    \node[draw=none,fill=none] (cdotsL) [right=of fourL]{$\cdots$};
    \node[main, label=below:$2p-2$] (2pminus2) [right=of cdotsL]{};
    \node[main, label=below:$2p$] (2pL) [right=of 2pminus2]{};
    \node[draw=none,fill=none] (cdotsR) [right=of 2pL]{$\cdots$};
    \node[main, label=below:$2p$] (2pR) [right=of cdotsR]{};
    \node[main, label=right:$p$] (pT) [above right=of 2pR]{};
    \node[main, label=right:$p$] (pB) [below right=of 2pR]{};
    \node[draw,label=left:$2$] (twoF) [above =of 2pL]{};
    \draw[-] (twoL)--(fourL)--(cdotsL)--(2pminus2)--(2pL)--(cdotsR)--(2pR);
    \draw[-] (pT)--(2pR)--(pB);
    \draw[-] (twoF)--(2pL);
    \draw [decorate, 
    decoration = {brace,
        raise=20pt,
        amplitude=5pt}] (2pR) --  (2pminus2) node[pos=0.5,below=25pt,black]{$k-(2p+1)$};
    \end{tikzpicture}}\right)\nonumber\\&\cup\mathcal C\left(\resizebox{!}{50pt}{\begin{tikzpicture}[baseline=(current bounding box.center), main/.style = {draw, circle},scale=0.25]
    \node[main, label=below:$2$] (twoL) []{};
    \node[main, label=below:$4$] (fourL) [right=of twoL]{};
    \node[draw=none,fill=none] (cdotsL) [right=of fourL]{$\cdots$};
    \node[main, label=below:$2p-2$] (2pminus2) [right=of cdotsL]{};
    \node[main, label=below:$2p-1$] (2pminus1)[right=of 2pminus2]{};
    \node[main, label=below:$2p$] (2pL) [right=of 2pminus1]{};
    \node[draw=none,fill=none] (cdotsR)[right=of 2pL]{$\cdots$};
    \node[main, label=below:$2p$] (2pR) [right=of cdotsR]{};
    \node[main, label=right:$p$] (pT) [above right=of 2pR]{};
    \node[main, label=right:$p$] (pB) [below right=of 2pR]{};
    \node[main, label=left:$1$] (one) [above =of 2pminus2]{};
    \node[draw,label=left:$2$] (twoF) [above =of one]{};
    \node[draw,label=left:$1$] (oneF) [above=of 2pL]{};
    \draw[-] (twoL)--(fourL)--(cdotsL)--(2pminus2)--(2pminus1)--(2pL)--(cdotsR)--(2pR);
    \draw[-] (twoF)--(one)--(2pminus2);
    \draw[-] (oneF)--(2pL);
    \draw[-] (pT)--(2pR)--(pB);
     \draw [decorate, 
    decoration = {brace,
        raise=20pt,
        amplitude=5pt}] (2pR) --  (2pminus2) node[pos=0.5,below=25pt,black]{$k-(2p+1)$};
    \end{tikzpicture}}\right)
\end{align}

The first quiver in the union is the affine Grassmannian slice $\overline{\left[\mathcal W_{D_{k-n}}\right]}^{[0,\cdots,0,1,0\cdots,0]]}_{[0,\cdots,0]}$ where the $1$ in the flavour vector is in the $p^{th}$ position. This moduli space has a global symmetry of $SO(2k-2n)$. There is no particular name for the  second moduli space but it has a global symmetry of $SO(2k-2n)\times \urm(1)$.

The case for $k=2p+2$ has three possible alignments of the $\surm(p+1)$ quotient quiver and so the HKQ is also a union according to the rules in Section \ref{sec:rules}. We find:

\begin{align}
    \overline{\mathcal O}^{D_{2p+2}}_{(2^{2p},1^{2})}///\surm(p+1)&=\mathcal C\left(\resizebox{!}{60pt}{\begin{tikzpicture}[baseline=(current bounding box.center), main/.style = {draw, circle},scale=0.25]
    \node[main, label=below:$2$] (2) []{};
    \node[main, label=below:$4$] (4) [right=of 2]{};
    \node[] (cdots) [right=of 4] {$\cdots$};
    \node[main, label=right:$2p-2$] (2pm2) [right=of cdots]{};
    \node[main, label=below:$p$,fill=black] (pt) [above right=of 2pm2]{};
    \node[main, label=below:$p$] (pb) [below right=of 2pm2]{};
    \node[draw, label=below:$2$] (2ft) [right=of pt]{};
    \node[draw, label=below:$2$] (2fb) [right=of pb]{};
    \draw[-] (2)--(4)--(cdots)--(2pm2)--(pt)--(2ft);
    \draw[-] (2fb)--(pb)--(2pm2);
    \end{tikzpicture}}\right)\nonumber\\&\cup\mathcal C\left(\resizebox{!}{85pt}{\begin{tikzpicture}[baseline=(current bounding box.center), main/.style = {draw, circle},scale=0.25]
    \node[main, label=below:$2$] (2) []{};
    \node[main, label=below:$4$] (4) [right=of 2]{};
    \node[] (cdots) [right=of 4] {$\cdots$};
    \node[main, label=right:$2p-2$] (2pm2) [right=of cdots]{};
    \node[main, label=below:$p-1$,fill=black] (pt) [above right=of 2pm2]{};
    \node[main, label=below:$p$] (pb) [below right=of 2pm2]{};
    \node[draw, label=below:$2$] (2fb) [right=of pb]{};
    \node[main, label=left:$1$] (1) [above=of 2pm2]{};
    \node[draw, label=left:$2$] (2f) [above=of 1]{};
    \draw[-] (2)--(4)--(cdots)--(2pm2)--(1)--(2f);
    \draw[-] (pt)--(2pm2)--(pb)--(2fb);
    \end{tikzpicture}}\right)\nonumber\\&\cup \mathcal C\left(\resizebox{!}{85pt}{\begin{tikzpicture}[baseline=(current bounding box.center), main/.style = {draw, circle},scale=0.25]
    \node[main, label=below:$2$] (2) []{};
    \node[main, label=below:$4$] (4) [right=of 2]{};
    \node[] (cdots) [right=of 4] {$\cdots$};
    \node[main, label=right:$2p-2$] (2pm2) [right=of cdots]{};
    \node[main, label=below:$p-1$] (pt) [above right=of 2pm2]{};
    \node[main, label=below:$p$,fill=black] (pb) [below right=of 2pm2]{};
    \node[draw, label=below:$2$] (2fb) [right=of pb]{};
    \node[main, label=left:$1$] (1) [above=of 2pm2]{};
    \node[draw, label=left:$2$] (2f) [above=of 1]{};
    \draw[-] (2)--(4)--(cdots)--(2pm2)--(1)--(2f);
    \draw[-] (pt)--(2pm2)--(pb)--(2fb);
    \end{tikzpicture}}\right).\label{eq:DUnionof3}
    \end{align}

The nodes of rank $p$ attached to the node of rank $2p-2$ are identical under an outer autmorphism and so we fill one of these nodes black and leave the other unfilled. The first Coulomb branch in the union \eqref{eq:DUnionof3} is the affine Grassmannian slice $\overline{\left[\mathcal W_{D_{p+1}}\right]}^{[0,\cdots,0,2,2]}_{[0,\cdots,0]}$ which has an $SO(2p+2)$ global symmetry. The final two moduli spaces in the union \eqref{eq:DUnionof3} are identical but have slightly different fugacities as made clear by the colouring of the nodes. This moduli space does not have any particular name but does have a global symmetry of $SO(2p+2)\times \urm(1)$.

\subsubsection{$\overline{\mathcal O}^{D_{2k}}_{(2^{2k})}///\surm(n), n\leq k$}
\begin{figure}[h!]
    \centering
    \begin{tikzpicture}[main/.style={draw,circle}]
        \node[main, label=below:$1$] (1) []{};
        \node[main, label=below:$2$] (2) [right=of 1]{};
        \node[] (cdots) [right=of 2]{$\cdots$};
        \node[main, label=right:$2k-2$] (2km2) [right=of cdots]{};
        \node[main, label=above:$k$] (k) [above right=of 2km2]{};
        \node[main, label=below:$k-1$] (km1) [below right=of 2km2]{};
        \node[main, label=right:$1$] (2F) [right=of k]{};

        \draw[-] (1)--(2)--(cdots)--(2km2)--(k);
        \draw[-] (2km2)--(km1);
        \draw[double,double distance=3pt,line width=0.4pt] (k)--(2F);
    \end{tikzpicture}
    \caption{Unframed magnetic quiver \Quiver{fig:OrbD2k} for $\overline{\mathcal O}^{D_{2k}}_{(2^{2k})}$.}
    \label{fig:OrbD2k}
\end{figure}
Next consider the possible $\urm(n)$ quotient quiver subtractions from the unframed magnetic quiver \Quiver{fig:OrbD2k} for $\overline{\mathcal O}^{D_{2k}}_{(2^{2k})}$, shown in \Figref{fig:OrbD2k}. From consideration of the length of the $\urm(n)$ quotient quiver, it is clear that $n\leq k$, with the general case for $n<k$ and the special case for $n=k$.

In the general case for $n<k$, there is only one possible alignment of the quotient quiver, resulting in: \begin{equation}
    \overline{\mathcal O}^{D_{2k}}_{(2^{2k})}///\surm(n)=\mathcal C\left(\resizebox{!}{50pt}{\begin{tikzpicture}[baseline=(current bounding box.center), main/.style = {draw, circle},scale=0.25]

    \node[main, label=below:$2$] (2) []{};
    \node[main, label=below:$4$] (4) [right=of 2]{};
    \node[draw=none,fill=none] (cdotsL) [right=of 4]{$\cdots$};
    \node[main, label=below:$2n-2$] (2nm2) [right=of cdotsL]{};
    \node[main ,label=below:$2n-1$] (2nm1) [right=of 2nm2]{};
    \node[draw=none,fill=none] (cdotsR) [right=of 2nm1]{$\cdots$};
    \node[main, label=right:$2k-1$] (2km1) [right=of cdotsR]{};
    \node[main, label=above:$k$] (k) [above right=of 2km1]{};
    \node[main, label=below:$k-1$] (km1) [below right=of 2km1]{};
    \node[draw, label=right:$2$] (2F) [right=of k]{};
    \node[main, label=left:$1$] (1T) [above=of 2nm2]{};
    
    \draw[-] (2)--(4)--(cdotsL)--(2nm2)--(2nm1)--(cdotsR)--(2km1)--(k)--(2F);
    \draw[-] (2km1)--(km1);
    \draw[-] (2nm2)--(1T);
    \end{tikzpicture}}\right).
\end{equation}
The global symmetry of the Coulomb branch is $SO(4k-2n)\times \urm(1)$.

Now consider the limiting case where $n=k$; the $\urm(n)$ quotient quiver has two possible alignments and so the HKQ is a union:
\begin{align}
    \overline{\mathcal O}^{D_{2k}}_{(2^{2k})}///\surm(n)&s
    =\mathcal C\left(\resizebox{!}{50pt}{\begin{tikzpicture}[baseline=(current bounding box.center), main/.style = {draw, circle},scale=0.25]
    \node[main, label=below:$2$] (2) []{};
    \node[main, label=below:$4$] (4) [right=of 2]{};
    \node[] (cdots) [right=of 4]{$\cdots$};
    \node[main, label=right:$2k-4$] (2km4) [right=of cdots]{};
    \node[main, label=above:$k-1$] (km1) [above right=of 2km4]{};
    \node[main, label=below:$k-1$] (km12) [below right=of 2km4]{};
    \node[draw, label=right:$2$] (2F2) [right=of km12]{};
    \node[main, label=below:$1$] (1) [right=of km1]{};
    \node[draw, label=right:$2$] (2F) [right=of 1]{};
    \draw[-] (2)--(4)--(cdots)--(2km4)--(km12)--(2F2);
    \draw[-] (2km4)--(km12);
    \draw[-] (2km4)--(km1);
    \draw[double,double distance=3pt,line width=0.4pt] (km1)--(1);
    \draw[-] (1)--(2F);
    \end{tikzpicture}}\right)\nonumber\\&\cup\mathcal C\left(\resizebox{!}{70pt}{\begin{tikzpicture}[baseline=(current bounding box.center), main/.style = {draw, circle},scale=0.25]
    \node[main, label=below:$2$] (2) []{};
    \node[main, label=below:$4$] (4) [right=of 2]{};
    \node[] (cdots) [right=of 4]{$\cdots$};
    \node[main, label=right:$2k-4$] (2km4) [right=of cdots]{};
    \node[main, label=below:$k-2$] (km2) [below right=of 2km4]{};
    \node[main, label=below:$k$] (k) [above right=of 2km4]{};
    \node[main, label=below:$1$] (1) [right=of k]{};
    \node[draw, label=right:$2$] (2F) [above=of k]{};
    \draw[-] (2)--(4)--(cdots)--(2km4)--(km2);
    \draw[-] (2km4)--(k)--(2F);
    \draw[double,double distance=3pt,line width=0.4pt] (k)--(1);
    \end{tikzpicture}}\right).
\end{align}
Both quivers in the union have an $SO(2k)\times \urm(1)$ global symmetry.

\subsubsection{$\overline{\mathcal O}^{D_{2k+1}}_{(2^{2k},1^2)}///\surm(n), n\leq k$}
\begin{figure}[h!]
    \centering
    \begin{tikzpicture}[main/.style={draw,circle}]
        \node[main, label=below:$1$] (1) []{};
        \node[main, label=below:$2$] (2) [right=of 1]{};
        \node[] (cdots) [right=of 2]{$\cdots$};
        \node[main, label=right:$2k-1$] (2km1) [right=of cdots]{};
        \node[main, label=above:$k$] (k) [above right=of 2km1]{};
        \node[main, label=below:$k$] (k2) [below right=of 2km1]{};
        \node[main, label=right:$1$] (12) [below right=of k]{};;

        \draw[-] (1)--(2)--(cdots)--(2km1)--(k)--(12);
        \draw[-] (2km1)--(k2)--(12);

    \end{tikzpicture}
    \caption{Unframed magnetic quiver \Quiver{fig:OrbD2kp1} for $\overline{\mathcal O}^{D_{2k}}_{(2^{2k},1^2)}$.}
    \label{fig:OrbD2kp1}
\end{figure}

Finally, consider the possible $\surm(n)$ quiver subtractions from the magnetic quiver \Quiver{fig:OrbD2kp1} for $\overline{\mathcal O}^{D_{2k}}_{(2^{2k},1^2)}$. Recalling that the junction, located at the node of rank $2k-1$ in \Quiver{fig:OrbD2kp1}, must occur on a node of 2 on the quotient quiver in accordance with the \hyperlink{rule:Junction}{Junction Rule}, we find that $n\leq k$.

Considering first the general case $n<k$, there is only one alignment of the $\urm(n)$ quotient quiver and we find \begin{equation}
    \overline{\mathcal O}^{D_{2k}}_{(2^{2k},1^2)}///\surm(n)=\mathcal C\left(\resizebox{!}{50pt}{\begin{tikzpicture}[baseline=(current bounding box.center), main/.style = {draw, circle},scale=0.25]
        \node[main, label=below:$2$] (2) []{};
        \node[main, label=below:$4$] (4) [right=of 2]{};
        \node[] (cdotsL) [right=of 4]{$\cdots$};
        \node[main, label=below:$2n-2$] (2nm2) [right=of cdotsL]{};
        \node[main, label=below:$2n-1$] (2nm1) [right=of 2nm2]{};
        \node[] (cdotsR) [right=of 2nm1] {$\cdots$};
        \node[main, label=right:$2k-1$] (2km1) [right=of cdotsR]{};
        \node[main, label=above:$k$] (k) [above right=of 2km1]{};
        \node[main, label=below:$k$] (k2) [below right=of 2km1]{};
        \node[main, label=right:$1$] (12) [below right=of k]{};
        \node[draw, label=left:$1$] (1F) [above=of 2nm2]{};

        \draw[-] (2)--(4)--(cdotsL)--(2nm2)--(2nm1)--(cdotsR)--(2km1)--(k)--(12);
        \draw[-] (2km1)--(k2)--(12);
        \draw[-] (2nm2)--(1F);

    \end{tikzpicture}}\right),
\end{equation}where the Coulomb branch has an $SO(4k+2-2n)\times \urm(1)$ global symmetry.

For the special case $n=k$ there is again only one alignment of the $\urm(k)$ quotient quiver which results in
\begin{equation}
    \overline{\mathcal O}^{D_{2k}}_{(2^{2k},1^2)}///\surm(k)=\mathcal C\left(\resizebox{!}{50pt}{\begin{tikzpicture}[baseline=(current bounding box.center), main/.style = {draw, circle},scale=0.25]
        \node[main, label=below:$2$] (2) []{};
        \node[main, label=below:$4$] (4) [right=of 2]{};
        \node[] (cdotsL) [right=of 4]{$\cdots$};
        \node[main, label=right:$2k-2$] (2km1) [right=of cdotsL]{};
        \node[main, label=above:$k$] (k) [above right=of 2km1]{};
        \node[main, label=below:$k$] (k2) [below right=of 2km1]{};
        \node[main, label=right:$1$] (12) [below right=of k]{};

        \node[draw, label=right:$1$] (1F) [right= of k]{};
        \node[draw, label=right:$1$] (1F2) [right= of k2]{};

        \draw[-] (2)--(4)--(cdotsL)--(2km1)--(k)--(12);
        \draw[-] (2km1)--(k2)--(12);
        \draw[-] (k)--(1F);
        \draw[-] (k2)--(1F2);
        \end{tikzpicture}}\right),
\end{equation}the global symmetry of the Coulomb branch here is $SO(2k+2)\times \urm(1)$.

\section{Exceptional Orbits}
\label{sec:ExceptionalHKQ}
We now consider the possible $\surm(n)$ HKQs that one can take from exceptional orbits using quiver subtraction. Unlike, the classical case, there are no families of exceptional orbits, so these orbits are considered on a case-by-case basis. We believe the following is an exhaustive list of the magnetic quivers for exceptional orbits amenable to this approach.

We examine the $\surm(2)$ HKQs of $\overline{sub.reg. G_2}$, $\overline{min. F_4}$, $\overline{min. E_6}$, $\overline{min. E_7}$,and $\overline{min. E_8}$. We also study the $\surm(3)$ HKQ of $\overline{min. E_7}$ and $\overline{min. E_8}$ (the latter was first computed in \cite{Hanany:2022itc}). Finally we consider the $\surm(4)$ HKQ of $\overline{min. E_8}$. 

The relevant embedding of $\surm(n)$ into the fundamental will be given when studying each example.
\subsection{Exceptional Orbit $\surm(2)$ HKQ}
\label{sec:ExceptionalSU2HKQ}
\subsubsection{$\overline{sub. reg. G_2}///\surm(2)$}
\label{sec:subregG2A1}
\begin{figure}[h!]
    \centering
    \begin{tikzpicture}[main/.style={draw,circle}]
    \node[main, label=below:$1$] (1) {};
    \node[main, label=below:$2$] (2) [right=of 1]{};
    \node[main, label=below:$3$] (3) [right=of 2]{};

    \draw[-] (1)--(2)--(3);
    \draw (3) to [out=135, in=45,looseness=8] (3);

    \node[main, label=below:$1$] (1subL) [below=of 1]{};
    \node[main, label=below:$2$] (2sub) [right=of 1subL]{};
    \node[main, label=below:$1$] (1subR) [right=of 2sub]{};
    \node[draw=none,fill=none] (minus) [left=of 1subL]{$-$};

    \draw[-] (1subL)--(2sub)--(1subR);

    \node[draw,label=left:$2$] (2F) [below=of 1subR]{};
    \node[main, label=left:$2$] (2res) [below=of 2F]{};

    \draw[-] (2F)--(2res);
    \draw (2res) to [out=225, in=-45,looseness=8] (2res);

     \node[draw=none,fill=none] (topghost) [right=of 3]{};
    \node[draw=none,fill=none] (bottomghost) [right=of 2res]{};
    
    \draw [->] (topghost) to [out=-30,in=30,looseness=1] (bottomghost);
    
    \end{tikzpicture}
    \caption{Quiver subtraction of the $\urm(2)$ quotient quiver from an unframed magnetic quiver for $\overline{sub. reg. G_2}$ to produce quiver \Quiver{fig:subregG2A1Sub}.}
    \label{fig:subregG2A1Sub}
\end{figure}
An unframed magnetic quiver for the $\overline{sub.reg. G_2}$ is drawn in at the top of \Figref{fig:subregG2A1Sub}, which shows the subtraction of the $\urm(2)$ quotient quiver to result in the quiver \Quiver{fig:subregG2A1Sub}. The Coulomb branch of \Quiver{fig:subregG2A1Sub} is $Sym^2\left(\overline{min.A_1}\right)$ \cite{Cremonesi:2014xha} (it is also the orbifold $\mathbb C^4/\mathcal{D}_4$, where $\mathcal D_4$ is the dihedral group of order 8.). We conclude:
\begin{equation}
    \overline{sub. reg. G_2}///\surm(2)=Sym^2\left(\overline{min.A_1}\right).\label{eq:subregG2A1Sub}
\end{equation}

The result \eqref{eq:subregG2A1Sub} can be verified using the embedding of $G_2\hookleftarrow A_1\times A_1$, which decomposes the fundamental of $G_2$ as: \begin{equation}
    [0,1]_{G_2}\rightarrow [0;2]_{A_1\times A_1}+[1;1]_{A_1\times A_1}.
    \label{eq:G2A1Embed}
\end{equation}
Note that the two factors of $A_1$ are not symmetric in this embedding. It is clear from the quiver subtraction that we must preserve the fugacity label associated to the node of rank 3 in the magnetic quiver for $\overline{sub. reg. G_2}$. This corresponds to performing the HKQ with respect to the first factor of $A_1$ in \eqref{eq:G2A1Embed}.

\subsubsection{$\overline{min. F_4}///\surm(2)$}
\label{sec:minF4A1}
\begin{figure}[h!]
    \centering
       \begin{tikzpicture}[main/.style = {draw, circle}]
    \node[main, label=below:$1$] (1) {};
    \node[main,label=below:$2$] (2L) [right=of 1] {};
    \node[main,label=below:$3$] (3) [right=of 2L] {};
    \node[main,label=below:$2$] (2R) [right=of 3] {};
    \node[main,label=below:$1$] (1R) [right=of 2R] {};
    \node[main,label=below:$1$] (1subL) [below=of 1]{};
    \node[main, label=below:$2$] (2sub) [right=of 1subL]{};
    \node[main, label=below:$1$] (1subR) [right=of 2sub]{};
    \node[draw=none,fill=none] (-) [left=of 1subL]{$-$};
    \node[draw=none,fill=none] (ghost) [below=of 1R]{};

    \node[draw,label=right:$1$] (1fresult) [below right=of 1subR]{};
    \node[main,label=below:$2$] (2result) [below=of 1fresult]{};
    \node[main,label=below:$2$] (2lresult) [left=of 2result]{};
    \node[main,label=below:$1$] (1result)[right=of 2result]{};
    
    \node[draw=none,fill=none] (topghost) [right=of 1R]{};
    \node[draw=none,fill=none] (bottomghost) [right=of 1result]{};
    
    \draw[-](1fresult)--(2result)--(1result);
    \draw[line width=1pt, double distance=3pt,
     arrows = {-Latex[length=0pt 3 0]}] (2lresult) -- (2result);
    
     \draw[-] (1)--(2L)--(3);
    \draw[-] (2R)--(1R);
    \draw [line width=1pt, double distance=3pt,
             arrows = {-Latex[length=0pt 3 0]}] (3) -- (2R);
    
    \draw [->] (topghost) to [out=-30,in=30,looseness=1] (bottomghost);
    
    \draw[-] (1subL)--(2sub)--(1subR);
 
    \end{tikzpicture}
    \caption{Quiver subtraction of the $\urm(2)$ quotient quiver from an unframed magnetic quiver for $\overline{min. F_4}$ to produce to produce quiver \Quiver{fig:minF4A1Sub}.}
    \label{fig:minF4A1Sub}
\end{figure}
An unframed magnetic quiver for $\overline{min. F_4}$ is shown at the top of \Figref{fig:minF4A1Sub}, together with the subtraction of the $\urm(2)$ quotient quiver to result in quiver \Quiver{fig:minF4A1Sub}. Note that the subtraction occurs on the long side, in accordance with the \hyperlink{rule:ExternalLeg}{External Leg Rule}. The Coulomb branch of \Quiver{fig:minF4A1Sub} is $\overline{n.min. C_3}$. We conclude that:
\begin{equation}
    \overline{min. F_4}///\surm(2)=\overline{n. min. C_3}.
\end{equation}
We have checked this explicitly using the embedding of $F_4\hookleftarrow C_3\times A_1$, where the fundamental of $F_4$ decomposes in the following way:
\begin{equation}
    [0,0,0,1]_{F_4}\rightarrow [0,1,0;0]_{C_3\times A_1}+[1,0,0;1]_{C_3\times A_1}.
\end{equation}
%
\subsubsection{$\overline{min. E_6}///\surm(2)$}
\label{sec:minE6A1}
\begin{figure}[h!]
    \centering
    \begin{tikzpicture}[main/.style = {draw, circle}]
    \node[main, label=below:$1$] (oneL) []{};
    \node[main, label=below:$2$] (twoL) [right=of oneL]{};
    \node[main, label=below:$3$] (three) [right=of twoL]{};
    \node[main, label=below:$2$] (twoR) [right=of three]{};
    \node[main, label=below:$1$] (oneR) [right=of twoR]{};
    \node[main, label=left:$2$] (twoT) [above=of three]{};
    \node[main, label=left:$1$] (oneT) [above=of twoT]{};
    
    \draw[-] (oneL)--(twoL)--(three);
    \draw[-] (oneT)--(twoT)--(three)--(twoR)--(oneR);
    
    \node[main, label=below:$1$] (1sub) [below=of oneL]{};
    \node[main, label=below:$2$] (2sub) [right=of 1sub]{};
    \node[main, label=below:$1$] (1subR) [right=of 2sub]{};
    \node[draw=none,fill=none] (minus) [left=of 1sub]{$-$};
    \draw[-] (1sub)--(2sub)--(1subR);
    
    \node[draw, label=left:$1$] (1FL) [below=of 2sub]{};
    \node[main, label=below:$2$] (2L) [below=of 1FL]{};
    \node[main, label=below:$1$] (1L) [left=of 2L]{};
    \node[main, label=below:$2$] (2M) [right=of 2L]{};
    \node[main, label=below:$2$] (2R) [right=of 2M]{};
    \node[draw,label=right:$1$] (1FR) [above=of 2R]{};
    \node[main, label=below:$1$] (1R) [right=of 2R]{};
    
    \draw[-] (1L)--(2L)--(2M)--(2R)--(1R);
    \draw[-] (1FL)--(2L);
    \draw[-] (1FR)--(2R);
    
    \node[draw=none,fill=none] (topghost) [right=of oneR]{};
    \node[draw=none,fill=none] (bottomghost) [right=of 1R]{};
    \draw [->] (topghost) to [out=-30,in=30,looseness=1] (bottomghost);
    \end{tikzpicture}
    \caption{Quiver subtraction of the $\urm(2)$ quotient quiver from an unframed magnetic quiver for $\overline{min. E_6}$ to produce quiver \Quiver{fig:minE6A1QuivSub}.}
    \label{fig:minE6A1QuivSub}
\end{figure}

A magnetic quiver for $\overline{min. E_6}$ is shown at the top of \Figref{fig:minE6A1QuivSub}, together with the subtraction of the $\urm(2)$ quotient quiver to result in the quiver \Quiver{fig:minE6A1QuivSub}. The Coulomb branch of \Quiver{fig:minE6A1QuivSub} is $\overline{n. min. A_5}$. We conclude that: \begin{equation}
    \overline{min. E_6}///\surm(2)=\overline{n. min. A_5}.
\end{equation}
There is an embedding of $E_6\hookleftarrow A_5\times A_1$ where the fundamental of $E_6$ decomposes in the following way:
\begin{equation}
    [1,0,0,0,0,0]_{E_6}\rightarrow [0,1,0,0,0;0]_{A_5\times A_1}+[0,0,0,0,1;1]_{A_5\times A_1}.
\end{equation}
We have used this to compute the HKQ explicitly by Weyl integration and find agreement with quiver subtraction.

\subsubsection{$\overline{min. E_7}///\surm(2)$}
\label{sec:minE7A1}
\begin{figure}[h!]
    \centering
    \begin{tikzpicture}[main/.style = {draw, circle}]
    \node[main, label=below:$1$] (oneL) []{};
    \node[main, label=below:$2$] (twoL) [right=of oneL]{};
    \node[main, label=below:$3$] (threeL) [right=of twoL]{};
    \node[main, label=below:$4$] (four) [right=of threeL]{};
    \node[main, label=below:$3$] (threeR) [right=of four]{};
    \node[main, label=below:$2$] (twoR) [right=of threeR]{};
    \node[main, label=below:$1$] (oneR) [right=of twoR]{};
    \node[main, label=left:$2$] (twoT) [above=of four]{};
    
    \draw[-] (twoT)--(four);
    \draw[-] (oneL)--(twoL)--(threeL)--(four)--(threeR)--(twoR)--(oneR);
    
    \node[main, label=below:$1$] (1sub) [below=of oneL]{};
    \node[main, label=below:$2$] (2sub) [right=of 1sub]{};
    \node[main, label=below:$1$] (1subR) [right=of 2sub]{};
    \node[draw=none,fill=none] (minus) [left=of 1sub]{$-$};
    \draw[-] (1sub)--(2sub)--(1subR);
    
    \node[main, label=right:$2$] (2) [below right=of 1subR]{};
    \node[main, label=below:$4$] (4) [below=of 2]{};
    \node[main,label=below:$2$] (2L) [left=of 4]{};
    \node[main,label=below:$3$] (3) [right=of 4]{};
    \node[main, label=below:$2$] (2R) [right=of 3]{};
    \node[main, label=below:$1$] (1) [right=of 2R]{};
    \node[draw,label=left:$1$] (1F) [above left=of 4]{};
    
    \draw[-] (2L)--(4)--(3)--(2R)--(1);
    \draw[-] (1F)--(4)--(2);

    \node[draw=none,fill=none] (topghost) [right=of oneR]{};
    \node[draw=none,fill=none] (bottomghost) [right=of 1]{};
    \draw [->] (topghost) to [out=-30,in=30,looseness=1] (bottomghost);
    \end{tikzpicture}
    \caption{Quiver subtraction of the $\urm(2)$ quotient quiver from an unframed magnetic quiver for $\overline{min. E_7}$ to produce quiver \Quiver{fig:minE7A1QuivSub}.}
    \label{fig:minE7A1QuivSub}
\end{figure}
An unframed magnetic quiver for $\overline{min. E_7}$ is shown at the top of \Figref{fig:minE7A1QuivSub}, followed by the subtraction of the $\urm(2)$ quotient quiver to give \Quiver{fig:minE7A1QuivSub}. The Coulomb branch of \Quiver{fig:minE7A1QuivSub} is $\overline{n. n. min. D_6}$. We conclude that:
\begin{equation}
    \overline{min. E_7}///\surm(2)=\overline{n. n. min. D_6}.
\end{equation}
We have checked this explicitly using the embedding of $E_7\hookleftarrow D_6\times A_1$ which decomposes the fundamental as:
\begin{equation}
    [0,0,0,0,0,1,0]_{E_7}\rightarrow [0,0,0,0,0,1;0]_{D_6\times A_1}+[1,0,0,0,0,0;1]_{D_6\times A_1}.
\end{equation}

This example verifies the Higgs branch side of the $4d$ $\mathcal N=2$ duality proposed in \cite{Argyres:2007cn}, which says that $\sprm(2)$ with $6$ flavours at infinite coupling is dual to the $\surm(2)$ gauging of the rank 1 $E_7$ SCFT.

The $3d$ mirror of the latter theory is the quiver given at the top of \Figref{fig:minE7A1QuivSub} and the $\surm(2)$ gauging is implemented by the $\urm(2)$ quotient quiver subtraction. The resulting quiver \Quiver{fig:minE7A1QuivSub} is the $3d$ mirror of $\sprm(2)$ with 6 flavours.

\subsubsection{$\overline{min. E_8}///\surm(2)$}
\label{sec:minE8A1}
\begin{figure}[h!]
    \centering
       \begin{tikzpicture}[main/.style = {draw, circle}]
   \node[main,label=below:$1$] (1L) {};
    \node[main,label=below:$2$] (2L) [right=of 1L]{};
    \node[main,label=below:$3$] (3L) [right=of 2L]{};
    \node[main,label=below:$4$] (4L) [right=of 3L]{};
    \node[main,label=below:$5$] (5L) [right=of 4L]{};
    \node[main,label=below:$6$] (6) [right=of 5L]{};
    \node[main,label=below:$4$] (4R) [right=of 6]{};
    \node[main,label=below:$2$] (2R) [right=of 4R]{};
    \node[main,label=right:$3$] (3T) [above=of 6]{};
    
    \draw[-](1L)--(2L)--(3L)--(4L)--(5L)--(6)--(4R)--(2R);
    \draw[-](3T)--(6);
    
    \node[main, label=below:$1$] (1subL) [below=of 1L]{};
    \node[main, label=below:$2$] (2sub) [right=of 1subL]{};
    \node[main, label=below:$1$] (1sub) [right=of 2sub]{};
    \node[draw=none,fill=none] (-) [left=of 1subL]{$-$};
    
    \draw[-] (1subL)--(2sub)--(1sub);
    
    \node[draw,label=right:$1$] (oneF) [below right=of 1sub]{};
    \node[main, label=below:$4$] (four) [below =of oneF]{};
    \node[main, label=below:$2$] (twoL) [left=of four]{};
    \node[main, label=below:$5$] (five) [right=of four]{};
    \node[main, label=below:$6$] (six) [right=of five]{};
    \node[main, label=below:$4$] (fourR) [right=of six]{};
    \node[main,label=below:$2$] (twoR) [right=of fourR]{};
    \node[main,label=right:$3$] (three) [above=of six]{};
    
    \draw[-] (twoL)--(four)--(five)--(six)--(fourR)--(twoR);
    \draw[-] (three)--(six);
    \draw[-] (oneF)--(four);
    
    \node[draw=none,fill=none] (topghost) [right=of 2R]{};
    \node[draw=none,fill=none] (bottomghost) [right=of twoR]{};
    
    \draw [->] (topghost) to [out=-30,in=30,looseness=1] (bottomghost);

   \end{tikzpicture}
    \caption{Quiver subtraction of the $\urm(2)$ quotient quiver from an unframed magnetic quiver for $\overline{min. E_8}$ to produce quiver \Quiver{fig:minE8A1Sub}.}
    \label{fig:minE8A1Sub}
\end{figure}
 An unframed magnetic quiver for $\overline{min. E_8}$ is given at the top of \Figref{fig:minE8A1Sub} followed by the subtraction of the $\urm(2)$ quotient quiver to produce \Quiver{fig:minE8A1Sub}. The Coulomb branch of  \Quiver{fig:minE8A1Sub} is $\overline{n. min. E_7}$. We conclude that: \begin{equation}
    \overline{min. E_8}///\surm(2)=\overline{n. min. E_7}.
\end{equation}
We have checked this explicitly using the Dynkin-type embedding of $E_8\hookleftarrow E_7\times A_1$ under which the fundamental of $E_8$ decomposes as:
\begin{equation}
    [0,0,0,0,0,0,1,0]_{E_8}\rightarrow [1,0,0,0,0,0,0;0]_{E_7\times A_1}+[0,0,0,0,0,1,0;1]_{E_7\times A_1}+[0,0,0,0,0,0,0;2]_{E_7\times A_1}.
\end{equation}

\subsection{Exceptional Orbit $\surm(3)$ HKQ}
\label{sec:ExceptionalSU3HKQ}
It is important to note that $\overline{min. E_6}$ is not amenable to subtraction by the $\urm(3)$ quotient quiver as the junction occurs on a node of rank 3 on the quotient quiver as opposed to a node of rank 2 as required by the \hyperlink{rule:Junction}{Junction Rule}.

\subsubsection{$\overline{min. E_7}///\surm(3)$}
\label{sec:minE7A2}
\begin{figure}[h!]
    \centering
    \begin{subfigure}{0.45\textwidth}
    \centering\resizebox{0.6\width}{!}{\begin{tikzpicture}[main/.style = {draw, circle}]
   \node[main,label=below:$1$] (1L) {};
    \node[main,label=below:$2$] (2L) [right=of 1L]{};
    \node[main,label=below:$3$] (3L) [right=of 2L]{};
    \node[main,label=below:$4$] (4) [right=of 3L]{};
    \node[main,label=below:$3$] (3R) [right=of 4]{};
    \node[main,label=below:$2$] (2R) [right=of 3R]{};
    \node[main,label=below:$1$] (1R) [right=of 2R]{};
    \node[main,label=right:$2$] (2T) [above=of 4]{};
    
    \draw[-](1L)--(2L)--(3L)--(4)--(3R)--(2R)--(1R);
    \draw[-](2T)--(4);
    
    \node[main, label=right:$1$] (1subT) [below=of 4]{};
    \node[main, label=below:$2$] (2subR) [below=of 1subT]{};
    \node[main, label=below:$3$] (3sub) [left=of 2subR]{};
    \node[main, label=below:$2$] (2subL) [left=of 3sub]{};
    \node[main, label=below:$1$] (1subL) [left=of 2subL]{};
    
    \draw[-] (1subT)--(2subR)--(3sub)--(2subL)--(1subL);
    
    \node[draw=none,fill=none] (-) [left=of 1subL]{$-$};
    
    \node[main, label=right:$1$] (1resT) [below=of 2subR]{};
    \node[main, label=below:$2$] (2resL) [below=of 1resT]{};
    \node[main, label=below:$3$] (3res) [right=of 2resL]{};
    \node[main, label=below:$2$] (2resR) [right=of 3res]{};
    \node[main, label=below:$1$] (1resR) [right=of 2resR]{};
    \node[draw,label=right:$2$] (2resF) [above=of 3res]{};
    
    \draw[-] (1resT)--(2resL)--(3res)--(2resR)--(1resR);
    \draw[-] (3res)--(2resF);
    
    \node[draw=none,fill=none] (topghost) [right=of 1R]{};
    \node[draw=none,fill=none] (bottomghost) [right=of 1resR]{};
    
    \draw [->] (topghost) to [out=-30,in=30,looseness=1] (bottomghost);

   \end{tikzpicture}}
    \caption{}
    \label{fig:minE7subQ1}
    \end{subfigure}
    \hfill
    \begin{subfigure}{0.45\textwidth}
    \centering\resizebox{0.6\width}{!}{\begin{tikzpicture}[main/.style = {draw, circle}]
   \node[main,label=below:$1$] (1L) {};
    \node[main,label=below:$2$] (2L) [right=of 1L]{};
    \node[main,label=below:$3$] (3L) [right=of 2L]{};
    \node[main,label=below:$4$] (4) [right=of 3L]{};
    \node[main,label=below:$3$] (3R) [right=of 4]{};
    \node[main,label=below:$2$] (2R) [right=of 3R]{};
    \node[main,label=below:$1$] (1R) [right=of 2R]{};
    \node[main,label=right:$2$] (2T) [above=of 4]{};
    
    \draw[-](1L)--(2L)--(3L)--(4)--(3R)--(2R)--(1R);
    \draw[-](2T)--(4);
    
    \node[main, label=right:$1$] (1subR) [below=of 3R]{};
    \node[main, label=below:$2$] (2subR) [left=of 1subR]{};
    \node[main, label=below:$3$] (3sub) [left=of 2subR]{};
    \node[main, label=below:$2$] (2subL) [left=of 3sub]{};
    \node[main, label=below:$1$] (1subL) [left=of 2subL]{};
    
    \draw[-] (1subR)--(2subR)--(3sub)--(2subL)--(1subL);
    
    \node[draw=none,fill=none] (-) [left=of 1subL]{$-$};
    
    \node[main, label=left:$2$] (2resT) [below=of 2subR]{};
    \node[main, label=below:$2$] (2resL) [below=of 2resT]{};
    \node[main, label=below:$2$] (2resM) [right=of 2resL]{};
    \node[main, label=below:$2$] (2resR) [right=of 2resM]{};
    \node[main, label=below:$1$] (1resR) [right=of 2resR]{};
    \node[draw,label=right:$2$] (2resF) [right=of 2resT]{};
    \node[draw,label=above:$1$] (1resF) [above=of 2resR]{};
    
    \draw[-] (2resF)--(2resT)--(2resL)--(2resM)--(2resR)--(1resR);
    \draw[-] (1resF)--(2resR);
    
    \node[draw=none,fill=none] (topghost) [right=of 1R]{};
    \node[draw=none,fill=none] (bottomghost) [right=of 1resR]{};
    
    \draw [->] (topghost) to [out=-30,in=30,looseness=1] (bottomghost);

   \end{tikzpicture}}
   \caption{}
    \label{fig:minE7subQ2}\end{subfigure}
    \centering
    \begin{subfigure}{\textwidth}
        \centering
        \begin{tikzpicture}[main/.style={draw,circle}]
        \node[main, label=below:$1$] (1L) []{};
        \node[main, label=below:$2$] (2L) [right=of 1L]{};
        \node[main, label=below:$2$] (2M) [right=of 2L]{};
        \node[main, label=below:$2$] (2R) [right=of 2M]{};
        \node[main, label=below:$1$] (1R) [right=of 2R]{};
        \node[draw, label=left:$1$] (1FL) [above=of 2L]{};
        \node[draw, label=left:$1$] (1FR) [above=of 2R]{};
        \draw[-] (1L)--(2L)--(2M)--(2R)--(1R);
        \draw[-] (1FL)--(2L);
        \draw[-] (1FR)--(2R);
        \end{tikzpicture}
        \caption{}
        \label{fig:minE7subInt}
    \end{subfigure}
    \caption{Both alignments of the $\urm(3)$ quotient quiver against $\overline{min. E_7}$ to produce quivers \Quiver{fig:minE7subQ1} and \Quiver{fig:minE7subQ2}. Their intersection, reached via $A_1$ Kraft-Procesi transitions, is \Quiver{fig:minE7subInt}.}
    \label{fig:minE7A2SubBoth}
\end{figure}

The unframed magnetic quiver for the $\overline{min. E_7}$ permits two possible alignments for subtraction of an $\urm(3)$ quotient quiver, as shown in \Figref{fig:minE7A2SubBoth}. The resulting quivers \Quiver{fig:minE7subQ1} and \Quiver{fig:minE7subQ2} are magnetic quivers for the nilpotent orbits $\overline{\mathcal O}^{A_5}_{(2^3)}$ and $\overline{\mathcal O}^{A_5}_{(3,1^3)}$, respectively. We conclude that:
\begin{equation}
    \overline{min. E_7}///\surm(3)=\overline{\mathcal O}^{A_5}_{(2^3)}\cup\overline{\mathcal O}^{A_5}_{(3,1^3)}.
\end{equation}
The intersection, reached in both cases by an $A_1$ KP transition, is the quiver \Quiver{fig:minE7subInt}, which is a magnetic quiver for $\overline{n.min. A_5}$.
There is an embedding of $E_7\hookleftarrow A_5\times A_2$ which decomposes the fundamental of $E_7$ as:
\begin{align}    [0,0,0,0,0,1,0]_{E_7}&\rightarrow[0,0,1,0,0;0,0]_{A_5\times A_2}+[1,0,0,0,0;0,1]_{A_5\times A_2}\nonumber\\&+[0,0,0,0,1;1,0]_{A_5\times A_2},
\end{align}
and we have used this to compute the HKQ using Weyl integration and to verify the above result.

\subsubsection{$\overline{min. E_8}///\surm(3)$}
\label{sec:minE8A2}
\begin{figure}[h!]
    \centering
       \begin{tikzpicture}[main/.style = {draw, circle}]
   \node[main,label=below:$1$] (1L) {};
    \node[main,label=below:$2$] (2L) [right=of 1L]{};
    \node[main,label=below:$3$] (3L) [right=of 2L]{};
    \node[main,label=below:$4$] (4L) [right=of 3L]{};
    \node[main,label=below:$5$] (5L) [right=of 4L]{};
    \node[main,label=below:$6$] (6) [right=of 5L]{};
    \node[main,label=below:$4$] (4R) [right=of 6]{};
    \node[main,label=below:$2$] (2R) [right=of 4R]{};
    \node[main,label=right:$3$] (3T) [above=of 6]{};
    
    \draw[-](1L)--(2L)--(3L)--(4L)--(5L)--(6)--(4R)--(2R);
    \draw[-](3T)--(6);
    
    \node[main, label=below:$1$] (1subL) [below=of 1L]{};
    \node[main, label=below:$2$] (2sub) [right=of 1subL]{};
    \node[main, label=below:$3$] (3sub) [right=of 2sub]{};
    \node[main, label=below:$2$] (2subR) [right=of 3sub]{};
    \node[main, label=below:$1$] (1subR) [right=of 2subR]{};
    
    \node[draw=none,fill=none] (-) [left=of 1subL]{$-$};
    
    \draw[-] (1subL)--(2sub)--(3sub)--(2subR)--(1subR);

    \node[draw, label=left:$1$] (1F) [below=of 1subR]{};
    \node[main, label=below:$6$] (6r) [below right=of 1F] {};
    \node[main, label=below:$4$] (4resl) [left=of 6r] {};
    \node[main, label=below:$2$] (2resl) [left=of 4resl]{};
    \node[main, label=below:$4$] (4resr) [right=of 6r]{};
    \node[main, label=below:$2$] (2resr) [right=of 4resr]{};
    \node[main, label=right:$3$] (3res) [above=of 6r]{};

    \draw[-] (2resl)--(4resl)--(6r)--(4resr)--(2resr);
    \draw[-] (1F)--(6r)--(3res);
    
    \node[draw=none,fill=none] (topghost) [right=of 2R]{};
    \node[draw=none,fill=none] (bottomghost) [right=of 2resr]{};
    
    \draw [->] (topghost) to [out=-30,in=30,looseness=1] (bottomghost);

   \end{tikzpicture}
    \caption{Quiver subtraction of the $\urm(3)$ quotient quiver from an unframed magnetic quiver for $\overline{min. E_8}$ to produce quiver \Quiver{fig:minE8A2Sub}.}
    \label{fig:minE8A2Sub}
\end{figure}
The computation of this $\surm(3)$ HKQ was first done in \cite{Hanany:2022itc} both using a diagrammatic quiver subtraction of the $\urm(3)$ quotient quiver from the magnetic quiver for $\overline{min. E_8}$, as reproduced in \Figref{fig:minE8A2Sub}, and explicitly using Weyl integration. The resulting diagram is not a magnetic quiver for an $E_6$ orbit, but turns out to be a magnetic quiver for the double cover of the 21-dimensional orbit of $E_6$. The quiver \Quiver{fig:minE8A2Sub} is also a slice in the affine Grassmannian of $E_6$ and so we find:\begin{equation}
    \overline{min. E_8}///\surm(3)=\left[\overline{\mathcal W_{E_6}}\right]^{[0,0,1,0,0,0]}_{[0,0,0,0,0,0]}.
\end{equation}

\subsection{Exceptional Orbit $\surm(4)$ HKQ}
\label{sec:ExceptionalSU4HKQs}
\subsubsection{$\overline{min. E_8}///\surm(4)$}
\label{sec:minE8A3}
\begin{figure}[h!]
     \centering
    \begin{subfigure}{.45\textwidth}
    \centering
    \resizebox{\textwidth}{!}{
        \begin{tikzpicture}[main/.style = {draw, circle}]
   \node[main,label=below:$1$] (1L) {};
    \node[main,label=below:$2$] (2L) [left=of 1L]{};
    \node[main,label=below:$3$] (3L) [left=of 2L]{};
    \node[main,label=below:$4$] (4L) [left=of 3L]{};
    \node[main,label=below:$5$] (5L) [left=of 4L]{};
    \node[main,label=below:$6$] (6) [left=of 5L]{};
    \node[main,label=below:$4$] (4R) [left=of 6]{};
    \node[main,label=below:$2$] (2R) [left=of 4R]{};
    \node[main,label=left:$3$] (3T) [above=of 6]{};
    
    \draw[-](1L)--(2L)--(3L)--(4L)--(5L)--(6)--(4R)--(2R);
    \draw[-](3T)--(6);
    
    \node[main, label=left:$1$] (1subT) [below=of 6]{};
    \node[main, label=below:$2$] (2subR) [below=of 1subT]{};
    \node[main, label=below:$3$] (3subR) [right=of 2subR]{};
    \node[main, label=below:$4$] (4sub) [right=of 3subR]{};
    \node[main, label=below:$3$] (3subL) [right=of 4sub]{};
    \node[main, label=below:$2$] (2subL) [right=of 3subL]{};
    \node[main, label=below:$1$] (1subL) [right=of 2subL]{};
    
    \draw[-] (1subT)--(2subR)--(3subR)--(4sub)--(3subL)--(2subL)--(1subL);
    
    \node[main, label=right:$2$] (2resT) [below=of 2subR]{};
    \node[main, label=below:$4$] (4resL) [below=of 2resT]{};
    \node[main, label=below:$2$] (2resL) [right=of 4resL]{};
    \node[main, label=below:$4$] (4resR) [left=of 4resL]{};
    \node[main, label=below:$2$] (2resR)[left=of 4resR]{};
    \node[draw,label=left:$2$] (2flav) [above=of 4resR]{};
    
    \node[draw=none,fill=none] (-) [left=of 2subR]{$-$};
    
    \draw[-] (2resL)--(4resL)--(4resR)--(2resR);
    \draw[-] (2flav)--(4resR);
    \draw[-] (2resT)--(4resL);
    
    \node[draw=none,fill=none] (topghost) [left=of 2R]{};
    \node[draw=none,fill=none] (bottomghost) [left=of 2resR]{};
    
    \draw [->] (topghost) to [out=-150,in=150,looseness=1] (bottomghost);

   \end{tikzpicture}}
    
    \caption{}
    \label{fig:minE8subQ1}
    \end{subfigure}
    \hfill
     \centering
    \begin{subfigure}{.45\textwidth}
    \centering
    \resizebox{\textwidth}{!}{\begin{tikzpicture}[main/.style = {draw, circle}]
   \node[main,label=below:$1$] (1L) {};
    \node[main,label=below:$2$] (2L) [left=of 1L]{};
    \node[main,label=below:$3$] (3L) [left=of 2L]{};
    \node[main,label=below:$4$] (4L) [left=of 3L]{};
    \node[main,label=below:$5$] (5L) [left=of 4L]{};
    \node[main,label=below:$6$] (6) [left=of 5L]{};
    \node[main,label=below:$4$] (4R) [left=of 6]{};
    \node[main,label=below:$2$] (2R) [left=of 4R]{};
    \node[main,label=left:$3$] (3T) [above=of 6]{};
    
    \draw[-](1L)--(2L)--(3L)--(4L)--(5L)--(6)--(4R)--(2R);
    \draw[-](3T)--(6);
    
    \node[main, label=below:$1$] (1subR) [below=of 1L]{};
    \node[main, label=below:$2$] (2subR) [left=of 1subR]{};
    \node[main, label=below:$3$] (3subR) [left=of 2subR]{};
    \node[main, label=below:$4$] (4sub) [left=of 3subR]{};
    \node[main, label=below:$3$] (3subL) [left=of 4sub]{};
    \node[main, label=below:$2$] (2subL) [left=of 3subL]{};
    \node[main, label=below:$1$] (1subL) [left=of 2subL]{};
    
     \node[draw=none,fill=none] (-) [left=of 1subL]{$-$};
    
    \draw[-] (1subR)--(2subR)--(3subR)--(4sub)--(3subL)--(2subL)--(1subL);
    
    \node[main, label=left:$3$] (3resT) [below=of 2subL]{};
    \node[main, label=below:$4$] (4res) [below=of 3resT]{};
    \node[main, label=below:$2$] (2resL) [right=of 4res]{};
    \node[main, label=below:$3$] (3resR)[left=of 4res]{};
    \node[main, label=below:$2$] (2resR)[left=of 3resR]{};
    \node[draw, label=right:$2$] (2resflav) [right=of 3resT]{};
    \node[draw, label=right:$1$] (1resflav) [above=of 2resR]{};
    
    \draw[-] (1resflav)--(2resR)--(3resR)--(4res)--(3resT)--(2resflav);
    \draw[-] (2resL)--(4res);
    
    \node[draw=none,fill=none] (topghost) [left=of 2R]{};
    \node[draw=none,fill=none] (bottomghost) [left=of 2resR]{};
    
    \draw [->] (topghost) to [out=-150,in=150,looseness=1] (bottomghost);

   \end{tikzpicture}}
    
    \caption{}
    \label{fig:minE8subQ2}
    \end{subfigure}
    \centering
    \begin{subfigure}{\textwidth}
    \centering
    \begin{tikzpicture}[main/.style={draw,circle}]
        \node[main, label=below:$2$] (2L) []{};
        \node[main, label=below:$3$] (3L) [right=of 2L]{};
        \node[main, label=below:$4$] (4) [right=of 3L]{};
        \node[main, label=right:$2$] (2T) [above right=of 4]{};
        \node[main, label=right:$2$] (2B) [below right=of 4]{};
        \node[draw, label=left:$1$] (1FL) [above=of 2L]{};
        \node[draw, label=left:$1$] (1FR) [above=of 4]{};
        \draw[-] (1FL)--(2L)--(3L)--(4)--(2T);
        \draw[-] (1FR)--(4)--(2B);
    \end{tikzpicture}
        \caption{}
        \label{fig:minE8subInt}
    \end{subfigure}
    \caption{Both alignments of the $\urm(4)$ quotient quiver against $\overline{min. E_8}$ to produce quivers \Quiver{fig:minE8subQ1} and \Quiver{fig:minE8subQ2}. Their intersection, reached via $A_1$ KP transitions, is \Quiver{fig:minE8subInt}.}
    \label{fig:minE8A3SubBoth}
\end{figure}

Subtraction of the $\urm(4)$ quotient quiver from the unframed magnetic quiver for $\overline{min. E_8}$ has two possible alignments, as shown in \Figref{fig:minE8A3SubBoth}, which produce quivers \Quiver{fig:minE8subQ1} and \Quiver{fig:minE8subQ2}. These are both slices in the affine Grassmannian of $D_5$, being $\overline{[\mathcal W_{D_5}]}^{[0,2,0,0,0]}_{[0,0,0,0,0]}$ and  $\overline{[\mathcal W_{D_5}]}^{[1,0,0,2,0]}_{[0,0,0,0,0]}$, respectively. The latter can also be identified as the normalisation of the orbit $\overline{\mathcal O}^{D_5}_{(3^2,2^2)}$, calculated using the Nilpotent Orbit Normalisation formula \cite{Hanany:2017ooe}. We conclude that:
\begin{equation}
\overline{min. E_8}///\surm(4)=\overline{[\mathcal W_{D_5}]}^{[0,2,0,0,0]}_{[0,0,0,0,0]}\cup \overline{[\mathcal W_{D_5}]}^{[1,0,0,2,0]}_{[0,0,0,0,0]}
\end{equation}
The intersection \Quiver{fig:minE8subInt}, related in both cases by an $A_1$ KP transition, is a magnetic quiver whose Coulomb branch is also a slice in the affine Grassmannian of $D_5$, $\overline{[\mathcal W_{D_5}]}^{[1,0,1,0,0]}_{[0,0,0,0,0]}$. Comparison of the volumes of the Coulomb branches indicates it is also the double cover of the orbit $\overline{\mathcal O}^{D_5}_{(3^2,1^4)}$. This is in accordance with the Springer correspondence and geometric Satake equivalence, as discussed in \cite{2011arXiv1108.4999A}.

This identity can be checked explicitly using Weyl integration under the embedding of $E_8\hookleftarrow D_5\times A_3$ which decomposes the fundamental of $E_8$ as:
\begin{align}
    [0,0,0,0,0,0,1,0]_{E_8}&\rightarrow [0,1,0,0,0;0,0,0]_{D_5\times A_3}+[0,0,0,0,1;0,0,1]_{D_5\times A_3}\nonumber\\&+[0,0,0,1,0;1,0,0]_{D_5\times A_3}+[1,0,0,0,0;0,1,0]_{D_5\times A_3}+[0,0,0,0,0;1,0,1]_{D_5\times A_3}.
\end{align}

\section{Slices in Exceptional Affine Grassmannians}
\label{sec:ExceptionalAGHKQ}
Here we consider magnetic quivers for some slices in the affine Grassmannian of E-type algebras. The quivers are good by construction and many satisfy the \hyperlink{rules:ExternalLeg}{External Leg Rule} since E-type Dynkin diagrams have a long leg suitable for quotient quiver subtraction.

The cases presented are not chosen in a systematic way since a comprehensive study of affine Grassmannians is beyond the scope of this paper. Nevertheless, we draw upon these examples in the affine Grassmannian to demonstrate the generality of our rules for quivers whose Coulomb branches are challenging to compute.

\subsection{$\surm(2)$ HKQ}
\label{sec:ExceptionalAGSU2HKQ}
\subsubsection{$\overline{\left[\mathcal W_{E_6}\right]}^{[1,0,0,0,0,1]}_{[1,0,0,0,0,0]}///\surm(2)$}
\begin{figure}[h!]
    \centering
    \begin{subfigure}[h!]{\linewidth}
    \centering
    \begin{tikzpicture}[main/.style = {draw, circle}]
    \node[main, label=below:$1$] (1L) [] {};
    \node[main, label=below:$2$] (2L) [right= of 1L]{};
    \node[main, label=below:$3$] (3) [right=of 2L]{};
    \node[main, label=below:$2$] (2R) [right=of 3]{};
    \node[main, label=below:$1$] (1R) [right=of 2R]{};
    \node[main, label=left:$2$] (2T) [above=of 3]{};
    \node[draw, label=left:$1$] (1FL) [above=of 1L]{};
    \node[draw,label=left:$1$] (1FT) [above=of 2T]{};
    
    \draw[-] (1FL)--(1L)--(2L)--(3)--(2R)--(1R);
    \draw[-] (3)--(2T)--(1FT);
       
    \end{tikzpicture}
    \caption{}
    \label{fig:quivE6100000Framed}
    \end{subfigure}
    \begin{subfigure}[h!]{\linewidth}
    \centering
    \begin{tikzpicture}[main/.style = {draw, circle}]
    \node[main, label=below:$1$] (1L) [] {};
    \node[main, label=below:$2$] (2L) [right= of 1L]{};
    \node[main, label=below:$3$] (3) [right=of 2L]{};
    \node[main, label=below:$2$] (2R) [right=of 3]{};
    \node[main, label=below:$1$] (1R) [right=of 2R]{};
    \node[main, label=right:$2$] (2T) [above=of 3]{};
    \node[main, label=left:$1$] (1E) [above right=of 1L]{};
    
    \draw[-] (1R)--(2R)--(3)--(2L)--(1L)--(1E)--(2T)--(3);
    \end{tikzpicture}
    \caption{{}}
    \label{fig:quivE6100000Unframed}
    \end{subfigure}
   \caption{Framed quiver \Quiver{fig:quivE6100000Framed} and unframed quiver \Quiver{fig:quivE6100000Unframed} for $\overline{\left[\mathcal W_{E_6}\right]}^{[1,0,0,0,0,1]}_{[1,0,0,0,0,0]}$}
    \label{fig:my_label}
\end{figure}
\begin{figure}[h!]
    \centering
    \begin{tikzpicture}[main/.style = {draw, circle}]
    \node[main, label=below:$1$] (1L) [] {};
    \node[main, label=below:$2$] (2L) [right= of 1L]{};
    \node[main, label=below:$3$] (3) [right=of 2L]{};
    \node[main, label=below:$2$] (2R) [right=of 3]{};
    \node[main, label=below:$1$] (1R) [right=of 2R]{};
    \node[main, label=right:$2$] (2T) [above=of 3]{};
    \node[main, label=left:$1$] (1E) [above right=of 1L]{};
    
    \draw[-] (1R)--(2R)--(3)--(2L)--(1L)--(1E)--(2T)--(3);
    
    \node[main, label=below:$1$] (1subR) [below =of 1R]{};
    \node[main, label=below:$2$] (2sub) [left=of 1subR]{};
    \node[main, label=below:$1$] (1subL) [left=of 2sub]{};
    \node[draw=none,fill=none] (minus) [left=of 1subL]{$-$};
    
    \draw[-] (1subR)--(2sub)--(1subL);
    
    \node[main, label=right:$2$] (twoT) [below =of 1subL]{};
    \node[main, label=below:$2$] (twoM) [below=of twoT]{};
    \node[main, label=below:$2$] (twoL) [left=of twoM]{};
    \node[main, label=below:$1$] (oneL) [left=of twoL]{};
    \node[main, label=left:$1$] (oneT) [above right=of oneL]{};
    \node[draw, label=below:$1$] (oneF) [above right= 0.5cm and 0.5cm of twoL]{};
    
    \draw[-] (twoT)--(twoM)--(twoL)--(oneL)--(oneT)--(twoT)--(oneF)--(twoL);
    
    \node[draw=none,fill=none] (topghost) [right=of 1R]{};
    \node[draw=none,fill=none] (bottomghost) [right=of twoM]{};
    
    \draw [->] (topghost) to [out=-30,in=30,looseness=1] (bottomghost);

    \end{tikzpicture}
    \caption{Subtraction of the $\urm(2)$ quotient quiver from \Quiver{fig:quivE6100000Unframed}
    to yield \Quiver{fig:E6100000A1Sub}.}
    \label{fig:E6100000A1Sub}
\end{figure}

Framed and unframed magnetic quivers \Quiver{fig:quivE6100000Framed} and \Quiver{fig:quivE6100000Unframed} for $\overline{\left[\mathcal W_{E_6}\right]}^{[1,0,0,0,0,1]}_{[1,0,0,0,0,0]}$ are shown in \Figref{fig:my_label}.

The Coulomb branch global symmetry is $SO(10)\times \urm(1)$. By inspection of \Quiver{fig:quivE6100000Unframed}, we see there is only one possible alignment for an $\urm(2)$ quotient quiver that obeys all of the selection rules. This subtraction yields the quiver \Quiver{fig:E6100000A1Sub} shown in \Figref{fig:E6100000A1Sub}. From consideration of the balance of gauge nodes of \Quiver{fig:E6100000A1Sub} one would expect a Coulomb branch global symmetry of at least $\surm(4)\times \urm(1)^2$. When we compute the Hilbert series of its Coulomb branch using the monopole formula we find a global symmetry enhancement to $\surm(4)\times \surm(2)\times \urm(1)$.

The unrefined HS is:
\begin{align}
    HS\left[\mathcal C\left(\mathcal Q_{\ref{fig:E6100000A1Sub}}\right)\right]&=\frac{\left(\begin{aligned}1&+6 t^{2}+33 t^{3}+136 t^{4}+541 t^{5}+1862 t^{6}+5913 t^{7}+16736 t^{8}+43638 t^{9}\\&+103724 t^{10}+228506 t^{11}+464616 t^{12}+880496 t^{13}+1553400 t^{14}\\&+2568079 t^{15}+3976634 t^{16}+5792877 t^{17}+7935748 t^{18}+10255355 t^{19}\\&+12496990 t^{20}+14392503 t^{21}+15653956 t^{22}+16104580 t^{23}\\&+\text{palindrome}+t^{44}\end{aligned}\right)}{\left(1-t\right)^{-6} \left(1-t^2\right)^7\left(1-t^3\right)^8\left(1-t^4\right)^7}
\end{align}

Noting that the $\urm(2)$ quotient quiver is subtracted from the balanced $D_5$ Dynkin diagram contained within \Quiver{fig:quivE6100000Framed}, we have used the embedding \eqref{eq:DkA1Embed} to compute the HKQ using Weyl integration and thereby to confirm the result from quotient quiver subtraction.

\subsubsection{$\overline{\left[\mathcal W_{E_6}\right]}^{[0,0,1,0,0,0]}_{[0,0,0,0,0,1]}///\surm(2)$}
\begin{figure}[h!]
    \centering
    \begin{tikzpicture}[main/.style = {draw, circle}]
    \node[main, label=below:$1$] (1L) [] {};
    \node[main, label=below:$2$] (2L) [right= of 1L]{};
    \node[main, label=below:$3$] (3) [right=of 2L]{};
    \node[main, label=below:$2$] (2R) [right=of 3]{};
    \node[main, label=below:$1$] (1R) [right=of 2R]{};
    \node[main, label=left:$1$] (1TL) [above left=of 3]{};
    \node[main, label=right:$1$] (1TR) [above right=of 3]{};
    
    \draw[-] (1R)--(2R)--(3)--(2L)--(1L);
    \draw[-] (1TL)--(3)--(1TR);
    
    \node[main, label=below:$1$] (1subR) [below =of 1R]{};
    \node[main, label=below:$2$] (2sub) [left=of 1subR]{};
    \node[main, label=below:$1$] (1subL) [left=of 2sub]{};
    \node[draw=none,fill=none] (minus) [left=of 1subL]{$-$};
    
    \draw[-] (1subR)--(2sub)--(1subL);
    
    \node[draw,label=left:$1$] (oneFTL) [below left=of 1subL]{};
    \node[draw,label=right:$1$] (oneFTR) [below right=of 1subL]{};
    \node[main, label=left:$1$] (oneTL) [below=of oneFTL]{};
    \node[main, label=right:$1$] (oneTR) [below=of oneFTR]{};
    \node[main, label=below:$2$] (twoR) [below right=of oneTL]{};
    \node[main, label=below:$2$] (twoL) [left=of twoR]{};
    \node[main, label=below:$1$] (oneL) [left=of twoL]{};
    \node[draw,label=left:$1$] (oneF) [above left=of twoL]{};
    
    \draw[-] (oneFTL)--(oneTL)--(twoR)--(oneTR)--(oneFTR);
    \draw[-] (twoR)--(twoL)--(oneL);
    \draw[-] (twoL)--(oneF);
    
    \node[draw=none,fill=none] (topghost) [right=of 1R]{};
    \node[draw=none,fill=none] (bottomghost) [right=of twoR]{};
    
    \draw [->] (topghost) to [out=-30,in=30,looseness=1] (bottomghost);

    \end{tikzpicture}
    \caption{Subtraction of the $\urm(2)$ quotient quiver from the unframed magnetic quiver for  $\overline{\left[\mathcal W_{E_6}\right]}^{[0,0,1,0,0,0]}_{[0,0,0,0,0,1]}$  to yield \Quiver{fig:E6000001A1Sub}.}
\label{fig:E6000001A1Sub}
\end{figure}

The magnetic quiver for $\overline{\left[\mathcal W_{E_6}\right]}^{[0,0,1,0,0,0]}_{[0,0,0,0,0,1]}$ is shown at the top of \Figref{fig:E6000001A1Sub}, together with the subtraction of the $\urm(2)$ quotient quiver to produce \Quiver{fig:E6000001A1Sub}. The global symmetry of $\overline{\left[\mathcal W_{E_6}\right]}^{[0,0,1,0,0,0]}_{[0,0,0,0,0,1]}$ is $\surm(6)\times \urm(1)$. The Coulomb branch of \Quiver{fig:E6000001A1Sub} is $\overline{\left[\mathcal W_{D_5}\right]}^{[0,1,0,1,1]}_{[0,0,0,1,1]}$, and has a global symmetry of $\surm(4)\times \urm(1)^2$. We conclude that \begin{equation}
    \overline{\left[\mathcal W_{E_6}\right]}^{[0,0,1,0,0,0]}_{[0,0,0,0,0,1]}///\surm(2)=\overline{\left[\mathcal W_{D_5}\right]}^{[0,1,0,1,1]}_{[0,0,0,1,1]}
\end{equation} and confirm this from the unrefined Hilbert series:
\begin{align} 
\hs[\mathcal C\left(\mathcal Q_{\ref{fig:E6000001A1Sub}}\right)]&=\frac{\left(\begin{aligned}1&+3 t+18 t^{2}+55 t^{3}+198 t^{4}+539 t^{5}+1445 t^{6}+3288 t^{7}+7052 t^{8}\\&+13416 t^{9}+23838 t^{10}+38390 t^{11}+57751 t^{12}+79667 t^{13}+102836 t^{14}\\&+122489 t^{15}+136705 t^{16}+141190 t^{17}+\text{palindrome}+t^{34}\end{aligned}\right)}
{\left(1-t^2\right)^2\left(1-t^3\right)^{12}\left(1-t^4\right)^5\left(1+t\right)^3}.
\end{align}

Noting that the $\urm(2)$ quotient quiver is subtracted from a leg that forms a balanced Dynkin diagram of $A_5$ within \Quiver{fig:E6000001A1Sub}, we have used the embedding \eqref{eq:AkAnEmbed} to compute the HKQ using Weyl integration, in agreement with the result from quotient quiver subtraction.
\subsubsection{$\overline{\left[\mathcal W_{E_6}\right]}^{[0,0,0,0,0,2]}_{[0,0,0,0,0,1]}///\surm(2)$}
\begin{figure}[h!]
    \centering
    \begin{tikzpicture}[main/.style = {draw, circle}]
    \node[main, label=below:$1$] (1L) []{};
    \node[main, label=below:$2$] (2L) [right=of 1L]{};
    \node[main, label=below:$3$] (3) [right=of 2L]{};
    \node[main, label=below:$2$] (2R) [right=of 3]{};
    \node[main, label=below:$1$] (1R) [right=of 2R]{};
    \node[main, label=left:$2$] (2T) [above=of 3]{};
    \node[main, label=left:$1$] (1TF) [above=of 2T]{};
    
    \draw[-] (1L)--(2L)--(3)--(2R)--(1R);
    \draw[-] (3)--(2T);
    \draw[double,double distance=3pt,line width=0.4pt] (1TF)--(2T);
    
    \node[main, label=below:$1$] (1subL) [below=of 1L]{};
    \node[main, label=below:$2$] (2sub) [right=of 1subL]{};
    \node[main, label=below:$1$] (1subR) [right=of 2sub]{};
    \node[draw=none,fill=none] (minus) [left=of 1subL]{$-$};
    \draw[-] (1subL)--(2sub)--(1subR);
    
    \node[draw, label=left:$1$] (oneF) [below=of 2sub]{};
    \node[main, label=below:$2$] (twoL) [below=of oneF]{};
    \node[main, label=below:$1$] (oneL) [left=of twoL]{};
    \node[main, label=below:$2$] (twoM) [right=of twoL]{};
    \node[main, label=below:$2$] (twoR) [right=of twoM]{};
    \node[main, label=below:$1$] (oneR) [right=of twoR]{};
    \node[draw,label=right:$1$] (oneFR) [above=of twoR]{};
    
    \draw[-] (twoL)--(twoM)--(twoR)--(oneR);
    \draw[-] (oneF)--(twoL);
    \draw[-] (oneFR)--(twoR);
    \draw[double,double distance=3pt,line width=0.4pt] (twoL)--(oneL);
    
    \node[draw=none,fill=none] (topghost) [right=of 1R]{};
    \node[draw=none,fill=none] (bottomghost) [right=of oneR]{};
    
    \draw [->] (topghost) to [out=-30,in=30,looseness=1] (bottomghost);

    \end{tikzpicture}
    \caption{Subtraction of the $\urm(2)$ quotient quiver from the unframed magnetic quiver for $\overline{\left[\mathcal W_{E_6}\right]}^{[0,0,0,0,0,2]}_{[0,0,0,0,0,1]}$  to yield \Quiver{fig:E6AG000002SU2Sub}.}
    \label{fig:E6AG000002SU2Sub}
\end{figure}
The unframed magnetic quiver for $\overline{\left[\mathcal W_{E_6}\right]}^{[0,0,0,0,0,2]}_{[0,0,0,0,0,1]}$ is shown at the top of \Figref{fig:E6AG000002SU2Sub}, together with the subtraction of the $\urm(2)$ quotient quiver to produce \Quiver{fig:E6AG000002SU2Sub}. The global symmetry of  $\overline{\left[\mathcal W_{E_6}\right]}^{[0,0,0,0,0,2]}_{[0,0,0,0,0,1]}$ is $\surm(6)\times \urm(1)$. The Coulomb branch of \Quiver{fig:E6AG000002SU2Sub} has an $\surm(4)\times \urm(1)^2$ Coulomb branch global symmetry. The monopole formula gives the unrefined HS:
\begin{align}
    \hs[\mathcal C\left(\mathcal Q_{\ref{fig:E6AG000002SU2Sub}}\right)]&=\frac{\left(\begin{aligned}1&+4 t^{2}+19 t^{3}+64 t^{4}+212 t^{5}+608 t^{6}+1609 t^{7}+3788 t^{8}+8198 t^{9}\\&+16140 t^{10}+29403 t^{11}+49352 t^{12}+77074 t^{13}+111680 t^{14}\\&+151113 t^{15}+190540 t^{16}+224913 t^{17}+248056 t^{18}+256504 t^{19}\\&+\text{palindrome}+t^{36}\end{aligned}\right)}{\left(1-t^2\right)^4\left(1-t^3\right)^8\left(1-t^4\right)^4\left(1+t\right)^4}
\end{align}
Noting the $\urm(2)$ quotient quiver is subtracted from an external leg which forms a balanced Dynkin diagram of $A_5$, we have used the embedding \eqref{eq:AkAnEmbed} to compute the HKQ using Weyl integration, in agreement with the result from quotient quiver subtraction.

\subsubsection{$\overline{\left[\mathcal W_{E_6}\right]}^{[0,0,0,1,1,0]}_{[0,0,0,0,0,1]}///\surm(2)$}
\begin{figure}[h!]
    \centering
    \begin{tikzpicture}[main/.style = {draw, circle}]
    \node[main, label=below:$1$] (1L) [] {};
    \node[main, label=below:$2$] (2L) [right= of 1L]{};
    \node[main, label=below:$3$] (3) [right=of 2L]{};
    \node[main, label=below:$3$] (3R) [right=of 3]{};
    \node[main, label=below:$2$] (2R) [right=of 2R]{};
    \node[main, label=left:$1$] (1T) [above=of 3]{};
    \node[main, label=left:$1$] (1TR) [above right= 1cm and 0.5cm of 3R]{};
    
    \draw[-] (1L)--(2L)--(3)--(3R)--(2R)--(1TR)--(3R);
    \draw[-] (1T)--(3);
    
    \node[main, label=below:$1$] (1subR) [below =of 3]{};
    \node[main, label=below:$2$] (2sub) [left=of 1subR]{};
    \node[main, label=below:$1$] (1subL) [left=of 2sub]{};
    \node[draw=none,fill=none] (minus) [left=of 1subL]{$-$};
    
    \draw[-] (1subR)--(2sub)--(1subL);
    
    \node[main, label=left:$1$] (oneL) [below=of 1subR]{};
    \node[main, label=below:$2$] (twoL) [below=of oneL]{};
    \node[main, label=below:$3$] (three) [right=of twoL]{};
    \node[main, label=below:$2$] (twoR) [right=of three]{};
    \node[main, label=right:$1$] (oneT) [above right= 1cm and 0.5cm of three]{};
    \node[draw, label=right:$1$] (oneF) at (three|-oneL){};
    
    \draw[-] (three)--(oneF)--(oneL)--(twoL)--(three)--(twoR)--(oneT)--(three);

    \node[draw=none,fill=none] (topghost) [right=of 2R]{};
    \node[draw=none,fill=none] (bottomghost) [right=of twoR]{};
    
    \draw [->] (topghost) to [out=-30,in=30,looseness=1] (bottomghost);

    \end{tikzpicture}
    \caption{Subtraction of the $\urm(2)$ quotient quiver from the unframed magnetic quiver for $\overline{\left[\mathcal W_{E_6}\right]}^{[0,0,0,1,1,0]}_{[0,0,0,0,0,1]}$ to yield \Quiver{fig:E6000001twoA1Sub}.}
    \label{fig:E6000001twoA1Sub}
\end{figure}
The unframed magnetic quiver for $\overline{\left[\mathcal W_{E_6}\right]}^{[0,0,0,1,1,0]}_{[0,0,0,0,0,1]}$ is drawn at the top of \Figref{fig:E6000001twoA1Sub}, together with the subtraction of the $\urm(2)$ quotient quiver to produce \Quiver{fig:E6000001twoA1Sub}. Note that there is only one possible alignment which respects all of the selection rules given in \ref{sec:rules}. The global symmetry of $\overline{\left[\mathcal W_{E_6}\right]}^{[0,0,0,1,1,0]}_{[0,0,0,0,0,1]}$ is $\surm(6)\times \urm(1)$, and the global symmetry of the Coulomb branch of \Quiver{fig:E6000001twoA1Sub} is $\surm(4)\times \urm(1)^2$. The latter is confirmed by the computation of the HS using the monopole formula:
\begin{equation}
    \hs[\mathcal C\left(\mathcal Q_{\ref{fig:E6000001twoA1Sub}}\right)]=\frac{\left(\begin{aligned}1&+7 t+39 t^2+170 t^3+668 t^4+2372 t^5+7780 t^6+23590 t^7+66604 t^8\\&+175712 t^9+435301 t^{10}+1016083 t^{11}+2242404 t^{12}+4691381 t^{13}\\&+9328137 t^{14}+17665378 t^{15}+31926140 t^{16}+55157964 t^{17}\\&+91240107 t^{18}+144699995 t^{19}+220286604 t^{20}+322261559 t^{21}\\&+453465718 t^{22}+614264933 t^{23}+801601813 t^{24}+1008376706 t^{25}\\&+1223428820 t^{26}+1432240058 t^{27}+1618409691 t^{28}+1765687065 t^{29}\\&+1860269901 t^{30}+1892885098 t^{31}+\text{palindrome}+t^{62}\end{aligned}\right)}{\left(1-t\right)^{-1}\left(1-t^3\right)^7\left(1-t^4\right)^6\left(1-t^5\right)^6\left(1+t\right)^6}.
\end{equation}
The $\urm(2)$ quotient quiver is subtracted from a leg which forms the balanced Dynkin diagram of $A_5$. We have used the embedding in \eqref{eq:AkAnEmbed} to compute the HKQ using Weyl integration and find agreement with quiver subtraction.
\subsubsection{$\left[\overline{\mathcal W_{E_6}}\right]^{[1,0,0,1,0,0]}_{[0,2,0,0,0,0]}///\surm(2)$}
\begin{figure}[h!]
    \centering
    \begin{tikzpicture}[main/.style = {draw, circle}]
        \node[main, label=below:$1$] (1L) []{};
        \node[main, label=below:$1$] (1M) [right=of 1L]{};
        \node[main, label=below:$2$] (2L) [right=of 1M]{};
        \node[main, label=below:$2$] (2R) [right=of 2L]{};
        \node[main, label=below:$1$] (1R) [right=of 2R]{};
        \node[main, label=left:$1$] (1T) [above=of 2L]{};
        \node[main, label=below:$1$] (1B) [below right= 1cm and 0.5cm of 1M]{};
        \draw[-] (1L)--(1M)--(2L)--(2R)--(1R);
        \draw[-] (2R)--(1B)--(1L);
        \draw[-] (1T)--(2L);
        
    \end{tikzpicture}
    \caption{Unframed magnetic quiver \Quiver{fig:E6100100} for $\left[\overline{\mathcal W_{E_6}}\right]^{[1,0,0,1,0,0]}_{[0,2,0,0,0,0]}$. }
    \label{fig:E6100100}
\end{figure}
\begin{figure}[h!]
 \centering
    \begin{subfigure}{0.4\textwidth}
    \centering
    \resizebox{0.75\width}{!}{\begin{tikzpicture}[main/.style = {draw, circle}]

    \node[main, label=below:$1$] (1L) []{};
    \node[main, label=below:$1$] (1M) [right=of 1L]{};
    \node[main, label=below:$2$] (2L) [right=of 1M]{};
    \node[main, label=below:$2$] (2R) [right=of 2L]{};
    \node[main, label=below:$1$] (1R) [right=of 2R]{};
    \node[main, label=left:$1$] (1T) [above=of 2L]{};
    \node[main, label=below:$1$] (1B) [below right= 1cm and 0.5cm of 1M]{};
    \draw[-] (1L)--(1M)--(2L)--(2R)--(1R);
    \draw[-] (2R)--(1B)--(1L);
    \draw[-] (1T)--(2L);

    \node[main, label=below:$1$] (1subR) [below=of 1R]{};
    \node[main, label=below:$2$] (2sub) [left=of 1subR]{};
    \node[main, label=below:$1$] (1subL) [below =of 1B]{};

    \draw[-] (1subR)--(2sub)--(1subL);

    \node[] (minus) [right=of 1subR]{$-$};

    \node[draw, label=right:$2$] (twof) [below right= 1cm and 0.5cm of 1subL]{};
    \node[main, label=below:$2$] (two) [below =of twof]{};
    \node[main, label=below:$1$] (onem) [left=of two]{};
    \node[main, label=below:$1$] (onel) [left=of onem]{};
    \node[main, label=below:$1$] (oner) [right=of two]{};
    \node[draw, label=left:$1$] (onef) [above=of onel]{};
    \draw[-] (onef)--(onel)--(onem)--(two)--(oner);
    \draw[-] (twof)--(two);

    \node[] (topghost) [left=of 1L]{};
    \node[] (bottomghost) [left=of onel]{};

    \draw [->] (topghost) to [out=-150,in=150,looseness=1] (bottomghost);
    \end{tikzpicture}}\caption{}
    \label{fig:E6100100Sub1}\end{subfigure}
    \hfill
    \begin{subfigure}{0.4\textwidth}
    \centering
    \resizebox{0.75\width}{!}{\begin{tikzpicture}[main/.style = {draw, circle}]

    \node[main, label=below:$1$] (1L) []{};
    \node[main, label=below:$1$] (1M) [right=of 1L]{};
    \node[main, label=below:$2$] (2L) [right=of 1M]{};
    \node[main, label=below:$2$] (2R) [right=of 2L]{};
    \node[main, label=below:$1$] (1R) [right=of 2R]{};
    \node[main, label=left:$1$] (1T) [above=of 2L]{};
    \node[main, label=below:$1$] (1B) [below right= 1cm and 0.5cm of 1M]{};
    \draw[-] (1L)--(1M)--(2L)--(2R)--(1R);
    \draw[-] (2R)--(1B)--(1L);
    \draw[-] (1T)--(2L);

    \node[main, label=below:$1$] (1subR) [below=of 1R]{};
    \node[main, label=below:$2$] (2sub) [left=of 1subR]{};
    \node[main, label=below:$1$] (1subL) [left =of 2sub]{};

    \draw[-] (1subR)--(2sub)--(1subL);

    \node[] (minus) [right=of 1subR]{$-$};

    \node[draw, label=right:$1$] (onefr) [below=of 1subR]{};
    \node[main, label=below:$1$] (oner) [below=of onefr]{};
    \node[main, label=below:$1$] (onemr) [left=of oner]{};
    \node[main, label=below:$1$] (onem) [left=of onemr]{};
    \node[main, label=below:$1$] (oneml) [left=of onem]{};
    \node[main, label=below:$1$] (onel) [left=of oneml]{};
    \node[draw, label=left:$1$] (onefm) [above=of onem]{};
    \node[draw, label=left:$2$] (twof) [above=of onel]{};

    \draw[-] (twof)--(onel)--(oneml)--(onem)--(onemr)--(oner)--(onefr);
    \draw[-] (onefm)--(onem);

    \node[] (topghost) [left=of 1L]{};
    \node[] (bottomghost) [left=of onel]{};

    \draw [->] (topghost) to [out=-150,in=150,looseness=1] (bottomghost);
    \end{tikzpicture}}\caption{}\label{fig:E6100100Sub2}\end{subfigure}
    \centering
    \begin{subfigure}{\textwidth}
    \centering
    \begin{tikzpicture}[main/.style = {draw, circle}]

    \node[draw, label=right:$1$] (onefr) []{};
    \node[main, label=below:$1$] (oner) [below=of onefr]{};
    \node[main, label=below:$1$] (onemr) [left=of oner]{};
    \node[main, label=below:$1$] (onem) [left=of onemr]{};
    \node[main, label=below:$1$] (oneml) [left=of onem]{};
    \node[draw, label=left:$1$] (onefm) [above=of onem]{};
    \node[draw, label=left:$1$] (onefl)[above=of oneml]{};
    
    \draw[-] (onefr)--(oner)--(onemr)--(onem)--(oneml)--(onefl);
    \draw[-] (onefm)--(onem);

    \end{tikzpicture}\caption{}\label{fig:E6100100SubInt}
    \end{subfigure}
    \caption{Both alignments of the $\urm(2)$ quotient quiver against the unframed magnetic quiver \Quiver{fig:E6100100} for $\left[\overline{\mathcal W_{E_6}}\right]^{[1,0,0,1,0,0]}_{[0,2,0,0,0,0]}$, giving the quivers \Quiver{fig:E6100100Sub1} and \Quiver{fig:E6100100Sub2}. Their intersection, via $A_1$ KP transitions, is quiver \Quiver{fig:E6100100SubInt}.}
    
\end{figure}
The unframed magnetic quiver \Quiver{fig:E6100100} for $\left[\overline{\mathcal W_{E_6}}\right]^{[1,0,0,1,0,0]}_{[0,2,0,0,0,0]}$ is given in \Figref{fig:E6100100}. The global symmetry of $\left[\overline{\mathcal W_{E_6}}\right]^{[1,0,0,1,0,0]}_{[0,2,0,0,0,0]}$ is $\surm(5)\times \urm(1)^2$. There are two equivalent external legs for $\urm(2)$ quotient quiver subtraction respecting the rules given in \ref{sec:rules}, so either may be chosen for the subtraction. There are two possible alignments of the $\urm(2)$ quotient quiver, which produce the magnetic quivers \Quiver{fig:E6100100Sub1} and \Quiver{fig:E6100100Sub2}, for the affine Grassmannian slices $\left[\overline{\mathcal W_{A_4}}\right]^{[1,0,2,0]}_{[0,1,0,0]}$and$ \left[\overline{\mathcal W_{A_5}}\right]^{[2,0,1,0,1]}_{[1,0,1,0,0]}$, respectively. We conclude that:
\begin{equation}
    \left[\overline{\mathcal W_{E_6}}\right]^{[1,0,0,1,0,0]}_{[0,2,0,0,0,0]}///\surm(2) = \left[\overline{\mathcal W_{A_4}}\right]^{[1,0,2,0]}_{[0,1,0,0]}\cup \left[\overline{\mathcal W_{A_5}}\right]^{[2,0,1,0,1]}_{[1,0,1,0,0]}\label{eq:E6100100A1}
\end{equation}
The intersection of \Quiver{fig:E6100100Sub1} and \Quiver{fig:E6100100Sub2} is \Quiver{fig:E6100100SubInt}, which is a magnetic quiver for $\left[\overline{\mathcal W_{A_4}}\right]^{[1,1,0,1]}_{[0,1,0,0]}$. Proceeding as before, we find the HS of the union \eqref{eq:E6100100A1}:
\begin{align}
    &HS\left[\mathcal C\left(\left[\overline{\mathcal W_{A_4}}\right]^{[1,0,2,0]}_{[0,1,0,0]}\cup \left[\overline{\mathcal W_{A_5}}\right]^{[2,0,1,0,1]}_{[1,0,1,0,0]}\right)\right]\nonumber\\
    &=\frac{(1+t^2) \left(1+5 t^2+8 t^3+14 t^4+16 t^5+23 t^6+16 t^7+14 t^8+8 t^9+5 t^{10}+t^{12}\right)}
    {\left(1-t^2\right)^{6} \left(1-t^3\right)^4}\nonumber\\&+\frac{\left(\begin{aligned}1&+t+11 t^2+22 t^3+70 t^4+128 t^5+266 t^6+398 t^7+623 t^8+761 t^9+935 t^{10}+938 t^{11}\\&+935 t^{12}+761 t^{13}+623 t^{14}+398 t^{15}+266 t^{16}+128 t^{17}+70 t^{18}+22 t^{19}+11 t^{20}+t^{21}+t^{22}\end{aligned}\right)}{\left(1-t\right)^{-1}\left(1-t^2\right)^3\left(1-t^3\right)^5\left(1-t^4\right)^3}\nonumber\\&-\frac{\left(\begin{aligned}1&+2 t+9 t^2+24 t^3+50 t^4+76 t^5+108 t^6+120 t^7\\&+108 t^8+76 t^9+50 t^{10}+24 t^{11}+9 t^{12}+2 t^{13}+t^{14}\end{aligned}\right)}{\left(1-t\right)^{-2}\left(1-t^2\right)^6\left(1-t^3\right)^4}\label{eq:HSE61001001A1Union}\\\nonumber\\
    &=\frac{\left(\begin{aligned}1&+t+11 t^2+22 t^3+78 t^4+160 t^5+353 t^6+601 t^7+980 t^8+1335 t^9+1682 t^{10}\\&+1827 t^{11}+1800 t^{12}+1529 t^{13}+1151 t^{14}+722 t^{15}+372 t^{16}+138 t^{17}+13 t^{18}\\&-25 t^{19}-28 t^{20}-15 t^{21}-8 t^{22}-t^{23}-t^{24}\end{aligned}\right)}
    {\left(1-t\right)^{-1}\left(1-t^2\right)^3\left(1-t^3\right)^5\left(1-t^4\right)^3}
\end{align}
 
The HS in the signed sum \eqref{eq:HSE61001001A1Union} are for $\left[\overline{\mathcal W_{A_4}}\right]^{[1,0,2,0]}_{[0,1,0,0]}$,$ \left[\overline{\mathcal W_{A_5}}\right]^{[2,0,1,0,1]}_{[1,0,1,0,0]}$, and $\left[\overline{\mathcal W_{A_4}}\right]^{[1,1,0,1]}_{[0,1,0,0]}$, respectively.

The $\urm(2)$ quotient quiver is subtracted from a leg which forms the Dynkin diagram of $A_4$. One can use the embedding in \eqref{eq:AkAnEmbed} to compute the HKQ using Weyl integration and to find agreement with quiver subtraction.
\subsubsection{$\left[\overline{\mathcal W_{E_7}}\right]^{[0,0,0,1,0,0,0]}_{[0,0,0,0,0,0,1]}///\surm(2)$}
\begin{figure}[h!]
    \centering
    \begin{tikzpicture}[main/.style = {draw, circle}]
        \node[main, label=below:$1$] (1L) []{};
        \node[main, label=below:$2$] (2L) [right=of 1L]{};
        \node[main, label=below:$3$] (3L) [right=of 2L]{};
        \node[main, label=below:$3$] (3R) [right=of 3L]{};
        \node[main, label=below:$2$] (2R) [right=of 3R]{};
        \node[main, label=below:$1$] (1R) [right=of 2R]{};
        \node[main, label=left:$1$] (1TL) [above=of 3L]{};
        \node[main,label=right:$1$] (1F) [above=of 3R]{};

        \draw[-] (1L)--(2L)--(3L)--(3R)--(2R)--(1R);
        \draw[-] (1TL)--(3L);
        \draw[-] (1F)--(3R);

        \node[main, label=below:$1$] (1subL) [below=of 1L]{};
        \node[main, label=below:$2$] (2sub) [right=of 1subL]{};
        \node[main, label=below:$1$] (1subR) [right=of 2sub]{};
        \node[draw=none,fill=none] (minus) [left=of 1subL] {$-$};

        \draw[-] (1subL)--(2sub)--(1subR);

        \node[draw, label=left:$1$] (1fres) [below=of 2sub]{};
        \node[main, label=below:$1$] (1resL) [below=of 1fres]{};
        \node[main,label=below:$2$] (2resL) [right=of 1resL]{};
        \node[main, label=below:$3$] (3res) [right=of 2resL]{};
        \node[main, label=below:$2$] (2resR) [right=of 3res]{};
        \node[main, label=below:$1$] (1resR) [right=of 2resR]{};
        \node[main, label=right:$1$] (1resT) [above=of 3res]{};
        \node[draw, label=right:$1$] (1resF) [above right=of 3res]{};

        \draw[-] (1fres)--(1resL)--(2resL)--(3res)--(2resR)--(1resR);
        \draw[-] (1resT)--(3res)--(1resF);

         \node[draw=none,fill=none] (topghost) [right=of 1R]{};
        \node[draw=none,fill=none] (bottomghost) [right=of 1resR]{};
    
        \draw [->] (topghost) to [out=-30,in=+30,looseness=1] (bottomghost);

    \end{tikzpicture}
    \caption{Subtraction of the $\urm(2)$ quotient quiver from the unframed magnetic quiver for $\left[\overline{\mathcal W_{E_7}}\right]^{[0,0,0,1,0,0,0]}_{[0,0,0,0,0,0,1]}$ to produce the magnetic quiver \Quiver{fig:E7affGoveronlast}.}
    \label{fig:E7affGoveronlast}
\end{figure}
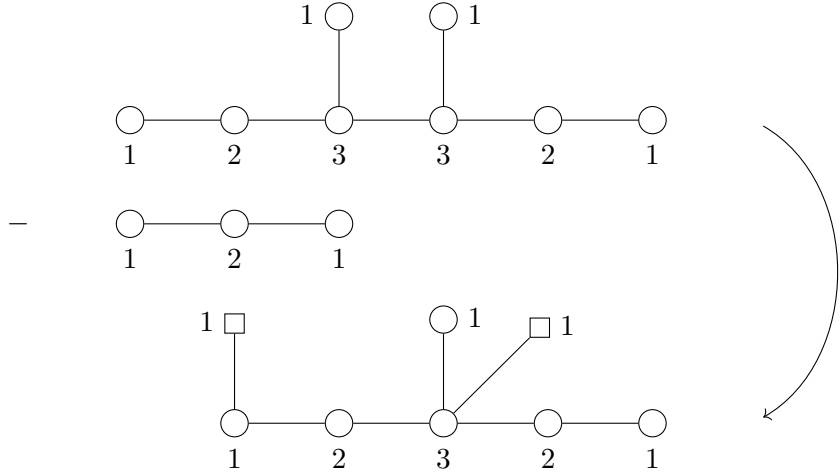
The magnetic quiver for $\left[\overline{\mathcal W_{E_7}}\right]^{[0,0,0,1,0,0,0]}_{[0,0,0,0,0,0,1]}$ is shown at the top of \Figref{fig:E7affGoveronlast}, followed by subtraction of the $\urm(2)$ quotient quiver to produce \Quiver{fig:E7affGoveronlast}, which is a magnetic quiver for $\left[\overline{\mathcal W_{E_6}}\right]^{[1,0,1,0,0,0]}_{[1,0,0,0,0,1]}$. The global symmetry of $\left[\overline{\mathcal W_{E_7}}\right]^{[0,0,0,1,0,0,0]}_{[0,0,0,0,0,0,1]}$ is $\surm(7)\times \urm(1)$. Note that there is only one external leg which respects the selection rules for quiver subtraction. We obtain the result: \begin{equation}
    \left[\overline{\mathcal W_{E_7}}\right]^{[0,0,0,1,0,0,0]}_{[0,0,0,0,0,0,1]}///\surm(2)=\left[\overline{\mathcal W_{E_6}}\right]^{[1,0,1,0,0,0]}_{[1,0,0,0,0,1]}.
\end{equation}
The global symmetry of $\left[\overline{\mathcal W_{E_6}}\right]^{[1,0,1,0,0,0]}_{[1,0,0,0,0,1]}$ is $\surm(5)\times \urm(1)^2$, which is verified through the computation of the HS using the monopole formula. 
\begin{align}
    HS\left[\mathcal C\left(\mathcal Q_{\ref{fig:E7affGoveronlast}}\right)\right]=\frac{\left(\begin{aligned}1&+4 t+29 t^2+116 t^3+507 t^4+1764 t^5+5897 t^6+17316 t^7+47598 t^8\\&+119000 t^9+277876 t^{10}+600000 t^{11}+1213902 t^{12}+2292834 t^{13}\\&+4073898 t^{14}+6800152 t^{15}+10715088 t^{16}+15929986 t^{17}\\&+22418427 t^{18}+29855900 t^{19}+37713801 t^{20}+45170786 t^{21}\\&+51383981 t^{22}+55484436 t^{23}+56938432 t^{24}+\text{palindrome}+t^{48}\end{aligned}\right)}{(1-t)^{-4}(1-t^2)^{7}(1-t^3)^{10} (1-t^4)^7}
\end{align}

As the $\urm(2)$ quotient quiver is subtracted from a balanced subdiagram of the magnetic quiver for $\left[\overline{\mathcal W_{E_7}}\right]^{[0,0,0,1,0,0,0]}_{[0,0,0,0,0,0,1]}$, which forms the $A_6$ Dynkin diagram, we employ the embedding given in \eqref{eq:AkAnEmbed} to verify the result.
\subsubsection{$\overline{\left[\mathcal W_{E_7}\right]}^{[0,1,0,0,0,0,0]}_{[1,0,0,0,0,0,0]}///\surm(2)$}
\begin{figure}[h!]
    \centering
    \begin{tikzpicture}[main/.style = {draw, circle}]
        \node[main, label=below:$1$] (1L) []{};
        \node[main, label=below:$3$] (3L) [right=of 1L]{};
        \node[main, label=below:$4$] (4) [right=of 3L]{};
        \node[main, label=below:$3$] (3R) [right=of 4]{};
        \node[main, label=below:$2$] (2R) [right=of 3R]{};
        \node[main, label=below:$1$] (1R) [right=of 2R]{};
        \node[main,label=left:$1$] (1T) [above=of 3L]{};
        \node[main, label=right:$2$] (2T) [above=of 4]{};

        \draw[-] (1L)--(3L)--(4)--(3R)--(2R)--(1R);
        \draw[-] (1T)--(3L);
        \draw[-] (2T)--(4);

        \node[main, label=below:$1$] (1subR) [below=of 1R]{};
        \node[main, label=below:$2$] (2sub) [left=of 1subR]{};
        \node[main, label=below:$1$] (1subL) [left=of 2sub]{};
        \node[draw=none,fill=none] (minus) [left=of 1subL]{$-$};

        \draw[-] (1subR)--(2sub)--(1subL);

        \node[main, label=right:$2$] (twoT) [below=of minus]{};
        \node[main, label=below:$4$] (four) [below=of twoT]{};
        \node[main, label=below:$3$] (three) [left=of four]{};
        \node[main, label=below:$1$] (oneL) [left=of three]{};
        \node[main, label=below:$2$] (two) [right=of four]{};
        \node[main, label=left:$1$] (oneT) [above=of three]{};
        \node[draw,label=right:$1$] (oneF) [above right=of four]{};

        \draw[-] (two)--(four)--(three)--(oneL);
        \draw[-] (twoT)--(four)--(oneF);
        \draw[-] (oneT)--(three);

        \node[draw=none,fill=none] (topghost) [left=of 1L]{};
        \node[draw=none,fill=none] (bottomghost) [left=of oneL]{};
    
        \draw [->] (topghost) to [out=150,in=-150,looseness=1] (bottomghost);
    
    \end{tikzpicture}
    \caption{Subtraction of the $\urm(2)$ quotient quiver from an unframed magnetic quiver for $\overline{\left[\mathcal W_{E_7}\right]}^{[0,1,0,0,0,0,0]}_{[1,0,0,0,0,0,0]}$ to give \Quiver{fig:E7AG0100000SU2Sub}.}
    \label{fig:E7AG0100000SU2Sub}
\end{figure}
The unframed magnetic quiver for $\overline{\left[\mathcal W_{E_7}\right]}^{[0,1,0,0,0,0,0]}_{[1,0,0,0,0,0,0]}$, which has global symmetry $SO(12)\times \urm(1)$, is shown at the top of \Figref{fig:E7AG0100000SU2Sub}. This is followed by the subtraction of the $\urm(2)$ quotient quiver to produce the magnetic quiver \Quiver{fig:E7AG0100000SU2Sub}. This has a global symmetry of $SO(8)\times \surm(2)\times \urm(1)$, as can be determined from the Hilbert series:

\begin{equation}
    HS\left[\mathcal C\left(\mathcal Q_{\ref{fig:E7AG0100000SU2Sub}}\right)\right]=\frac{\left(\begin{aligned}1&+9 t+66 t^2+373 t^3+1914 t^4+8701 t^5+36329 t^6+138552 t^7\\&+489648 t^8+1605300 t^9+4918751 t^{10}+14118695 t^{11}+38138780 t^{12}\\&+97185445 t^{13}+234361303 t^{14}+535948910 t^{15}+1165046145 t^{16}\\&+2411578527 t^{17}+4762032274 t^{18}+8983378907 t^{19}+16213188569 t^{20}\\&+28027708562 t^{21}+46461483210 t^{22}+73925927636 t^{23}\\&+113003617343 t^{24}+166076737959 t^{25}+234831061465 t^{26}\\&+319661125610 t^{27}+419130743340 t^{28}+529576399650 t^{29}\\&+645063463512 t^{30}+757718368534 t^{31}+858546072189 t^{32}\\&+938541874231 t^{33}+990021037865 t^{34}+1007788178974 t^{35}\\&+\text{palindrome}+t^{70}\end{aligned}\right)}{\left(1-t\right)^{-9}\left(1-t^2\right)^{11}\left(1-t^3\right)^{13}\left(1-t^4\right)^{11}}.
\end{equation}

Note that the pattern of balanced gauge nodes in \Quiver{fig:E7AG0100000SU2Sub} requires a Coulomb branch global symmetry of at least $SO(8)\times \urm(1)^2$, and so an enhancement has occurred.

The HKQ can be computed explicitly with Weyl integration using the embedding \eqref{eq:DkA1Embed} to find agreement with quotient quiver subtraction.

\subsection{$\surm(3)$ HKQ}
\label{sec:ExceptionalAGSU3HKQs}
\subsubsection{$\overline{\left[\mathcal W_{E_7}\right]}^{[0,1,0,0,0,0,0]}_{[1,0,0,0,0,0,0]}///\surm(3)$}
\begin{figure}[h!]
    \centering
\begin{subfigure}{0.4\textwidth}
    \centering\resizebox{0.8\width}{!}{
    \begin{tikzpicture}[main/.style = {draw, circle}]
       \node[main, label=below:$1$] (1L) []{};
        \node[main, label=below:$3$] (3L) [right=of 1L]{};
        \node[main, label=below:$4$] (4) [right=of 3L]{};
        \node[main, label=below:$3$] (3R) [right=of 4]{};
        \node[main, label=below:$2$] (2R) [right=of 3R]{};
        \node[main, label=below:$1$] (1R) [right=of 2R]{};
        \node[main,label=left:$1$] (1T) [above=of 3L]{};
        \node[main, label=right:$2$] (2T) [above=of 4]{};

        \draw[-] (1L)--(3L)--(4)--(3R)--(2R)--(1R);
        \draw[-] (1T)--(3L);
        \draw[-] (2T)--(4);

        \node[main, label=left:$1$] (1subL) [below=of 4]{};
        \node[main, label=below:$2$] (2subL) [below=of 1subL]{};
        \node[main, label=below:$3$] (3sub) [right=of 2subL]{};
        \node[main, label=below:$2$] (2subR) [right=of 3sub]{};
        \node[main, label=below:$1$] (1subR) [right=of 2subR]{};

        \node[draw=none,fill=none] (minus) [left=of 2subL]{$-$};

        \draw[-] (1subL)--(2subL)--(3sub)--(2subR)--(1subR);

        \node[main, label=right:$1$] (1resT) [below=of minus]{};
        \node[main, label=below:$3$] (3res) [below=of 1resT]{};
        \node[main, label=below:$1$] (1resL) [left=of 3res]{};
        \node[main, label=below:$2$] (2res) [right=of 3res]{};
        \node[main, label=below:$1$] (1resR) [right=of 2res]{};
        \node[draw, label=left:$2$] (2resF) [above left=of 3res]{};

        \draw[-] (1resL)--(3res)--(2res)--(1resR);
        \draw[-] (2resF)--(3res)--(1resT);
        
         \node[draw=none,fill=none] (topghost) [left=of 1L]{};
        \node[draw=none,fill=none] (bottomghost) [left=of 1resL]{};
    
    \draw [->] (topghost) to [out=180,in=150,looseness=0.5] (bottomghost);
        
    \end{tikzpicture}}
    \caption{}
    \label{fig:E71000000A2Sub1}
\end{subfigure}
\hfill
\begin{subfigure}{0.4\textwidth}
    \centering\resizebox{0.8\width}{!}{
    \begin{tikzpicture}[main/.style = {draw, circle}]
     \node[main, label=below:$1$] (1L) []{};
        \node[main, label=below:$3$] (3L) [right=of 1L]{};
        \node[main, label=below:$4$] (4) [right=of 3L]{};
        \node[main, label=below:$3$] (3R) [right=of 4]{};
        \node[main, label=below:$2$] (2R) [right=of 3R]{};
        \node[main, label=below:$1$] (1R) [right=of 2R]{};
        \node[main,label=left:$1$] (1T) [above=of 3L]{};
        \node[main, label=right:$2$] (2T) [above=of 4]{};

        \draw[-] (1L)--(3L)--(4)--(3R)--(2R)--(1R);
        \draw[-] (1T)--(3L);
        \draw[-] (2T)--(4);

        \node[main, label=below:$1$] (1subL) [below=of 3L]{};
        \node[main, label=below:$2$] (2subL) [right=of 1subL]{};
        \node[main, label=below:$3$] (3sub) [right=of 2subL]{};
        \node[main, label=below:$2$] (2subR) [right=of 3sub]{};
        \node[main, label=below:$1$] (1subR) [right=of 2subR]{};

        \node[draw=none,fill=none] (minus) [left=of 1subL]{$-$};

        \draw[-] (1subL)--(2subL)--(3sub)--(2subR)--(1subR);

        \node[main, label=right:$1$] (1resT) [below=of 1subL]{};
        \node[main, label=below:$2$] (2resL) [below=of 1resT]{};
        \node[main, label=below:$1$] (1resL) [left=of 2resL]{};
        \node[main, label=below:$2$] (2resM) [right=of 2resL]{};
        \node[main, label=below:$2$] (2resR) [right=of 2resM]{};
        \node[draw, label=right:$2$] (2resF) [above=of 2resR]{};
        \node[draw, label=left:$1$] (1resF) at (1resL|-1resT){};

        \draw[-] (2resF)--(2resR)--(2resM)--(2resL)--(1resL)--(1resF)--(1resT)--(2resL);

         \node[draw=none,fill=none] (topghost) [right=of 1R]{};
        \node[draw=none,fill=none] (bottomghost) [right=of 2resR]{};
    
    \draw [->] (topghost) to [out=-30,in=0,looseness=0.5] (bottomghost);

    \end{tikzpicture}}
    \caption{}
    \label{fig:E71000000A2Sub2}
\end{subfigure}
\centering
    \begin{subfigure}{\linewidth}
    \centering
        \begin{tikzpicture}[main/.style = {draw, circle}]
        \node[main, label=right:$1$] (1resT) {};
        \node[main, label=below:$2$] (2resL) [below=of 1resT]{};
        \node[main, label=below:$1$] (1resL) [left=of 2resL]{};
        \node[main, label=below:$2$] (2resM) [right=of 2resL]{};
        \node[main, label=below:$1$] (2resR) [right=of 2resM]{};
        \node[draw, label=left:$1$] (1resF) at (1resL|-1resT){};

        \draw[-] (2resR)--(2resM)--(2resL)--(1resL)--(1resF)--(1resT)--(2resL);

        \end{tikzpicture}
        \caption{}
        \label{fig:E6000001A2SubInt}
        \end{subfigure}
    \caption{Both alignments of the $\urm(3)$ quotient quiver against the unframed magnetic quiver for $\overline{\left[\mathcal W_{E_7}\right]}^{[0,1,0,0,0,0,0]}_{[1,0,0,0,0,0,0]}$ to give the quivers \Quiver{fig:E71000000A2Sub1} and \Quiver{fig:E71000000A2Sub2}. Their intersection, via $A_1$ KP transitions, is quiver \Quiver{fig:E6000001A2SubInt}. }
    \label{fig:E6000001A2SubBoth}
\end{figure}
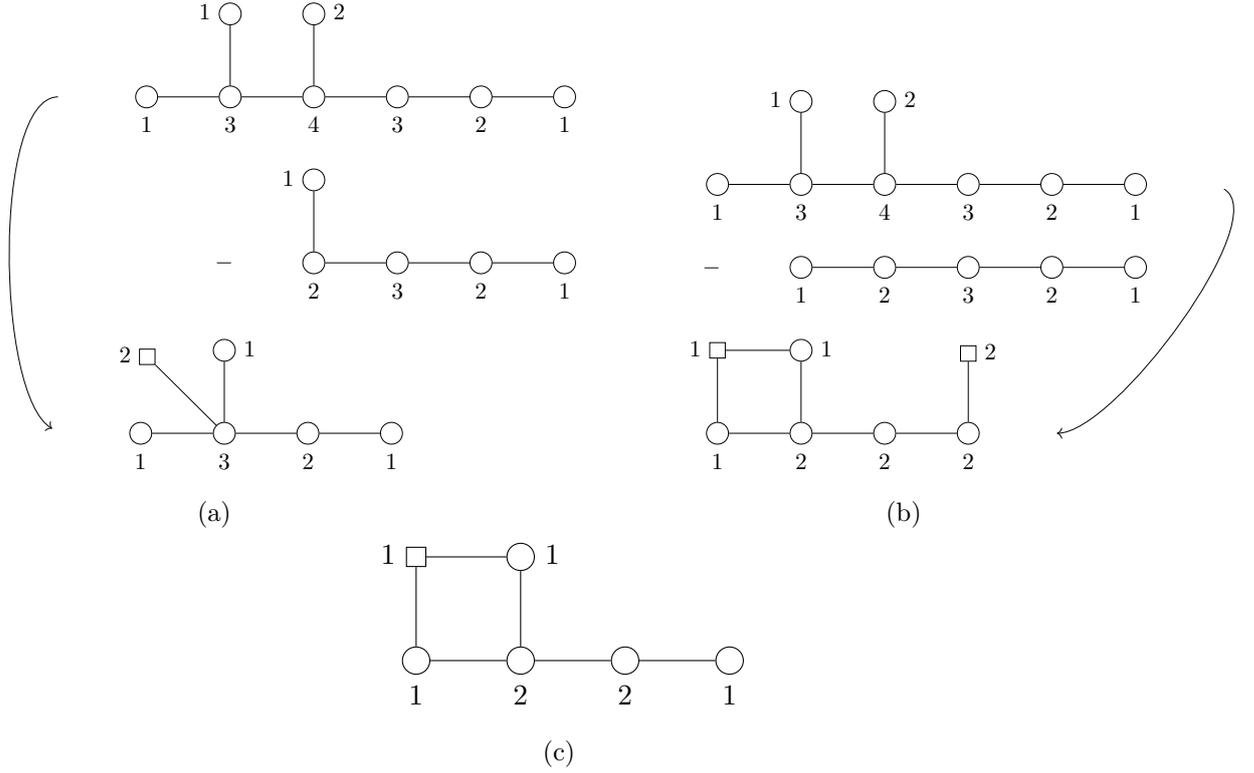

We can also take the unframed magnetic quiver for $\overline{\left[\mathcal W_{E_7}\right]}^{[0,1,0,0,0,0,0]}_{[1,0,0,0,0,0,0]}$ and consider the subtraction of the $\urm(3)$ quotient quiver. There are two possible alignments shown in \Figref{fig:E6000001A2SubBoth}, producing quivers \Quiver{fig:E71000000A2Sub1} and \Quiver{fig:E71000000A2Sub2}. These are magnetic quivers for $\overline{\left[\mathcal W_{D_5}\right]}^{[0,0,2,0,0]}_{[0,0,0,1,1]}$ and $ \overline{\left[\mathcal W_{D_5}\right]}^{[2,0,0,1,1]}_{[2,0,0,1,1]}$ respectively. We conclude that, \begin{equation}
    \overline{\left[\mathcal W_{E_7}\right]}^{[0,1,0,0,0,0,0]}_{[1,0,0,0,0,0,0]}///\surm(3)=\overline{\left[\mathcal W_{D_5}\right]}^{[0,0,2,0,0]}_{[0,0,0,1,1]}\cup \overline{\left[\mathcal W_{D_5}\right]}^{[2,0,0,1,1]}_{[2,0,0,1,1]}.
\end{equation}

To compute the union we note that there is an $A_1$ KP transition from each of \Quiver{fig:E71000000A2Sub1} and \Quiver{fig:E71000000A2Sub2} to their intersection, which is a magnetic quiver for $\overline{\left[\mathcal W_{D_5}\right]}^{[0,1,0,1,1]}_{[0,0,0,1,1]}$. The monopole formula yields the HS of the union as:
\begin{align}
   & HS\left[\overline{\left[\mathcal W_{D_5}\right]}^{[0,0,2,0,0]}_{[0,0,0,1,1]}\cup \overline{\left[\mathcal W_{D_5}\right]}^{[2,0,0,1,1]}_{[2,0,0,1,1]}\right]\nonumber\\
   &=\frac{\left(\begin{aligned}1&+3 t+15 t^2+46 t^3+148 t^4+386 t^5+954 t^6+2064 t^7+4183 t^8+7649 t^9+13081 t^{10}\\&+20490 t^{11}+30060 t^{12}+40738 t^{13}+51804 t^{14}+61138 t^{15}+67790 t^{16}+69920 t^{17}\\&+\text{palindrome}+t^{34}\end{aligned}\right)}{\left(1-t\right)^{-3}\left(1-t^2\right)^{8}\left(1-t^3\right)^7\left(1-t^4\right)^4}\nonumber\\
   &+\frac{\left(\begin{aligned}1&+4 t+21 t^2+72 t^3+264 t^4+792 t^5+2276 t^6+5728 t^7+13478 t^8+28656 t^9+56852 t^{10}\\&+103680 t^{11}+177030 t^{12}+280752 t^{13}+418307 t^{14}+582484 t^{15}+764001 t^{16}+940120 t^{17}\\&+1091630 t^{18}+1191688 t^{19}+1228644 t^{20}+\text{palindrome}+t^{40}\end{aligned}\right)}{\left(1-t\right)^{-4}\left(1-t^2\right)^6\left(1-t^3\right)^8\left(1-t^4\right)^6}\nonumber\\&-\frac{\left(\begin{aligned}1&+3 t+18 t^2+55 t^3+198 t^4+539 t^5+1445 t^6+3288 t^7+7052 t^8+13416 t^9+23838 t^{10}\\&+38390 t^{11}+57751 t^{12}+79667 t^{13}+102836 t^{14}+122489 t^{15}+136705 t^{16}+141190 t^{17}\\&+\text{palindrome}+t^{34}\end{aligned}\right)}{\left(1-t\right)^{-3}\left(1-t^3\right)^7\left(1-t^4\right)^{5}}\\&=\frac{\left(\begin{aligned}1&+4 t+21 t^2+72 t^3+264 t^4+792 t^5+2311 t^6+5948 t^7+14510 t^8+32172 t^9\\&+67045 t^{10}+128688 t^{11}+231618 t^{12}+386696 t^{13}+605330 t^{14}+882664 t^{15}\\&+1207465 t^{16}+1542388 t^{17}+1848901 t^{18}+2071096 t^{19}+2175454 t^{20}\\&+2132696 t^{21}+1955542 t^{22}+1666800 t^{23}+1321273 t^{24}+964732 t^{25}+646981 t^{26}\\&+391080 t^{27}+209934 t^{28}+94416 t^{29}+31585 t^{30}+2532 t^{31}-6746 t^{32}-7412 t^{33}\\&-5153 t^{34}-2904 t^{35}-1378 t^{36}-568 t^{37}-206 t^{38}-64 t^{39}-19 t^{40}-4 t^{41}-t^{42}\end{aligned}\right)}{\left(1-t\right)^{-4}\left(1-t^2\right)^6\left(1-t^3\right)^8\left(1-t^4\right)^6}
   ,\label{eq:E6000001A2SubHS}
\end{align}

where in \eqref{eq:E6000001A2SubHS} the HS components are those for $\overline{\left[\mathcal W_{D_5}\right]}^{[0,0,2,0,0]}_{[0,0,0,1,1]}$, $\overline{\left[\mathcal W_{D_5}\right]}^{[2,0,0,1,1]}_{[2,0,0,1,1]}$, and their intersection $\overline{\left[\mathcal W_{D_5}\right]}^{[0,1,0,1,1]}_{[0,0,0,1,1]}$, respectively.

There is an embedding of $D_6\hookleftarrow A_3\times D_3\hookleftarrow A_3\times A_2\times \urm(1)$. We have performed the $\surm(3)$ HKQ analytically and find agreement with quiver subtraction.

\section{Miscellaneous Moduli Spaces}
\label{sec:miscellaneous}

Here we consider some miscellaneous moduli spaces and their HKQs using quiver subtraction on their magnetic quivers. 

First we consider $\surm(2)$ HKQs of the generic star shaped quivers $\surm(2)^N$, which for cases $N=3,4$ are the 4 (quaternionic) dimensional moduli space of free fields (Section \ref{sec:freetheories}) and the 5 dimensional $\overline{min. D_4}$ (Section \ref{sec:DtypeHKQ}), respectively.

\subsection{Star shaped quiver $\surm(2)^{5} ///\surm(2)$}
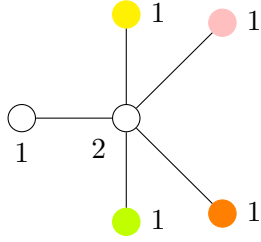
\begin{figure}[h!]
    \centering
    \begin{tikzpicture}[main/.style={draw,circle}]
    \node[main, label=below:$1$] (1) {};
    \node[main, label=below left:$2$] (2) [right=of 1]{};
    \node[main, label=right:$1$,yellow,fill] (y) [above=of 2]{};
    \node[main, label=right:$1$,pink,fill] (p) [above right=of 2]{};
    \node[main, label=right:$1$,orange,fill] (o) [below right=of 2]{};
    \node[main, label=right:$1$,lime,fill] (l) [below=of 2]{};
    \draw[-] (1)--(2)--(y);
    \draw[-] (p)--(2);
    \draw[-] (o)--(2)--(l);  
    \end{tikzpicture}
    \caption{Unframed star shaped magnetic quiver $\mathcal Q_{\ref{fig:T5SU2}}$ with global symmetry $\surm(2)^{ 5}$.}
    \label{fig:T5SU2}
\end{figure}

The Coulomb branch of the quiver $\mathcal Q_{\ref{fig:T5SU2}}$ is 6 (quaternionic) dimensional and has a global symmetry of $\surm(2)^5$, with each factor of $\surm(2)$ being equivalent under the outer automorphism. Quiver subtraction of the $\urm(2)$ quotient quiver is performed by aligning it with the uncoloured nodes and one coloured node in turn. The result of the HKQ is the union of the Coulomb branches of four possible magnetic quivers, each of which is $\left(\mathbb{C}_2/\mathbb{Z}_2\right)^3$. These have 6 two-way intersections, 4 three-way intersections and 1 four-way intersection at the origin. Summing the HWGs signed in accordance with \eqref{eq:UnionsofCones} leads to the result:
\begin{equation}
    HWG[\mathcal C(\mathcal Q_{\ref{fig:T5SU2}})///\surm(2)]=PE[(\mu^2+\nu^2+\rho^2+\sigma^2)t^2-\mu^2\nu^2\rho^2\sigma^2t^8],
\end{equation}
where we have associated the highest weight fugacities $\mu,\nu,\rho,\sigma$ to the coloured $\surm(2)$ symmetries. The moduli space is 3 (quaternionic) dimensional.

\subsection{Star shaped quiver $\surm(2)^{N+1} ///\surm(2), N>1$}
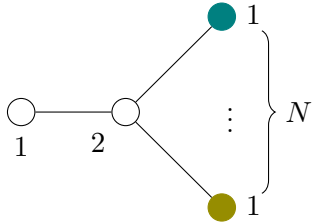
\begin{figure}[h!]
    \centering
    \begin{tikzpicture}[main/.style={draw,circle}]
    \node[main, label=below:$1$] (1) {};
    \node[main, label=below left:$2$] (2) [right=of 1]{};
    \node[main, label=right:$1$,teal,fill] (t) [above right=of 2]{};
    \node[draw=none, fill=none] (dots) [right=of 2]{$\vdots$};
    \node[main, label=right:$1$,olive,fill] (b) [below right=of 2]{};

    \draw[-] (t)--(2)--(b);
    \draw[-] (1)--(2);

     \draw [decorate, 
    decoration = {brace,
        raise=15pt,
        amplitude=5pt}] (t) --  (b) node[pos=0.5,right=20pt,black]{$N$};
        
    \end{tikzpicture}
    \caption{Unframed star shaped quiver $\mathcal Q_{\ref{fig:TNSU2}}$ with global symmetry $\surm(2)^{N+1}$.}
    \label{fig:TNSU2}
\end{figure}

The Coulomb branch of the quiver $\mathcal Q_{\ref{fig:TNSU2}}$ is $N+2$ (quaternionic) dimensional and has a global symmetry of $\surm(2)^{N+1}$, with each factor of $\surm(2)$ being equivalent under the outer automorphism.
Generalising the analyses for $\mathcal Q_{\ref{fig:4DimFree}}$, $\mathcal Q_{\ref{fig:minD4Quiv}}$ and $\mathcal Q_{\ref{fig:T5SU2}}$, we conjecture that the HWG for the  $N-1$ (quaternionic) dimensional $\surm(2)$ HKQ is:
\begin{equation}
    HWG[\mathcal C(\mathcal Q_{\ref{fig:TNSU2}})///\surm(2)]=PE\left[\sum_{i=1}^N\mu_{(i)}^2t^2-\left(\prod_{i=1}^N\mu_{(i)}^2\right)t^{2N}\right],
\end{equation}
where we have assigned highest weight fugacities $\mu_{(1),\cdots,(N)}$ to the $N$ coloured $\surm(2)$ nodes. The interpretation of the relation at order $t^{2N}$ in the HWG is as an invariant under the $S_N$ outer automorphism permuting the coloured nodes.

\subsection{$A_{k-1}$ Class ${\mathcal S}$ theory on a torus with 1 puncture}
Another moduli space we consider is the $A_{k-1}$ class $\mathcal S$ theory on a torus with 1 puncture labelled by the partition $(k-2,1^2)$. The magnetic quiver has an alternative description as an $\surm(2)$ node surrounded by a bouquet of $k+1$ gauge nodes subject to an $S_k$ discrete quotient. The case for $k=3$ is the $\overline{sub. reg. G_2}$. The $\surm(2)$ HKQs of these spaces can be found using quiver subtraction.

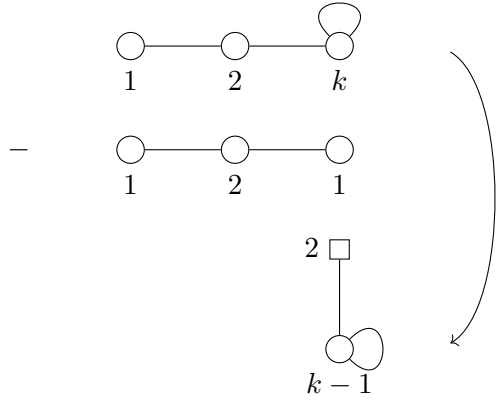
\begin{figure}[h!]
    \centering
    \begin{tikzpicture}[main/.style={draw,circle}]
    \node[main, label=below:$1$] (1) {};
    \node[main, label=below:$2$] (2) [right=of 1]{};
    \node[main, label=below:$k$] (k) [right=of 2]{};

     \draw (k) to [out=135, in=45,looseness=8] (k);

    \draw[-] (1)--(2)--(k);

    \node[main, label=below:$1$] (1subL) [below=of 1]{};
    \node[main, label=below:$2$] (2sub) [right=of 1subL]{};
    \node[main, label=below:$1$] (1subR) [right=of 2sub]{};

    \draw[-] (1subL)--(2sub)--(1subR);

    \node[draw=none,fill=none] (minus) [left=of 1subL]{$-$};

    \node[draw, label=left:$2$] (2F) [below=of 1subR]{};
    
    \node[main, label=below:$k-1$] (kminus) [below=of 2F]{};

    \draw (kminus) to [out=45, in=-45,looseness=8] (kminus);

    \draw[-] (2F)--(kminus);
    
     \node[draw=none,fill=none] (topghost) [right=of k]{};
    \node[draw=none,fill=none] (bottomghost) [right=of kminus]{};
    
    \draw [->] (topghost) to [out=-30,in=30,looseness=0.6] (bottomghost);
    
    \end{tikzpicture}
    \caption{Subtraction of the $\urm(2)$ quotient quiver from the unframed magnetic quiver $\mathcal Q_{\ref{fig:12kwithAdjA1Sub} \textcolor{blue}{a}}$ for a Class ${\mathcal S}$ torus, with fixture symmetry $SL(k)$ and 1 puncture $(k-2,1^{2})$, to produce the magnetic quiver $\mathcal Q_{\ref{fig:12kwithAdjA1Sub} \textcolor{blue}{b}}$ 
    .}
    \label{fig:12kwithAdjA1Sub}
\end{figure}

The unframed magnetic quiver $\mathcal Q_{\ref{fig:12kwithAdjA1Sub} \textcolor{blue}{a}}$ for a Class ${\mathcal S}$ theory of type $A_{k-1}$, on a torus (of genus 1), with 1 puncture described by the partition data $(k-2,1^{2})$, is drawn at the top of \Figref{fig:12kwithAdjA1Sub}. For the case $k=3$, where all nodes become balanced, there is an enhancement of the global symmetry of the Coulomb branch to ${G_2}$, and the space forms the orbit closure $\overline{sub. reg. G_2}$. Apart from this case, the global symmetry of the Coulomb branch is $\surm(2)^2$.

One can perform quiver subtraction by the $\urm(2)$ quotient quiver, which results in the magnetic quiver  $\mathcal Q_{\ref{fig:12kwithAdjA1Sub} \textcolor{blue}{b}}$. It is known that this is an ADHM quiver and is also a magnetic quiver for $\text{Sym}^{k-1}(\mathbb{C}^2/\mathbb{Z}_2)$. Thus, we conjecture that for this family of quivers:
\begin{equation}
    \mathcal C\left( \mathcal Q_{\ref{fig:12kwithAdjA1Sub} \textcolor{blue}{a}} \right)///\surm(2)=\text{Sym}^{k-1}\left(\mathbb{C}^2/\mathbb{Z}_2\right),\quad k\ge 2.
\end{equation}

We have checked the result by Weyl integration for small values of $k$, taking the HKQ
w.r.t the $\surm(2)$ associated to the node of rank 1 in $\mathcal Q_{\ref{fig:12kwithAdjA1Sub} \textcolor{blue}{a}}$.

\subsection{Extended $E_6$ quiver $\overline{min. E_6}\times \mathbb H^2///\surm(2)$}

\begin{figure}[h!]
    \centering
    \begin{tikzpicture}[main/.style={draw,circle}]
         \node[main, label=below:$1$] (1l) []{};
        \node[main, label=below:$2$] (2l) [right=of 1l]{};
        \node[main, label=below:$3$] (3l) [right=of 2l]{};
        \node[main, label=below:$2$] (2) [right=of 3l]{};
        \node[main, label=below:$2$,fill=black] (2r) [right=of 2]{};
        \node[main, label=below:$1$] (1r) [right=of 2r]{};
        \node[main, label=left:$2$] (2t) [above=of 3l]{};
        \node[main, label=left:$1$] (1t) [above=of 2t]{};

        \draw[-] (1l)--(2l)--(3l)--(2)--(2r)--(1r);
        \draw[-] (1t)--(2t)--(3l);
    \end{tikzpicture}
    \caption{Magnetic quiver \Quiver{fig:E6H2} for $\overline{min. E_6}\times \mathbb H^2$.}
    \label{fig:E6H2}
\end{figure}

\begin{figure}[h!]
    \centering
    \begin{tikzpicture}[main/.style={draw,circle}]
         \node[main, label=below:$1$] (1l) []{};
        \node[main, label=below:$2$] (2l) [right=of 1l]{};
        \node[main, label=below:$3$] (3l) [right=of 2l]{};
        \node[main, label=below:$2$] (2) [right=of 3l]{};
        \node[main, label=below:$2$] (2r) [right=of 2]{};
        \node[main, label=below:$1$] (1r) [right=of 2r]{};
        \node[main, label=left:$2$] (2t) [above=of 3l]{};
        \node[main, label=left:$1$] (1t) [above=of 2t]{};

        \draw[-] (1l)--(2l)--(3l)--(2)--(2r)--(1r);
        \draw[-] (1t)--(2t)--(3l);

        \node[main, label=below:$1$] (1subr) [below=of 1r]{};
        \node[main, label=below:$2$] (2sub) [left=of 1subr]{};
        \node[main, label=below:$1$] (1subl) [left=of 2sub]{};

        \draw[-] (1subr)--(2sub)--(1subl);
        \node[draw=none,fill=none] (minus) [right=of 1subr]{$-$};

        \node[] (ghost) [below=of 2sub]{};

        \node[main, label=below:$1$] (1R) [below=of ghost]{};
        \node[main, label=below:$2$] (2R) [left=of 1R]{};
        \node[main, label=below:$3$] (3) [left=of 2R]{};
        \node[main, label=below:$2$] (2L) [left=of 3]{};
        \node[main, label=below:$1$] (1L) [left=of 2L]{};
        \node[main, label=left:$1$] (1tL) [above left=of 3]{};
        \node[main, label=right:$1$] (1tR) [above right=of 3]{};

        \draw[-] (1L)--(2L)--(3)--(2R)--(1R) (1tL)--(3)--(1tR);

        \node[draw=none,fill=none] (topghost) [left=of 1l]{};
        \node[draw=none,fill=none] (bottomghost) [left=of 1L]{};
    
    \draw [->] (topghost) to [out=-150,in=150,looseness=0.6] (bottomghost);

    \end{tikzpicture}
    \caption{Subtraction of the $\urm(2)$ quotient quiver from the unframed magnetic quiver \Quiver{fig:E6H2} for $\overline{min. E_6}\times \mathbb H^2$ to produce the magnetic quiver \Quiver{fig:E6H2A1Sub}.}
    \label{fig:E6H2A1Sub}
\end{figure}

One can take the affine $E_6$ shown in the top of \Figref{fig:minE6A1QuivSub} and extend one leg by a further node of $\urm(2)$. This is shown in \Figref{fig:E6H2}, with the node in black having a balance of $-1$. The Hilbert series of the Coulomb branch of this quiver is computed as \begin{equation}
    HS\left[\mathcal C(\text{\Quiver{fig:E6H2}})\right]=\frac{\left(\begin{aligned}1 &+ 56 t^2 + 945 t^4 + 6776 t^6 + 23815 t^8 + 43989 t^{10} \\ &+ 
 43989 t^{12} + 23815 t^{14}+ 6776 t^{16} + 945 t^{18} + 
 56 t^{20} + t^{22}\end{aligned}\right)}{(1 - t)^{4} (1 - t^2)^{22}}.
\end{equation} 

The moduli space is $\overline{min. E_6}\times \mathbb H^2$ and the global symmetry is $E_6\times C_2$ which is confirmed from the $t^2$ coefficient of the above HS. 
One alignment and subtraction of the $\urm(2)$ quotient quiver from \Quiver{fig:E6H2} is shown in \Figref{fig:E6H2A1Sub} to produce quiver \Quiver{fig:E6H2A1Sub}. The Hilbert series of the Coulomb branch of \Quiver{fig:E6H2A1Sub} is\begin{equation}
    HS\left[\mathcal C(\text{\Quiver{fig:E6H2A1Sub}})\right]=\frac{\left(\begin{aligned}1 &+ 6 t + 41 t^2 + 206 t^3 + 900 t^4 + 3326 t^5 + 10846 t^6 + 
   31100 t^7 + 79677 t^8 \\ &+ 183232 t^9 + 381347 t^{10} + 720592 t^{11} + 
   1242416 t^{12} + 1959850 t^{13} + 2837034 t^{14}\\ &+ 3774494 t^{15} + 
   4624009 t^{16} + 5220406 t^{17} + 5435982 t^{18} +\text{palindrome}+
   t^{36}\end{aligned}\right)} {(1 - t^2)^{16} (1 - t^3)^{4} (1 + t + t^2)^{6}},
\end{equation}the coefficient of the $t^2$ term indicates that the global symmetry of $\mathcal C(\text{\Quiver{fig:E6H2A1Sub}})$ is $A_5\times \urm(1)$.

We conclude that \begin{equation}
    \left(\overline{min. E_6}\times \mathbb H^2\right)///\surm(2)=\mathcal C(\text{\Quiver{fig:E6H2A1Sub}}).
\end{equation}
The above result is verified from Weyl integration by integrating over the $\surm(2)$ in the global symmetry of $\mathcal C(\text{\Quiver{fig:E6H2}})$.

This result realises the Higgs branch side of the other example (alongside \ref{sec:minE7A1}) contained in \cite{Argyres:2007cn}. In this case the proposed duality is between $\surm(3)$ gauge theory with $6$ flavours at infinite coupling and the gauging of the $\surm(2)$ in a rank 1 $E_6$ SCFT with a free fundamental hyper of $\surm(2)$. 

The Coulomb branch of \Quiver{fig:E6H2} is exactly the Higgs branch of the latter theory. The gauging of $\surm(2)$ is implemented through the quiver subtraction to produce \Quiver{fig:E6H2A1Sub}. The quiver \Quiver{fig:E6H2A1Sub} is indeed the $3d$ mirror for $\surm(3)$ with $6$ flavours. 

\section{Discussion and Conclusions}
\label{sec:Conclusions}
\paragraph{Quiver subtraction for $\surm(n)$ HKQ}{In this work we have presented a new diagrammatic technique to compute the hyper-Kähler quotient by $\surm(n)$ of the Coulomb branch of a quiver gauge theory. This technique works for all unitary magnetic quivers, including both "good" and "ugly" quivers, that respect the selection rules presented in Section \ref{sec:rules}. The procedure involves the subtraction of a $\urm(n)$ quotient quiver of the form $(1)-(2)-\cdots-(n)-\cdots-(2)-(1)$ from a target quiver which has been unframed w.r.t. an overall $\urm(1)$. Such a quotient quiver is "bad", as its central gauge node of rank $n$ has a balance of $-2$, and this method of quiver subtraction differs from the usual subtraction of "good" quivers for Kraft-Procesi transitions.}

Quotient quiver subtraction requires that the target quiver should contain an external leg that forms a balanced Dynkin diagram of $\surm(n)$, with nodes matching at least $(1)-(2)-\cdots-(n)-\cdots$.

The sum of the ranks of the gauge nodes in the $\urm(n)$ quotient quiver entails that the combination of this subtraction and unframing is dimensionally consistent with an $\surm(n)$ HKQ.

For cases where there are multiple permissible alignments of the quotient quiver against the target quiver, the subtraction generates the HKQ as the union of the Coulomb branches of the alternative resulting quivers. The selection rules entail that these alternative quivers are related to their intersection by KP transitions, being Kleinian singularities of type $A$. (This follows due to the quiver subtraction being taken from a long external leg of the target quiver, along with the requirement that junctions only occur on a node of rank 2 on the quotient quiver.) The HS of the union is computed using the unions of cones formula \eqref{eq:UnionsofCones}.

Compared with HKQ via Weyl integration, the diagrammatic technique of quotient quiver subtraction is able to give additional insight into the structure of the resulting HS and/or HWG. This is because of the direct identification of the corresponding magnetic quiver(s). In cases where the HKQ is a union, this identification of the component magnetic quivers is particularly helpful and often simplifies calculation of the HS.

We have verified the validity of $\surm(n)$ HKQs by quotient quiver subtraction for various classical and exceptional nilpotent orbits, and some free field moduli spaces, by comparison with the results from Weyl integration. Also, based on diagrammatic analyses, we have been able to conjecture general results for certain families whose magnetic quivers satisfy the selection rules.

\paragraph{$\surm(n)$ HKQs of Free Fields}{We summarise some of the results from Section \ref{sec:freetheories} in Table \ref{tab:FreeSU2}. These establish relationships between $\mathbb H^{n}$ for $n\geq 4$ and the minimal nilpotent orbits of type $B$ and $D$ for odd and even $n$ respectively. This is in accordance with \cite{Kraft1982OnGroups}.}

\begin{table}[h!]
    \centering
    \begin{tabular}{|c|c|c|}
    \hline
        Target Moduli Space & $\surm(2)$ HKQ & Rank\\\hline\hline
        $\mathbb H^{2k}$ & $\overline{min. D_k}$ & $k\geq2$\\\hline
        $\mathbb H^{2k+1}$ & $\overline{min. B_k}$ & $k\geq 2$\\
        \hline
    \end{tabular}
    \caption{The moduli spaces $\mathbb H^n$ and their $\surm(2)$ HKQs. These fall into two families for even and odd $n$. These families are labelled by an integer $k$, with the appropriate limits on $k$ shown.}
    \label{tab:FreeSU2}
\end{table}

\begin{table}[h!]
    \centering
    \begin{tabular}{|c|c|c|}
    \hline
       Target Moduli Space  & $\surm(2)$ HKQ & Rank\\
       \hline\hline
        $\overline{n.min. A_4}$ & $\overline{\mathcal O}^{A_2}_{(3)}\cup\overline{[\mathcal W_{D_3}]}^{[0,1,3]}_{[0,0,2]}$ &\\
        $\overline{n.min. A_k}$ & $\overline{\mathcal O}_{(3,1^{k-4})}^{A_{k-2}}\cup \overline{\left[\mathcal W_{D_{k-1}}\right]}^{[0,1,\cdots,2]}_{[0,0,\cdots,2]}$& $k\geq5$\\
        \hline
        $\overline{max. A_3}$&$\mathcal{S}^{A_3}_{\mathcal N,(2^2)}\otimes\mathbb{C}^2/\mathbb{Z}_4 $&\\
        $\overline{max. A_k}$ &$\left[\overline{\mathcal W_{D_{k-1}}}\right]^{[k+1,0,\cdots,0]}_{[0,0,\cdots,2]}$&$k\geq 4$\\
        \hline
        $\overline{min. B_k}$&$\overline{n.min B_{k-2}}\cup \left(\overline{min. A_1}\otimes \overline{min. B_{k-2}}\right)$& $k\geq 4$\\
        \hline
        $\overline{n.min. B_k}$&$\overline{\mathcal O}^{D_{k-1}}_{(3,1^{2k-7})}/\mathbb Z_2\cup \left(\overline{min. A_1}\otimes \overline{\mathcal O}^{B_{k-2}}_{(3,1^{2k-6})}\right)$&$k\geq 3$\\\hline
        $\overline{min. D_k}$&$\overline{n.min D_{k-2}}\cup \left(\overline{min. A_1}\otimes \overline{min. D_{k-2}}\right)$& $k\geq 4$\\\hline
        $\overline{n. min. D_k}$&$\overline{n. min. B_{k-2}}/\mathbb{Z}_2\cup (\overline{min. A_1}\otimes \overline{n. min. D_{k-2}}$)&$k\geq 4$\\
        \hline
        $\overline{sub. reg. G_2}$&$Sym^2\left(\overline{min.A_1}\right)$&\\\hline
        $\overline{min. F_4}$&$\overline{n. min. C_3}$&\\\hline
        $\overline{min. E_6}$&$\overline{n. min. A_5}$&\\\hline
        $\overline{min. E_7}$&$\overline{n. n. min. D_6}$&\\\hline
        $\overline{min. E_8}$&$\overline{n. min. E_7}$&\\
         \hline
    \end{tabular}
    \caption{Some classical and exceptional nilpotent orbits and their $\surm(2)$ HKQs. For families of orbits, labelled by their rank $k$, the limits on $k$ are shown.}
    \label{tab:OrbSU2}
\end{table}

\begin{table}[h!]
    \centering
    \begin{tabular}{|c|c|c|}
    \hline
    Target Moduli Space & $\surm(3)$ HKQ & Rank\\
    \hline
        $\overline{max. A_5}$&$\mathcal{S}^{A_5}_{\mathcal N,(2^3)}\otimes \mathbb{C}^2/\mathbb{Z}_6$ & \\
         $\overline{max. A_k}$& $\overline{\left[\mathcal W_{E_{k-2}}\right]}^{[0,\cdots,0,k+1,0]}_{[0,\cdots,0,4]}$& $k=6,7,8,9,10$\\
        \hline
        $\overline{\mathcal O}^{B_k}_{(2^4,1^{2k-7})}$&$\overline{\left[\mathcal W_{B_{k-3}}\right]}^{[0,2,0,\cdots,0]}_{[0,0,0,\cdots,0]}\cup\overline{\left[\mathcal W_{B_{k-2}}\right]}^{[2,0,0,1,0,\cdots,0]}_{[2,0,0,0,0,\cdots,0]}$& $k\geq 6$\\
        \hline
         $\overline{min. E_7}$&$\overline{\mathcal O}^{A_5}_{(2^3)}\cup\overline{\mathcal O}^{A_5}_{(3,1^3)}$&\\
        \hline
         $\overline{min. E_8}$&$\left[\overline{\mathcal W_{E_6}}\right]^{[0,0,1,0,0,0]}_{[0,0,0,0,0,0]}$&\\
        \hline
    \end{tabular}
    \caption{Some classical and exceptional nilpotent orbits and their $\surm(3)$ HKQs. For families of orbits, labelled by their rank $k$, the limits on $k$ are shown.}
    \label{tab:OrbSU3}
\end{table}

\begin{table}[h!]
    \centering
    \begin{tabular}{|c|c|}
    \hline
    Target Moduli Space & $\surm(4)$ HKQ \\
    \hline
        $\overline{min. E_8}$ & $\overline{[\mathcal W_{D_5}]}^{[0,2,0,0,0]}_{[0,0,0,0,0]}\cup \overline{[\mathcal W_{D_5}]}^{[1,0,0,2,0]}_{[0,0,0,0,0]}$\\
         \hline
    \end{tabular}
    \caption{Exceptional nilpotent orbits and their $\surm(4)$ HKQs.}
    \label{tab:OrbSU4}
\end{table}

\paragraph{$\surm(n)$ HKQs of Nilpotent Orbits}{We summarise some of the results from Sections \ref{sec:AType}, \ref{sec:BtypeHKQ}, \ref{sec:DtypeHKQ}, and \ref{sec:ExceptionalHKQ} in Tables \ref{tab:OrbSU2}, \ref{tab:OrbSU3}, and \ref{tab:OrbSU4}. These establish interesting relationships between classical and exceptional nilpotent orbits under $\surm(n)$ HKQ and other orbits or affine Grassmannian slices, and include many results that we have not found in the Literature.}

In particular this work extends the work of Kobak-Swann \cite{Kobak1996CLASSICALNO}, who find discrete quotients between nilpotent orbit closures, to include relationships under continous hyper-Kähler quotients between nilpotent orbit closures and/or affine Grassmannian slices. Several aspects of these examples merit discussion.

Firstly, while a number of the HKQs of orbits studied form 1-parameter families based on the rank of the target group, the first member of a family may take an irregular form. For example, consider the quotients $\overline{max. A_k}///\surm(2)$. The first member of this family, consistent with the selection rules, is for $k=3$. However, application of the general form for $k\geq 4$, which is given by the affine Grassmannian slice $\left[\overline{\mathcal W_{D_{k-1}}}\right]^{[k+1,0,\cdots,0]}_{[0,0,\cdots,2]}$, would require a slice of the semi-simple $D_2$ algebra, $\left[\overline{\mathcal W_{D_{2}}}\right]^{[4,0]}_{[0,2]}$. This description as a slice is not valid since the flavour co-weight minus the balance co-weight does not lie in the positive co-root lattice. The result for $k=3$ is instead described by a product space with $A_1 \times \urm(1)$ global symmetry.

Or, consider the results in Table \ref{tab:OrbSU3} for $\overline{max. A_k}///\surm(3)$, which takes a different form for $k=5$ compared with $k=6,7,8,9,10$, the latter being an E-sequence of affine Grassmannian slices $\overline{\left[\mathcal W_{E_{k-2}}\right]}^{[0,\cdots,0,k+1,0]}_{[0,\cdots,0,4]}$. If we were to apply the more general form to $k=5$, we would obtain $\overline{\left[\mathcal W_{E_{3}}\right]}^{[0,6,0]}_{[0,0,4]}$.

Interpreting the Dynkin diagram of $E_3$ as $A_2\times A_1$ suggests this is the product of two affine Grassmannian slices $\overline{\left[\mathcal W_{A_{2}}\right]}^{[0,6]}_{[0,0]}\times \overline{\left[\mathcal W_{A_{1}}\right]}^{[0]}_{[4]}$. However, in the second term, the flavour co-weight minus the balance co-weight 

is not in the positive coroot lattice of $A_1$, and so, once again, this is not a valid affine Grassmannian slice.

\paragraph{Non-commutativity of KP transitions and $\surm(n)$ HKQ}{There are other aspects of these quotient quiver subtractions that merit further comment, including their commutation relations with other diagrammatic operations. Here, we examine the commutation between quiver subtraction for KP transitions and for $\surm(n)$ HKQ. Just as the order of KP transitions is important, we find that the operations of quotient quiver subtraction and KP transition quiver subtraction do not commute.

\begin{figure}[h!]
    \centering
    \begin{tikzpicture}
    \filldraw[black] (0,0) circle (2pt) node[anchor=north]{$0$};
    \filldraw[black] (0,10) circle (2pt) node[anchor=south]{$29$};
    \filldraw[black] (-5,340/58) circle (2pt) node[anchor=east]{$17$};
    \filldraw[black] (-5,520/58) circle (2pt) node[anchor=east]{$26$};
    \filldraw[black] (-2,220/58) circle (2pt) node[anchor=west]{$11$};
    \filldraw[black] (-2,320/58) circle (2pt) node[anchor=west]{$16$};
    \filldraw[black] (-2,420/58) circle (2pt) node[anchor=west]{$21$};
    \filldraw[black] (5,140/58) circle (2pt) node[anchor=west]{$7$};
    \filldraw[black] (5,200/58) circle (2pt) node[anchor=west]{$10$};
    \filldraw[black] (5,260/58) circle (2pt) node[anchor=west]{$13$};
    \filldraw[black] (4,280/58) circle (2pt) node[anchor=east]{$14$};
    \filldraw[black] (6,280/58) circle (2pt) node[anchor=west]{$14$};
    \draw[-] (0,0) -- (-5,340/58) node[pos=0.5,left]{$e_7$} -- (-5,520/58) node[pos=0.5,left]{$d_6$};
    \draw[-] (0,0) -- (-2,220/58) node[pos=0.5,right]{$e_6$} -- (-2,320/58) node[pos=0.5,left]{$a_5$}-- (-2,420/58)node[pos=0.5,left]{$d_4$};
    \draw[-] (0,0) -- (5,140/58) node[pos=0.5,right]{$d_5$} -- (5,200/58) node[pos=0.5,right]{$d_3$}-- (5,260/58) node[pos=0.5,right]{$a_3$}-- (4,280/58) node[pos=0.5,below]{$A_1$};
    \draw[-] (5,260/58) -- (6,280/58) node[pos=0.5,right]{$A_1$};

    \draw[-] (0,10)--(0,0)node[pos=0.5,right]{$e_8$};

    \draw[dashed] (0,10)--(-5,520/58) node[pos=0.5,sloped,above]{$\surm(2)$};
    \draw[dashed] (0,10)--(-2,420/58) node[pos=0.5,left]{$\surm(3)$};
    \draw[dashed] (0,10)--(4,280/58) node[pos=0.5,sloped,above]{$\surm(4)$};
    \draw[dashed] (0,10)--(6,280/58);
    \end{tikzpicture}
    \caption{Hasse diagram of the $\overline{min. E_8}$, subject to $\surm(2)$, $\surm(3)$, and $\surm(4)$ HKQs by quotient quiver subtraction, shown by the dashed lines. The unbroken lines below each moduli space decompose their Hasse diagrams, labelled by KP transitions in the usual manner. The number next to each leaf is the quaternionic dimension of the moduli space.}
    \label{fig:E8Hasse}
\end{figure}
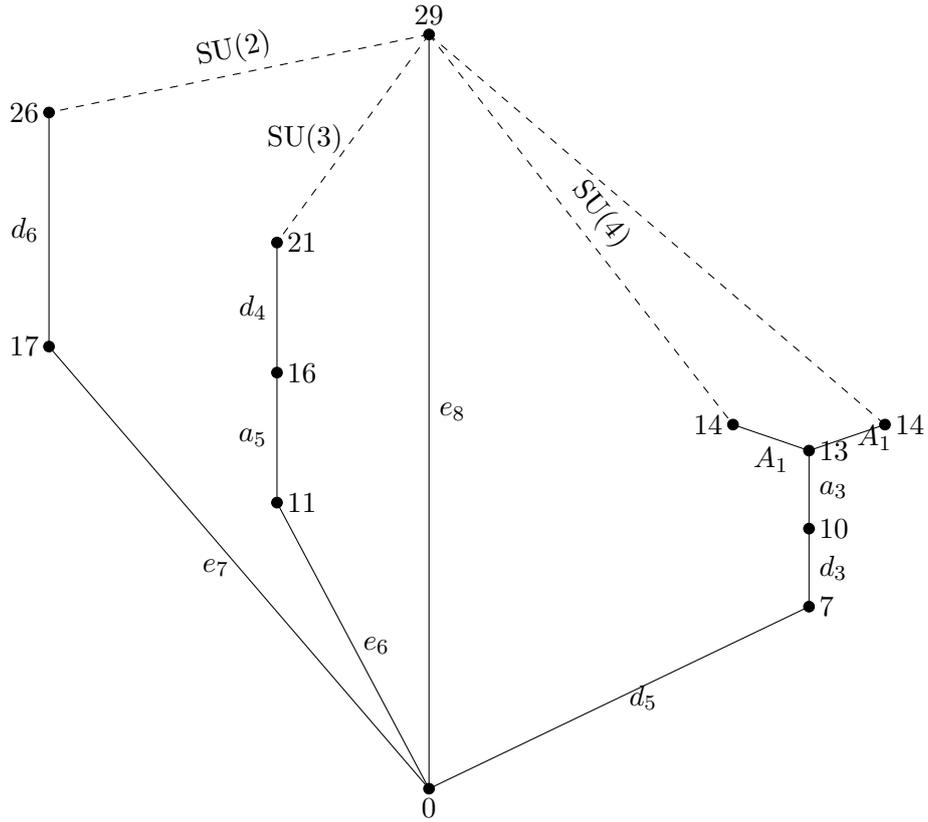
\FloatBarrier

By way of illustration, the Hasse diagram in \Figref{fig:E8Hasse} shows the permissible KP transitions from $\overline{min. E_8}$ before and after an initial quotient quiver subtraction for $n=2,3,4$.

Without any quotient quiver subtraction and hence using KP transitions alone, only an $e_8$ slice can be subtracted – to the trivial leaf. Suppose instead that an $\surm(n)$ HKQ, for $n = 2,3$, or $4$, is performed first. This can then be followed by subsequent KP transitions down to the trivial leaf. Each initial $\surm(n)$ HKQ takes us to a different branch of the Hasse diagram and, although the magnetic quivers at each leaf may be amenable to further $\surm(n)$ HKQs, these do not generally link between the different branches. We conclude that quiver subtractions, whether for HKQs or KP transitions, do not commute.}

A final comment we make is that KP transitions on nilpotent orbits of an algebra lead to other moduli spaces within the nilcone of the algebra. The difference for quotient quiver subtraction on nilpotent orbits is that one moves to a different algebra, being that of the commutant of the embedding of $\surm(n)$ inside the original algebra.

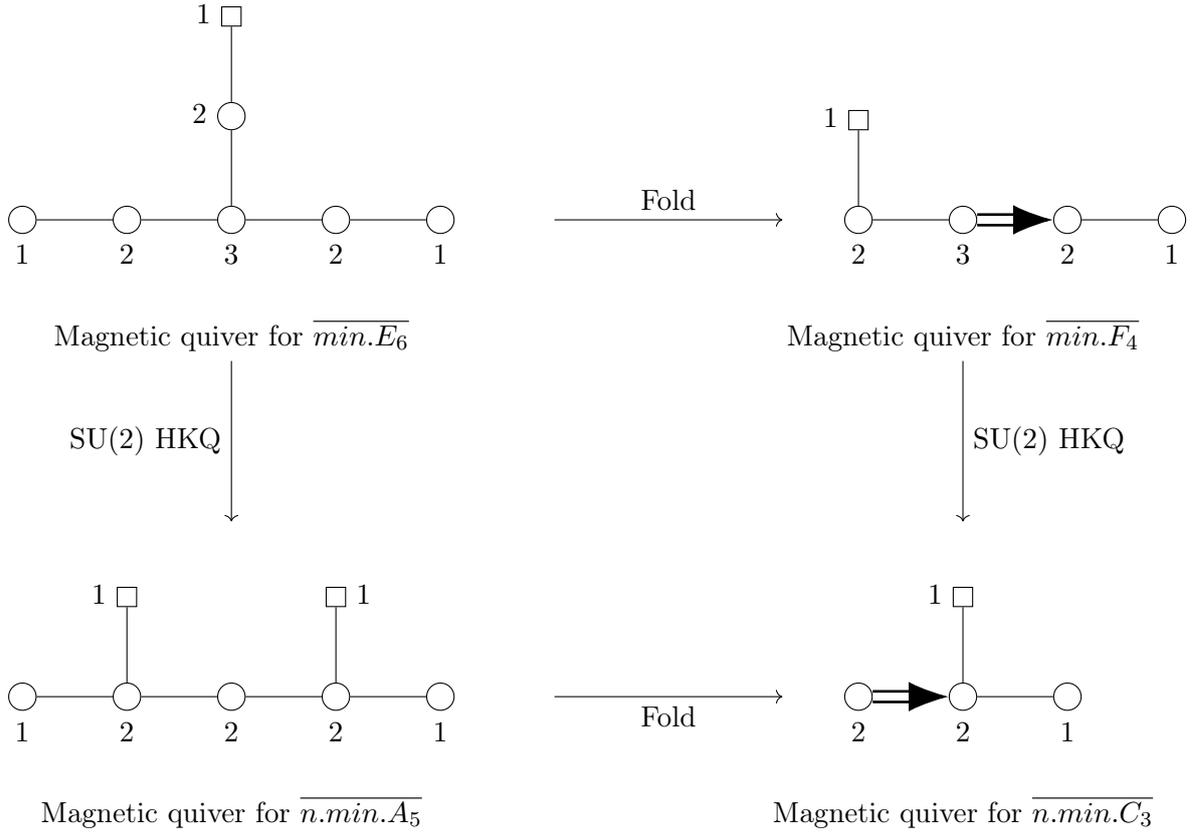
\begin{figure}[h!]
    \centering
    \begin{tikzpicture}[main/.style={draw,circle}]
    \node[main, label=below:$1$] (oneL) []{};
    \node[main, label=below:$2$] (twoL) [right=of oneL]{};
    \node[main, label=below:$3$] (three) [right=of twoL]{};
    \node[main, label=below:$2$] (twoR) [right=of three]{};
    \node[main, label=below:$1$] (oneR) [right=of twoR]{};
    \node[main, label=left:$2$] (twoT) [above=of three]{};
    \node[draw, label=left:$1$] (oneT) [above=of twoT]{};
    \node[draw=none,fill=none](labelE6) [below=of three]{Magnetic quiver for $\overline{min. E_6}$};
    
    \draw[-] (oneL)--(twoL)--(three);
    \draw[-] (oneT)--(twoT)--(three)--(twoR)--(oneR);
    
    \draw[->] (7,0)--(10,0) node[pos=0.5,above]{Fold};

    \node[main,label=below:$2$] (2L) at (11,0) {};
    \node[draw, label=left:$1$] (1) [above=of 2L] {};
    \node[main,label=below:$3$] (3) [right=of 2L] {};
    \node[main,label=below:$2$] (2R) [right=of 3] {};
    \node[main,label=below:$1$] (1R) [right=of 2R] {};
    \node[draw=none,fill=none](labelF4) [below=of 3]{Magnetic quiver for $\overline{min. F_4}$};

    \draw[-] (1)--(2L)--(3);
    \draw[-] (2R)--(1R);
    \draw [line width=1pt, double distance=3pt,
             arrows = {-Latex[length=0pt 3 0]}] (3) -- (2R);

    \draw[->] (labelE6)-- (labelE6 |-,-4) node[pos=0.5,left]{$\surm(2)$ HKQ};

    \node[draw, label=left:$1$] (oneFLres) at (twoL |-,-5) {};
    \node[main, label=below:$2$] (twoLres) [below=of oneFLres]{};
    \node[main, label=below:$1$] (oneLres) [left=of twoLres]{};
    \node[main, label=below:$2$] (twoMres) [right=of twoLres]{};
    \node[main, label=below:$2$] (twoRres) [right=of twoMres]{};
    \node[draw,label=right:$1$] (oneFRres) [above=of twoRres]{};
    \node[main, label=below:$1$] (oneRres) [right=of twoRres]{};
    \node[draw=none,fill=none] (labelA5) [below=of twoMres]{Magnetic quiver for $\overline{n. min. A_5}$};
    
    \draw[-] (oneLres)--(twoLres)--(twoMres)--(twoRres)--(oneRres);
    \draw[-] (oneFLres)--(twoLres);
    \draw[-] (oneFRres)--(twoRres);

    \draw[->] (7,  |-  oneRres)--(10,  |-  oneRres)node[pos=0.5, below]{Fold};

    \node[main, label=below:$2$] (2Lres) at (2L |- oneRres){};
    \node[main, label=below:$2$] (2Mres) [right=of 2Lres]{};
    \node[main, label=below:$1$] (1res) [right=of 2Mres]{};
    \node[draw, label=left:$1$] (1Fres) [above=of 2Mres]{};
    \node[draw=none,fill=none] (labelC3) [below=of 2Mres]{Magnetic quiver for $\overline{n. min. C_3}$};

    \draw[-] (1Fres)--(2Mres)--(1res);
    \draw [line width=1pt, double distance=3pt,
             arrows = {-Latex[length=0pt 3 0]}] (2Lres) -- (2Mres);

    \draw[->] (labelF4)-- (labelF4 |-,-4) node[pos=0.5,right]{$\surm(2)$ HKQ};
    
    \end{tikzpicture}
    \caption{Commutative diagram showing folding and $\surm(2)$ HKQ by quotient quiver subtraction from a magnetic quiver for $\overline{min. E_6}$ to obtain $\overline{n. min. C_3}$.}
    \label{fig:E6Commute}
\end{figure}

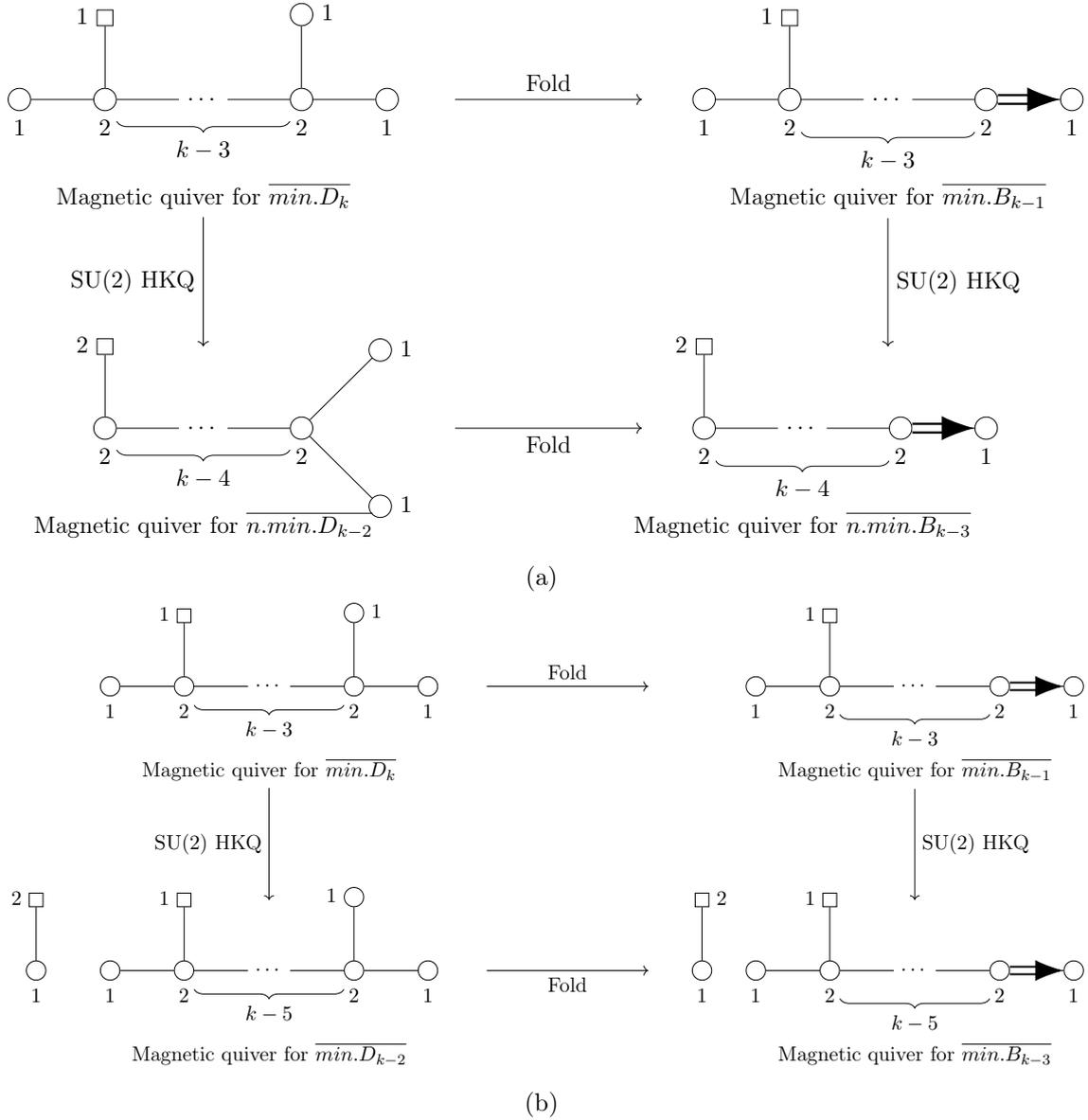
\begin{figure}[h!]
    \centering
    \begin{subfigure}{\textwidth}\resizebox{\textwidth}{!}{
    \begin{tikzpicture}[main/.style={draw,circle}]
    \node[main, label=below:$1$] (a) {};
    \node[main,label=below:$2$] (b) [right=of a] {};
    \node[] (c) [right=of b] {$\cdots$};
    \node[main,label=below:$2$] (d) [right=of c] {};
    \node[main,label=below:$1$] (e) [right=of d] {};
    \node[draw, label=left:$1$] (f) [above=of b]{};
    \node[main, label=right:$1$] (g) [above=of d]{};
    \draw[-] (a)--(b);
    \draw[-] (e)--(d);
    \draw[-] (f)--(b)--(c)--(d)--(g);
    \draw [decorate, 
    decoration = {brace,
        raise=10pt,
        amplitude=5pt}] (d) --  (b) node[pos=0.5,below=15pt,black]{$k-3$};
    \node[draw=none,fill=none](labelD) [below=of c]{Magnetic quiver for $\overline{min. D_k}$};

    \draw[->] (7,0)--(10,0) node[pos=0.5,above]{Fold};

    \node[main, label=below:$1$] (A) at (11,0) {};
    \node[main, label=below:$2$] (C) [right=of A]{};
    \node[] (D) [right=of C] {$\cdots$};
    \node[main,label=below:$2$] (E) [right=of D] {};
    \node[main,label=below:$1$] (F) [right=of E] {};
    \node[draw, label=left:$1$] (G) [above=of C]{};
    
    \draw[-] (A)--(C)--(D)--(E)--(F);
    \draw[-] (C)--(G);
    \draw [line width=1pt, double distance=3pt,
             arrows = {-Latex[length=0pt 3 0]}] (E) -- (F);
    \draw [decorate, 
    decoration = {brace,
        raise=15pt,
        amplitude=5pt}] (E) --  (C) node[pos=0.5,below=20pt,black]{$k-3$};
    \node[draw=none,fill=none](labelB) [below=of D]{Magnetic quiver for $\overline{min. B_{k-1}}$};

    \draw[->] (labelD)-- (labelD |-,-4) node[pos=0.5,left]{$\surm(2)$ HKQ};

    \node[draw, label=left:$2$] (aa) at (b |-, -4){};
    \node[main, label=below:$2$] (bb) [below=of aa]{};
    \node[draw=none,fill=none] (cdots2) [right=of bb]{$\cdots$};
    \node[main, label=below:$2$] (cc) [right=of cdots2]{};
    \node[main, label=right:$1$] (dd)[above right=of cc]{};
    \node[main, label=right:$1$] (ee)[below right=of cc]{};
    
    \draw[-] (ee)--(cc)--(dd);
    \draw[-] (cc)--(cdots2)--(bb)--(aa);
    \draw [decorate, 
    decoration = {brace,
        raise=10pt,
        amplitude=5pt}] (cc) --  (bb) node[pos=0.5,below=15pt,black]{$k-4$};
    \node[draw=none,fill=none](labelDres1) [below=of cdots2]{Magnetic quiver for $\overline{n. min. D_{k-2}}$};
    
    \draw[->] (7,  |-  cc)--(10,  |-  cc)node[pos=0.5, below]{Fold};

    \node[main, label=below:$2$] (CC) at (A |- cc){};
    \node[] (DD) [right=of CC] {$\cdots$} ;
    \node[main,label=below:$2$] (EE) [right=of DD] {};
    \node[main,label=below:$1$] (FF) [right=of EE] {};
    \node[draw, label=left:$2$] (GG) [above=of CC]{};
    
    \draw[-] (CC)--(DD)--(EE)--(FF);
    \draw[-] (CC)--(GG);
    \draw [line width=1pt, double distance=3pt,
             arrows = {-Latex[length=0pt 3 0]}] (EE) -- (FF);
    \draw [decorate, 
    decoration = {brace,
        raise=15pt,
        amplitude=5pt}] (EE) --  (CC) node[pos=0.5,below=20pt,black]{$k-4$};
    \node[draw=none,fill=none](labelB2) [below=of DD]{Magnetic quiver for $\overline{n. min. B_{k-3}}$};

    \draw[->] (labelB)-- (labelB |-,-4) node[pos=0.5,right]{$\surm(2)$ HKQ};
    \end{tikzpicture}}
    \caption{}
    \label{fig:minDkCommute1}
    \end{subfigure}
    \centering
    \begin{subfigure}{\textwidth}\resizebox{\textwidth}{!}{
    \begin{tikzpicture}[main/.style={draw,circle}]
    \node[main, label=below:$1$] (a) {};
    \node[main,label=below:$2$] (b) [right=of a] {};
    \node[] (c) [right=of b] {$\cdots$};
    \node[main,label=below:$2$] (d) [right=of c] {};
    \node[main,label=below:$1$] (e) [right=of d] {};
    \node[draw, label=left:$1$] (f) [above=of b]{};
    \node[main, label=right:$1$] (g) [above=of d]{};
    \draw[-] (a)--(b);
    \draw[-] (e)--(d);
    \draw[-] (f)--(b)--(c)--(d)--(g);
    \draw [decorate, 
    decoration = {brace,
        raise=10pt,
        amplitude=5pt}] (d) --  (b) node[pos=0.5,below=15pt,black]{$k-3$};
    \node[draw=none,fill=none](labelD) [below=of c]{Magnetic quiver for $\overline{min. D_k}$};

    \draw[->] (7,0)--(10,0) node[pos=0.5,above]{Fold};

    \node[main, label=below:$1$] (A) at (12,0) {};
    \node[main, label=below:$2$] (C) [right=of A]{};
    \node[] (D) [right=of C] {$\cdots$};
    \node[main,label=below:$2$] (E) [right=of D] {};
    \node[main,label=below:$1$] (F) [right=of E] {};
    \node[draw, label=left:$1$] (G) [above=of C]{};
    
    \draw[-] (A)--(C)--(D)--(E)--(F);
    \draw[-] (C)--(G);
    \draw [line width=1pt, double distance=3pt,
             arrows = {-Latex[length=0pt 3 0]}] (E) -- (F);
    \draw [decorate, 
    decoration = {brace,
        raise=15pt,
        amplitude=5pt}] (E) --  (C) node[pos=0.5,below=20pt,black]{$k-3$};
    \node[draw=none,fill=none](labelB) [below=of D]{Magnetic quiver for $\overline{min. B_{k-1}}$};

    \draw[->] (labelD)-- (labelD |-,-4) node[pos=0.5,left]{$\surm(2)$ HKQ};
    
    \node[draw, label=left:$1$] (gg) at (b |-,-4){};
    \node[main,label=below:$2$] (bb) [below=of gg] {};
    \node[main, label=below:$1$] (aa) [left=of bb] {};
    \node[] (cc) [right=of bb] {$\cdots$};
    \node[main,label=below:$2$] (dd) [right=of cc] {};
    \node[main,label=below:$1$] (ee) [right=of dd] {};
    \node[main, label=left:$1$] (ff) [above=of dd]{};
    
    \draw[-] (aa)--(bb)--(gg);
    \draw[-] (ee)--(dd);
    \draw[-] (bb)--(cc)--(dd)--(ff);
    \draw [decorate, 
    decoration = {brace,
        raise=10pt,
        amplitude=5pt}] (dd) --  (bb) node[pos=0.5,below=15pt,black]{$k-5$};
    \node[draw=none,fill=none](labelD1) [below=of cc]{Magnetic quiver for $\overline{min. D_{k-2}}$};

    \node[main, label=below:$1$] (one) [left=of aa]{};
    \node[draw, label=left:$2$] (twoF) [above=of one]{};
    
    \draw[-] (twoF)--(one);

    \draw[->] (7,  |-  cc)--(10,  |-  cc)node[pos=0.5, below]{Fold};

    \node[main, label=below:$1$] (ONE) at (11,|-dd) {};
    \node[draw, label=right:$2$] (TWOF) [above=of ONE]{};
    
    \draw[-] (TWOF)--(ONE);

    \node[main, label=below:$1$] (AA) at (A|-dd){};
    \node[main, label=below:$2$] (CC) [right=of AA]{};
    \node[] (DD) [right=of CC] {$\cdots$};
    \node[main,label=below:$2$] (EE) [right=of DD] {};
    \node[main,label=below:$1$] (FF) [right=of EE] {};
    \node[draw, label=left:$1$] (GG) [above=of CC]{};
    
    \draw[-] (AA)--(CC)--(DD)--(EE)--(FF);
    \draw[-] (CC)--(GG);
    \draw [line width=1pt, double distance=3pt,
             arrows = {-Latex[length=0pt 3 0]}] (EE) -- (FF);
    \draw [decorate, 
    decoration = {brace,
        raise=15pt,
        amplitude=5pt}] (EE) --  (CC) node[pos=0.5,below=20pt,black]{$k-5$};
    \node[draw=none,fill=none](labelB2) [below=of DD]{Magnetic quiver for $\overline{min. B_{k-3}}$};
    
    \draw[->] (labelB)-- (labelB |-,-4) node[pos=0.5,right]{$\surm(2)$ HKQ};

    \end{tikzpicture}}
    
    \caption{}
    \label{fig:minDkCommute2}
    \end{subfigure}
    \caption{The quotient quiver subtraction from $\overline{min. D_k}$ has two alignments. For each alignment we draw the commutative diagram of folding and $\surm(2)$ HKQ.}
    \label{fig:minDkCommute}
\end{figure}

\begin{figure}[h!]
    \centering
     \centering
    \begin{subfigure}{\textwidth}\resizebox{\textwidth}{!}{
    \begin{tikzpicture}[main/.style={draw,circle}]
    \node[main, label=below:$2$] (a) []{};
    \node[] (b) [right=of a]{$\cdots$};
    \node[main, label=below:$2$] (c) [right=of b]{};
    \node[main, label=right:$1$] (d) [above right=of c]{};
    \node[main, label=right:$1$] (e) [below right=of c]{};
    \node[draw, label=left:2] (f) [above=of a]{};
    \draw[-] (f)--(a)--(b)--(c);
    \draw[-] (d)--(c)--(e);

    \draw [decorate, 
    decoration = {brace,
        raise=10pt,
        amplitude=5pt}] (c) --  (a) node[pos=0.5,below=15pt,black]{$k-2$};
    \node[draw=none,fill=none](labelD) [below=of b]{Magnetic quiver for $\overline{n. min. D_k}$};

    \draw[->] (5,0)--(8,0) node[pos=0.5,above]{Fold*};

    \node[] (A) at (9,0) {$\mathcal Q_{\ref{fig:minBKUsual}}$};
    \node[] (G) [right=0.1cm of A]{$\sim$};
    \node[main, label=below:$1$] (B) [right=of A]{};
    \node[main, label=below:$2$] (C) [right=of B]{};
    \node[] (D) [right=of C]{$\cdots$};
    \node[main, label=below:$2$] (E) [right=of D]{};

    \node[draw, label=left:$1$] (F) [above=of C]{};

    \draw[-] (B)--(C)--(D)--(E);
    \draw[-] (F)--(C);

    \draw (E) to [out=135, in=45,looseness=8] (E);
    \draw [decorate, 
    decoration = {brace,
        raise=15pt,
        amplitude=5pt}] (E) --  (B) node[pos=0.5,below=20pt,black]{$k-1$};
    \node[draw=none,fill=none](labelB) [below=of D]{Magnetic quiver for $\overline{n.min. B_{k-1}}$};

    \draw[->] (labelD)-- (labelD |-,-4) node[pos=0.5,left]{$\surm(2)$ HKQ};

    \node[] (cc) at (b |-, -5.5){$\cdots$};
    \node[main, label=below:$2$] (bb) [left=of cc]{};
    \node[main, label=below:$1$] (aa) [left=of bb]{};
    \node[main, label=below:$2$] (dd) [right=of cc]{};
    \node[draw, label=right:$2$] (ee) [above=of dd]{};

    \draw[-] (bb)--(cc)--(dd)--(ee);
    \draw[double,double distance=3pt,line width=0.4pt] (aa)--(bb);
     \draw [decorate, 
    decoration = {brace,
        raise=15pt,
        amplitude=5pt}] (dd) --  (aa) node[pos=0.5,below=20pt,black]{$k-2$};
    \node[draw=none,fill=none](labelB2) [below=of cc]{Magnetic quiver for $\overline{n. min. B_{k-2}}/\mathbb{Z}_2$};

    \draw[->] (5,  |-  dd)--(8,  |-  dd)node[pos=0.5, below]{Fold**};

    \node[main, label=below:$2$] (CC) at (E|- dd){};
    \node[] (BB) [left=of CC]{$\cdots$};
    \node[main, label=below:$2$] (AA) [left=of BB] {};
    \node[draw, label=right:$2$] (DD) [above=of AA]{};

    \draw[-] (DD)--(AA)--(BB)--(CC);
    \draw (CC) to [out=135, in=45,looseness=8] (CC);
    \draw [decorate, 
    decoration = {brace,
        raise=15pt,
        amplitude=5pt}] (CC) --  (AA) node[pos=0.5,below=20pt,black]{$k-3$};
    \node[draw=none,fill=none](labelD2) [below=of BB]{Magnetic quiver for $\overline{n. min. D_{k-3}}/\mathbb Z_2$};
    
    \draw[->] (labelB)-- (labelB |-,-4) node[pos=0.5,right]{$\surm(2)$ HKQ};
    
    \end{tikzpicture}}\caption{}\label{fig:nminDkCommute1}\end{subfigure}
    \centering
    \begin{subfigure}{\textwidth}\resizebox{\textwidth}{!}{
    \begin{tikzpicture}[main/.style={draw,circle}]
    \node[main, label=below:$2$] (a) []{};
    \node[] (b) [right=of a]{$\cdots$};
    \node[main, label=below:$2$] (c) [right=of b]{};
    \node[main, label=right:$1$] (d) [above right=of c]{};
    \node[main, label=right:$1$] (e) [below right=of c]{};
    \node[draw, label=left:2] (f) [above=of a]{};
    \draw[-] (f)--(a)--(b)--(c);
    \draw[-] (d)--(c)--(e);

    \draw [decorate, 
    decoration = {brace,
        raise=10pt,
        amplitude=5pt}] (c) --  (a) node[pos=0.5,below=15pt,black]{$k-2$};
    \node[draw=none,fill=none](labelD) [below=of b]{Magnetic quiver for $\overline{n. min. D_k}$};

    \draw[->] (5,0)--(8,0) node[pos=0.5,above]{Fold*};

    \node[] (A) at (9,0) {$\mathcal Q_{\ref{fig:minBKUsual}}$};
    \node[main, label=below:$1$] (B) [right=of A]{};
    \node[main, label=below:$2$] (C) [right=of B]{};
    \node[] (D) [right=of C]{$\cdots$};
    \node[main, label=below:$2$] (E) [right=of D]{};
    \node[] (G) [right=0.1cm of A]{$\sim$};
    
    \node[draw, label=left:$1$] (F) [above=of C]{};

    \draw[-] (B)--(C)--(D)--(E);
    \draw[-] (F)--(C);

    \draw (E) to [out=135, in=45,looseness=8] (E);
    \draw [decorate, 
    decoration = {brace,
        raise=15pt,
        amplitude=5pt}] (E) --  (B) node[pos=0.5,below=20pt,black]{$k-1$};
    \node[draw=none,fill=none](labelB) [below=of D]{Magnetic quiver for $\overline{n.min. B_{k-1}}$};

    \draw[->] (labelD)-- (labelD |-,-4) node[pos=0.5,left]{$\surm(2)$ HKQ};
    
    \node[] (bb) at (b|-, -5.5){$\cdots$};
    \node[main, label=below:$2$] (aa) [left=of bb]{};
    \node[main, label=below:$2$] (cc) [right=of bb]{};
    \node[main, label=right:$1$] (dd) [above right=of cc]{};
    \node[main, label=right:$1$] (ee) [below right=of cc]{};
    \node[draw, label=left:2] (ff) [above=of aa]{};
    \draw[-] (ff)--(aa)--(bb)--(cc);
    \draw[-] (dd)--(cc)--(ee);

    \draw [decorate, 
    decoration = {brace,
        raise=10pt,
        amplitude=5pt}] (cc) --  (aa) node[pos=0.5,below=15pt,black]{$k-4$};
    \node[draw=none,fill=none](labelD2) [below=of bb]{Magnetic quiver for $\overline{n. min. D_{k-2}}$};

    \node[main, label=below:$1$] (one) [left=of aa]{};
    \node[draw, label=left:$2$] (twof) [above=of one]{};
    \draw[-] (one)--(twof);

    \draw[->] (5,  |-  cc)--(8,  |-  cc)node[pos=0.5, below]{Fold*};

    \node[main, label=below:$2$] (EE) at  (E|- cc){};
    \node[] (DD) [left=of EE]{$\cdots$};
    \node[main, label=below:$2$] (CC) [left=of DD]{};
    \node[main, label=below:$1$] (BB) [left=of CC]{};
    \node[draw, label=left:$1$] (FF) [above=of CC]{};

    \draw[-] (BB)--(CC)--(DD)--(EE);
    \draw[-] (FF)--(CC);

    \node[main, label=below:$1$] (ONE) at (A|-cc){};
    \node[draw, label=left:$2$] (TWOF) [above=of ONE]{};
    \draw[-] (ONE)--(TWOF);

    \draw (EE) to [out=135, in=45,looseness=8] (EE);
    \draw [decorate, 
    decoration = {brace,
        raise=15pt,
        amplitude=5pt}] (EE) --  (BB) node[pos=0.5,below=20pt,black]{$k-3$};
    \node[draw=none,fill=none](labelB2) [below=of DD]{Magnetic quiver for $\overline{n.min. B_{k-3}}$};

     \node[] (AA) [right=of EE] {$\mathcal Q_{\ref{fig:minBKUsual}}$};
    \node[] (GG) [left=0.1cm of AA]{$\sim$};

    \draw[->] (labelB)-- (labelB |-,-4) node[pos=0.5,right]{$\surm(2)$ HKQ};
    
    \end{tikzpicture}}\caption{}\label{fig:nminDkCommute2}\end{subfigure}
    \caption{The $\surm(2)$ quiver subtraction from $\overline{n. min. D_k}$ has two alignments. For each alignment we draw the commutative diagram of folding and $\surm(2)$ HKQ. The folding operation is accompanied by a quiver loop transformation, denoted * or **, as discussed in the text.}
    \label{fig:nminDkCommute}
\end{figure}
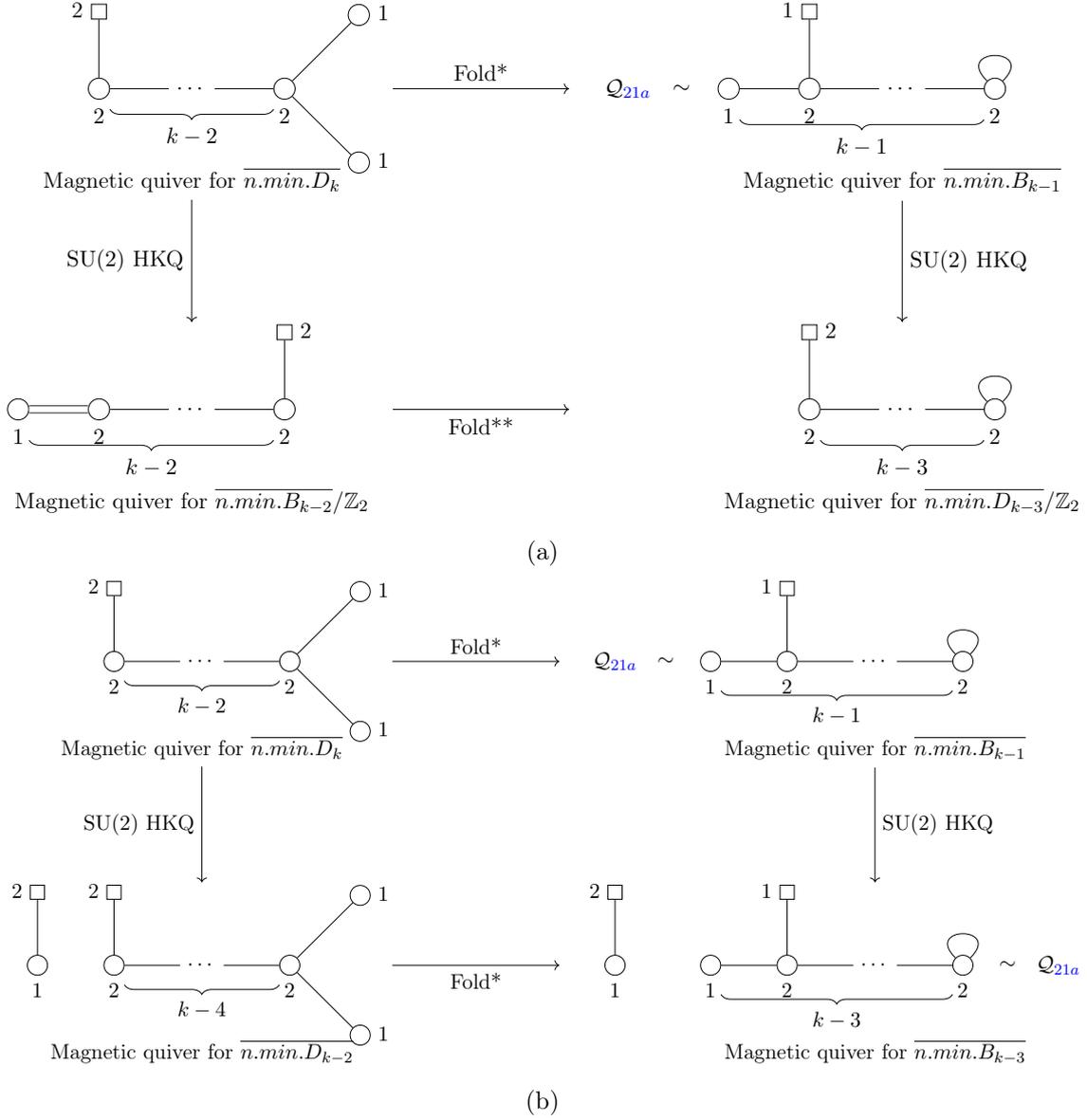

\begin{figure}[h!]
    \centering
    \begin{subfigure}{\textwidth}\resizebox{\textwidth}{!}{
    \begin{tikzpicture}[main/.style={draw,circle}]
    \node[main, label=below:$1$] (a) {};
    \node[main,label=below:$2$] (b) [right=of a] {};
    \node[] (c) [right=of b] {$\cdots$};
    \node[main,label=below:$2$] (d) [right=of c] {};
    \node[main,label=below:$1$] (e) [right=of d] {};
    \node[draw, label=left:$1$] (f) [above=of b]{};
    \node[main, label=right:$1$] (g) [above=of d]{};
    \draw[-] (a)--(b);
    \draw[-] (e)--(d);
    \draw[-] (f)--(b)--(c)--(d)--(g);
    \draw [decorate, 
    decoration = {brace,
        raise=10pt,
        amplitude=5pt}] (d) --  (b) node[pos=0.5,below=15pt,black]{$k-3$};
    \node[draw=none,fill=none](labelD) [below=of c]{Magnetic quiver for $\overline{min. D_k}$};

    \draw[->] (7,0)--(10,0) node[pos=0.5,above]{$\mathbb{Z}_2$ Gauge};

    \node[main, label=below:$1$] (A) at (11,0) {};
    \node[main, label=below:$2$] (C) [right=of A]{};
    \node[] (D) [right=of C] {$\cdots$};
    \node[main,label=below:$2$] (E) [right=of D] {};
    \node[main,label=below:$2$] (F) [right=of E] {};
    \node[draw, label=left:$1$] (G) [above=of C]{};
    
    \draw[-] (A)--(C)--(D)--(E)--(F);
    \draw[-] (C)--(G);
    \draw (F) to [out=135, in=45,looseness=8] (F);
    \draw [decorate, 
    decoration = {brace,
        raise=15pt,
        amplitude=5pt}] (E) --  (C) node[pos=0.5,below=20pt,black]{$k-3$};
    \node[draw=none,fill=none](labelB) [below=of D]{Magnetic quiver for $\overline{n. min. B_{k-1}}$};

    \draw[->] (labelD)-- (labelD |-,-4) node[pos=0.5,left]{$\surm(2)$ HKQ};

    \node[draw, label=left:$2$] (aa) at (b |-, -4){};
    \node[main, label=below:$2$] (bb) [below=of aa]{};
    \node[draw=none,fill=none] (cdots2) [right=of bb]{$\cdots$};
    \node[main, label=below:$2$] (cc) [right=of cdots2]{};
    \node[main, label=right:$1$] (dd)[above right=of cc]{};
    \node[main, label=right:$1$] (ee)[below right=of cc]{};
    
    \draw[-] (ee)--(cc)--(dd);
    \draw[-] (cc)--(cdots2)--(bb)--(aa);
    \draw [decorate, 
    decoration = {brace,
        raise=10pt,
        amplitude=5pt}] (cc) --  (bb) node[pos=0.5,below=15pt,black]{$k-4$};
    \node[draw=none,fill=none](labelDres1) [below=of cdots2]{Magnetic quiver for $\overline{n. min. D_{k-2}}$};
    
    \draw[->] (7,  |-  cc)--(10,  |-  cc)node[pos=0.5, below]{$\mathbb{Z}_2$ Gauge};

    \node[main, label=below:$2$] (CC) at (A |- cc){};
    \node[] (DD) [right=of CC] {$\cdots$} ;
    \node[main,label=below:$2$] (EE) [right=of DD] {};
    \node[main,label=below:$2$] (FF) [right=of EE] {};
    \node[draw, label=left:$2$] (GG) [above=of CC]{};
    
    \draw[-] (CC)--(DD)--(EE)--(FF);
    \draw[-] (CC)--(GG);
   \draw (FF) to [out=135, in=45,looseness=8] (FF);
    \draw [decorate, 
    decoration = {brace,
        raise=15pt,
        amplitude=5pt}] (EE) --  (CC) node[pos=0.5,below=20pt,black]{$k-4$};
    \node[draw=none,fill=none](labelB2) [below=of DD]{Magnetic quiver for $\overline{n. min. D_{k-2}}/\mathbb{Z}_2$};

    \draw[->] (labelB)-- (labelB |-,-4) node[pos=0.5,right]{$\surm(2)$ HKQ};
    \end{tikzpicture}}
    \caption{}
    \label{fig:minDkCommute3}
    \end{subfigure}
    \centering
    \begin{subfigure}{\textwidth}\resizebox{\textwidth}{!}{
    \begin{tikzpicture}[main/.style={draw,circle}]
    \node[main, label=below:$1$] (a) {};
    \node[main,label=below:$2$] (b) [right=of a] {};
    \node[] (c) [right=of b] {$\cdots$};
    \node[main,label=below:$2$] (d) [right=of c] {};
    \node[main,label=below:$1$] (e) [right=of d] {};
    \node[draw, label=left:$1$] (f) [above=of b]{};
    \node[main, label=right:$1$] (g) [above=of d]{};
    \draw[-] (a)--(b);
    \draw[-] (e)--(d);
    \draw[-] (f)--(b)--(c)--(d)--(g);
    \draw [decorate, 
    decoration = {brace,
        raise=10pt,
        amplitude=5pt}] (d) --  (b) node[pos=0.5,below=15pt,black]{$k-3$};
    \node[draw=none,fill=none](labelD) [below=of c]{Magnetic quiver for $\overline{min. D_k}$};

    \draw[->] (7,0)--(10,0) node[pos=0.5,above]{$\mathbb{Z}_2$ Gauge};

    \node[main, label=below:$1$] (A) at (11,0) {};
    \node[main, label=below:$2$] (C) [right=of A]{};
    \node[] (D) [right=of C] {$\cdots$};
    \node[main,label=below:$2$] (E) [right=of D] {};
    \node[main,label=below:$2$] (F) [right=of E] {};
    \node[draw, label=left:$1$] (G) [above=of C]{};
    
    \draw[-] (A)--(C)--(D)--(E)--(F);
    \draw[-] (C)--(G);
    \draw (F) to [out=135, in=45,looseness=8] (F);
    \draw [decorate, 
    decoration = {brace,
        raise=15pt,
        amplitude=5pt}] (E) --  (C) node[pos=0.5,below=20pt,black]{$k-3$};
    \node[draw=none,fill=none](labelB) [below=of D]{Magnetic quiver for $\overline{n.min. B_{k-1}}$};

    \draw[->] (labelD)-- (labelD |-,-4) node[pos=0.5,left]{$\surm(2)$ HKQ};
    
    \node[draw, label=left:$1$] (gg) at (b |-,-4){};
    \node[main,label=below:$2$] (bb) [below=of gg] {};
    \node[main, label=below:$1$] (aa) [left=of bb] {};
    \node[] (cc) [right=of bb] {$\cdots$};
    \node[main,label=below:$2$] (dd) [right=of cc] {};
    \node[main,label=below:$1$] (ee) [right=of dd] {};
    \node[main, label=right:$1$] (ff) [above=of dd]{};
    
    \draw[-] (aa)--(bb)--(gg);
    \draw[-] (ee)--(dd);
    \draw[-] (bb)--(cc)--(dd)--(ff);
    \draw [decorate, 
    decoration = {brace,
        raise=10pt,
        amplitude=5pt}] (dd) --  (bb) node[pos=0.5,below=15pt,black]{$k-5$};
    \node[draw=none,fill=none](labelD1) [below=of cc]{Magnetic quiver for $\overline{min. D_k}$};

    \node[main, label=below:$1$] (one) [left=of aa]{};
    \node[draw, label=left:$2$] (twoF) [above=of one]{};
    
    \draw[-] (twoF)--(one);

    \draw[->] (7,  |-  cc)--(10,  |-  cc)node[pos=0.5, below]{$\mathbb{Z}_2$ Gauge};

    \node[main, label=below:$1$] (ONE)  at (A|- dd){};;
    \node[draw, label=left:$2$] (TWOF) [above=of ONE]{};
    
    \draw[-] (TWOF)--(ONE);
    
    \node[main, label=below:$1$] (AA) [right=of ONE]{};
    \node[main, label=below:$2$] (CC) [right=of AA]{};
    \node[] (DD) [right=of CC] {$\cdots$};
    \node[main,label=below:$2$] (EE) [right=of DD] {};
    \node[main,label=below:$2$] (FF) [right=of EE] {};
    \node[draw, label=left:$1$] (GG) [above=of CC]{};
    
    \draw[-] (AA)--(CC)--(DD)--(EE)--(FF);
    \draw[-] (CC)--(GG);
    \draw (FF) to [out=135, in=45,looseness=8] (FF);
    \draw [decorate, 
    decoration = {brace,
        raise=15pt,
        amplitude=5pt}] (EE) --  (CC) node[pos=0.5,below=20pt,black]{$k-5$};
    \node[draw=none,fill=none](labelB2) [below=of DD]{Magnetic quiver for $\overline{n.min. B_{k-3}}$};
    
    \draw[->] (labelB)-- (labelB |-,-4) node[pos=0.5,right]{$\surm(2)$ HKQ};

    \end{tikzpicture}}
    
    \caption{}
    \label{fig:minDkCommute4}
    \end{subfigure}
    \caption{The $\surm(2)$ quiver subtraction from $\overline{min. D_k}$ has two alignments. For each alignment we draw the commutative diagram between $\mathbb{Z}_2$ gauging and $\surm(2)$ HKQ.}
    \label{fig:minDkCommuteDiscreteGauge}
\end{figure}
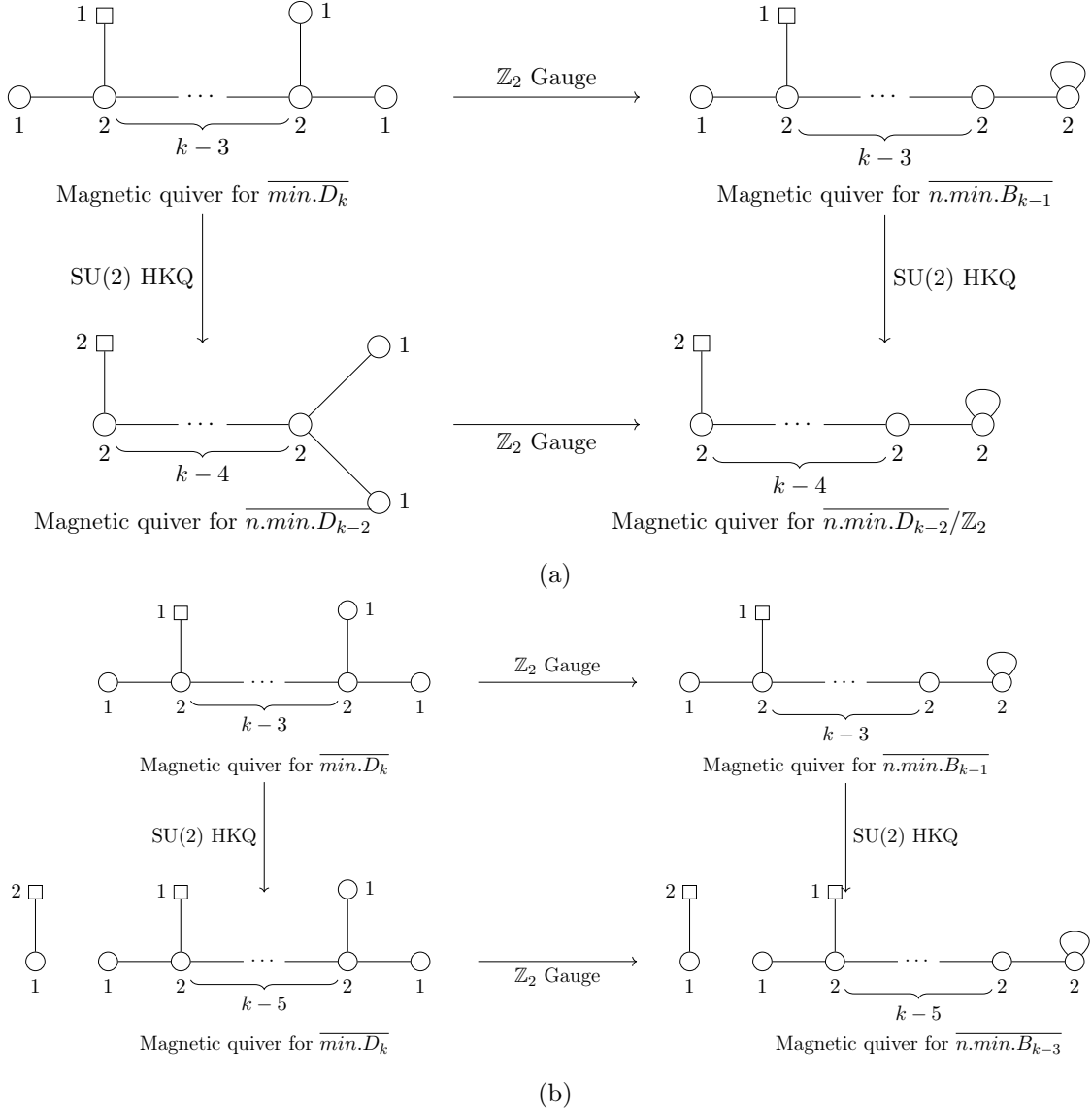
    
\paragraph{Commutativity of Folding with $\surm(n)$ HKQ}
{A commutative diagram of the folding and $\surm(2)$ HKQ of a magnetic quiver for $\overline{min. E_6}$ is shown in \Figref{fig:E6Commute}, based on results from Table \ref{tab:OrbSU2}. Applying a $\mathbb{Z}_2$ folding to the magnetic quiver for $\overline{min. E_6}$ gives a magnetic quiver for $\overline{min. F_4}$, and a subsequent $\surm(2)$ HKQ by quotient quiver subtraction gives a magnetic quiver for $\overline{n. min. C_3}$. If the quotient quiver subtraction is carried out first, a magnetic quiver for $\overline{n. min. A_5}$ is obtained, but subsequent folding once again yields the magnetic quiver for $\overline{n.min. C_3}$.

A second example, shown in \Figref{fig:minDkCommute}, is the commutation of folding and $\surm(2)$ HKQ by quotient quiver subtraction on the magnetic quiver \Quiver{fig:quiv_min_Dk} for  $\overline{min. D_k}$ for $k\geq 5$. Recall that there are two possible alignments of the quotient quiver, resulting in the quivers $\mathcal Q_{\ref{fig:minDKA1Sub1}}$ and $\mathcal Q_{\ref{fig:minDKA1Sub2}}$; so we draw commuting diagrams for each alternative. In each case, an initial folding of the magnetic quiver $\mathcal Q_{\ref{fig:quiv_min_Dk}}$ for $\overline{min. D_k}$ gives $\mathcal Q_{\ref{fig:quiv_min_Bk}}$ for $\overline{min. B_{k-1}}$. The subsequent quotient quiver subtraction gives magnetic quivers $\mathcal Q_{\ref{fig:minBKA1Sub1}}$ and $\mathcal Q_{\ref{fig:minBKA1Sub2}}$ for $\overline{n.min B_{k-3}}$ and $ \left(\overline{min. A_1}\otimes \overline{min. B_{k-3}}\right)$, respectively. The other way around is to perform the quotient quiver subtraction first, which gives the magnetic quivers $\mathcal Q_{\ref{fig:minDKA1Sub1}}$ and $\mathcal Q_{\ref{fig:minDKA1Sub2}}$ for $\overline{n.min D_{k-2}}$ and $ \left(\overline{min. A_1}\otimes \overline{min. D_{k-2}}\right)$, respectively. The subsequent folding once again gives $\mathcal Q_{\ref{fig:minBKA1Sub1}}$ and $\mathcal Q_{\ref{fig:minBKA1Sub2}}$.

We note that the $\overline{min. A_1}$  in $\mathcal Q_{\ref{fig:minDKA1Sub2}}$ does not participate in the folding.

A more subtle example is provided by the folding and $\surm(2)$ HKQ by quotient quiver subtraction on the magnetic quiver for $\overline{n.min. D_k}$ shown in \Figref{fig:nminDkCommute}. In this case, the initial folding is accompanied by a transformation to an alternative quiver for $\overline{n.min. B_{k-1}}$, as discussed in Section \ref{sec:nminBkA1}. The consistency of the operation marked Fold** in \Figref{fig:nminDkCommute1} has been confirmed by analysis of the HWGs of the two magnetic quivers involved. (There is an identification of highest weight fugacities in the respective HWGs that indicates an operation on the Coulomb branch similar to folding.) However, we do not yet have a full understanding of the Fold** diagrammatic transformation.

The fact that folding and quotient quiver subtraction often commute is not surprising. The folding of legs introduces a non-simply laced edge, with the folded nodes becoming short roots. Any subsequent quotient quiver subtraction is always taken by subtraction from the long roots and so is independent of those roots that have participated in the folding.

It should be noted that a similar commutative diagram of folding and $\urm(2)$ quotient quiver subtraction may be drawn for the magnetic quivers for the moduli space of free fields \Quiver{fig:free2kp2} \Quiver{fig:free2kp2}, which take the form of finite Dynkin diagrams of type $D$ and $B$ respectively.}

\paragraph{Commutativity of Discrete Quotients with $\surm(n)$ HKQ}
A similar approach can be applied to explore the commutation between discrete gauging and $\surm(n)$ HKQs. Discrete gauging and quotient quiver subtraction often commute because the nodes that are discretely gauged cannot subsequently be involved in quiver subtraction and vice versa.

Consider the example in Figure \ref{fig:minDkCommuteDiscreteGauge}. The magnetic quiver $\mathcal Q_{\ref{fig:quiv_min_Dk}}$ for $\overline{min. D_k}$ for $k\geq 4$ contains a bouquet of two $\urm(1)$ gauge nodes that can be discretely gauged to give a gauge node of rank 2 with an adjoint link. This implements the Kostant-Brylinski relation  $\overline{min. D_k}/\mathbb Z_2=\overline{n.min. B_{k-1}}$ \cite{1992math......4227B} in a diagrammatic way, and gives a magnetic quiver $\mathcal Q_{\ref{fig:minBkAdj}}$ for $\overline{n.min. B_{k-1}}$. Its $\surm(2)$ HKQ can be computed using quiver subtraction to yield two magnetic quivers $\mathcal Q_{\ref{fig:minBKA1Sub1}}$ and $\mathcal Q_{\ref{fig:minBKA1Sub2}}$, whose union is $\overline{n. min. D_{k-3}/\mathbb{Z}_2}\cup \overline{min. A_1}\otimes \overline{n. min. B_{k-3}}$. Alternatively, performing the $\surm(2)$ quiver subtraction first, we find the magnetic quivers $\mathcal Q_{\ref{fig:minDKA1Sub1}}$ and $\mathcal Q_{\ref{fig:minDKA1Sub2}}$, whose union gives $\overline{n. min. D_{k-2}} \cup \overline{min. A_1} \otimes \overline{min. D_{k-2}}$. The subsequent discrete gauging of the bouquet of rank 1 nodes in these quivers once again produces $\mathcal Q_{\ref{fig:minBKA1Sub1}}$ and $\mathcal Q_{\ref{fig:minBKA1Sub2}}$.

Here, the discrete gauging does not affect the $\overline{min. A_1}$ part of the moduli space.

It should be noted that a similar commutative diagram of discrete gauging by $\mathbb Z_2$ and $\urm(2)$ quotient quiver subtraction may be drawn for the magnetic quivers for the moduli space of free fields \Quiver{fig:free2kp2} \Quiver{fig:free2kp2}, which take the form of finite Dynkin diagrams of type $D$ and $B$ respectively.

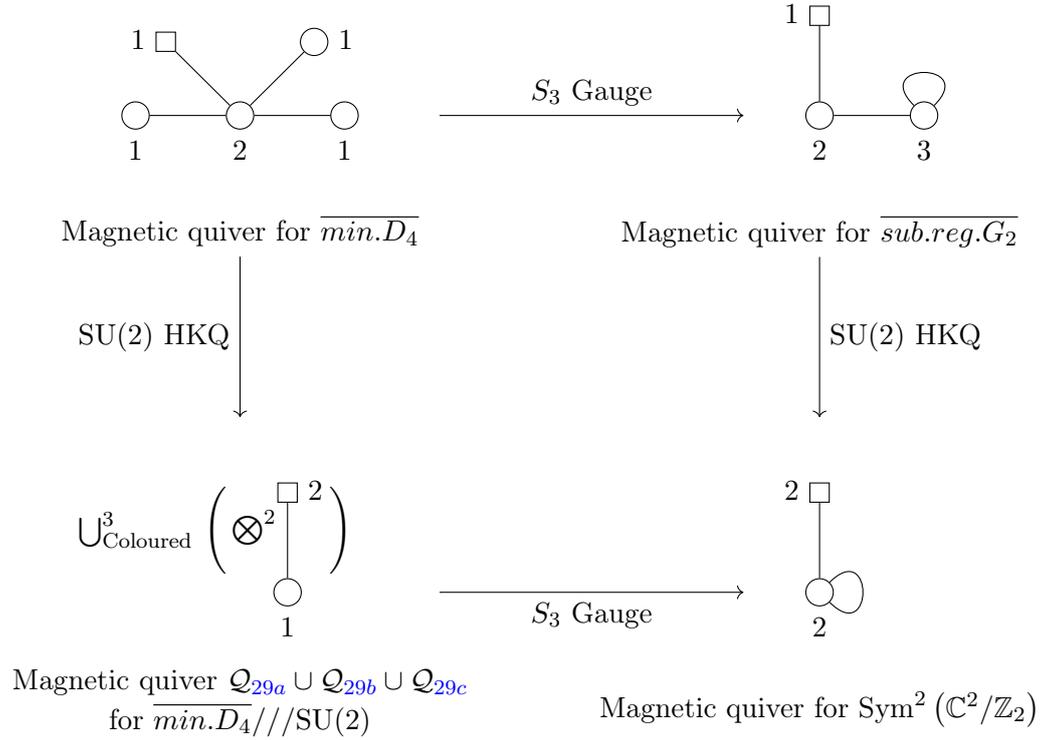
\begin{figure}[h!]
    \centering
    \begin{tikzpicture}[main/.style={draw,circle}]
    \node[main, label=below:$1$] (a) []{};
    \node[main, label=below:$2$] (b) [right=of a]{};
    \node[main, label=below:$1$] (c) [right=of b]{};
    \node[draw, label=left:$1$] (d) [above left=0.71 and 0.71 of b]{};
    \node[main, label=right:$1$] (e) [above right=0.71 and 0.71 of b]{};

    \draw[-] (a)--(b)--(c);
    \draw[-] (d)--(b)--(e);
    \node[draw=none,fill=none](labelD) [below=of b]{Magnetic quiver for $\overline{min. D_4}$};

    \draw[->] (4,0)--(8,0) node[pos=0.5,above]{$S_3$ Gauge};

    \node[main, label=below:$2$] (B) at (9,0){};
    \node[draw, label=left:$1$] (A) [above=of B] {};
    
    \node[main,label=below:$3$] (C) [right=of B] {};
    
    \draw[-] (A)--(B)--(C);
    \draw (C) to [out=135, in=45,looseness=8] (C);
    \node[draw=none,fill=none](labelG) [below=of B]{Magnetic quiver for $\overline{sub. reg. G_2}$};

    \draw[->] (labelD)-- (labelD |-,-4) node[pos=0.5,left]{$\surm(2)$ HKQ};

    \draw[->] (labelG)-- (labelG |-,-4) node[pos=0.5,right]{$\surm(2)$ HKQ};

    \node[draw, label=left:$2$] (AA) at (labelG |-, -5){};
    \node[main, label=below:$2$] (BB) [below=of AA]{};

    \draw[-] (AA)--(BB);
    \draw (BB) to [out=45, in=-45,looseness=8] (BB);

    \node[draw=none,fill=none](labelC) [below=of BB]{Magnetic quiver for $\text{Sym}^2\left(\mathbb C^2/\mathbb Z_2\right)$};

    \node[] (aa) at (labelD|-,-5.5){$\Bigg(\bigotimes ^2$};
    \node[main, label=below:$1$] (bb) at (2,|-BB){};
    \node[draw, label=right:$2$] (cc) [above=of bb]{};
    \node[] (ee) at (2.7,|-aa){$\Bigg)$};
    \draw[-] (bb)--(cc);
    \node[] (dd) at (aa-|a) {$\bigcup^3_{\text{Coloured}}$};
    
    \node[draw=none,fill=none](labelquot) [below=of aa]{\stackanchor{Magnetic quiver $\mathcal{Q}_{\ref{fig:minD4Quiv} \textcolor{blue}{a}}\cup\mathcal{Q}_{\ref{fig:minD4Quiv} \textcolor{blue}{b}}\cup\mathcal{Q}_{\ref{fig:minD4Quiv} \textcolor{blue}{c}}$}{for $\overline{min. D_4}///\surm(2)$}};

    \draw[->] (4,  |-  BB)--(8,  |-  BB)node[pos=0.5, below]{$S_3$ Gauge};
    
    \end{tikzpicture}
    \caption{Commutative diagram showing $S_3$ discrete gauging and $\surm(2)$ HKQ by quotient quiver subtraction from a magnetic quiver for $\overline{min. D_4}$ to obtain $\text{Sym}^2\left(\mathbb C_2/\mathbb Z_2\right)$.}
    \label{fig:minD4Commute}
    \end{figure}

A further non-trivial example is the commutativity of $\surm(2)$ HKQ and $S_3$ gauging of $\overline{min. D_4}$ shown in \Figref{fig:minD4Commute}. If the $S_3$ gauging is performed on a magnetic quiver $\mathcal Q_{\ref{fig:minD4Quiv}}$ for $\overline{min. D_4}$ a magnetic quiver for $\overline{sub. reg. G_2}$ is obtained. The quotient quiver subtraction then gives a magnetic quiver for $\text{Sym}^2\left(\mathbb C^2/\mathbb Z_2\right)$. Alternatively, carrying out the $\surm(2)$ HKQ of $\overline{min. D_4}$ first, we find the union of coloured $\overline{min. A_1}\otimes \overline{min. A_1}$, as described in Section \ref{sec:DtypeSU2HKQ}. Performing the $S_3$ gauging then recovers $\text{Sym}^2\left(\mathbb C^2/\mathbb Z_2\right)$. 

There is no simple bouquet of $\urm(1)$ gauge nodes in this last step, so it is difficult to see this $S_3$ quotient diagrammatically. We can however check by comparing the volumes of the Coulomb branches. We do this by taking the ratio of unrefined HS in the limit $t\rightarrow 1$, to find:
\begin{equation}
    \frac{\text{Vol}\left(\overline{min. D_4}///\surm(2)\right)}{\text{Vol}\left(\text{Sym}^2\left(\mathbb C^2/\mathbb Z_2\right)\right)}=\lim_{t\rightarrow 1}\frac{\left(1+t^2\right)^2 \left(1+5t^2+12t^4-7t^6+t^8\right)}{1+t^2+4t^4+t^6+t^8}=6 ,
\end{equation}
and since $|S_3|=6$, this is consistent with an $S_3$ gauging.

Moreover, we can explicitly recover the $S_3$ quotient using the Burnside lemma \cite{Bourget:2020bxh}. The HS of $\overline{min. D_4}///\surm(2)$ can be mapped to $\surm(2)$ as $[2;0;0]_{A_1\times A_1\times A_1}+[0;2;0]_{A_1\times A_1\times A_1}+[0;0;2]_{A_1\times A_1\times A_1}\rightarrow 3[2]_{A_1}$. We find the HWG:
\begin{align}
    HWG\left[\overline{min. D_4}///\surm(2)\right]_{\rightarrow{A_1}}
    &=\left(1 + \mu^2 t^2 + 2 t^4 + 3 \mu^2 t^4 + \mu^4 t^4 - \mu^2 t^6 - \mu^4 t^8\right)PE\left[2\mu^2t^2+t^4\right],
\end{align}where $\mu$ is an $\surm(2)$ highest weight fugacity. We now use the symbols $\mathbf{1},\varepsilon$, and $\mathbf{2}$ to refer to the trivial, alternating, and fundamental representations of $S_3$ respectively, choose the following assignment of irreps, and perform the group average to find:
\begin{align}
   &\left(\mathbf{1}1 + \mathbf{1}\mu^2 t^2 + \mathbf{2} t^4 + \mathbf{2} \mu^2 t^4 +\varepsilon\mu^2t^4 +\mathbf{1}\mu^4 t^4 - \mathbf{1}\mu^2 t^6 - \mathbf{1}\mu^4 t^8\right)PE\left[\mathbf{2}\mu^2t^2+\mathbf{1}t^4\right]\nonumber\\&\rightarrow PE\left[\mu^2 t^2+\left(\mu^4+1\right) t^4+\mu^4 t^6-\mu^8 t^{12}\right].
\end{align}
The result is the HWG for $\text{Sym}^2\left(\mathbb C^2/\mathbb Z_2\right)$, which confirms the construction.

Based on such examples, including the star-shaped quivers of $\surm(2)^n$ studied in Section \ref{sec:miscellaneous}, we conjecture that $\surm(2)$ HKQ and $S_{k<n}$ discrete gauging are commuting operations for quivers with bouquets of $\surm(2)^n$.

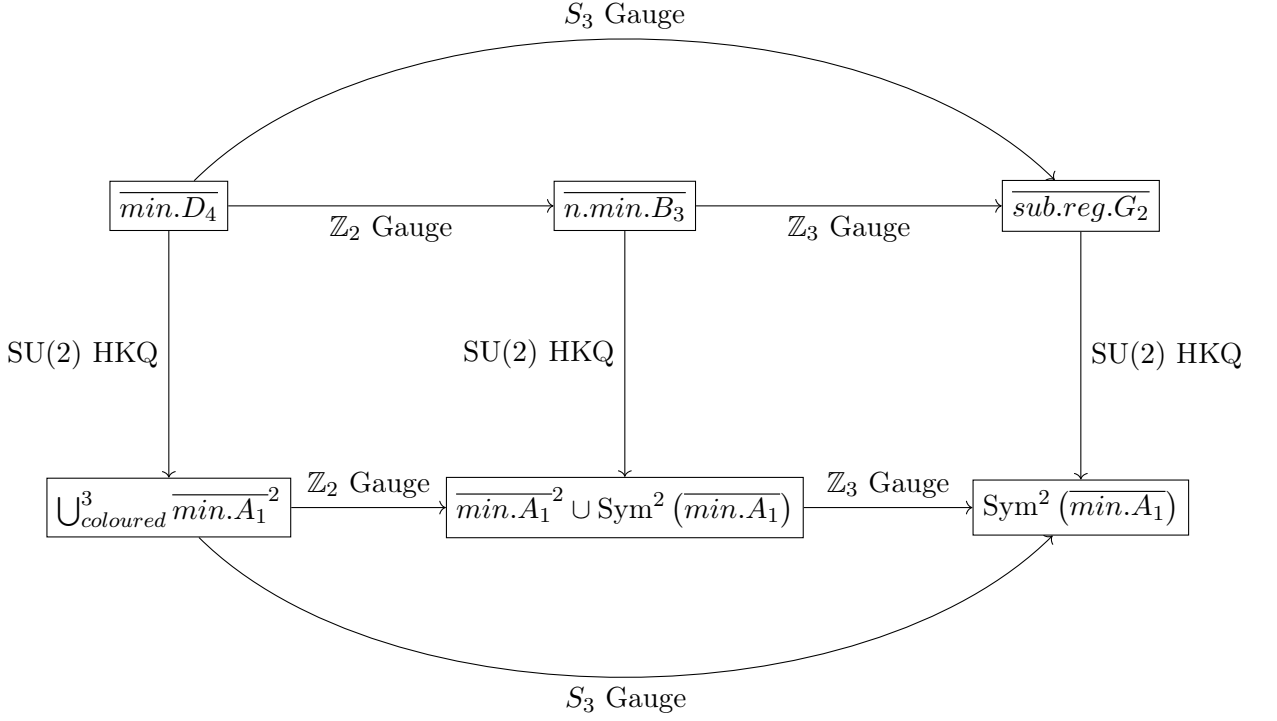
\begin{figure}
    \centering
    \begin{tikzpicture}[main/.style={draw,circle}]
    \node[draw] (minD4) at (0,0) {$\overline{min. D_4}$};
    \node[draw] (nminB3) at (6,0) {$\overline{n.min. B_3}$};
    \node[draw] (srG2) at (12,0) {$\overline{sub. reg. G_2}$};
    \node[draw] (minD4SU2) at (0,-4) {$\bigcup^3_{coloured}\overline{min. A_1}^2$};
    \node[draw] (nminB3SU2) at (6,-4) {$\overline{min. A_1}^2\cup \text{Sym}^2\left(\overline{min. A_1}\right)$};
    \node[draw] (subregG2SU2) at (12,-4) {$\text{Sym}^2\left(\overline{min. A_1}\right)$};

    \draw[->] (minD4)--(nminB3)node[pos=0.5, below]{$\mathbb{Z}_2$ Gauge};
    
    \draw[->](nminB3)--(srG2) node[pos=0.5, below]{$\mathbb{Z}_3$ Gauge};

    \draw[->] (minD4SU2)--(nminB3SU2)node[pos=0.5, above]{$\mathbb{Z}_2$ Gauge};
    \draw[->](nminB3SU2)--(subregG2SU2) node[pos=0.5, above]{$\mathbb{Z}_3$ Gauge};

    \draw[->] (minD4)--(minD4SU2) node[pos=0.5, left
    ]{$\surm(2)$ HKQ};
    \draw[->] (nminB3)--(nminB3SU2) node[pos=0.5, left
    ]{$\surm(2)$ HKQ};
    \draw[->] (srG2)--(subregG2SU2) node[pos=0.5, right
    ]{$\surm(2)$ HKQ};

     \draw [->,out=45,in=135,looseness=0.8] (minD4) to node[above]{$S_3$ Gauge} (srG2) ;

     \draw [->,out=-45,in=-135,looseness=0.8] (minD4SU2) to node[below]{$S_3$ Gauge} (subregG2SU2) ;
    \end{tikzpicture}
    \caption{Commutative diagram of the discrete gauging relationships between $\overline{min. D_4}$, $\overline{n.min. B_3}$, and $\overline{sub. reg. G_2}$ and their $\surm(2)$ HKQ.}
    \label{fig:D4CombinedCommute}
\end{figure}

One particular example of a star-shaped quiver of $\surm(2)^n$ that we have studied throughout this work is for $n=4$. This has an enhancement of the global symmetry of its Coulomb branch from $\surm(2)^4\rightarrow SO(8)$, giving a magnetic quiver for $\overline{min. D_4}$. This quiver admits two obvious discrete gauging actions of $S_2\simeq \mathbb Z_2$, and $S_3$, which give magnetic quivers for $\overline{n. min. B_{3}}$ and $\overline{sub. reg. G_2}$, respectively. These actions form part of commutative diagrams, as shown in \Figref{fig:minDkCommuteDiscreteGauge} (specialised to the case of $\overline{min. D_4}$) and \Figref{fig:minD4Commute}. There is also a discrete gauging of $\overline{n.min. B_3}/\mathbb Z_3=\overline{sub. reg. G_2}$ (another Kostant-Brylinski relation) \cite{1992math......4227B} which also commutes with quotient quiver subtraction. All of these results can be collected into a larger commutative diagram as shown in \Figref{fig:D4CombinedCommute}.

In general, we conjecture that if two magnetic quivers are related by a discrete action, and both quivers are amenable to quotient quiver subtraction, then we expect that the quivers resulting from the HKQ are also related through the same discrete action \footnote{As we have seen in \Figref{fig:nminDkCommute}, this discrete action may not be obvious diagrammatically and, in such circumstances, this conjecture may assist in its identification.}.

\paragraph{$\surm(n)$ HKQs of Exceptional Affine Grassmannian Slices}{Further non-trivial tests of $\surm(n)$ HKQs by quotient quiver subtraction are provided by target quivers which are magnetic quivers for affine Grassmannian slices of exceptional algebras.

Summaries of some of the examples computed in Section \ref{sec:ExceptionalAGHKQ} are shown in Tables \ref{tab:AGSU2} and \ref{tab:AGSU3}. The relationships between different affine Grassmannian slices are particularly interesting, especially when a union of affine Grassmannian slices emerges under HKQ. The quotient quiver subtraction on viable magnetic target quivers may allow one to move between affine Grassmannian slices of different algebras.}
\begin{table}[h!]
    \centering
    \begin{tabular}{|c|c|}
    \hline
    Target Moduli Space & $\surm(2)$ HKQ\\
    \hline\hline
        $ \overline{\left[\mathcal W_{E_6}\right]}^{[0,0,1,0,0,0]}_{[0,0,0,0,0,1]}$&$\overline{\left[\mathcal W_{D_5}\right]}^{[0,1,0,1,1]}_{[0,0,0,1,1]} $\\
        \hline
         $\left[\overline{\mathcal W_{E_6}}\right]^{[1,0,0,1,0,0]}_{[0,2,0,0,0,0]}$&$\left[\overline{\mathcal W_{A_4}}\right]^{[1,0,2,0]}_{[0,1,0,0]}\cup \left[\overline{\mathcal W_{A_5}}\right]^{[2,0,1,0,1]}_{[1,0,1,0,0]}$\\
         \hline$\left[\overline{\mathcal W_{E_7}}\right]^{[0,0,0,1,0,0,0]}_{[0,0,0,0,0,0,1]}$&$\left[\overline{\mathcal W_{E_6}}\right]^{[1,0,1,0,0,0]}_{[1,0,0,0,0,1]}$\\
         \hline
    \end{tabular}
    \caption{Exceptional affine Grassmannian slices and their $\surm(2)$ HKQs.}
    \label{tab:AGSU2}
\end{table}

\begin{table}[h!]
    \centering
    \begin{tabular}{|c|c|}
    \hline
    Target Moduli Space & $\surm(3)$ HKQ\\\hline\hline
        $\overline{\left[\mathcal W_{E_7}\right]}^{[0,1,0,0,0,0,0]}_{[1,0,0,0,0,0,0]}$&$\overline{\left[\mathcal W_{D_5}\right]}^{[0,0,2,0,0]}_{[0,0,0,1,1]}\cup \overline{\left[\mathcal W_{D_5}\right]}^{[2,0,0,1,1]}_{[2,0,0,1,1]}$\\\hline
    \end{tabular}
    \caption{Exceptional affine Grassmannian slices and their $\surm(3)$ HKQs.}
    \label{tab:AGSU3}
\end{table}

\paragraph{$\urm(n)$ HKQ of Dynkin type}{Although not the focus of our study, the possibility of combining $\surm(n)$ and $\urm(1)$ HKQs into a $\urm(n)$ HKQ merits some comment. Often there are many ways of taking a $\urm(1)$ HKQ from a unitary magnetic quiver, while the $\surm(n)$ HKQ is limited by the selection rules \ref{sec:rules}. However, in the natural case, where the embedding of $\urm(n)$ follows from a Dynkin embedding into an external leg of form $(1)- \cdots - (n) - \cdots $, then the magnetic quiver for the $\urm(n)$ HKQ can be conjectured to follow as a simple extension of the $\surm(n)$ HKQ rules herein, by omitting the step of unframing prior to the quotient quiver subtraction. This is illustrated in the examples in Appendix \ref{sec:unex}, wherein care needs to be taken regarding application of the Junction Rule to avoid HKQs where incomplete Higgsing arises.}

\paragraph{Open problems}{

Our method of quotient quiver subtraction imposes selection rules on the types of magnetic quiver to which it can be applied. Further work could provide diagrammatic techniques for the HKQs of the Coulomb branches of quivers that lie outside these rules.

In particular, quotient quiver subtraction requires a Dynkin type embedding of $\surm(n)$ into the global symmetry of the Coulomb branch in accordance with the External Leg rule. However, there may be other embeddings of $\surm(n)$ into the global symmetry $G$ of the magnetic quiver. Is it possible to find rules that diagrammatically produce $\surm(n)$ HKQs for such different embeddings?

We have explored $\surm(n)$ HKQs. Are there diagrammatic techniques for HKQs by other classical groups?

We have only considered unitary magnetic quivers. Can similar diagrammatic techniques be developed for orthosymplectic quivers?

Although we have seen how to perform $\surm(n)$ HKQs at the level of the quiver, these quotient quiver subtractions remain to be interpreted in the context of brane systems.}

\acknowledgments

We would like to thank Julius Grimminger, Tudor Dimofte, Travis Schedler, and Marcus Sperling for discussions. The work of AH, RK, and GK is partially supported by STFC grant ST/T000791/1. The work of GK is supported by STFC DTP research studentship grant ST/X508433/1.

\appendix

\section{$\urm(1)$ HKQs}
\label{sec:U1HKQ}

\begin{table}[h!]
    \centering
    \begin{tabular}{|c|c|c|c|c|}
\hline
    Target Orbit&Resulting Orbit&Target Quiver & Resulting Quiver & HKQ \\\hline
   $\overline{\mathcal O}_{(2^2)}^{A_3}$ & $\overline{\mathcal O}_{(3)}^{A_2}$&
     \begin{tikzpicture}[main/.style={draw,circle},baseline=0]
     \node[main, label=below:$1$] (1L) []{};
     \node[main, label=below:$2$] (2) [right=of 1L]{};
     \node[main, label=below:$1$] (1R) [right=of 2]{};
     \node[draw, label=left:$2$] (2F) [above=of 2]{};
    
     \draw[-] (1L)--(2)--(1R);
     \draw[-] (2)--(2F);
     \end{tikzpicture}
     & \begin{tikzpicture}[main/.style={draw,circle},baseline=0]
     \node[main, label=below:$1$] (1L) []{};
     \node[main, label=below:$2$] (2) [right=of 1L]{};
     \node[draw, label=right:$3$] (3F) [above=of 2]{};
     \draw[-] (1L)--(2)--(3F);
     \end{tikzpicture}  &$\urm(1)$
     \\\hline
   $\overline{\mathcal O}_{(2^2,1^4)}^{D_4}$ & $\overline{\mathcal O}_{(2^2)}^{A_3}$&
     \begin{tikzpicture}[main/.style={draw,circle},baseline=0]
     \node[main, label=below:$1$] (1L) []{};
     \node[main, label=below:$2$] (2) [right=of 1L]{};
     \node[main, label=below:$1$] (1R) [right=of 2]{};
     \node[draw, label=left:$1$] (1F) [above left= 0.7 and 0.7 of 2]{};
     \node[main, label=right:$1$] (1T) [above right= 0.7 and 0.7 of 2]{};
    
     \draw[-] (1L)--(2)--(1R);
     \draw[-] (1F)--(2)--(1T);
     \end{tikzpicture}
     & \begin{tikzpicture}[main/.style={draw,circle},baseline=0]
     \node[main, label=below:$1$] (1L) []{};
     \node[main, label=below:$2$] (2) [right=of 1L]{};
     \node[main, label=below:$1$] (1R) [right=of 2]{};
     \node[draw, label=left:$2$] (2F) [above=of 2]{};
    
     \draw[-] (1L)--(2)--(1R);
     \draw[-] (2)--(2F);
     \end{tikzpicture}  &$\urm(1)$
     \\\hline
      $\overline{\mathcal O}_{(2^2,1)}^{A_4}$ & $\overline{\mathcal O}_{(3,1)}^{A_3}$ &
      \begin{tikzpicture}[main/.style={draw,circle},baseline=0]
      \node[main, label=below:$1$] (1L) []{};
      \node[main, label=below:$2$] (2L) [right=of 1L]{};
      \node[main, label=below:$2$] (2R) [right=of 2L]{};
      \node[main, label=below:$1$] (1R) [right=of 2R]{};
      \node[draw, label=left:$1$] (1FL)  [above=of 2L]{};
      \node[draw, label=right:$1$] (1FR) [above=of 2R]{};

      \draw[-] (1L)--(2L)--(2R)--(1R);
      \draw[-] (1FL)--(2L);
      \draw[-] (1FR)--(2R);
          
      \end{tikzpicture}
      &\begin{tikzpicture}[main/.style={draw,circle},baseline=0]
      \node[main, label=below:$1$] (1L) []{};
      \node[main, label=below:$2$] (2L) [right=of 1L]{};
      \node[main, label=below:$2$] (2R) [right=of 2L]{};
      \node[draw, label=left:$1$] (1FL)  [above=of 2L]{};
      \node[draw, label=right:$2$] (2FR) [above=of 2R]{};

      \draw[-] (1L)--(2L)--(2R);
      \draw[-] (1FL)--(2L);
      \draw[-] (2FR)--(2R);
          
      \end{tikzpicture}&$\urm(1)$
      \\\hline
     $\overline{\mathcal O}_{(2^2,1^4)}^{D_4}$ & $\overline{\mathcal O}_{(3)}^{A_2}$&
     \begin{tikzpicture}[main/.style={draw,circle},baseline=0]
     \node[main, label=below:$1$] (1L) []{};
     \node[main, label=below:$2$] (2) [right=of 1L]{};
     \node[main, label=below:$1$] (1R) [right=of 2]{};
     \node[draw, label=left:$1$] (1F) [above left= 0.7 and 0.7 of 2]{};
     \node[main, label=right:$1$] (1T) [above right= 0.7 and 0.7 of 2]{};
    
     \draw[-] (1L)--(2)--(1R);
     \draw[-] (1F)--(2)--(1T);
     \end{tikzpicture}
     & \begin{tikzpicture}[main/.style={draw,circle},baseline=0]
     \node[main, label=below:$1$] (1L) []{};
     \node[main, label=below:$2$] (2) [right=of 1L]{};
     \node[draw, label=right:$3$] (3F) [above=of 2]{};
     \draw[-] (1L)--(2)--(3F);
     \end{tikzpicture}  &$\urm(1)\times \urm(1)$
       \\\hline
        $\overline{\mathcal O}_{(2^2,1^6)}^{D_5}$ & $\overline{\mathcal O}_{(3,1)}^{A_3}$&
    \begin{tikzpicture}[main/.style={draw,circle},baseline=0]
    \node[main, label=below:$1$] (1L) []{};
    \node[main, label=below:$2$] (2L) [right=of 1L]{};
    \node[main, label=below:$2$] (2R) [right=of 2L]{};
    \node[main, label=right:$1$] (1TR) [above right=0.7 and 0.7 of 2R]{};
    \node[main, label=right:$1$] (1BR) [below right=0.7 and 0.7 of 2R]{};
    \node[draw, label=left:$1$] (1F) [above= of 2L]{};

    \draw[-] (1L)--(2L)--(2R)--(1TR);
    \draw[-] (1BR)--(2R);
    \draw[-] (1F)--(2L);
        
    \end{tikzpicture}
    &\begin{tikzpicture}[main/.style={draw,circle},baseline=0]
      \node[main, label=below:$1$] (1L) []{};
      \node[main, label=below:$2$] (2L) [right=of 1L]{};
      \node[main, label=below:$2$] (2R) [right=of 2L]{};
      \node[draw, label=left:$1$] (1FL)  [above=of 2L]{};
      \node[draw, label=right:$2$] (2FR) [above=of 2R]{};

      \draw[-] (1L)--(2L)--(2R);
      \draw[-] (1FL)--(2L);
      \draw[-] (2FR)--(2R);
          
      \end{tikzpicture}&$\urm(1)\times \urm(1)$
      \\\hline
       
\end{tabular}
    
    \caption{Selection of $\urm(1)$ and $\urm(1)^2$ HKQs of classical orbits with their magnetic quivers taken from Kobak-Swann \cite{Kobak1996CLASSICALNO}.}
    \label{tab:U1Quots}
\end{table}

Here we review the $\urm(1)$ and $\urm(1)^2$ HKQ of nilpotent orbits with a Dynkin type embedding given in Kobak-Swann \cite{Kobak1996CLASSICALNO}. 

The diagrammatic technique of quiver subtraction for $\urm(1)$ HKQ differs subtly from the diagrammatic method of non-abelian $\surm(n)$ quotient quiver subtraction presented in Section \ref{sec:rules}.
The $\urm(1)$ HKQ involves the subtraction of the quiver $(1)$, which is the first member of the sequence of quotient quivers we use for $\surm(n)$. The difference lies in that $\urm(1)$ gauging does not require the initial step of transforming the framed target quiver into an unframed quiver. Otherwise the subtraction procedure is similar, albeit simpler, because alternatives do not arise at junctions.

In fact, it is already known in the Literature that the gauging of a $\urm(1)$ can be done diagrammatically by taking a framed magnetic quiver and turning a $\urm(1)$ gauge node into a flavour node \cite{Crawley-Boevey2001GeometryQuivers}.
%
%
A selection of examples taken from Kobak-Swann are shown with target and resulting magnetic quivers in Table \ref{tab:U1Quots}. This diagrammatic procedure for the $\urm(1)$ HKQ is obvious from the quivers.

\section{$\urm(n)$ HKQ Examples}
\label{sec:unex}

Given the diagrammatic method of performing a $\urm(1)$ HKQ in Appendix \ref{sec:U1HKQ}, and noting the isomorphism $\urm(n)\simeq \surm(n)\times \urm(1)$, we can ask if a $\urm(n)$ HKQ can be carried out diagrammatically. Complications can arise because the diagrammatic quotient quiver subtractions required by $\urm(1)$ and $\surm(n)$ HKQs do not necessarily commute.

However, for cases where the target quiver possesses an appropriate external leg with $\urm(n)$ symmetry, it is possible to combine the two HKQs into a single well-defined operation, as shown in the following examples.


\subsection{$\urm(n)$ HKQ of $\overline{max. A_k}, k\geq 3$}
\label{sec:maxAkUnHKQ}

For a straightforward example of $\urm(n)$ HKQ we will take the family of $\overline{max. A_k}$ orbits. The first step is the $\surm(n)$ quotient quiver subtraction which is valid for $n\leq (k+1)/2$ with the value of $n=(k+1)/2$ saturating the inequality being a special case. First for the non-saturating cases, it is simple to show that \begin{equation}
    \overline{max. A_k}///\surm(n)=\mathcal C\left(\resizebox{0.7\textwidth}{!}{\begin{tikzpicture}[baseline=(current bounding box.center),main/.style={draw,circle}]

    \node[main, label=below:$2$] (2) []{};
    \node[main, label=below:$4$] (4) [right=of 2]{};
    \node[] (cdotsl) [right=of 4]{$\cdots$};
    \node[main, label=below:$2n-2$] (2nm2) [right=of cdotsl]{};
    \node[main, label=below:$2n$] (2n) [right=of 2nm2]{};
    \node[] (cdotsr) [right=of 2n]{$\cdots$};
    \node[main, label=below:$k$] (k)[right=of cdotsr]{};
    \node[draw, label=right:$k+1$] (flav) [above=of k]{};
    \node[main, label=left:$1$] (1) [above=of 2n]{};

    \draw[-] (2)--(4)--(cdotsl)--(2nm2)--(2n)--(cdotsr)--(k)--(flav);
    \draw[-] (1)--(2n);

    \draw [decorate, 
    decoration = {brace,
        raise=10pt,
        amplitude=5pt}] (k) --  (2n) node[pos=0.5,below=15pt,black]{$k-2n+1$};
    \end{tikzpicture}}\right)\label{eq:maxAKSUn},
\end{equation}
this moduli space has no particular name, however the gauge nodes along the bottom are balanced and form the Dynkin diagram of $A_{k-n}$ the node of rank 1 at the top is overbalanced and hence the global symmetry is $\surm(k-n+1)\times \urm(1)$.

The $\urm(1)$ HKQ is easily implemented by subtracting the $\surm(1)$ quotient quiver from the top node of rank 1 in the result of \label{eq:maxAKSUn} we find that \begin{equation}
    \overline{max. A_k}///\urm(n)=\overline{\left[\mathcal W_{A_{k-n}}\right]}^{[0,\cdots,0,1,0,\cdots,0,k+1]}_{[0,\cdots,0]},\label{eq:maxAKSUn}
\end{equation}where the 1 in the flavour vector of the affine Grassmannian slice in \eqref{eq:maxAKSUn} is located in the $n^{th}$ position.

The saturating case for $n=2k-1$ was studied in Section \ref{sec:MaxAkSUk}. Using the labelling as in Section \ref{sec:MaxAkSUk} we found that $\overline{max. A_{2k-1}}///\surm(k)=\mathbb C^2/\mathbb Z_{2k}\otimes\mathcal{S}^{A_{2k-1}}_{\mathcal N,(2^k)}$, where $\mathbb C^2/\mathbb Z_{2k}$ has a $\urm(1)$ global symmetry and $\mathcal{S}^{A_{2k-1}}_{\mathcal N,(2^k)}$ has a global symmetry of $\surm(k)$. The $\urm(1)$ HKQ is then performed on the $\mathbb C^2/\mathbb Z_{2k}$ part of the product $\mathbb C^2/\mathbb Z_{2k}\otimes\mathcal{S}^{A_{2k-1}}_{\mathcal N,(2^k)}$, the former has a magnetic quiver $(1)-[2k]$ and thus the $\urm(1)$ HKQ results in the quiver $[2k+1]$ which has a trivial moduli space. Hence we conclude that \begin{equation}
    \overline{max. A_{2k-1}}///\urm(k)=\mathcal{S}^{A_{2k-1}}_{\mathcal N,(2^k)}.
\end{equation}

\subsection{$\urm(k)$ HKQ of $\overline{\left[\mathcal W_{A_{n+k-1}}\right]}^{[0,\cdots,0,n-k,0,0,\cdots,0]}_{[0,\cdots,0,0,n-k-2,0,\cdots,0]}$}
\label{sec:UkHKQ}

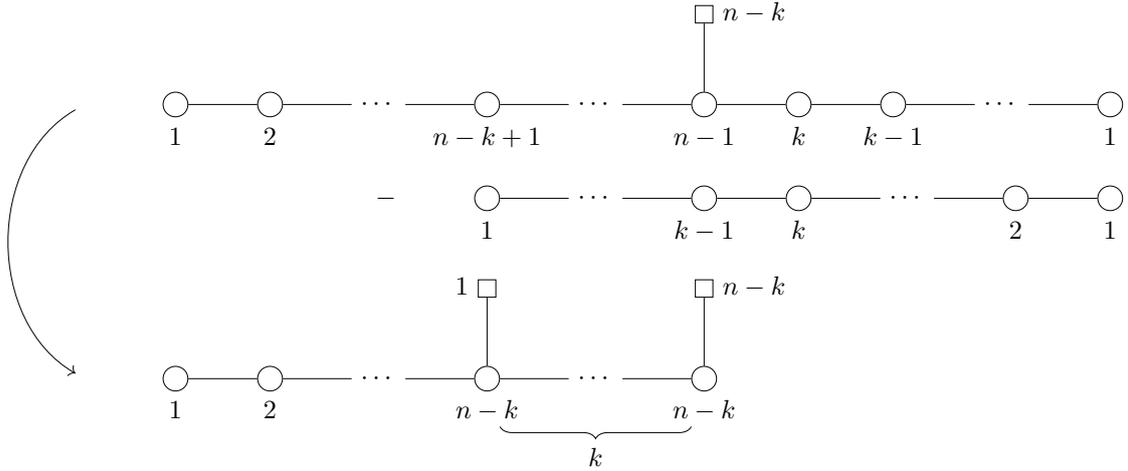
\begin{figure}[h!]
    \centering
     \resizebox{\textwidth}{!}{
    \begin{tikzpicture}[main/.style={draw,circle}]

    \node[main, label=below:$1$] (1) []{};
    \node[main, label=below:$2$] (2) [right=of 1]{};
    \node[] (cdotsl) [right=of 2]{$\cdots$};
    \node[main, label=below:$n-k+1$] (nmkp1) [right=of cdotsl]{};
    \node[] (cdots) [right=of nmkp1]{$\cdots$};
    \node[main, label=below:$n-1$] (nm1) [right=of cdots]{};
    \node[main, label=below:$k$] (k) [right=of nm1]{};
    \node[main, label=below:$k-1$] (km1) [right=of k]{};
    \node[] (cdotsr) [right=of km1]{$\cdots$};
    \node[main, label=below:$1$] (1r) [right=of cdotsr]{};
    \node[draw, label=right:$n-k$] (flav) [above=of nm1]{};

    \draw[-] (1)--(2)--(cdotsl)--(nmkp1)--(cdots)--(nm1)--(k)--(km1)--(cdotsr)--(1r);
    \draw[-] (nm1)--(flav);

    \node[main, label=below:$1$] (1subr) [below=of 1r]{};
    \node[main, label=below:$2$] (2subr) [left=of 1subr]{};
    \node[] (cdotsrsub) [left=of 2subr] {$\cdots$};
    \node[main,label=below:$k$] (ksub) [left=of cdotsrsub]{};
    \node[main, label=below:$k-1$] (km1sub) [left=of ksub]{};
    \node[] (cdotslsub) [left=of km1sub]{$\cdots$};
    \node[main, label=below:$1$] (1lsub) [left=of cdotslsub]{};
    \node[] (minus) [left=of 1lsub]{$-$};

    \draw[-] (1subr)--(2subr)--(cdotsrsub)--(ksub)--(km1sub)--(cdotslsub)--(1lsub);

    \node[draw, label=right:$n-k$] (nmkf) [below=of km1sub]{};
    \node[main, label=below:$n-k$] (nmkr) [below=of nmkf]{};
    \node[] (cdotsres) [left=of nmkr] {$\cdots$};
    \node[main, label=below:$n-k$] (nmkl) [left=of cdotsres]{};
    \node[] (cdotsresl) [left=of nmkl] {$\cdots$};
    \node[main, label=below:$2$] (2res) [left=of cdotsresl]{};
    \node[main, label=below:$1$] (1res) [left=of 2res]{};
    \node[draw, label=left:$1$] (1f) [above=of nmkl]{};

    \draw[-] (nmkf)--(nmkr)--(cdotsres)--(nmkl)--(cdotsresl)--(2res)--(1res);
    \draw[-] (1f)--(nmkl);

    \draw [decorate, 
    decoration = {brace,
        raise=20pt,
        amplitude=5pt}] (nmkr) --  (nmkl) node[pos=0.5,below=25pt,black]{$k$};

    \node[draw=none,fill=none] (topghost) [left=of 1]{};
    \node[draw=none,fill=none] (bottomghost) [left=of 1res]{};
    
    \draw [->] (topghost) to [out=-150,in=150,looseness=1] (bottomghost);

    \end{tikzpicture}}
    \caption{Quiver subtraction of the $\urm(k)$ quotient quiver from the magnetic quiver for $\overline{\left[\mathcal W_{A_{n+k-1}}\right]}^{[0,\cdots,0,n-k,0,0,\cdots,0]}_{[0,\cdots,0,0,n-k-2,0,\cdots,0]}$ to produce the magnetic quiver for $\overline{\mathcal O}^{A_{n-1}}_{(n-k+1,1^{k-1})}$.}
    \label{fig:implosionex}
\end{figure}

As a more technical example, a $\urm(k)$ HKQ for $\overline{\left[\mathcal W_{A_{n+k-1}}\right]}^{[0,\cdots,0,n-k,0,0,\cdots,0]}_{[0,\cdots,0,0,n-k-2,0,\cdots,0]}$, shown in the top of \Figref{fig:implosionex}, was computed in \cite{Bourget:2021qpx}, by utilising $3d$ mirror symmetry and carrying out the HKQ by gauging on the Higgs branch. The outcome of this $\urm(k)$ HKQ may be viewed in terms of a quotient quiver subtraction between magnetic quivers. For the $\urm(k)$ HKQ we start with the framed target magnetic quiver and subtract the $\urm(k)$ quotient quiver as shown in \Figref{fig:implosionex}. This results in \Quiver{fig:implosionex}, which is a magnetic quiver for $\overline{\mathcal O}^{A_{n-1}}_{(n-k+1,1^{k-1})}$, as given in \cite{Bourget:2021qpx}.

We note that the use of a framed quiver means that the gauge nodes are linear and thus the opportunity to apply the \hyperlink{rule:JunctionRule}{Junction Rule} is  missed. For values of $k \leq 2$, the Junction Rule is of no consequence, and we obtain the same results as those from the rules for an $\surm(k)$ quotient quiver subtraction followed by a $\urm(1)$ quotient quiver subtraction. However, for values of $k \geq 3$, the Junction Rule prevents the subtraction of an $\surm(k)$ quotient quiver and we are unable to replicate the results of \cite{Bourget:2021qpx} by using $\surm(k)$ quotient quiver subtraction. Also, we find that for $k\geq 3$ the Weyl integration method identifies incomplete Higgsing.

It appears that the solution in \cite{Bourget:2021qpx} for avoiding incomplete Higgsing involves making a particular choice around the breaking of the $\urm(k)$ gauge group, such that the HKQ implemented diagrammatically is no longer strictly that of $\urm(k)$.
\section{Violation of the Junction Rule}
\label{sec:JunctionRule}
\begin{figure}[h!]
    \centering
    \begin{subfigure}{0.4\textwidth}
    \centering
    \resizebox{!}{0.5\textwidth}{\begin{tikzpicture}[baseline=(current bounding box.center), main/.style = {draw, circle}]

    \node[main, label=below:$1$] (1l) {};
    \node[main, label=below:$2$] (2l) [right=of 1l]{};
    \node[main, label=below:$3$] (3) [right=of 2l]{};
    \node[main, label=below:$2$,fill=gray] (2r) [right=of 3]{};
    \node[main, label=below:$1$,fill=gray] (1r) [right=of 2r]{};
    \node[main, label=left:$2$,fill=red] (2a) [above=of 3]{};
    \node[main, label=left:$1$,fill=red] (1a) [above=of 2a]{};

    \draw[-] (1l)--(2l)--(3)--(2r)--(1r);
    \draw[-] (3)--(2a)--(1a);

    \node[main, label=below:$1$] (1subl) [below =of 1l]{};
    \node[main, label=below:$2$] (2subl) [right=of 1subl]{};
    \node[main, label=below:$3$] (3sub) [right=of 2subl]{};
    \node[main, label=below:$2$] (2subr) [right=of 3sub]{};
    \node[main, label=below:$1$] (1subr) [right=of 2subr]{};

    \draw[-] (1subl)--(2subl)--(3sub)--(2subr)--(1subr);

    \node[draw, label=right:$3$] (3f) [below=of 2subl]{};
    \node[main, label=below:$2$,fill=red] (2res) [below=of 3f]{};
    \node[main, label=below:$1$,fill=red] (1res) [left=of 2res]{};

    \draw[-] (3f)--(2res)--(1res);

    \node[] (minus) [right=of 1r]{$-$};

    \node[] (topghost) [left=of 1l]{};
    \node[] (bottomghost) [left=of 1res]{};

    \draw [->] (topghost) to [out=-150,in=150,looseness=1] (bottomghost);

    \end{tikzpicture}}\caption{}\label{fig:minE6A2Sub1}\end{subfigure}
    \hfill
    \centering
    \begin{subfigure}{0.4\textwidth}
    \centering
    \resizebox{!}{0.5\textwidth}{\begin{tikzpicture}[baseline=(current bounding box.center), main/.style = {draw, circle}]

    \node[main, label=below:$1$] (1l) {};
    \node[main, label=below:$2$] (2l) [right=of 1l]{};
    \node[main, label=below:$3$] (3) [right=of 2l]{};
    \node[main, label=below:$2$,fill=red] (2r) [right=of 3]{};
    \node[main, label=below:$1$,fill=red] (1r) [right=of 2r]{};
    \node[main, label=left:$2$,fill=gray] (2a) [above=of 3]{};
    \node[main, label=left:$1$,fill=gray] (1a) [above=of 2a]{};

    \draw[-] (1l)--(2l)--(3)--(2r)--(1r);
    \draw[-] (3)--(2a)--(1a);

    \node[main, label=below:$1$] (1subl) [below =of 1l]{};
    \node[main, label=below:$2$] (2subl) [right=of 1subl]{};
    \node[main, label=below:$3$] (3sub) [right=of 2subl]{};
    \node[main, label=below:$2$] (2subr) [right=of 3sub]{};
    \node[main, label=below:$1$] (1subr) [right=of 2subr]{};

    \draw[-] (1subl)--(2subl)--(3sub)--(2subr)--(1subr);

    \node[draw, label=right:$3$] (3f) [below=of 2subl]{};
    \node[main, label=below:$2$,fill=gray] (2res) [below=of 3f]{};
    \node[main, label=below:$1$,fill=gray] (1res) [left=of 2res]{};

    \draw[-] (3f)--(2res)--(1res);

    \node[] (minus) [right=of 1r]{$-$};

    \node[] (topghost) [left=of 1l]{};
    \node[] (bottomghost) [left=of 1res]{};

    \draw [->] (topghost) to [out=-150,in=150,looseness=1] (bottomghost);

    \end{tikzpicture}}\caption{}\label{fig:minE6A2Sub2}\end{subfigure}
    \caption{Both (incorrect) alignments of the $\surm(3)$ quotient quiver against the unitary unframed magnetic quiver for $\overline{min. E_6}$ to give quivers \Quiver{fig:minE6A2Sub1} and \Quiver{fig:minE6A2Sub2}. Their intersection is trivial.}
    \label{fig:my_label}
\end{figure}
Here we give an example of the incorrect application of quiver subtraction for $\surm(n)$ HKQ by explicitly breaking the \hyperlink{rule:JunctionRule}{Junction Rule}.

For our example we will take the magnetic quiver for $\overline{min. E_6}$ and try and take the $\surm(3)$ HKQ using quotient quiver subtraction. We ignore the \hyperlink{rule:JunctionRule}{Junction Rule} and permit the junction for the alignment of the $\surm(3)$ quotient quiver and target quiver to be located at the central node of rank 3 of the quotient quiver instead of a node of rank 2. There are two alignments of the $\surm(3)$ quotient quiver and these are related by an outer automorphism, thus we use colours for each leg of the target quiver to distinguish them. This procedure gives quivers \Quiver{fig:minE6A2Sub1} and \Quiver{fig:minE6A2Sub2} whose Coulomb branch is $\overline{max. A_2}$. The incorrect application of the rules of quiver subtraction leads us to believe\begin{equation}
    \overline{min. E_6}///\surm(3)\stackrel{?}{=}\textcolor{red}{\overline{max. A_2}}\cup \textcolor{darkgray}{\overline{max. A_2}}\label{eq:minE6A2}.
\end{equation}

It is simple to compute the refined HS and HWG of the union \eqref{eq:minE6A2} however, to show inconsistency between Weyl integration and ignorance of the \hyperlink{rule:JunctionRule}{Junction Rule} only the unrefined HS will suffice. We find:
\begin{align}
    HS\left[\textcolor{red}{\overline{max. A_2}}\cup \textcolor{darkgray}{\overline{max. A_2}}\right]&=2\times\frac{1+2t^2+2t^4+t^6}{\left(1-t^2\right)^6}-1\\&=\frac{1+10t^2-11t^4+22t^6-15t^8+6t^{10}-t^{12}}{\left(1-t^2\right)^6}
\end{align}

There is an embedding of $A_2\times A_2\times A_2$ into $E_6$ which decomposes the fundamental as \begin{equation}
    [0,0,0,0,0,1]_{E_6}\rightarrow [1,0;0,0;1,0]_{A_2\times A_2\times A_2}+[0,0;1,0;0,1]_{A_2\times A_2\times A_2}+[0,1;0,1;0,0]_{A_2\times A_2\times A_2}.
\end{equation}
This embedding is also a Dynkin type embedding. After applying this embedding, Weyl integration can be used to compute the $\surm(3)$ HKQ of $\overline{min. E_6}$ whose HS we find to be \begin{equation}
    HS\left[\overline{min. E_6}///\surm(3)\right]=\frac{1 + 9 t^2 + 43 t^4 + 81 t^6 + 20 t^8 - 89 t^{10} + 53 t^{12} - 
 11 t^{14} + t^{16}}{(1 - t^2)^7},
\end{equation}
where the quaternionic dimension of the moduli space is 3.5, indicating that incomplete Higgsing has occurred. This does not match the prediction from quotient quiver subtraction if we fail to apply the \hyperlink{rule:JunctionRule}{Junction Rule}. We find that the application of the \hyperlink{rule:JunctionRule}{Junction Rule} is necessary to avoid such mismatches that result from incomplete Higgsing.

\section{Violation of the External Leg Rule}
\label{sec:ExternalLeg}

\begin{figure}[h!]
    \centering
    \begin{subfigure}{0.45\textwidth}
    \centering\resizebox{0.75\width}{!}{\begin{tikzpicture}[main/.style = {draw, circle}]
    \node[main, label=below:$1$] (a) {};
    \node[main,label=below:$2$] (b) [right=of a] {};
    \node[main, label=below:$2$] (c) [right=of b]{};
    \node[] (d) [right=of c] {$\cdots$};
    \node[main,label=below:$2$] (e) [right=of d] {};
    \node[main,label=below:$2$] (f) [right=of e] {};
    \node[main, label=left:$1$] (g) [above=of b]{};
    
    \draw[-] (a)--(b)--(c)--(d)--(e)--(f);
    \draw[-] (b)--(g);
    \draw [line width=1pt, double distance=3pt,
             arrows = {-Latex[length=0pt 3 0]}] (f) -- (e);
    \draw [decorate, 
    decoration = {brace,
        raise=20pt,
        amplitude=5pt}] (e) --  (b) node[pos=0.5,below=25pt,black]{$k-2$};

    \node[main, label=left:$1$] (1subT) [below=of b]{};
    \node[main, label=below:$2$] (2sub) [below=of 1subT]{};
    \node[main, label=below:$1$] (1subL) [left=of 2sub]{};
    \node[draw=none,fill=none] (-) [left=of 1subL]{$-$};
    \node[draw=none,fill=none] (ghost) [right=of 2sub]{};
    \draw[-] (1subT)--(2sub)--(1subL);
    
    \node[main,label=left:$2$] (2resF) [below=of ghost]{};
    \node[draw=none,fill=none] (dotsres) [right=of 2resF] {$\cdots$};
    \node[main,label=below:$2$] (eres) [right=of dotsres] {};
    \node[main,label=below:$2$] (fres) [right=of eres] {};
    \node[draw, label=left:$2$] (gres) [above=of 2resF]{};
    
    \draw[-] (gres)--(2resF)--(dotsres)--(eres);
    \draw [line width=1pt, double distance=3pt,
             arrows = {-Latex[length=0pt 3 0]}] (fres) -- (eres);
    \draw [decorate, 
    decoration = {brace,
        raise=10pt,
        amplitude=5pt}] (eres) --  (2resF) node[pos=0.5,below=15pt,black]{$k-3$};

    \node[draw=none,fill=none] (topghost) [right=of f]{};
    \node[draw=none,fill=none] (bottomghost) [right=of fres]{};
    
    \draw [->] (topghost) to [out=-30,in=30,looseness=1] (bottomghost);

   \end{tikzpicture}}
  \caption{}
    \label{fig:nminCKA1Sub1}
        
    \end{subfigure}
    \hfill
    \begin{subfigure}{0.45\textwidth}
    \centering\resizebox{0.75\width}{!}{\begin{tikzpicture}[main/.style = {draw, circle}]
    \node[main, label=below:$1$] (a) {};
    \node[main,label=below:$2$] (b) [right=of a] {};
    \node[main, label=below:$2$] (c) [right=of b]{};
    \node[] (d) [right=of c] {$\cdots$};
    \node[main,label=below:$2$] (e) [right=of d] {};
    \node[main,label=below:$2$] (f) [right=of e] {};
    \node[main, label=left:$1$] (g) [above=of b]{};
    
    \draw[-] (a)--(b)--(c)--(d)--(e)--(f);
    \draw[-] (b)--(g);
    \draw [line width=1pt, double distance=3pt,
             arrows = {-Latex[length=0pt 3 0]}] (f) -- (e);
    \draw [decorate, 
    decoration = {brace,
        raise=20pt,
        amplitude=5pt}] (e) --  (b) node[pos=0.5,below=25pt,black]{$k-2$};

    \node[main, label=left:$1$] (1subT) [below=of b]{};
    \node[main, label=below:$2$] (2sub) [below=of 1subT]{};
    \node[main, label=below:$1$] (1subR) [right=of 2sub]{};
    \node[draw=none,fill=none] (-) [left=of 2sub]{$-$};
    \draw[-] (1subT)--(2sub)--(1subR);
    
    \node[draw,label=left:$2$] (2resF) [below left=of 2sub]{};
    \node[main, label=below:$1$] (1) [below=of 2resF] {};
    
    \draw[-] (2resF)--(1);
    
    \node[draw,label=left:$1$] (1F) [below right=of 1subR]{};
    \node[main, label=below:$2$] (2) [below =of 1F]{};
    \node[main, label=below:$1$] (1) [left=of 2]{};
    \node[draw=none,fill=none] (cdots) [right=of 2]{$\cdots$};
    \node[main,label=below:$2$] (2R) [right=of cdots]{};
    \node[main, label=below:$2$] (1R)[right=of 2R]{};
    
    \draw[-] (1F)--(2);
    \draw[-] (1)--(2)--(cdots)--(2R);
    \draw [line width=1pt, double distance=3pt,
             arrows = {-Latex[length=0pt 3 0]}] (1R) -- (2R);
    \draw [decorate, 
    decoration = {brace,
        raise=20pt,
        amplitude=5pt}] (2R) --  (1) node[pos=0.5,below=25pt,black]{$k-3$};

    \node[draw=none,fill=none] (topghost) [right=of f]{};
    \node[draw=none,fill=none] (bottomghost) [right=of 1R]{};
    
    \draw [->] (topghost) to [out=-30,in=30,looseness=1] (bottomghost);

   \end{tikzpicture}}
   \caption{}
    \label{fig:nminCKA1Sub2}
    \end{subfigure}
   \centering
\begin{subfigure}{\textwidth}
\centering
\begin{tikzpicture}[main/.style = {draw, circle}]
    \node[draw,label=left:$1$] (1F) {};
    \node[main, label=below:$2$] (2) [below =of 1F]{};
    \node[main, label=below:$1$] (1) [left=of 2]{};
    \node[draw=none,fill=none] (cdots) [right=of 2]{$\cdots$};
    \node[main,label=below:$2$] (2R) [right=of cdots]{};
    \node[main, label=below:$2$] (1R)[right=of 2R]{};
    
    \draw[-] (1F)--(2);
    \draw[-] (1)--(2)--(cdots)--(2R);
    \draw [line width=1pt, double distance=3pt,
             arrows = {-Latex[length=0pt 3 0]}] (1R) -- (2R);
    \draw [decorate, 
    decoration = {brace,
        raise=20pt,
        amplitude=5pt}] (2R) --  (1) node[pos=0.5,below=25pt,black]{$k-3$};
    
\end{tikzpicture}
\caption{}
\label{fig:nminCKA1SubInt}
\end{subfigure}
    \caption{Both alignments of the $\surm(2)$ quotient quiver against the magnetic quiver for $\overline{min. C_k}$ for $k\geq4$ giving quivers \Quiver{fig:nminCKA1Sub1} and \Quiver{fig:nminCKA1Sub2}. Their intersection, reached via $A_1$ KP transitions, is quiver \Quiver{fig:nminCKA1SubInt}.}
    \label{fig:minCkA1SubBoth}
\end{figure}

The \hyperlink{rule:ExternalLeg}{External Leg Rule} requires that the $\surm(n)$ quotient quiver must be subtracted from nodes of the target quiver corresponding to long roots. Here we give an example where this rule is ignored. 

We will attempt to subtract the $\surm(2)$ quotient quiver from the family of $\overline{n.min. C_k}$ quivers. In order to prevent violation of the \hyperlink{rule:SingleEdge}{Single Edge Rule} we require $k\geq 4$. The "subtraction" is shown in \Figref{fig:minCkA1SubBoth} and produces quivers \Quiver{fig:nminCKA1Sub1} and \Quiver{fig:nminCKA1Sub2}. The respective Coulomb branches are identified as $\overline{n.min. C_{k-2}}$ and $\overline{min. C_{k-2}}\otimes \overline{min. A_1}$. So we are lead to conjecture that:
\begin{equation}
    \overline{min. C_k}///\surm(2)\stackrel{?}{=}\overline{n.min. C_{k-2}}\cup\overline{min. C_{k-2}}\otimes \overline{min. A_1}, 
    \label{eq:minCkA1}
\end{equation}
where the global symmetry of the union \eqref{eq:minCkA1} is $\sprm(k-2)\times \surm(2)$.

Although it is straightforward to compute the refined HS and HWG of the union \eqref{eq:minCkA1} for particular values of $k\geq 4$, calculation of the unrefined HS will suffice to demonstrate the inconsistency with Weyl integration that arises if the \hyperlink{rule:ExternalLeg}{External Leg Rule} is ignored. We present the calculation using Weyl integration and quiver subtraction for the case $k=5$, however, we have also computed up to $k=6$. 

We find from quiver subtraction that:
\begin{align}
    &HS\left[\overline{n.min. C_{3}}\cup\overline{min. C_{3}}\otimes \overline{min. A_1}\right]\nonumber\\&=\frac{\left(\begin{aligned}1 &+ 15 t^2 + 135 t^4 + 625 t^6 + 1725 t^8 + 3048 t^{10} + 3686 t^{12} \\&+ 
 3048 t^{14} + 1725 t^{16} + 625 t^{18} + 135 t^{20} + 
 15 t^{22} + t^{24}\end{aligned}\right)}{(1 - t^2)^{12} (1 + t^2)^6)}\nonumber\\&+ \frac{(1-t^4)}{(1-t^2)^3}\frac{(1 + t^2) (1 + 10 t^2 + 41 t^4 + 10 t^6 + t^8)}{(1 - t^2)^{10}}\nonumber\\&- \frac{(1 + t^2) (1 + 10 t^2 + 41 t^4 + 10 t^6 + t^8)}{(1 - t^2)^{10}}\\&=\frac{\left(\begin{aligned}1 &+ 18 t^2 + 185 t^4 + 1004 t^6 + 3219 t^8 + 6457 t^{10} + 8438 t^{12}\\& + 
 7094 t^{14} + 3653 t^{16} + 934 t^{18} - 11 t^{20} - 66 t^{22} - 
 13 t^{24} - t^{26}\end{aligned}\right)}{(1 - t^2)^{12} (1 + t^2)^6}
\end{align}

There is a Dynkin type embedding of $C_k\hookleftarrow C_{k-1}\times A_1$ which branches the vector of $C_k$ as:
\begin{align}    [1,0,\cdots,0]_{C_k}&\rightarrow[1,0\cdots,0]_{C_{k-1}}+[1]_{A_1}\label{eq:CkDynkA1}.
\end{align}
If the embedding \eqref{eq:CkDynkA1} is chosen for Weyl integration then the expected global symmetries do not match as Weyl integration would give a result with $\sprm(k-1)$ global symmetry as opposed to the union \eqref{eq:minCkA1}. A second problem is that this embedding gives incomplete Higgsing as seen from the HS: \begin{equation}
    HS\left[\overline{n. min. C_5}///\surm(2)\right]=\frac{(1 + t^2) (1 + 22 t^2 + 253 t^4 + 812 t^6 + 1058 t^8 + 392 t^{10} + 
    36 t^{12})}{(1 - t^2)^{13}}.
\end{equation}

This is clearly in disagreement with \eqref{eq:minCkA1} which was obtained using $\surm(2)$ quotient quiver subtraction but ignoring the \hyperlink{rule:ExternalLegRule}{External Leg Rule}. We find that the application of the \hyperlink{rule:ExternalLegRule}{External Leg Rule} is necessary to avoid such pathological results where incomplete Higgsing occurs.


\bibliographystyle{JHEP}
\bibliography{references}

\end{document}